\documentclass[seceq]{ptptex}
\usepackage{graphicx}
\usepackage{wrapft}
\usepackage{multirow}

\notypesetlogo 
\title{Minimal Supersymmetric ${\boldsymbol{SU}(5)}$ Model \\
with Nonrenormalizable Operators} 

\subtitle{Seesaw Mechanism and Violation of Flavour and $CP$}

\author{Francesca \textsc{Borzumati}$^{1,2}$ 
and Toshifumi \textsc{Yamashita}$^{3}$}

\inst{$^1$Department of Physics, National Taiwan University, 
Taipei 10617, Taiwan\\
$^2$Theory Division, KEK, Tsukuba 305-0801, Japan\\
$^3$Department of Physics, Nagoya University, Nagoya 464-8602, Japan}

\abst{Flavour and {\it CP} violations that the neutrino-seesaw couplings
 of types~I,~II, and~III induce radiatively in the soft massive
 parameters of the minimal supersymmetric {\it SU}(5) model, made
 realistic by nonrenormalizable operators, are analyzed. 
Effective couplings are used to parametrize the couplings of
 renormalizable operators and of the corrections that nonrenormalizable
 ones provide at the tree level. 
It is found that for a limited, but sufficient accuracy in the 
 calculations of such violations, it is possible to extend the picture
 of effective couplings to the quantum level, all the way to the cutoff
 scale.
The arbitrariness introduced by nonrenormalizable operators is analyzed
 in detail. 
It is shown that it can be drastically reduced in the Yukawa sector if
 the effective Yukawa couplings involving colored triplet Higgs bosons
 are tuned to suppress the decay rate of the proton. 
In the supersymmetry-breaking sector, the usual requirement of 
 independence of flavour and field type for the mechanism of mediation
 of supersymmetry breaking is not sufficient to forbid arbitrary flavour
 and {\it CP} violations at the tree level. 
Special conditions to be added to this requirement, under which such
 violations can be avoided, are identified. 
Depending on how and whether these conditions are implemented, different
 phenomenological scenarios emerge. 
Flavour and {\it CP} violations of soft massive parameters induced by
 neutrino-seesaw couplings are discussed explicitly for the simplest
 scenario, in which no such violations are present at the tree level. 
Guidelines for studying them in other, less simple scenarios are given.  
Lists of all renormalization group equations needed for their 
 calculations are provided for each of the three types of seesaw
 mechanism, at all energies between the TeV scale and Planck scale.}

\PTPindex{113, 132, 140, 142}

\begin{document}

\maketitle
\tableofcontents

\section{Introduction and motivation}
\label{sec:Intro}
%
Among the existing proposals to solve the hierarchy problem of the
 Standard Model~(SM), supersymmetry~(SUSY) is still one of the most
 compelling.
A solution to this problem without excessive tuning requires that the
 massive parameters that break SUSY softly are around the TeV scale,
 hereafter identified with the electroweak scale, $M_{\rm weak}$ or
 $\widetilde{m}$.

As is well known, if no restrictions are invoked for these parameters, 
 flavour violations in the sfermion mass matrices, or sfermion flavour
 violations~(sFVs), in general, exist. 
In particular, off-diagonal elements in the chirality-conserving 
 sectors of these matrices, and/or misalignments of the chirality-mixing
 sectors with the corresponding Yukawa matrices, may be nonnegligible.
Then, loop diagrams with exchange of superpartners can give large
 contributions to flavour-changing-neutral-current (FCNC) processes.
Since these are experimentally known to be rare, and not in
 disagreement with the SM predictions~\cite{FLVreview}, it follows that
 sFVs must be rather small~\cite{sFVsmall}, or altogether absent, at
 least, at one scale.

Even if vanishing at the cutoff scale, $M_{\rm cut}$, however, sFVs are
 in general nonvanishing at $M_{\rm weak}$, as they are induced at the
 quantum level by the SM Yukawa couplings~\cite{DNW,BKS,DGH,BBMR,MFV}.
They are suppressed by a loop factor, but enhanced by the large
 logarithms of the ratios of $M_{\rm cut}$ and $M_{\rm weak}$, usually
 resummed through renormalization group~(RG) techniques. 
Nevertheless, the squark flavour violations (sQFVs) obtained in this way
 in the minimal supersymmetric standard model (MSSM) are not
 particularly large~\cite{BKS,BBMR} (unless $\tan \beta$ is
 large~\cite{FB}), mainly because of the smallness of the off-diagonal
 elements of the Cabibbo--Kobayashi--Maskawa (CKM) matrix, $K_{\rm CKM}$,
 and the pattern of fermion masses.

Irrespective of the extensions made to solve the hierarchy problem,
 the leptonic sector requires also some enlargement of the originally
 proposed SM structure in order to accommodate neutrino masses.
One way to proceed is to introduce the well-known seesaw mechanism,
 which is classified into three different
 types,~I~\cite{SEESAW1},~II~\cite{SEESAW2}, and~III~\cite{SEESAW3},
 depending on which heavy fields are advocated for its
 mediation~\cite{STRUMIAVISSANI}.
Singlets of the SM, or right-handed neutrinos~(RHNs) are used for the
 type~I. 
All three types of seesaw mediators have couplings of ${\cal O}(1)$
 with the lepton doublets if their scale, $M_{\rm ssw}$, is large,
 considerably above the range of energies at which direct detection is
 possible.

It is therefore very important to search for signals that can give
 information on these heavy fields.
An obvious magnifying glass for them may be precisely their large Yukawa
 couplings to the left-handed leptons and the large mixing angles of the
 Maki--Nakagawa--Sakata~(MNS) matrix, $K_{\rm MNS}$. 
These affect the RG flow of the soft SUSY-breaking parameters for
 sleptons~\cite{NuY->LFV}, inducing slepton flavour violations (sLFVs),
 and consequently, flavour-violating effects in the charged-lepton fermion
 sector. 
Thus, information on the seesaw fields can be hopefully gleaned through
 the study of flavour-violating processes in models in which sFVs at
 $M_{\rm cut}$ are vanishing (or under control, as in models with a
 Yukawa mediation of SUSY breaking~\cite{YuMEDIAT}). 
The existence of arbitrary tree-level flavour violations in the slepton
 mass parameters at this scale, even if relatively small, for example of
 ${\cal O}(10^{-2}$), can completely obscure the loop effects induced by
 the seesaw mechanism.

In such studies, a minimal number of parameters is commonly used
 to specify the soft SUSY-breaking terms at $M_{\rm cut}$: a common
 gaugino mass, $M_{1/2}$, a mass common to all scalars, with vanishing
 intergenerational mixings, $\widetilde{m}_0$, and two, in general,
 uncorrelated parameters, the bilinear and trilinear couplings $B_0$ and
 $A_0$, which guarantee the alignment of the bilinear and trilinear soft
 terms to the corresponding mass and Yukawa terms in the superpotential. 
This short list of parameters encapsulates the concept of universal
 boundary conditions for soft masses, which may be obtained in the
 context of minimal supergravity~\cite{BFS,HLW}, or more generally,
 from a  field- and flavour-blind mediation of SUSY breaking, such as
 gravity mediation, which is assumed throughout this paper.

If the MSSM with the seesaw mechanism is embedded in a SUSY {\it SU}(5)
 grand-unification theory~(GUT), the seesaw mediators interact with
 large Yukawa couplings not only with the {\it SU}(2)-doublet leptons,
 but with all components of the multiplets to which these leptons belong,
 that is, also with {\it SU}(2)-singlet down quarks.
Thus, as pointed out by Baek et al.~\cite{BGOO1} and
 Moroi~\cite{NuY->QFV}, in the minimal SUSY~{\it SU}(5)~(MSSU(5)) model
 with the seesaw of type~I, these interactions can induce sQFVs at
 $M_{\rm weak}$ in the right--right down-squark sector, whereas those
 generated by the top Yukawa coupling are in the left--left sector.
Such sQFVs are related to sLFVs in a simple way.
In addition, various independent phases are present in this model, only
 one of which corresponds to the CKM phase of the SM, when evolved at
 low scale.
These GUT phases can also leave visible imprints in the superpartner
 mass matrices, during the RG flow from $M_{\rm cut}$ to the GUT scale,
 $M_{\rm GUT}$, if the soft SUSY-breaking parameters are real at
 $M_{\rm cut}$~\cite{NuY->QFV}.

Both these facts have raised hopes that combined studies of flavour- and
 {\it CP}-violating processes in the quark and lepton sectors may provide
 interesting avenues for detecting the presence of RHNs in the MSSU(5)
 model with universal soft terms at $M_{\rm cut}$. 
Indeed, considerable attention has been paid to the correlations of 
 sQFVs and sLFVs in this
 model~\cite{BGOO1,NuY->QFV,BGOO,ETbefMEAS,ETaftMEAS,ParkEt}, much
 less in the MSSU(5) models with the other two seesaw
 types.\footnote{Exceptions are 
  Refs.~\citen{AROSSIorig,JoaROSSI,MOHOKA}, in which specific boundary
  conditions for the soft masses, universal~\cite{AROSSIorig} or
  obtained from a combined gauge- and Yukawa-mediation mechanism of
  SUSY breaking~\cite{JoaROSSI,MOHOKA}, are imposed at $M_{\rm GUT}$.}

Soon after the observations made in Refs.~\citen{BGOO1}
 and~\citen{NuY->QFV}, it was argued that the MSSU(5) model with a
 seesaw of type~I, precisely because it induces sQFVs in the
 right--right down-squark sector, could accommodate values for the
 $B_s$-$\bar{B}_s$ mixing (at that time unmeasured) distinguishable from
 the typical SM ones~\cite{PatEmiShab}, without upsetting the observed
 agreement between experimental results and SM predictions for other
 FCNC processes.
In contrast, predictions for this mixing in the MSSM with universal
 boundary conditions of the soft parameters (usually assumed at
 $M_{\rm GUT}$) tend to deviate less from those obtained in the SM.

The now existing measurement of the $B_s$-$\bar{B}_s$
 mixing~\cite{TEVmix}, unfortunately, turns out to be inconclusive for
 searches of new physics.
The measurement of $\Delta M_s$ does not show disagreement with the SM,
 but, despite its outstanding precision, cannot exclude new-physics
 contributions either, because of the $\sim 30\%$ error that
 plagues the SM calculation~\cite{AlexUlli}.
In contrast, the SM calculation for the phase of this mixing is much
 less uncertain~\cite{RFleischer}, but the experimental results are
 less precise~\cite{TEVphase}. 
The most recent global fits for this phase made by the UTfit
 collaboration~\cite{UTfitnew} and the HFAG~\cite{HFavG} seem to be
 $2.9$ or $2.5\sigma$ away from the SM value; the fit from a combined
 CDF and D0 analysis based on $B_s\to J/\psi\phi$ data alone is only 
 $2 \sigma$ away~\cite{CDFDOcomb}.
A phase deviation from the SM prediction could easily be accommodated in 
 the MSSU(5) models with different types of seesaw mechanism.

In view of new and more extensive phenomenological studies, it is time
 to assess the state-of-the-art of the theoretical treatment of these
 models.
As is well known, the MSSU(5) model is not realistic as it predicts a
 too rapid proton decay~\cite{SakaiYanag} and wrong relations between 
 the down-quark and charged-lepton masses~\cite{FERMASSES}.
Suitable extensions of the field
 content~\cite{FERMASSES,missingPARTN,AFM}, or the inclusion of
 nonrenormalizable operators~(NROs) suppressed by $1/M_{\rm cut}$,
 are both considered sufficient to address these problems.
Since the former extensions are more obtrusive, as they bring larger
 modifications to the model, we opt here for the second possibility.

\subsection{Existing studies of NROs in MSSU(5) models}
%
The consequences that NROs may have for sFVs were first discussed in
 Ref.~\citen{NROsFLsAHCH} in the context of the SUSY~{\it SO}(10) model.
They were also studied in Ref.~\citen{NROsFLsHNOST} in the 
 SUSY~{\it SU}(5) model with vanishing neutrino masses, and were then 
 neglected until  Ref.~\citen{BGOO}, where the seesaw of type~I was 
 implemented to obtain massive neutrinos.
In Ref.~\citen{BGOO}, only one NRO of dimension five was included in the
 Yukawa sector (the minimal number sufficient to obtain a suitable
 fermion spectrum).\footnote{After Ref.~\citen{BGOO}, the only other 
  papers including NROs in the same context are those listed in 
  Ref.~\citen{ParkEt}, which follow quite closely the treatment of 
  Ref.~\citen{BGOO}. Within a SUSY {\it SO}(10) model, the same problem 
  is studied in Ref.~\citen{TWW}.}
 Nevertheless, the correlations between sLFVs and sQFVs induced by the
 seesaw Yukawa couplings were shown to be sizably altered.

The direct effect that NROs have through RGEs is at most of 
 ${\cal O}(10^{-4})$, for the usual hierarchy between $M_{\rm cut}$ and
 $M_{\rm GUT}$.
Although small, it may not be neglected if the seesaw couplings happen
 to be somewhat smaller than ${\cal O}(1)$. 
{}For this reason, the authors of Ref.~\citen{BGOO} aimed at collecting
 all contributions of  ${\cal O}(10^{-4})$.
The largest effect that NROs have, however, comes from the arbitrariness
 they introduce in the choice of the flavour rotations of the SM fields
 to be embedded in {\it SU}(5) matter multiplets.
This is expressed by the appearance of unitary matrices of mismatch in 
 the diagonalization (hereafter, mismatch matrices) of various Yukawa 
 couplings, in addition to the RGE-evolved CKM and MNS ones.
There are two such matrices in the analysis of Ref.~\citen{BGOO}.

Since the coefficient of the NRO in Ref.~\citen{BGOO} is tuned to
 provide corrections of ${\cal O}$(1) to the Yukawa couplings of first- 
 and second-generation fermions, it is somewhat obvious that these two
 mismatch matrices modify the pattern of sFVs in the 
 first- and second-generation sector of the down-squark mass matrix. 
What is perhaps less obvious is the fact that they can affect also, in a
 sizable way, the pattern of sFVs in which the third-generation 
 down squark is involved~\cite{ParkEt}.
Unfortunately, the authors of Ref.~\citen{BGOO} failed to emphasize this
 point, thereby implicitly substantiating the perception that the
 predictions for FCNC processes involving the bottom quark remain
 unaffected by the inclusion of NROs.

The authors of Refs.~\citen{NROsFLsAHCH} and~\citen{NROsFLsHNOST}
 included in their analyses all possible NROs in the Yukawa sector of
 the superpotential. 
To deal with such a complex situation, they made use of the picture of
 effective couplings.
{}For each renormalizable operator, these collect at the tree level the
 coupling of the operator itself and the corrections contributed by
 different NROs with superheavy fields replaced by their
 vacuum-expectation values~(VEVs).
This picture was assumed by these authors to be valid also at the
 quantum level.

No effort was made by any of the three groups of authors to verify
 whether the proton-decay rate can be suppressed in their scenarios.
In the case of Ref.~\citen{BGOO}, this seems hardly to be the case,
 given the presence of only one NRO. 
Enough {\it SU}(5)-breaking effects are induced by NROs in the analyses 
 presented in Refs.~\citen{NROsFLsAHCH} and~\citen{NROsFLsHNOST}.
These can disentangle~\cite{ZURAB,GORAN,DESYpeople} the Yukawa
 couplings giving rise to the fermion masses from the Yukawa couplings
 contributing to the coefficients of the effective operators
 responsible for proton decay.
The corresponding rate can then be tuned to be smaller than the existing
 experimental limits~\cite{ExperimentPD}, also for colored Higgs fields
 with mass of ${\cal O}(M_{\rm GUT})$.
Nevertheless, no mention of this issue is made in these papers, and no
 awareness appears in them of the fact that this tuning can have a
 substantial feedback into the sFV problem.

\subsection{This paper}
%
Although closer in method to the treatment of NROs proposed in 
 Ref.~\citen{NROsFLsAHCH}, this paper is the first step towards a
 generalization of Ref.~\citen{BGOO} for all three seesaw types, with
 no restrictions on the type and number of NROs to be included.
The three resulting models are often dubbed in this analysis as 
 nrMSSU(5) models, to distinguish them from the corresponding MSSU(5) 
 models without NROs. 
Special attention is paid to the possible shortcomings of the nrMSSU(5) 
 models, in particular, the loss of predictability, and the breaking
 of minimality in their K\"ahler potentials.
As in Ref.~\citen{BGOO}, we aim at collecting all the leading
 RGE-induced effects due to NROs, of ${\cal O}(10^{-4})$.

We plan a comprehensive analysis of flavour and {\it CP} violations in the
 sfermion and fermion sectors of general nrMSSU(5) models in which any
 of the three seesaw types is implemented.
In this particular paper, we lay out all theoretical aspects of the 
 problem, deriving partial phenomenological results in some cases.

We start by reviewing the MSSU(5) model (without seesaw mechanism) in
 \S\ref{sec:MatterSEC} and the three types of seesaw mechanism in
 \S\ref{sec:seesaw}.
We study in \S\ref{sec:HiggsSECandVACUA} the vacuum structure of
 the MSSU(5) model, which in a certain justified approximation,
 specified in \S\ref{sec:NROps}, is also the vacuum of the
 corresponding nrMSSU(5) model.
We provide analytic expressions for the scalar and auxiliary VEVs of the
 Higgs field in the adjoint representation, $24_H$.
To the best of our knowledge, the expression of the auxiliary VEV is 
 given here explicitly for the first time.

Still in \S\ref{sec:HiggsSECandVACUA}, we show how the tunings
 needed in the MSSU(5) model to induce light Higgs boson masses are
 stable under radiative corrections, irrespective of the type of soft
 parameters assumed at $M_{\rm cut}$.
The proof is already given in a compact form in Ref.~\citen{KMY}. 
Ours is more direct and explicit and uses the RGEs for the leading
 components of the VEVs of the field $24_H$ in the expansion in powers
 of $(\widetilde{m}/M_{\rm GUT})$. 
We derive them in this section in terms of the original parameters of
 the model, and in a practically model-independent way in
 Appendix~\ref{sec:New}.
In particular, this proof remains valid for the nrMSSU(5) models.

In \S\ref{sec:NROps}, we survey in full generality the different NROs
 that can appear in the SUSY-conserving sector of the nrMSSU(5) models,
 and we list explicitly the various effective couplings induced.
In \S\ref{sec:NROsBC-YUKAWA}, we show that, in general, the number of
 mismatch matrices, and therefore, the amount of arbitrariness,
 introduced in the Yukawa sector is considerably larger than in
 Ref.~\citen{BGOO}.
Despite this, only one mismatch matrix affects sizably the pattern of
 the seesaw-induced sFVs in the down-squark sector, as in the simpler
 case studied in Ref.~\citen{BGOO}.

We do not attempt an actual calculation of the proton-decay rate in this
 paper, limiting ourselves to recall
 old~\cite{ZURAB,GORAN,DESYpeople} and new~\cite{BMYtalks} possibilities
 to suppress it through NROs. 
Nevertheless, in \S\ref{sec:PDconstBC}, we illustrate the 
 consequences for sFVs of the two ans\"atze used in 
 Ref.~\citen{DESYpeople} for its suppression.
These correspond to specific choices of the effective Yukawa couplings 
 of operators involving the colored Higgs triplets in the large
 parameter space opened up by the introduction of NROs.
Interestingly, for these particular choices, the above mismatch matrices
 turn out to be well approximated by the unit matrix and their effects
 on the correlations between sQFVs and sLFVs are negligible.  
We argue, however, that this may not be a generic feature of these
 models and that a dedicated study should be devoted to this problem.

We survey in \S\ref{sec:SOFTNROS} the NROs that can appear in the 
 SUSY-breaking part of the superpotential and K\"ahler potential.
We emphasize the expressions of the effective trilinear couplings and
 effective soft masses squared, to which also the auxiliary VEV $F_{24}$
 contributes (a point which was missed in Ref.~\citen{BGOO}, and
 presumably also in Refs.~\citen{NROsFLsAHCH} and~\citen{NROsFLsHNOST}).

As a consequence, tree-level sFVs for effective couplings cannot be
 avoided, even if the boundary conditions for the original soft
 couplings of the nrMSSU(5) models are consistent with a flavour- and 
 field-independent mechanism of mediation of SUSY breaking. 
We show this in \S\ref{sec:NROsBC-SOFT}.
We give in this section, and later in \S\ref{sec:EmerScen}, explicit
 guidelines on how to deal with this complicated situation.

Nevertheless, after some guesses made in this section, we show in a
 systematic way in \S\ref{sec:UNIVERSALITY} how to avoid these
 tree-level sFVs by restricting further the type of couplings between 
 the mediator of SUSY breaking and the various operators of the
 K\"ahler potential and the superpotential.
Thanks to these restrictions, it is possible to recover for the
 effective soft coupling at $M_{\rm cut}$ a four-parameter description
 in terms of $M_{1/2}$, $\widetilde{m}_0$, $A_0$, $B_0$, i.e., the same
 parameters that describe the soft couplings of the MSSM obtained in 
 the flat limit of minimal supergravity~\cite{BY-SUSY09}.

We are then faced with the problem of whether it is possible to upgrade
 this picture of effective couplings to the quantum level. 
In \S\ref{sec:ValidityEffPic}, we examine the possible loop diagrams 
 that break this picture and we determine the level of accuracy in the 
 calculation of sFVs for which this upgrade is doable. 
We show, in a sketchy way in \S\ref{sec:NROinRGE} and in detail in
 Appendix~\ref{sec:proof}, that, within this level of accuracy, the RGEs
 for effective couplings are as those for renormalizable couplings, even
 above the GUT scale where the superheavy degrees of freedom are still
 active.

This statement has already been claimed in Ref.~\citen{NROsFLsAHCH}, and
 then used also in Ref.~\citen{NROsFLsHNOST} by assuming that it is
 possible to neglect the wave function renormalization of the adjoint
 Higgs field. 
In reality, within our target accuracy, this approximation is not valid
 for the effective trilinear couplings. 
Nevertheless, by making use of the evolution equation of the scalar and 
 auxiliary VEVs derived in \S\ref{sec:minSU5}, we show that the
 formal equality of the RGEs for effective couplings and
 renormalizable couplings holds exactly, still within our target
 accuracy. 
We also show how the neglect of the auxiliary VEV of the field 
 $24_H$ has led to an overestimation of sFVs in Ref.~\citen{BGOO}.

This discussion is supplemented by complete lists of RGEs, given in
 Appendix~\ref{sec:RGEsSU5} for all ranges of energies from
 $M_{\rm cut}$ to $M_{\rm weak}$, for each of the three seesaw types,
 correcting mistakes in some of the equations reported in the existing
 literature, and giving here for the first time those missing. 
In particular, the RGEs for the MSSM with seesaw of type~III, those 
 for the MSSU(5) models with seesaw of types~II and~III, and the RGEs
 for the nrMSSU(5) models with all seesaw types appear here for the
 first time.

{}Finally, in \S\ref{sec:Results}, we give approximated analytic 
 expressions for low-energy off-diagonal terms in the squark and slepton
 mass matrices induced by the seesaw couplings, for vanishing and 
 nonvanishing NROs. 
Those for vanishing NROs are still predictions for the flawed
 MSSU(5) model, which are, however, missing in the existing literature, 
 or can be interpreted as predictions for the points of parameter
 space of nrMSSM(5) models in which the proton-decay constraints
 reduce the mismatch matrices to be trivial. 
We show explicitly how NROs can modify the seesaw-induced transitions
 between squarks of first and second generations, as well as first and
 third and second and third generations.
We summarize our analysis in \S\ref{sec:summary}.

\subsection{Notation issues}
%
We do not distinguish in this paper fundamental and antifundamental
 indices as upper and lower ones, as usually done.
We give, however, in Appendix~\ref{sec:emb-GROUPfact} explicit
 multiplication rules for each term of the superpotential and K\"ahler
 potential, clarifying our notation.
In the same Appendix, we give also explanatory details for many of the
 equations appearing in the text.
In particular, the {\it SU}(5) generators $T_i$ listed there are assumed 
 to be acting on the antifundamental representation. 
Thus, they act as $-T_i^T$ on the fundamental representation $5$.
Throughout this paper, we use the same symbol for Higgs superfields and
 their scalar components. 
An exception is made only for the field $24_H$ in
 \S\ref{sec:HiggsSECandVACUA} and Appendix~\ref{sec:proof}.
Finally, a dot on a parameter $P(Q)$ denotes the partial derivative
 $(16 \pi^2)\partial P/\partial \ln (Q/Q_0)$, with $Q_0$ an arbitrary 
 reference scale.

\section{The MSSU(5) model with seesaw}
\label{sec:minSU5}
%
In the MSSU(5) model~\cite{SUSYSU5}, the supersymmetric version of the
 Georgi-Glashow {\it SU}(5) model~\cite{SU5FIRST}, the superpotential 
 can be split as
\begin{equation}
  W^{\rm MSSU(5)} =W^{\rm MSSU(5)}_{\rm M} +W^{\rm MSSU(5)}_{\rm H},
\label{eq:WminSU5}
\end{equation}
 where $W^{\rm MSSU(5)}_{\rm M}$ contains all matter interactions:
\begin{equation}  
 W^{\rm MSSU(5)}_{\rm M} =   
   \sqrt{2} \, \bar{5}_M  Y^5    \, 10_M \bar{5}_H  
 - \frac{1}{4} 10_M       Y^{10} \, 10_M 5_H  ,
\label{eq:WminSU5m}
\end{equation}
 and $W^{\rm MSSU(5)}_{\rm H}$ all Higgs interactions:
\begin{equation}
 W^{\rm MSSU(5)}_{\rm H} =        
  M_5 5_H \bar{5}_H +
  \lambda_{5} 5_H 24_H \bar{5}_H + 
  \frac{1}{2} M_{24} (24_H)^2 +
  \frac{1}{6} \lambda_{24} (24_H)^3 .
\label{eq:WminSU5H}
\end{equation}
Here,{{{}}} $Y^{10}$ is a symmetric matrix, whereas $Y^5$ is a generic one.
As is well known, the two irreducible representations $\bar{5}_M$ and
 $10_M$ collect, generation by generation, all matter superfields of the
 MSSM: $10_M = \{Q, U^c, E^c \}$ and $\bar{5}_M =\{D^c, L\}$. 
The two Higgs fields $5_H$ and $\bar{5}_H$ contain the two weak Higgs
 doublets of the MSSM, $H_u$ and $H_d$, and the two colour triplets,
 $H^C_U$ and $H^C_D$: $5_H =\{H^C_U,H_u\}$ and
 $\bar{5}_H=\{H^C_D,H_d\}$.
The explicit form of the Higgs field $24_H$ is given in
 Appendix~\ref{sec:emb-GROUPfact}, where also the SM decomposition of
 most superpotential terms can be found. 
It is sufficient here to say that its components are: $G_H$ and $W_H$, 
 respectively in the adjoint representations of {\it SU}(3) and 
 {\it SU}(2); $X_H$, an {\it SU}(2) doublet and {\it SU}(3) antitriplet; 
 $\bar{X}_H$, an {\it SU}(2) antidoublet and {\it SU}(3) triplet; and 
 $B_H$, a SM singlet.

In this form, the MSSU(5) model predicts vanishing neutrino masses. 
The extension usually made to obviate this problem consists in the
 introduction of a heavy seesaw sector.
Depending on the particle content of this sector, three realizations of
 the seesaw mechanism are possible, denoted as seesaw of types~I,~II,
 and~III, which will be discussed in \S\ref{sec:seesaw}. 
We also postpone a discussion of the vacuum structure of the model 
 to \S\ref{sec:HiggsSECandVACUA}, whereas we concentrate on the quark
 and lepton interactions in the next section.

\subsection{Matter sector}
\label{sec:MatterSEC}
%
After performing some rotations in flavour space discussed in 
 Appendix~\ref{sec:emb-SMfields}, it is possible to recast
 $W^{\rm MSSU(5)}_{\rm M}$ in the form
\begin{equation}  
 W^{\rm MSSU(5)}_{\rm M} =  
  \sqrt{2} \, \bar{5}_M \widehat{Y}^5 \, 10_M \bar{5}_H  
 -\frac{1}{4} 10_M 
  \left( K_{10}^T P_{10}\, \widehat{Y}^{10}\,K_{10}\right)
              10_M 5_H ,
\label{eq:WminSU5mRot}
\end{equation}
 which shows explicitly the physical parameter of this superpotential
 term: $K_{10}$ and $P_{10}$~\cite{EllGaill}, which like all $K$- and
 $P$-matrices throughout this paper, are respectively a unitary matrix
 with three mixing angles and one phase, and a diagonal phase matrix
 with determinant equal to one, and the real and positive entries in the
 diagonal matrices of Yukawa couplings, $\widehat{Y}^5$ and
 $\widehat{Y}^{10}$.

By decomposing the two terms in Eq.~(\ref{eq:WminSU5mRot}) in SM
 representations, we can split $ W^{\rm MSSU(5)}_{\rm M}$ as
\begin{equation}
  W^{\rm MSSU(5)}_{\rm M} =
  W^{\prime\,{\rm MSSM}} + W^{\rm MSSU(5)}_{\mbox{\scriptsize{M-H$^{\rm C}$}}} ,
\label{eq:WminSU5mDec}
\end{equation}
 with $W^{\prime\,{\rm MSSM}}$ containing only MSSM fields,
 $W^{\rm MSSU(5)}_{\mbox{\scriptsize{M-H$^{\rm C}$}}}$ collecting all the Yukawa
 interactions involving Higgs triplets.
The following identification of the MSSM fields $Q$, $D^c$, $U^c$, $L$
 and $E^c$ among the components of the MSSU(5) fields $\bar{5}_M$ and
 $10_M$: 
\begin{equation}
 \bar{5}_M \to \{D^c, e^{-i\phi_l} P_l^\dagger L\},
 \hspace*{1truecm}
 10_M  \to \{Q,K^\dagger_{10}P_{10}^\dagger U^c,e^{i\phi_l}P_l E^c\},
\label{eq:SMfieldsID}
\end{equation}
 removes the so-called GUT phases $P_{10}$ from the MSSM-like part, 
 and reduces the up-quark Yukawa coupling to be nonsymmetric. 
After the renaming: 
 \begin{equation}
  \widehat{Y}^{10}\to \widehat{Y}_U,   \hspace*{1truecm}
  \widehat{Y}^{5} \to \widehat{Y}_D,   \hspace*{1truecm}
  K_{10} \to K_{\rm CKM}  ,
 \label{eq:YukawaID}
 \end{equation}
 where $\widehat{Y}^5_D$, $\widehat{Y}^{10}_U$ and $K_{\rm CKM}$ are,
 respectively, the diagonal matrices of down- and up-quark Yukawa
 couplings, ${\rm diag}(y_d,y_s,y_b)$, ${\rm diag}(y_u,y_c,y_t)$, and
 the CKM matrix, all at $M_{\rm GUT}$, the MSSM-like becomes
\begin{equation}
  W^{\prime\,{\rm MSSM}} =
  U^c  \left(\widehat{Y}_U K_{\rm CKM} \right) Q H_u 
- D^c  \,\widehat{Y}_D                       \,Q H_d 
- E^c  \,\widehat{Y}_D                       \,L H_d, 
\label{eq:WMSSMprime}
\end{equation}
 with the down-quark diagonal Yukawa matrix also in the leptonic term.
In its place, the MSSM superpotential has a different diagonal matrix
 $\widehat{Y}_E$, with three additional independent parameters,
 $\widehat{Y}_E\equiv {\rm diag}(y_e,y_\mu,y_\tau)$.

Note that the above identification is not unique, and so are the
 expressions of $Y^5$ and $Y^{10}$ in terms of the physical parameters
 of $W^{\rm MSSU(5)}_{\rm M}$, and of the MSSM Yukawa matrices (see
 Appendix~\ref{sec:emb-SMfields}). 
The appearance of the three phases $e^{i\phi_l}P_l$ shows the
 arbitrariness that still remains in choosing the {\it SU}(2) doublets
 and singlet lepton fields, $L$ and $E^c$. 
This will be used to remove three phases in the seesaw sector.

With this identification and the form of $W^{\prime\,{\rm MSSM}}$ in
 Eq.~(\ref{eq:WMSSMprime}), $W^{\rm MSSU(5)}_{\mbox{\scriptsize{M-H$^{\rm C}$}}}$
 looks as
\begin{eqnarray}
 W^{\rm MSSU(5)}_{\mbox{\scriptsize{M-H$^{\rm C}$}}}                 & = & 
 \frac{1}{2}
 Q \left(K_{\rm CKM}^T \widehat{Y}_U P_{10} K_{\rm CKM}\right) Q H_U^C
+ e^{+i\phi_l} \,U^c 
 \left(\widehat{Y}_U K_{\rm CKM} P_l \right) E^c H_U^C 
\nonumber\\                                               &   & 
- e^{-i\phi_l} \,L \left(P_l^\dagger \widehat{Y}_D \right) Q H_D^C 
- D^c \left(\widehat{Y}_D K_{\rm CKM}^\dagger P_{10}^\dagger
      \right) U^c H_D^C .  
\label{eq:WmattHc}
\end{eqnarray}
It contains all the phases $P_{10}$ and $e^{i\phi_l}P_l$ rotated away
 from $W^{\prime\,{\rm MSSM}}$. 
Consistently, once the Higgs triplets are integrated out, their
 dependence disappears{{{}}} from the superpotential below $M_{\rm GUT}$.
A trace of these phases, however, remains in the modification that the
 soft SUSY-breaking parameters undergo through renormalization from
 $M_{\rm cut}$ down to $M_{\rm GUT}$.
They remain also in higher-dimension operators generated by integrating
 out the colored Higgs triplets, such as the proton-decay operators
 discussed in \S\ref{sec:PDconstBC}.

The term $W^{\prime\,{\rm MSSM}}$ still differs from the conventional MSSM
 superpotential as it lacks the bilinear terms in the two Higgs
 doublets, with massive coupling $\mu$. 
Up to soft SUSY-breaking terms, this parameter and the dynamics leading
 to the breaking of {\it SU}(5) are determined by $W^{\rm MSSU(5)}_{\rm H}$.

\subsection{Higgs sector and {\it SU}(5)-breaking vacuum}
\label{sec:HiggsSECandVACUA}
%
We assume that the field $24_H$ acquires a nonvanishing~VEV of
 ${\cal O}(M_{\rm GUT})$ only in the hypercharge direction,
 $\left\langle 24_H^{24} \right\rangle$, since $24_H^{24}$ is the only
 SM singlet among its components.\footnote{The neutral component of
  $W_H$ in $24_H$, colorless, could also acquire a~VEV once the
  electroweak  symmetry is broken. A dimensional analysis shows that its
  scalar component, with mass squared of ${\cal O}(M_{\rm GUT}^2)$, can
  induce a tadpole term at most of
  ${\cal O}(\widetilde{m}^2 M_{\rm GUT})$, and therefore, a~VEV at most
  of  ${\cal O}(\widetilde{m}^2/ M_{\rm GUT})$, i.e., completely
  negligible. 
  We set to zero the~VEVs of $5_H$ and $\bar 5_H$, which are also
  negligible and completely irrelevant to this analysis.
  Since all the components of $24_H$ have a mass of
  ${\cal O}(M_{\rm GUT}^2)$, the vacuum $\langle 24_H^{24}\rangle$ is
  bound to remain a vacuum, at least locally, even when the
  SUSY-breaking terms are included, which have only the effect of
  producing tiny shifts in its value.}
Elsewhere, this field is denoted by the symbol $B_H$, which indicates
 both the complete superfield and its scalar component, like all
 symbols used for Higgs fields in this paper.
In this section and in Appendix~\ref{sec:proof}, where the double
 use of the same symbol may generate some confusion, we reserve the
 symbol $24_H^{24}$ for the superfield, whereas $B_H$ and $F_{B_H}$ are
 used for its scalar and auxiliary components.

The~VEV $24_H^{24}$ is decomposed as 
\begin{equation}
 \langle 24_H^{24} \rangle = 
 \langle B_H       \rangle + \theta^2
 \langle F_{B_H}\rangle,    
\label{eq:V24decomp}
\end{equation} 
 and the part of the superpotential relevant to its determination is a
 subset of $W^{\rm MSSU(5)}_H$ in Eq.~(\ref{eq:WminSU5H}):
\begin{equation}
 W_{24_H^{24}} = 
  \frac{1}{2} M_{24} \left(24_H^{24}\right)^2
 -\frac{1}{6\sqrt{30}}\lambda_{24} \left(24_H^{24}\right)^3. 
\label{eq:Wvacuum}
\end{equation}
{}For vanishing soft SUSY-breaking terms, the scalar potential determining
 the vacuum is simply given by $\vert F_{B_H}\vert^2$, with
\begin{equation}
 F_{B_H}^* 
 = \frac{\partial W(B_H)}{\partial B_H}
 = M_{24} B_H - \frac{1}{2\sqrt{30}} \lambda_{24} B_H^2. 
\label{eq:auxBH-vacuumCOND}
\end{equation}
{}For $F_{B_H} =0$, a nonvanishing~VEV for $B_H$ is generated:
\begin{equation}
\langle B_H \rangle  =  v_{24} \, \equiv\,  
   2 \sqrt{30} \, \frac{M_{24}}{\lambda_{24}}.
\quad \quad  
 \left(\left \langle F_{B_H} \right \rangle = 0 \right)
\label{eq:v24noSOFT}
\end{equation}
The VEV $v_{24}$, which coincides with $\langle B_H \rangle$ only
 in the limit of exact SUSY, splits the masses of the doublet and
 triplet components of the $5_H$ and $\bar{5}_H$ Higgs fields, the 
 parameters $\mu_2$ and $\mu_3$ defined in
 Appendix~\ref{sec:emb-GROUPfact}:
\begin{eqnarray}
 \mu_2 \, = \,
 M_5 -\displaystyle{\frac{1}{2}}\sqrt{\frac{6}{5}} \lambda_5 v_{24}
       & = &
 M_5 - 6 \lambda_5 \,\displaystyle{\frac{M_{24}}{\lambda_{24}}},
\label{eq:mu2noSOFT}
\\
 \mu_3 \, = \,
 M_5 +\displaystyle{\frac{1}{3}}\sqrt{\frac{6}{5}} \lambda_5 v_{24}
       & = &
 M_5 + 4 \lambda_5 \,\displaystyle{\frac{M_{24}}{\lambda_{24}}}
       \, = \,
 \mu_2 + 10 \lambda_5 \,\displaystyle{\frac{M_{24}}{\lambda_{24}}}.
\label{eq:mu3noSOFT}
\end{eqnarray}

It is clear that $M_5$ and $M_{24}$, both of ${\cal O}(M_{\rm GUT})$,
 must be fine tuned to render $\mu_2$ of ${\cal O}(M_{\rm weak})$, 
 whereas $\mu_3$ is naturally of ${\cal O}(M_{\rm GUT})$, i.e.,
 heavy enough not to disturb the success of the gauge coupling
 unification.  
This fine-tuning problem, or doublet--triplet splitting problem, is
 typical of GUT models and has spurred many proposals for its
 solution~\cite{missingPARTN,MPnonMSSU5,SS,flipped,DW,pseudoNG,5dDTS}. 
The known remedies in the case of {\it SU}(5) in four-dimensional
 space-time~\cite{missingPARTN} require a departure from the
 assumption of minimality encoded in $W^{\rm MSSU(5)}_{\rm H}$.
Here, we keep this assumption and we agree to tolerate the severe fine
 tuning of Eq.~(\ref{eq:mu2noSOFT}), which is stable under radiative
 corrections, thanks to the nonrenormalization theorem for
 superpotentials.  
The soft SUSY-breaking scalar potential is not equally protected and, in
 general, we expect that the fine tuning to be performed also in this
 sector is potentially destabilized by radiative corrections.
This turns out not to be the case, and both tunings are stable~\cite{KMY}.

To show this, we start by giving the form of the soft SUSY-breaking
 scalar potential:
\begin{equation}
 \widetilde{V}^{\rm MSSU(5)} =  \widetilde{V}^{\rm MSSU(5)}_{\rm M}
                  + \widetilde{V}^{\rm MSSU(5)}_{\rm H}
                  + \widetilde{V}^{\rm MSSU(5)}_{\rm gaug},
\label{eq:VminSU5}
\end{equation}
 split into a matter part,\footnote{Note the different conventions
  for the soft masses of the fields $10_M$ and $\bar{5}_M$, discussed
  below Eq.~(\ref{eq:SOFTmassDOUBL}) in Appendix~\ref{sec:RGEsSU5}.} a
 purely Higgs-boson part, and a gaugino part:
\begin{eqnarray}
\widetilde{V}^{\rm MSSU(5)}_{\rm M}              & = & 
\left\{\!-\frac{1}{4}\widetilde{10}_M A^{10}\widetilde{10}_M   5_H  
  +\!\sqrt{2} \,\widetilde{\bar{5}}_M A^5   \widetilde{10}_M\bar{5}_H 
+\! {\rm H.c.} \!\right\}
+\!\widetilde{10}_M^\ast \widetilde{m}_{10_M}^2 \widetilde{10}_M
\nonumber\\                               &   & 
\hspace*{8truecm}
+ \widetilde{\bar{5}}_M \widetilde{m}_{\bar{5}_M}^2\, 
                         \widetilde{\bar{5}}_M^\ast,  
\nonumber\\    
\widetilde{V}^{\rm MSSU(5)}_{\rm H}             & = &  
\left\{
   B_5 5_H \bar{5}_H       + A_{\lambda_{5}} 5_H 24_H \bar{5}_H 
+\frac{1}{2} B_{24} (24_H)^2 +\frac{1}{6} A_{\lambda_{24}} (24_H)^3 
+\! {\rm H.c.} \right\}
\nonumber\\                               &   &  
+ \widetilde{m}_{5_H}^2        5_H^\ast       5_H
+ \widetilde{m}_{\bar{5}_H}^2 \bar{5}_H^\ast \bar{5}_H     
+ \widetilde{m}_{24_H}^2       24_H^\ast      24_H ,  
\nonumber\\    
\widetilde{V}^{\rm MSSU(5)}_{\rm gaug}            & = & 
 \frac{1}{2} M_G \widetilde{G}_5 \widetilde{G}_5 .
\label{eq:VminSU5terms}
\end{eqnarray}

The inclusion of the soft SUSY-breaking terms shifts the scalar 
 potential relevant to the determination of the MSSU(5) vacuum as 
 follows:
\begin{equation}
 V_{B_H} =
 \vert F_{B_H} \vert^2 + \widetilde{V}_{B_H},
\label{eq:VBHcomplete}
\end{equation}
 where $\widetilde{V}_{B_H}$ is a subset of the potential  
 $\widetilde{V}^{\rm MSSU(5)}_{\rm H}$ in Eq.~(\ref{eq:VminSU5terms}): 
\begin{equation}
 \widetilde{V}_{B_H} =
  \widetilde{m}^2_{24_H} B_H^\ast B_H 
+ \left\{ \frac{1}{2} B_{24} B_H^2 
         -\frac{1}{6\sqrt{30}}A_{\lambda_{24}} B_H^3 + {\rm H.c.}
 \right\}. 
\label{eq:VBHsoft}
\end{equation}
Once the soft terms are introduced, in general, also the auxiliary
 component of $24_H^{24}$ acquires a~VEV and the value of the
 scalar-component~VEV gets modified~\cite{HLW}.  
The expressions for both these~VEVs can be obtained perturbatively,
 using $(\widetilde{m}/M_{\rm GUT})$ as 
 expansion parameter, where $\widetilde{m}$ is a typical soft mass: 
\begin{eqnarray}
 \left\langle B_H\right\rangle       & = &  v_{24}  
+ \delta v_{24}  
+ \delta^2 v_{24} 
+ {\cal O}\left(\displaystyle{\frac{\widetilde m^3}{M_{\rm GUT}^2}}
          \right), 
\nonumber\\
 \left\langle F_{B_H} \right\rangle  & = & \  0 \ 
+\,F_{24} \, 
+\,\delta F_{24} 
+ {\cal O}\left(\displaystyle{\frac{\widetilde m^3}{M_{\rm GUT}}}
           \right).
\label{eq:VEVS24expansion}
\end{eqnarray}
Here, $v_{24}$, $\delta v_{24}$, and $\delta^2 v_{24}$ are respectively
 of ${\cal O}\left(M_{\rm GUT}\right)$,
 ${\cal O}\left(\widetilde m\right)$, and
 ${\cal O}\left(\widetilde m^2/M_{\rm GUT}\right)$, whereas 
 $F_{24}$ and $\delta F_{24}$ are of
 ${\cal O}\left(\widetilde m M_{\rm GUT}\right)$ and
 ${\cal O}\left(\widetilde m^2\right)$.
A straightforward calculation yields
\begin{eqnarray}
 \delta v_{24} \  & = &  
-\left( \displaystyle{\frac{B_{24}}{M_{24}}} - 
        \displaystyle{\frac{A_{\lambda_{24}}}{\lambda_{24}}}
 \right)^{\!\ast}  
        \displaystyle{\frac{v^\ast_{24}}{M_{24}}},
\nonumber\\
  F_{24}  \       & = &  
\phantom{-}
 \left( \displaystyle{\frac{B_{24}}{M_{24}}} - 
        \displaystyle{\frac{A_{\lambda_{24}}}{\lambda_{24}}}
 \right) v_{24},
\nonumber\\
\delta F_{24}   & = & 
\left[
 \widetilde{m}_{24_H}^2 + 
  \displaystyle{\frac{B_{24}}{M_{24}}} 
 \left( \displaystyle{\frac{B_{24}}{M_{24}}} -
        \displaystyle{\frac{A_{\lambda_{24}}}{\lambda_{24}}}
 \right)^{\!\ast}
\right] \displaystyle{\frac{v^\ast_{24}}{M_{24}}}.
\label{eq:VEVS24shifts}
\end{eqnarray}

The MSSM $\mu$ and $B$ parameters are then expressed in terms of 
 $\langle B_H \rangle$ and $\langle F_{B_H} \rangle$:\footnote{The
  importance of the correct matching of $\mu$ and $B$ to the original
  GUT parameters is highlighted in Ref.~\citen{EMO} in the context of
  no-scale supergravity.}
\begin{eqnarray}
\mu  &=&    M_5 -
 \displaystyle{\frac{1}{2} \sqrt{\frac{6}{5}}}\, \lambda_5
                             \left\langle B_H    \right\rangle 
\equiv
\mu_2  +\delta \mu_2 
+ {\cal O}\left(\frac{\widetilde{m}^2}{M_{\rm GUT}}\right), 
\nonumber\\[1.5ex]
 B   &=&  
  B_5 -
 \displaystyle{\frac{1}{2} \sqrt{\frac{6}{5}}}
          \left(A_{\lambda_5}\left\langle B_H    \right\rangle
                  +\lambda_5 \left\langle F_{B_H}\right\rangle \right)
 \equiv  
 B_2 +\delta B_2 
+{\cal O}\left(\frac{\widetilde{m}^3}{M_{\rm GUT}}\right),
\label{eq:MUBparam1}
\end{eqnarray}
 where $\mu_2$ is already defined in Eq.~(\ref{eq:mu2noSOFT}),
 and $B_2$ is
\begin{equation}
  B_2 
 =  B_5 -
 \displaystyle{\frac{1}{2} \sqrt{\frac{6}{5}}}
 \left(A_{\lambda_5} v_{24} +\lambda_5 F_{24}\right)
 = \frac{B_5}{M_5}\mu_2    
   +6 \frac{\lambda_5}{\lambda_{24}} M_{24}\,\Delta,
\label{eq:B2}
\end{equation}
 with $\Delta$: 
\begin{equation}
\Delta  \equiv 
 \frac{B_5}   {M_5}   -\frac{A_{\lambda_5}}   {\lambda_5}
-\frac{B_{24}}{M_{24}}+\frac{A_{\lambda_{24}}}{\lambda_{24}}. 
\label{eq:DeltaB}
\end{equation}
Finally, $\delta \mu_2 $ and $\delta B_2$ are
\begin{eqnarray}
 \delta \mu_2 &=& 
 -\frac{1}{2} \sqrt{\frac{6}{5}}\lambda_5 \delta v_{24}
 =
\  6 \frac{\lambda_5}{\lambda_{24}^\ast}
   \frac{M_{24}^\ast}{M_{24}} 
 \left( \displaystyle{\frac{B_{24}}{M_{24}}} - 
        \displaystyle{\frac{A_{\lambda_{24}}}{\lambda_{24}}} \right)^\ast,
\nonumber \\
 \delta B_2  & = & 
 -\frac{1}{2} \sqrt{\frac{6}{5}}
 \left[A_{\lambda_5} \delta v_{24} + \lambda_5 \delta F_{24}\right] 
\nonumber \\ 
 &=&
-  6 \frac{\lambda_5}{\lambda_{24}^\ast}
   \frac{M_{24}^\ast}{M_{24}} 
\left[ 
\left( \displaystyle{\frac{B_{24}}{M_{24}}} - 
        \displaystyle{\frac{A_{\lambda_{24}}}{\lambda_{24}}} \right)^\ast
\left( \displaystyle{\frac{B_{24}}{M_{24}}} - 
        \displaystyle{\frac{A_{\lambda_5}}{\lambda_5}} \right)
  + \widetilde{m}_{24_H}^2 
\right].
\label{eq:dmu2db2}
\end{eqnarray}
The quantities to be fine tuned in these equations are $\mu_2$
 and $B_2$, and the tuning conditions to be imposed are
\begin{equation}
 \mu_2 = {\cal O}(\widetilde{m}), \qquad
 \Delta= {\cal O}\left(\!\frac{\widetilde{m}^2}{M_{\rm GUT}}\!\right).
\label{eq:tunings}
\end{equation}
That both these tuning conditions are stable against radiative
 corrections, as claimed above, can be seen by making use of the RGEs
 listed in Appendix~\ref{sec:RGEGUTtoPL} for the parameters of the 
 Higgs sector entering the second equalities for $\mu_2$ and
 $\Delta$ in Eqs.~(\ref{eq:mu2noSOFT}) and~(\ref{eq:DeltaB}).

The calculation becomes particularly simple in terms of the RGE for
 $v_{24}$, which we give here together with that for $F_{24}$:
\begin{eqnarray}
\dot v_{24}  &=& -\gamma_{24} \, v_{24},
\nonumber\\
\dot F_{24}  &=& -\gamma_{24} F_{24}-\tilde{\gamma}_{24}\,v_{24}. 
\label{eq:VEVS24evol}
\end{eqnarray}
These evolution equations, expressed in terms of the anomalous
 dimension of the field $24_H$, $\gamma_{24}$, and its soft counterpart
 $\tilde{\gamma}_{24}$, are obtained in a very general way in
 Appendix~\ref{sec:New}.
Both VEVs exist above $M_{\rm GUT}$.
The~VEV $v_{24}$, in particular, determined by superpotential
 parameters, is very different, for example, from the MSSM~VEVs
 $v_u$ and $v_d$, determined by K\"ahler potential parameters and whose
 existence is not guaranteed at all scales.
The RGE for $F_{24}$ shows that a nonvanishing value for this VEV
 is generated through radiative corrections, even when starting with a
 vanishing one at some scale. 
Since it will be of use later on, we define here also the quantity
 $f_{24}$, i.e., the ratio of the two VEVs:
\begin{equation}
 f_{24}   \equiv   
 \frac{ F_{24}} {v_{24}} 
   = \left(\frac{B_{24}}{M_{24}}-\frac{A_{\lambda_{24}}}{\lambda_{24}}
       \right), 
\label{eq:f24DEF}
\end{equation}
 with the simple evolution equation:
\begin{equation}
\dot f_{24}  = -\tilde{\gamma}_{24}. 
\label{eq:f24evol}
\end{equation}
The RGEs for $\delta v_{24}$, $\delta F_{24}$, and $\delta^2 v_{24}$, are
 very different as these quantities depend on the K\"ahler
 potential~(see Appendix~\ref{sec:New}), which gets vertex corrections
 at the loop level.

A consequence of the first RGE in Eq.~(\ref{eq:VEVS24evol}) is that
 $\lambda_5 v_{24}$ evolves as $M_5$, and so does $\mu_2$: 
\begin{equation}
\dot{\mu}_2  = \left(\gamma_{5_H} +\gamma_{\bar{5}_H}\right) \mu_2 .
\label{eq:MUevol}
\end{equation}
Since the quantity $\Delta$ turns out to be scale invariant:
\begin{equation}
  \dot{\Delta}=0, 
\label{eq:RGEDelta}
\end{equation}
 $B_2$ evolves also like $B_5$:
\begin{equation}
\dot{B}_2     = 
 \left(\gamma_{5_H} +\gamma_{\bar{5}_H}\right) B_2 + 
 \left(\tilde{\gamma}_{5_H}+\tilde{\gamma}_{\bar{5}_H}\right)\mu_2 .  
\label{eq:Bevol}
\end{equation}
This is sufficient to prove that the tuning for $\mu_2$ and $B_2$ is
 not destabilized by quantum corrections. 
Note that no specific ansatz was made for the boundary conditions
 assigned to the trilinear couplings $A_{\lambda_5}$ and $A_{\lambda_{24}}$,
 and the bilinear couplings $B_{5}$ and $B_{24}$, except that they are
 of ${\cal O}(\widetilde{m})$ and ${\cal O}(\widetilde{m} M_{\rm GUT})$,
 respectively.

On the other hand, the specific values chosen for $\mu_2$ and $\Delta$
 in Eq.~(\ref{eq:tunings}) can affect the values of the boundary
 conditions for the MSSM parameters $\mu$ and $B$. 
We take as an{{{}}} example minimal supergravity, in which at $M_{\rm cut}$, the
 soft masses for all scalar fields are $\widetilde{m}_0$, all ratios of
 trilinear couplings over the corresponding Yukawa couplings are
 $(A_{\lambda_i}/\lambda_i)(M_{\rm cut})= A_0$, and all ratios of heavy
 bilinear parameters over the corresponding superpotential masses are
 $(B_{i}/M_i)(M_{\rm cut}) = B_0$, with
 \begin{equation}
 B_0 = A_0 -\widetilde{m}_0.
\label{eq:B0sugra}
\end{equation}
In this context, at $M_{\rm cut}$, it is $\Delta=0$, and if the value of
 $\mu_2(M_{\rm cut})=0$ is imposed, we obtain for the boundary condition
 of the MSSM $\mu$ parameter:
 $\mu(M_{\rm cut}) =\delta \mu_2(M_{\rm cut})$, with 
\begin{equation}
 \delta \mu_2(M_{\rm cut}) = 
  -6 \widetilde{m}_0
\left[\left(\frac{\lambda_5}{\lambda_{24}^\ast}\right)
      \left(\frac{M_{24}^\ast}{M_{24}}\right)\right](M_{\rm cut}) ,
\label{eq:deltamu2BC}
\end{equation}
 and for the boundary condition of the MSSM $B$ parameter~\cite{HLW}:
\begin{equation}
 B(M_{\rm cut}) =  2\widetilde{m}_0 \delta \mu_2 (M_{\rm cut}).
\label{eq:BparBC1}
\end{equation}
In contrast, if $\mu_2(M_{\rm cut})$ (and therefore, $\mu(M_{\rm cut})$)
 is a nonvanishing arbitrary quantity of ${\cal O}(\widetilde{m})$, the
 relation between the boundary conditions of $B$ and $\mu$ is also
 arbitrary:
\begin{equation}
 B(M_{\rm cut}) =
 \left[B_0\mu_2(M_{\rm cut}) +2\widetilde{m}_0\delta\mu_2(M_{\rm cut})\right],
\label{eq:BparBC2}
\end{equation}
 with $B_0$ in Eq.~(\ref{eq:B0sugra}) and $\delta\mu_2(M_{\rm cut})$ in
 Eq.~(\ref{eq:deltamu2BC}).

{}For completeness, we give here also the tree-level mass of the
 triplet Higgs fields including the small correction $\delta \mu_3$, 
 of ${\cal O}(\widetilde{m})$, due to the term $\delta v_{24}$ of the
 VEV $\langle B_H\rangle$:
\begin{equation}
 M_{H^C}  \equiv   \mu_3  + \delta \mu_3   = 
  \mu  +10 \lambda_5\frac{M_{24}}{\lambda_{24}}
       -10 \frac{\lambda_5}{\lambda_{24}}\frac{M_{24}^\ast}{M_{24}}
           \left(\frac{B_{24}}{M_{24}} - \frac{A_{\lambda_{24}}}{\lambda_{24}}
           \right)^{\!\!\!\ast}.
\label{eq:McHREWRITE}
\end{equation}
We also collect together the various pieces of the parameters $\mu$ and
 $B$:
\begin{eqnarray}
  \mu  & =&
 \left[M_5 -6\lambda_5 \frac{M_{24}}{\lambda_{24}}\right]
+ 6\frac{\lambda_5}{\lambda_{24}^\ast}
   \frac{M_{24}^\ast}{M_{24}}
   \left(\frac{B_{24}}{M_{24}} - \frac{A_{\lambda_{24}}}{\lambda_{24}}
            \right)^{\!\!\ast},
\nonumber\\
B  & = &
 \frac{B_5}{M_5}\mu
  -6 \frac{\lambda_5}{\lambda_{24}^\ast}\frac{M_{24}}{M_{24}^\ast}
 \left[\left(\frac{B_{24}}{M_{24}} -\frac{A_{\lambda_{24}}}{\lambda_{24}}\right)^\ast 
     \left(2 \frac{B_{24}}{M_{24}} -\frac{A_{\lambda_{24}}}{\lambda_{24}}\right)
         +\tilde{m}^2\right]. \quad
\label{eq:muBcomplete}
\end{eqnarray}
Note that, differently from $\mu_2$, the quantities $\delta\mu_2$
 and $\delta\mu_3$ do not evolve like all the superpotential
 dimensionful parameters.  
In a similar way, $\delta B_2$ does not evolve like a typical bilinear
 soft parameter, whereas $B_2$ does.
This is because their expressions contain $\delta v_{24}$ and 
 $\delta F_{24}$.
Thus, above $M_{\rm GUT}$, the MSSM $\mu$ and $B$ parameters
 as well as the small correction to $M_{H^C}$ of
 ${\cal O}(\widetilde{m})$ also have nonconventional RGEs.

The derivation of $\langle B_H \rangle$ and $\langle F_{B_H}\rangle$
 in this section allows for independent phases of $M_{24}$ and 
 $\lambda_{24}$. 
It is possible, however, to redefine the field $24_H$ in such a way to
 align them.
That is, without any loss of generality, it is possible to take
 $v_{24}$ to be real, which is what will be assumed hereafter.
In contrast, the phases of the shifts $\delta v_{24}$, $F_{24}$, and
 $\delta F_{24}$ depend on those of the bilinear and trilinear terms 
 in $\widetilde{V}^{\rm MSSU(5)}_{\rm H}$.

\subsection{Seesaw sector}    
\label{sec:seesaw}
\begin{wrapfigure}{r}{6.6cm}
\centerline{\includegraphics[width=0.49\textwidth]{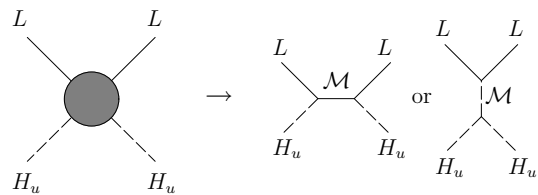}}
\caption{The seesaw mechanism.}
\label{fig:seesaw}
\end{wrapfigure}

%
The seesaw mechanism is a mechanism that generates the effective
 dimension-five operator for Majorana neutrino masses:
\begin{equation}
  W_{\nu} \ = \ -\frac{1}{2} L H_u \kappa L H_u,
\label{eq:nu}
\end{equation}
 where the symmetric matrix $\kappa$ is a dimensionful parameter of
 $O(M_{\rm ssw}^{-1})$.
This operator is obtained by integrating out the heavy seesaw degrees
 of freedom at the scale $M_{\rm ssw}$.
The mechanism is depicted in Fig.~\ref{fig:seesaw}, where a
 solid~(broken) line indicates a fermion (boson) or, in a supersymmetric
 context, a superfield with a $Z_2$ odd~(even) $R$-parity.
At the tree level, there are only two diagrams that can give rise to the
 above effective operator, one mediated by a solid line and one by a
 broken line.

Taking into account the gauge structure, it may seem at first glance 
 that in both cases, the inner line that represents the mediator
 ${\cal M}$ can be a singlet or triplet of {\it SU}(2).
In reality, the possibility of the singlet scalar mediator
 ${\cal M}$ is forbidden by the multiplication rule of {\it SU}(2):
$  2\times2=1_A+3_S $,
 where the indices $A$ and $S$ indicate antisymmetric and {{{}}} symmetric
 products, respectively.
Thus, there are only three types of seesaw mechanism, with mediators:

\begin{center}
\begin{tabular}{lcl}
{\large singlet fermions}  & \quad : \quad & {\large type~I,}  \\[1.2ex]
{\large a triplet scalar}  & \quad : \quad & {\large type~II,} \\[1.2ex]
{\large triplet fermions}  & \quad : \quad & {\large type~III,} 
\end{tabular}\vspace{2mm}
\end{center}
 i.e., the RHNs $N^c$, a triplet Higgs boson $T$, and matter triplets
 $W_M$, respectively.
They interact with the lepton doublets through the large Yukawa
 couplings $Y_\nu^i $ $(i={\rm I,II,III})$:
\begin{equation}
 N^c  Y_\nu^{\rm I} L H_u, 
\qquad
 \frac1{\sqrt2} L Y_\nu^{\rm II} T L,  
\qquad
 H_u W_M Y_\nu^{\rm III} L.
\label{eq:NUYukawaInteracs}
\end{equation}
With a minimal completion, we obtain the seesaw
 superpotentials:\footnote{We distinguish by a prime the second and
 third superpotentials from $W_{\rm ssw\,II}$ and $W_{\rm ssw\,III}$
 listed in Appendix~\ref{sec:RGEsSU5}, which contain more operators and
 fields, not needed for the implementation of the seesaw mechanism in
 the MSSM context, but necessary for the completion of the seesaw
 sectors of types~II and~III in the MSSU(5) model.}
\begin{eqnarray}
 W_{\rm ssw\,I}                     & = & 
    N^c Y_\nu^{\rm I} L H_u
  +\displaystyle{\frac{1}{2}} N^c M_N N^c ,
\nonumber\\
 W_{\rm ssw\,II}^\prime             & = & 
   \displaystyle{\frac{1}{\sqrt{2}}} LY_\nu^{\rm II} T L 
 + \frac{1}{\sqrt{2}} \lambda_{\bar{T}} H_u \bar{T} H_u +M_T T\bar{T} ,
\nonumber\\
 W_{\rm ssw\,III}^\prime            & = & 
   H_u  W_M Y_\nu^{\rm III} L  
 +\displaystyle{\frac{1}{2}} W_M M_{W_M} W_M , 
\label{eq:sswSUPERPOT}
\end{eqnarray}
 where, of the three superheavy masses $M_N$, $M_T$, and $M_{W_M}$, the
 first and the third are two $3\times 3$ matrices, $M_T$ is just a
 number.
{}Following Appendix~\ref{sec:emb-SMfields}, it is easy to see that
 $Y_\nu^{\rm I}$ and $Y_\nu^{\rm III}$ are complex matrices with fifteen
 parameters, whereas $Y_\nu^{\rm II}$ is a nine-parameter symmetric
 complex matrix.

By integrating out the mediators in the superpotentials of 
 Eq.~(\ref{eq:sswSUPERPOT}), and replacing $H_u$ by its~VEV $v_u$,
 we obtain relations between the neutrino mass matrix, $\kappa v_u^2$,
 and the seesaw Yukawa couplings:
\begin{equation}
 m_\nu = \kappa v_u^2  \ =  
\left\{
 \begin{array}{l@{\qquad}l}
\phantom{\frac{1}{2}}
  \left(Y_\nu^{\rm I}\right)^T
\!\displaystyle{\frac{1}{M_N}}
  \left(Y_\nu^{\rm I}\right)          v_u^2   & \mbox{for type I},    
\\[1.8ex]
\phantom{\frac{1}{2}} \ \,
   Y_\nu^{\rm II}
\ \displaystyle{\frac{1}{M_T}}\,\lambda_{\bar{T}}\,
                                   v_u^2   & \mbox{for type II},    
\\[1.5ex]
\displaystyle{\frac{1}{2}}
  \left(Y_\nu^{\rm III}\right)^T 
\!\displaystyle{\frac{1}{M_{W_M}}}
  \left(Y_\nu^{\rm III}\right)        v_u^2   & \mbox{for type III}.    
 \end{array}
 \right.
\label{eq:NUmassmatrix}
\end{equation}
(See Appendix~\ref{sec:emb-GROUPfact} for the normalization of the
 field $W_M$, which is in the adjoint representation of {\it SU}(2).)

The neutrino mass matrix is usually expressed as
\begin{equation} 
  m_\nu = V_{\rm MNS}^\ast \, \widehat{m}_\nu V_{\rm MNS}^\dagger.
\label{eq:mnuVMNS}
\end{equation}
In the basis in which the Yukawa coupling for the charged leptons is 
 diagonal, the MNS matrix $V_{\rm MNS}$, a unitary matrix with three
 mixings and three phases, is the diagonalization matrix of $m_\nu$.
It is often parametrized as
 $V_{\rm MNS}^\ast = K_{\rm MNS} P_{\rm MNS}$, where $K_{\rm MNS}$ is a
 $K_{\rm CKM}$-like matrix, with three mixing angles and one phase,
 $P_{\rm MNS}$ is a two-phase matrix, the so-called Majorana phases.

In the same basis, it is possible to invert the relations in 
 Eq.~(\ref{eq:NUmassmatrix}),  solve for the couplings
 $Y_\nu^i$ ($i$= I,II,III) in terms of $m_\nu$, and therefore, obtain the
 high-energy neutrino parameters in terms of those that can be tested at
 low energy, $V_{\rm MNS}$ and $\widehat{m}_\nu$.  
In the case of the seesaw of type~II, because the mediator has no
 flavour, the high-energy input in the neutrino mass matrix is just a
 number, i.e., the ratio $\lambda_{\bar{T}}/M_{\cal M}^{\rm II}$,
 and the flavour structure of $Y_\nu^{\rm II}$ is the same as that of
 the neutrino mass $m_\nu$. 
Thus, the expression for $Y_\nu^{\rm II}$ is rather simple:
\begin{equation}
 Y^{\rm II}_\nu= \frac{1}{v_u^2}
   V_{\rm MNS}^* \,{\widehat m_\nu} V_{\rm MNS}^\dagger 
 \frac{M_T}{\lambda_{\bar{T}}}.
\label{eq:TypeIIrel}
\end{equation}
The corresponding ones for types~I and~III, in which the mediators carry
 flavour indices, are far more involved.
The neutrino mass matrix, which depends on a large number of high-energy
 parameters, has a flavour structure different from those of
 $Y_\nu^{\rm I}$ and $Y_\nu^{\rm III}$.
In a basis in which also the mediator masses are diagonal, these 
 couplings can be expressed as
\begin{eqnarray}
\left(Y_\nu^{\rm I}\right)^T   &=&
 \displaystyle{\frac{\,1\,}{v_u}}\,
   V_{\rm MNS}^* \sqrt{\widehat m_\nu}\,R\,\sqrt{\widehat{M}_N}\,,
\nonumber\\
\left(Y_\nu^{\rm III}\right)^T &=&
 \displaystyle{\frac{\sqrt{2}}{v_u}}\,
   V_{\rm MNS}^* \sqrt{\widehat m_\nu}\,R\,\sqrt{\widehat{M}_{W_M}}\, .
\label{eq:CASASrel}
\end{eqnarray}
Here, an arbitrary complex orthogonal matrix $R$~\cite{CASAS},
 parametrized by three (arbitrary) mixing angles and three phases,
 appears.
Note also that in these two cases, $m_\nu$ is quadratic in
 $Y_\nu^{\rm I}$ and $Y_\nu^{\rm III}$, whereas in the seesaw of type~II,{{{}}}
 it is linear in $Y_\nu^{\rm II}$.

In {\it SU}(5), the multiplets respectively containing the mediators of 
 the seesaw of types~I, II, and III are matter singlets, or RHNs $N^c$, 
 a 15plet Higgs field, $15_H$, and three adjoint matter fields, $24_M$,
 distinguished from the Higgs field $24_H$ by the index $M$, for matter.
The components of the fields $24_M$ are denoted as $G_M$, $W_M$, $X_M$,
 $\bar{X}_M$, and $B_M$, whereas,{{{}}} in addition to the triplet Higgs 
 $T$, $15_H$ contains also the fields $S$ and $Q_{15}$~(see
 Appendix~\ref{sec:emb-GROUPfact}).
The complete seesaw superpotentials are\footnote{The superpotentials
  $W_{\rm ssw\,II}$ and $W_{\rm ssw\,III}$ listed in
  Appendix~\ref{sec:RGEsSU5}, which differ from
  $W_{\rm ssw\,II}^\prime$ and $W_{\rm ssw\,III}^\prime$ of 
  Eq.~(\ref{eq:sswSUPERPOT}), are obtained from the following 
  $W_{\rm 15H}$ and $W_{\rm 24M}$ by integrating out the triplet Higgs
  fields $H_U^C$, and by taking the limit $Y_{24_M}^{S,A}=0$ in
  $W_{\rm ssw\,III}$.}
\begin{eqnarray}
  W_{\rm RHN}  & = & 
  - N^{c} Y_{N}^{\rm I} \bar{5}_M 5_H
  + \displaystyle{\frac{1}{2}} N^{c} M_N N^{c},    
\nonumber\\
  W_{\rm 15H}\ & = & 
 \ \displaystyle{\frac{1}{\sqrt{2}}}
 \bar{5}_M Y_{N}^{\rm II} 15_H \bar{5}_M 
   + M_{15} 15_H \bar{15}_H 
\nonumber\\  &   &            
 + \displaystyle{\frac{1}{\sqrt{2}}}
 \lambda_{D\,} \bar{5}_H 15_H \bar{5}_H
 + \displaystyle{\frac{1}{\sqrt{2}}}
 \lambda_{U\,}  5_H \bar{15}_H 5_H
 + \lambda_{15} 15_H 24_H \bar{15}_H ,
\nonumber\\
  W_{\rm 24M}\ & = & 
 \ 5_H 24_M Y_{N}^{\rm III} \bar{5}_M 
 + \displaystyle{\frac{1}{2}} 24_M M_{24_M} 24_M
\nonumber\\   & & 
\hspace*{3.5truecm}
 + \displaystyle{\frac{1}{2}} \sum_{x=S,A} 
   \left(24_M Y_{24_M}^x 24_M \right)_{x} 24_H,  
\label{eq:SU5sswSUPERP}
\end{eqnarray}
 where the index $x$ in the last term of the third of these equations 
 refers to the symmetric and antisymmetric products of the two
 $24_M$~(see Appendix~\ref{sec:emb-GROUPfact}). 
If not small, the couplings $\lambda_{15}$ and $Y^x_{24_M}$ $(x=S,A)$
 can easily spoil the gauge coupling unification. 
{}For this reason, in this analysis, we take them to be vanishing, and 
 therefore, the mass term for the mediators $15_H$ and $24_M$ is
 approximately given by the second term in $W_{\rm 15H}$ and
 $W_{\rm 24M}$, respectively.
The couplings $Y_{N}^{\rm I}$ and $Y_{N}^{\rm III}$ are generic complex
 matrices, $Y_N^{\rm II}$ is a symmetric complex one, with,
 respectively, eighteen and twelve irreducible parameters.

When expanding these terms in their SM representations (see
 Appendix~\ref{sec:emb-GROUPfact} for details), in the SM basis of
 Eq.~(\ref{eq:SMfieldsID}), these Yukawa interactions become
\begin{eqnarray}
 - N^c Y_{N}^{\rm I} \bar{5}_M 5_H        &\to&
              N^c Y_\nu^{\rm I} L H_u
 -e^{i\phi_l} N^c \left(Y_\nu^{\rm I} P_l \right) D^c H_U^C, 
\nonumber\\[1.5ex]
  \bar{5}_M Y_{N}^{\rm II} 15_H \bar{5}_M  &\to&
   L Y_\nu^{\rm II} T L 
 -\sqrt{2}e^{i\phi_l} D^c \left(P_l Y_\nu^{\rm II}\right) L Q_{15}
\nonumber\\                           &  & 
\qquad  + e^{i2\phi_l} D^c \left(P_lY_\nu^{\rm II}P_l\right) S D^c,
\nonumber\\[1.1ex]
  5_H 24_M Y_{N}^{\rm III} \bar{5}_M       &\to& 
   H_u W_M Y_\nu^{\rm III} L
 -\displaystyle{\sqrt{\frac65}\frac12}
   H_u B_M Y_\nu^{\rm III} L
\nonumber\\                           &   &
 + e^{i\phi_l} H_u\bar X_M Y_\nu^{\rm III} P_l D^c
 +  e^{i\phi_l}
   H_U^C   G_M Y_\nu^{\rm III} P_l D^c
\nonumber\\                          &   &
 +\,e^{i\phi_l}\displaystyle{\sqrt{\frac65}\frac13}
   H_U^C B_M Y_\nu^{\rm III}P_l D^c
 +\, H_U^C X_M Y_\nu^{\rm III} L.
\label{eq:YUKAWAsswRotat}
\end{eqnarray}
Note that the couplings $Y_N^i$ $(i={\rm I,II,III})$ are substituted
 by the couplings $Y_\nu^i$ of
 Eqs.~(\ref{eq:NUYukawaInteracs}),~(\ref{eq:TypeIIrel}), 
 and~(\ref{eq:CASASrel}), which have less independent parameters, as
 three phases were removed by using the three unfixed phases
 $e^{i\phi_l}P_l$ in Eq.~(\ref{eq:SMfieldsID}).
The dependence on these phases, however, is not totally eliminated.
They reappear in terms containing either the triplet Higgs field,
 $H_U^C$, or a component of the seesaw mediators, and, depending on the
 type of seesaw implemented, are to be identified with the phases
 $e^{i\phi_i}P_i$ $(i={\rm I,II,III})$ of
 Appendix~\ref{sec:emb-SMfields}.

One feature of the seesaw of type III becomes very evident through the 
 expansion of the term $5_H 24_M Y_{N}^{\rm III} \bar{5}_M$. 
The component $B_M$ of $24_M$ interacts with the components of the
 fields $5_H$ and $\bar{5}_M$ as the RHNs $N^c$, except for different
 group factors. 
Thus, the implementation of a seesaw of type~III in the MSSU(5) model is
 connected to that of type~I, and the relation of
 Eq.~(\ref{eq:NUmassmatrix}), in which the superscript label is now
 somewhat of a misnomer, must be modified: 
\begin{equation}
 m_\nu = \kappa v_u^2 = 
\left[\frac{1}{2}
  \left(Y_\nu^{\rm III}\right)^T \frac{1}{M_{W_M}}
  \left(Y_\nu^{\rm III}\right)      +
\frac{3}{10}
  \left(Y_\nu^{\rm III}\right)^T \frac{1}{M_{B_M}}
  \left(Y_\nu^{\rm III}\right) 
\right] v_u^2 .
\label{eq:MODmnuIII}
\end{equation}
{}For simplicity, the same symbol $Y_\nu^{\rm III}$ is used for the
 couplings in the two terms, but in reality, these differ by small 
 {\it SU}(5)-breaking corrections.

Last, but not least, these decompositions show clearly that the Yukawa
 couplings $Y_\nu^i$ $(i={\rm I,II,III})$ induce flavour violations in
 both the scalar sector of leptons and right-handed down quarks in a
 related way. 
Differently than in the seesaw of type~I, where the second interaction
 in the first decomposition decouples at $M_{\rm GUT}$, in the seesaw of 
 type~II, all three interactions in the second decomposition remain 
 active down to $M_{\rm ssw}$, which differ by at least a couple of
 orders of magnitude from $M_{\rm GUT}$.
The situation is more complicated in the case of the seesaw of type~III.
The consequences of this feature will be discussed in
 \S\ref{sec:Results}.

We close this section by giving also the soft part of the scalar
 potential for the three types of seesaw mechanism:\footnote{The
  conventions for the soft mass terms of the fields $N^c$ and $24_M$
  are discussed below Eq.~(\ref{eq:SOFTmassDOUBL}) in
  Appendix~\ref{sec:RGEsSU5}.}
\begin{eqnarray}
\widetilde{V}_{\rm RHN}                        & = & 
\Bigl\{
 - \widetilde{N^c} A_{N}^{\rm I} \widetilde{\bar{5}}_M  5_H
 + \displaystyle{\frac{1}{2}} \widetilde{N^c} B_N \widetilde{N^c} 
 + \ {\rm H.c.} 
\Bigr\}
 + \widetilde{N^c} \widetilde{m}_{N^c}^2 \widetilde{N^c}^\ast ,
\nonumber\\[1.1ex]
\widetilde{V}_{\rm 15H}                        & = & 
\Bigl\{
  \displaystyle{\frac{1}{\sqrt{2}}}
    \widetilde{\bar{5}}_M A_{N}^{\rm II} 15_H \widetilde{\bar{5}}_M 
 + B_{15} 15_H \bar{15}_H 
\nonumber\\                                 &   & 
 + \displaystyle{\frac{1}{\sqrt{2}}}
                      A_{\lambda_{D\,}} \bar{5}_H     15_H \bar{5}_H
 + \displaystyle{\frac{1}{\sqrt{2}}}
                      A_{\lambda_{U\,}}      5_H \bar{15}_H     5_H
 +                    A_{\lambda_{15}} 15_H  24_H\bar{15}_H 
 + \ {\rm H.c.} 
\Bigr\}  
\nonumber\\                                 &   & 
 + \widetilde{m}_{15_H}^2            15_H^\ast       15_H        
 + \widetilde{m}_{\bar{15}_H}^2 \bar{15}_H^\ast \bar{15}_H,
\phantom{\frac{1}{2}}
\nonumber\\[1.1ex]
\widetilde{V}_{\rm 24M}                        & = & 
\Bigl\{
   5_H \widetilde{24}_M A_{N}^{\rm III} \widetilde{\bar{5}}_M 
 + \displaystyle{\frac{1}{2}} \widetilde{24}_M B_{24_M} \widetilde{24}_M
\nonumber\\
                                               &   & 
 + \displaystyle{\frac{1}{2}} \sum_{x=S,A} \left(
        \widetilde{24}_M A_{24_M}^x \widetilde{24}_M \right)_{x} 24_H 
 \phantom{\frac{1}{2}}  + \ {\rm H.c.}
\Bigr\}
 + \widetilde{24}_M  \widetilde{m}_{24_M}^2 \widetilde{24}_M^\ast.
\label{eq:SU5sswSOFTPOT}
\end{eqnarray}

\section{Effective couplings at the tree level in the SUSY-conserving
         sector}
\label{sec:NROps}
%
Nonrenormalizable operators exist in every effective model, and in
 general,{{{}}} all NROs allowed by the symmetries of the model do appear.
The peculiarity of GUT models in this respect is that they postulate the
 existence of a large scale not far from $M_{\rm cut}$.
Therefore, NROs in these models can be far less suppressed than, say,
 in the SM.

Those that immediately come to mind in the context of the MSSU(5) model
 are the NROs obtained from renormalizable operators in which the field
 $24_H$ is inserted in all possible ways. 
Neglecting for a moment the dynamical part of $24_H$, it is easy to see
 that the large~VEVs of its component $B_H$, in this discussion
 safely approximated by $v_{24}$ and $F_{24}$, can partially compensate
 for the huge suppression coming from inverse powers of
 $M_{\rm cut}$.
The result is a milder suppression factor:
\begin{equation}
  s \ \equiv \, \frac{v_{24}}{M_{\rm cut}},
\label{eq:SU5suppFACTOR}
\end{equation}
 a quantity of order $10^{-2}$, or larger if the GUT scale is closer to
 $M_{\rm cut}$.
This is too large to assume that NROs can be neglected.
Since it is difficult to explain theoretically why their coefficients
 are exactly vanishing or very small, NROs should actually be
 considered as an integral part of the MSSU(5) model itself.

In this section, we discuss NROs of this type in the SUSY-conserving
 sector of our theory.
We postpone the study of those in the SUSY-breaking sector to
 \S\ref{sec:SOFTNROS}.
Nonrenormalizable operators that do not contain the adjoint field
 $24_H$ can also play an important role in low-energy physics.
An example is given by the following class of NROs, which may be used
 to suppress the decay rate of the proton:
\begin{equation}
 Op^{\rm PD}   =  \sum_n     \frac{1}{M_{\rm cut}}
  \left(C^{\rm PD}_n \right)^{hijk}
  (10_M)_h (10_M)_i (\bar5_M)_j (10_M)_k 
             \left(\!\frac{24_H}{M_{\rm cut}}\!\right)^{\!\!n}.
\label{eq:supNROforPD}
\end{equation}
(See details of the actual form of this class of operators in
 Appendix~\ref{sec:emb-GROUPfact}.)
A discussion on this issue is postponed to \S\ref{sec:PDconstBC}.

\subsection{Effective Yukawa couplings}
\label{sec:effYUK}
%
Nonrenormalizable operators in the Yukawa sector of the superpotential
 are obtained by inserting in all possible ways powers of the field
 $24_H$, for example, in the two Yukawa operators of
 Eq.~(\ref{eq:WminSU5m}).
We obtain the two classes of NROs $Op^{5}$ and $Op^{10}$, with $Op^{5}$
 given by
\begin{eqnarray}
 Op^{5}         = 
 \sum_{n+m=0}^k\,
 \sqrt{2}\ \bar{5}_M \,C_{n,m}^{5} 
 \left(\!\displaystyle{\frac{24_H^T}{M_{\rm cut}}}\!
 \right)^{\!\!n}  \! 10_M 
 \left(\!\displaystyle{\frac{24_H}{M_{\rm cut}}}\!
 \right)^{\!\!m}  \! \bar5_H,
\label{eq:NRO5}  
\end{eqnarray}
 where $24_H^T$ is the transpose of $24_H$.
{}For the explicit expression  of $Op^{10}$, with coefficients
 $C_{n_1,n_2,n_3,n_4,n_5}^{10}$, we refer the reader to
 Appendix~\ref{sec:emb-GROUPfact}.
The NROs with $n$, $m$, and $n_i$ simultaneously vanishing in $Op^{5}$
 and $Op^{10}$ are the renormalizable ones listed in
 Eq.~(\ref{eq:WminSU5m}), with $C_{0,0}^{5}\equiv Y^5$ and
 $C_{0,0,0,0,0}^{10} \equiv Y^{10}$.
We shall sometimes refer to them as $Op^{5}\vert_4$ and
 $Op^{10}\vert_4$.

In general, there exist other NROs that differ from these by the trace
 of a product of an arbitrary number of $24_H$ fields, duly suppressed
 by the same number of factors $1/M_{\rm cut}$:
$ Op^{i\,\prime}  =  Op^i \times \prod_j {\rm Tr} ({24_H}/{M_{\rm cut}}\!)^{k_j}  
 (i= 5, 10)$. 
The corrections that they induce to the various Yukawa couplings can be
 reabsorbed into those coming from $Op^{5}$ and $Op^{10}$, once the
 field $24_H$ is replaced by its~VEV.
Therefore, we neglect such NROs in the following discussion.

The dimension-five operators often discussed in the literature are
\begin{eqnarray}
Op^{5}\vert_5   &=& \phantom{-}
 \displaystyle{\frac{\sqrt{2}}{\, M_{\rm cut}}}
 \left(\bar{5}_M \,C_{1,0}^{5}\, 24_H^T 10_M \bar5_H  \ + \
       \bar{5}_M \,C_{0,1}^{5}\, 10_M 24_H \bar5_H    \right),
\nonumber\\
Op^{10}\vert_5  &=&
 -\displaystyle{\frac{1}{4 M_{\rm cut}}}
 \left( 10_M \,C_A^{10} 10_M \,24_H\, 5_H \ + \
        10_M \,C_S^{10} 10_M   5_H \,24_H\right),  
\label{eq:NROdim5}
\end{eqnarray}
where $C_A^{10}$ and $C_S^{10}$ correspond to linear combinations of
 some coefficients of $Op^{10}$:
\begin{eqnarray}
 C_A^{10}     &=&  
 \left[
   C_{1,0,0,0,0}^{10\,T} +C_{0,1,0,0,0}^{10\,T} +
   C_{0,0,1,0,0}^{10}    +C_{0,0,0,1,0}^{10}
 \right]_A ,
\nonumber\\
 C_S^{10}     &=&
 C_{0,0,0,0,1}^{10}  -
 \left[
   C_{1,0,0,0,0}^{10\,T} +C_{0,1,0,0,0}^{10\,T} +
   C_{0,0,1,0,0}^{10}    +C_{0,0,0,1,0}^{10}
 \right]_S ,
\label{eq:C10coeff}
\end{eqnarray}
 and the indices $S$ and $A$ outside square brackets denote,
 respectively, the symmetrized and antisymmetrized forms of the matrices
 contained inside.
We recall that
 $10\times 10 = \overline{5}_S + \overline{45}_A + \overline{50}_S$
 and $5 \times 24 = 5 + 45 + 70$.
Thus the antisymmetric matrix $C_A^{10}$ corresponds to the coupling for
 the irreducible interaction between the two fields $10_M$ combined into
 a $\overline{45}$ representation of {\it SU}(5) and the fields $5_H$
 and $24_H$, combined into a $45$.
The matrix $C_S^{10}$, which is symmetric because $C_{0,0,0,0,1}^{10}$
 (like $Y^{10}$) is also a symmetric matrix, corresponds to the coupling
 for the irreducible interaction in which the two fields $10_M$ and the
 fields $5_H$ and $24_H$ combine, respectively, into a $\bar{5}$ and a
 $5$ representation of {\it SU}(5).

As for the dimension-six NROs, $Op^{5}\vert_6$ can be easily read off
 from Eq.~(\ref{eq:NRO5}); $Op^{10}\vert_6$ is given in
 Appendix~\ref{sec:emb-GROUPfact}.

Neglecting the dynamical part of the field $24_H$, after the
 substitution $24_H^{24}\to v_{24}$, $Op^{5}$ and $Op^{10}$ decompose
 into effective renormalizable operators.
In particular, in the case of $Op^{5}$, we obtain:\footnote{At the 
  GUT threshold, the MSSM coupling $Y_E$ has to be identified with 
  $(\mbox{\boldmath{$Y$}}^{5}_{E\,})^T$.}
\begin{equation}
  -D^c \mbox{\boldmath{$Y$}}^{5}_{D\,} Q H_d,      \quad\,
  -E^c(\mbox{\boldmath{$Y$}}^{5}_{E\,})^T L H_d,   \quad\,
  -D^c \mbox{\boldmath{$Y$}}^{5}_{DU} U^c H_D^C,   \quad\,
  -L\, \mbox{\boldmath{$Y$}}^{5}_{LQ} Q H_D^C , 
\label{eq:5effSMdecomp}
\end{equation}
 with couplings $\mbox{\boldmath{$Y$}}^{5}_i$
 $(i={\scriptstyle D,E,DU,LQ})$:
\begin{equation}
  \mbox{\boldmath{$Y$}}^{5}_i                           \ = \ 
  \sum_{n+m=0}^k \!
  \mbox{\boldmath{$Y$}}_i^{5}\vert_{n+m}                \ = \
  \sum_{n+m=0}^k 
   C_{n,m}^{5} \, s^{(n+m)}  
   \left(\left(I_{\bar{5}_M}\right)_i\right)^n
   \left(\left(I_{\bar{5}_H}\right)_i\right)^m,  
\label{eq:effY5def}
\end{equation}
 where $(I_{\bar{5}_M})_i$ and $(I_{\bar{5}_H})_i$ collect the
 hypercharge values:
\begin{eqnarray}
  \left(I_{\bar{5}_M}\right)_i    & = &
  \displaystyle{\sqrt{\frac{6}{5}}}
 \left\{
 +\displaystyle{\frac{1}{3}},-\frac{1}{2},+\frac{1}{3},
               -\frac{1}{2}\right\} ,  
\nonumber\\
  \left(I_{\bar{5}_H}\right)_i    & = &
  \displaystyle{\sqrt{\frac{6}{5}}}
 \left\{
 -\displaystyle{\frac{1}{2}},-\frac{1}{2},+\frac{1}{3},
               +\frac{1}{3}\right\}.     
\qquad (i={\scriptstyle D,E,DU,LQ})   
 \label{eq:effY5hypercharges}
\end{eqnarray}
In these definitions, a boldface type is used to distinguish the
 effective Yukawa couplings from the usual ones, and all boldfaced
 quantities appearing hereafter are assumed to incorporate in some way
 the effect of NROs in which the substitution $24_H^{24}\to v_{24}$ is
 made. 
Indices are attached to these effective couplings consisting in the
 sequence of flavour fields in the effective renormalizable operators
 in which these couplings appear.
We have used, however,
 $\mbox{\boldmath{$Y$}}_D^5$ and
 $\mbox{\boldmath{$Y$}}_E^{5\,T}$, and not 
 $\mbox{\boldmath{$Y$}}_{DQ}^5$ and
 $\mbox{\boldmath{$Y$}}_{LE}^{5\,T}$, for the couplings that exist also
 in the MSSM, such as $Y_D$ and $Y_E$.
In a similar way, 
 $\mbox{\boldmath{$Y$}}_U^{10}$ will be used later instead of
 $\mbox{\boldmath{$Y$}}_{UQ}^{10}$.

Specifically, up to ${\cal O}(s^2)$, the couplings
 $\mbox{\boldmath{$Y$}}^{5}_i$ are
\begin{eqnarray}
 \mbox{\boldmath{$Y$}}^{5}_D \    &=&  Y^5 +
 \displaystyle{\sqrt{\frac{6}{5}}}
 \left(\displaystyle{\frac{1}{3}} C_{1,0}^{5} 
      -\displaystyle{\frac{1}{2}} C_{0,1}^{5}
 \right) s + {\cal O}(s^2),
\nonumber\\
 \mbox{\boldmath{$Y$}}^{5}_E \    &=&  Y^5 -
 \displaystyle{\sqrt{\frac{6}{5}}}
 \left(\displaystyle{\frac{1}{2}} C_{1,0}^{5} 
      +\displaystyle{\frac{1}{2}} C_{0,1}^{5}
 \right) s + {\cal O}(s^2),
\nonumber\\
 \mbox{\boldmath{$Y$}}^{5}_{DU}   &=&  Y^5 +
 \displaystyle{\sqrt{\frac{6}{5}}}
 \left(\displaystyle{\frac{1}{3}} C_{1,0}^{5} 
      +\displaystyle{\frac{1}{3}} C_{0,1}^{5}
 \right) s + {\cal O}(s^2),
\nonumber\\
 \mbox{\boldmath{$Y$}}^{5}_{LQ}   &=&  Y^5 -
 \displaystyle{\sqrt{\frac{6}{5}}}
 \left(\displaystyle{\frac{1}{2}} C_{1,0}^{5} 
      -\displaystyle{\frac{1}{3}} C_{0,1}^{5}
 \right) s + {\cal O}(s^2),
\label{eq:expandEFFYUK5}
\end{eqnarray}
 from which, being 
 $\mbox{\boldmath{$Y$}}^{5}_D-\mbox{\boldmath{$Y$}}^{5}_E =\sqrt{(5/6)}
  C_{1,0}^{5} s$, it becomes clear that only one NRO of dimension five,
 that with coefficient $C_{1,0}^{5}$, is sufficient to correct the wrong
 fermion spectrum of the minimal model.
Once this is introduced, however, it is difficult to explain why the
 other NRO of dimension five, as well as NROs of higher dimensions in
 $Op^{5}$, should be absent.
We keep our formalism as general as possible and postpone to later the
 issue of the level of precision in $s$ that it is reasonable to
 require.

It is easy to see from this expansion that, for each element $(h,k)$, 
 the effective couplings $\mbox{\boldmath{$Y$}}^{5}_i$ satisfy the
 conditions:
\begin{eqnarray}
\left\vert \left(
      \mbox{\boldmath{$Y$}}^{5}_i  -\mbox{\boldmath{$Y$}}^{5}_{i'}
           \right)_{(h,k)} \right\vert     &\leq&
 {\cal O}(s), \quad  (i,i^\prime={\scriptstyle D,E,DU,LQ})
\nonumber\\
\left\vert \left(
      \mbox{\boldmath{$Y$}}^{5}_{D} - \mbox{\boldmath{$Y$}}^{5}_{E} 
     +\mbox{\boldmath{$Y$}}^{5}_{LQ}- \mbox{\boldmath{$Y$}}^{5}_{DU}
           \right)_{(h,k)} \right\vert     &\leq&
 {\cal O}(s^2).
\label{eq:5Constraint}
\end{eqnarray}
This last combination vanishes exactly if only NROs of dimension five
 are introduced.
We refer to these relations as the ${\cal O}(s)$- and
 ${\cal O}(s^2)$-constraints.
They help fixing, respectively, the ${\cal O}(1)$ and ${\cal O}(s)$
 contributions to the effective couplings $\mbox{\boldmath{$Y$}}^{5}_i$,
 proving themselves a very useful tool in selecting the appropriate
 boundary conditions for the various $\mbox{\boldmath{$Y$}}^{5}_i$.
Note that there are no ${\cal O}(s^3)$-constraints.
These constraints, like all the others listed in this section, as well
 as those in \S\S\ref{sec:trilSOFTsecNROS} 
 and \ref{sec:NROsBC-TrilSOFT} for effective trilinear couplings, are
 derived in {\it SU}(5)-symmetric bases.

The effective renormalizable operators induced by $Op^{10}$, with
 effective couplings $\mbox{\boldmath{$Y$}}^{10}_j$ labelled by the
 index $j={\scriptstyle \{U,UE,QQ\}}$, are
\begin{equation}
 U^c \mbox{\boldmath{$Y$}}^{10}_U    Q H_u,     \quad \,
 U^c \mbox{\boldmath{$Y$}}^{10}_{UE} E^c H_U^C, \quad \,
   \frac{1}{2}
  Q \,\mbox{\boldmath{$Y$}}^{10}_{QQ} Q H_U^C.
\label{eq:10effSMdecomp}
\end{equation}
Because of its complexity, we refrain in this case from giving general
 expressions of the effective couplings $\mbox{\boldmath{$Y$}}^{10}_j$
 in terms of the coefficients appearing in $Op^{10}$.
We give explicitly only the expansion up to ${\cal O}(s^2)$, the
 counterpart of Eq.~(\ref{eq:expandEFFYUK5}) for the effective couplings
 $\mbox{\boldmath{$Y$}}^{5}_i$:
\begin{eqnarray}
 \mbox{\boldmath{$Y$}}^{10}_U  \, &=&  Y^{10} -
  \displaystyle{\sqrt{\frac{6}{5}}}
 \left(
  \displaystyle{\frac{1}{2}} C_S^{10}
 -\displaystyle{\frac{5}{6}} C_A^{10}
 \right) s  + {\cal O}(s^2),
\nonumber\\
 \mbox{\boldmath{$Y$}}^{10}_{QQ}   &=&  Y^{10} +
  \displaystyle{\sqrt{\frac{6}{5}}}
 \left(
  \displaystyle{\frac{1}{3}} C_S^{10}
 \right) s  + {\cal O}(s^2),
\nonumber\\ 
 \mbox{\boldmath{$Y$}}^{10}_{UE}   &=&  Y^{10} +
  \displaystyle{\sqrt{\frac{6}{5}}}
 \left(
  \displaystyle{\frac{1}{3}} C_S^{10} + 
  \displaystyle{\frac{5}{3}} C_A^{10}
 \right) s  + {\cal O}(s^2).
\label{eq:expandEFFYUK10}
\end{eqnarray}
Note that $\mbox{\boldmath{$Y$}}^{10}_{QQ}$, as expected, is symmetric,
 whereas $\mbox{\boldmath{$Y$}}^{10}_{U}$ and
 $\mbox{\boldmath{$Y$}}^{10}_{UE}$ are not.

The corresponding ${\cal O}(s)$- and ${\cal O}(s^2)$-constraints for
 these couplings are
\begin{equation}
\begin{array}{cll}
\left\vert \bigl(
       \mbox{\boldmath{$Y$}}^{10}_j -
       \mbox{\boldmath{$Y$}}^{10}_{j'}
           \bigr)_{(h,k)}  \right\vert                         &\leq& 
  {\cal O}(s), \quad (j,j^\prime={\scriptstyle U,UE,QQ})
\\[1.001ex]
\left\vert  \left(
       \mbox{\boldmath{$Y$}}^{10}_{QQ} -
 \bigl(\mbox{\boldmath{$Y$}}^{10}_{UE} \bigr)^S
            \right)_{(h,k)}\right\vert                         &\leq&
  {\cal O}(s^2), 
\\[1.001ex]
\left\vert  \left(
2\bigl(\mbox{\boldmath{$Y$}}^{10}_{U}\bigr)^A -
 \bigl(\mbox{\boldmath{$Y$}}^{10}_{UE}\bigr)^A
            \right)_{(h,k)}\right\vert                         &\leq&
  {\cal O}(s^2),
\label{eq:10Constraint}
\end{array}      
\end{equation}
 where, as in the case of the effective couplings
 $\mbox{\boldmath{$Y$}}^{5}_i$, the left-hand side of the two
 ${\cal O}(s^2)$-constraints vanishes identically if NROs of dimension
 six and higher in $Op^{10}$ are vanishing.
Also in this case, the ${\cal O}(s)$- and  ${\cal O}(s^2)$-constraints
 fix or help fix the contributions of ${\cal O}(1)$ and ${\cal O}(s)$
 to the effective couplings $\mbox{\boldmath{$Y$}}^{10}_j$, for which
 there are no ${\cal O}(s^3)$-constraints.

Nonrenormalizable operators in the seesaw Yukawa sector are less
 relevant than those discussed until now, which can give ${\cal O}(1)$
 corrections to the Yukawa couplings of first- and second-generation
 fermions. 
Since we expect the renormalizable Yukawa couplings in the seesaw
 sector to be rather large, NROs in this sector play subleading roles, 
 unless their coefficients are considerably larger than ${\cal O}(1)$.
That remains true, of course, as long as $s$ is $\sim 10^{-2}$ and the 
 seesaw scale is $\sim 10^{14}\,$GeV.
(The situation could, however, change if the values of $s$ and/or of
 $M_{\rm ssw}$ were modified even by a not too large amount.) 
Nevertheless, we consider NROs also in the seesaw sectors, in order to
 have a description of all NROs relevant to flavour violation at the
 same order in $s$.

By inserting powers of the field $24_H$ in the first term of
 $W_{\rm RHN}$ in Eq.~(\ref{eq:SU5sswSUPERP}), we obtain the class of
 operators $Op^{N_{\rm I}}$ for the seesaw of type~I:
\begin{equation}
  Op^{N_{\rm I}}  =  -\sum_{n=0}^k \ 
  N^c \,C^{(N_{\rm I})}_n \, \bar{5}_M
   \left(\!\displaystyle{\frac{24_H^T}{M_{\rm cut}}}\!\right)^{\!\!n}
           \! 5_H,
\label{eq:NROssw}
\end{equation}
 with $C^{(N_{\rm I})}_{0}\equiv Y_N^{\rm I}$.
We do not include NROs for the mass term $(1/2) N^c M_N N^c$.
After the substitution $24_H^{24}\to v_{24}$, $Op^{N_{\rm I}}$
 decomposes in the effective renormalizable operators:
\begin{equation}
   N^c \mbox{\boldmath{$Y$}}_{N}^{\rm I}  L H_u,  \quad\,
 - N^c \mbox{\boldmath{$Y$}}_{ND}^{\rm I} D^c H_U^C ,
\label{eq:SSWeffSMdecomp}
\end{equation}
 with effective couplings defined as
\begin{equation}
 \mbox{\boldmath{$Y$}}_{h}^{\rm I}
 \ =\  \sum_{n=0}^k \mbox{\boldmath{$Y$}}_h^{\rm I}\left\vert_n \right.
 \ =\  \sum_{n=0}^k C_{n}^{N_{\rm I}} \, s^{n}  
   \left(\left(-I_{5_H}\right)_h\right)^n, 
 \quad \quad (h={\scriptstyle N,ND}) 
\label{eq:effYsswIdef}
\end{equation}
 in which the hypercharge factors in $(I_{5_H})_h$ are 
\begin{equation}
 \quad  \left(-I_{5_H}\right)_h =
 \sqrt{\frac{6}{5}}
 \left\{-\frac12,\frac13\right\}. \qquad  (h={\scriptstyle N,ND}) 
\label{eq:effYNhypercharges}
\end{equation}
Thus, up to ${\cal O}(s^2)$, $\mbox{\boldmath{$Y$}}_{N}^{\rm I}$ and
 $\mbox{\boldmath{$Y$}}_{ND}^{\rm I}$ have the very simple expression:
\begin{eqnarray}
 \mbox{\boldmath{$Y$}}_{N}^{\rm I}\  & = &  Y_{N}^{\rm I} - 
  \displaystyle{\sqrt{\frac{6}{5}}}\,
  \displaystyle{\frac{1}{2}} C_1^{N_{\rm I}} \,s +{\cal O}(s^2) ,
\nonumber\\
 \mbox{\boldmath{$Y$}}_{ND}^{\rm I}  & = &  Y_{N}^{\rm I} + 
  \displaystyle{\sqrt{\frac{6}{5}}}\,
  \displaystyle{\frac{1}{3}} C_1^{N_{\rm I}} \,s +{\cal O}(s^2) , 
\label{eq:expandEFFYUKNI}
\end{eqnarray}
 and must satisfy only one constraint:
\begin{equation}
\left\vert  \left(
 \mbox{\boldmath{$Y$}}^{\rm I}_{N}  - 
 \mbox{\boldmath{$Y$}}^{\rm I}_{ND}
            \right)_{(h,k)}\right\vert   \ \, \leq \ \,
  {\cal O}(s).
\label{eq:sswIConstraint}
\end{equation}
There are no ${\cal O}(s^2)$- and ${\cal O}(s^3)$-constraints on the 
 effective couplings $\mbox{\boldmath{$Y$}}_{h}^{\rm I}$, while 
 ${\cal O}(s^2)$-constraints exist in the case of the other two seesaw
 types.

The treatment for the seesaw of types~II and~III is not explicitly 
 reported here, as it is completely straightforward. 
As in type~I, only the first interaction term in $W_{\rm 15H}$ and the
 first in $W_{\rm 24M}$ are generalized to the classes of operators 
 $Op^{N_{\rm II}}$ and $Op^{N_{\rm III}}$.
We neglect NROs for the other two terms in $W_{\rm 24M}$ because they
 involve only heavy fields.
The same is true for the terms $M_{15} 15_H \bar{15}_H$ and
 $\lambda_{15} 15_H 24_H \bar{15}_H$ in $W_{\rm 15H}$.
We actually neglect NROs for all terms of $W_{\rm 15H}$, except for the
 first one, as they do not contain flavour fields.
The small modifications that NROs would induce in the couplings of the
 terms involving the fields $5_H$ and $\bar{5}_H$ in $W_{\rm 15H}$ are
 of no consequence to our discussion of sFVs~(see also
 \S\ref{sec:HIGGsNROS}).

The picture of effective Yukawa couplings is particularly appealing
 because, for a given value $k$ taken as an upper limit of the sums in
 $Op^{5}$, $Op^{10}$, and $Op^{N_{\rm I}}$, only few linear combinations
 of the many coefficients, $C_{n,m}^{5}$,
 $C_{n_1,n_2,n_3,n_4,n_5}^{10}$, and $C_n^{N_{\rm I}}$, in these sums
 are physically relevant.
These are the four, three, and two combinations defining
 $\mbox{\boldmath{$Y$}}^{5}_i$, $\mbox{\boldmath{$Y$}}^{10}_j$, and
 $\mbox{\boldmath{$Y$}}^{\rm I}_h$, respectively.
{}For example, for the effective couplings
 $\mbox{\boldmath{$Y$}}^{5}_i$, the number of these combinations, four,
 is smaller than the number obtained summing the number of couplings of
 the minimal model, one, i.e., $Y^5$, plus the number of
 coefficients of the relevant NROs up to dimension $4+k$.
This is true except in the cases $k=0,1$.
{}For $k=0$, there is only the original coupling of the renormalizable
 operator, $Y^5$, versus the four effective couplings
 $\mbox{\boldmath{$Y$}}^{5}_i$, giving rise to three
 ${\cal O}(s)$-constraints.
{}Similarly, for $k=1$, there are three original couplings,
 $Y^5$, $C_{1,0}^{5}$,  $C_{0,1}^{5}$, versus four effective couplings,
 inducing one ${\cal O}(s^2)$-constraint.
Clearly, there are no ${\cal O}(s^3)$-constraints, as the number of
 original couplings exceeds already the number of effective couplings.
The same reasoning holds for the other two types of effective couplings
 $\mbox{\boldmath{$Y$}}^{10}_j$ and
 $\mbox{\boldmath{$Y$}}^{\rm I}_h$.

The possibility of incorporating the effects of NROs in the classes 
 $Op^{5}$, $Op^{10}$, and $Op^{N_{\rm I}}$ (as well as $Op^{N_{\rm II}}$
 and $Op^{N_{\rm III}}$) of dimensions large enough could be particularly
 important in models with $s>10^{-2}$.

The main question here is whether it is possible to promote this picture
 to the quantum level. 
This issue will be discussed in \S\ref{sec:ValidityEffPic}.
{}For the remainder of this section, we shall continue the survey of the
 different NROs in the SUSY-conserving part of our model, irrespective
 of how they should be treated at the quantum level.

\subsection{NROs in the superpotential Higgs sector}
\label{sec:HIGGsNROS}
%
In addition to the NROs listed above, there are also NROs in the Higgs
 sector:
\begin{eqnarray}
& Op^{5_H}  & =  M_{\rm cut} \sum_{n=0}^{k+1}
 C_{n}^{5_H} \,{5}_H 
   \left(\!\frac{24_H}{M_{\rm cut}}\!\right)^{\!\!n}  \! \bar5_H ,
\nonumber\\
& Op^{24_H} & =  M_{\rm cut}^3 \sum_{\sum n_j = 2}^{k+3}
 C_{n_1,\cdots,n_j,\cdots}^{24_H}
  \prod_j \frac{1}{n_j!}
  {\rm Tr} \left(\!\frac{24_H}{M_{\rm cut}}\!\right)^{\!\!n_j},
\label{eq:NROHiggs}
\end{eqnarray}
 where it is $C_{0}^{5_H}= M_5/M_{\rm cut}$, $C_{1}^{5_H} =\lambda_5$,
 in the first class of operators, and $C_{2}^{24_H}= M_{24}/M_{\rm cut}$,
 $C_{3}^{24_H} =\lambda_{24}$, in the second one. 
Effective couplings can be defined also in this case.
{}For example, there are two effective couplings 
 $(\mbox{\boldmath{$M$}}_5)_p$
 ($p ={\scriptstyle H_U^C H_D^C, H_u H_d}$) 
 for the two operators in Eq.~(\ref{eq:mu2mu3noSOFTredef}) in
 Appendix~\ref{sec:emb-GROUPfact}, and six couplings
 $(\mbox{\boldmath{$\lambda$}}_5)_q$
 ($q={\scriptstyle H_U^C G_H H_D^C, \cdots, 
                   H_u B_H H_d}$), 
 for the six operators in Eq.~(\ref{eq:524bar5DECOMP}).
We do not give here their expressions in terms of the original couplings
 $C_{n}^{5_H}$.
These can be easily obtained as follows:
\begin{eqnarray}
 \left(\mbox{\boldmath{$M$}}_5\right)_{H_u H_d} \  
&=& 
 \frac{\partial \, W^{\rm nrMSSU(5)}_H}
      {\partial H_u \partial H_d}
\Big \vert_{24_H^{24}\to v_{24}}
\quad = \ 
 \frac{\partial \, Op^{5_H} }
      {\partial H_u \partial H_d}
\Big \vert_{24_H^{24}\to v_{24}},
\nonumber \\[1.2ex]
  \left(\mbox{\boldmath{$\lambda$}}_5\right)_{H_u B_H H_d} 
& = & 
  \frac{\partial \, W^{\rm nrMSSU(5)}_H}
       {\partial H_u \partial B_H \partial H_d}
 \Big \vert_{24_H^{24}\to v_{24}}
\ = \ 
  \frac{\partial \, Op^{5_H} }
       {\partial H_u \partial B_H \partial H_d}
 \Big \vert_{24_H^{24}\to v_{24}},
\label{eq:EffCoupHIGGS}
\end{eqnarray}
 where $ W^{\rm nrMSSU(5)}_H$ is the Higgs sector superpotential
 completed with NROs.

In general, NROs in the Higgs sector induce small shifts in the vacuum
 condition discussed in \S\ref{sec:HiggsSECandVACUA}, unless the
 trilinear coupling $\lambda_{24}$ is small.
In this case, NROs can drastically modify the vacuum 
 structure~\cite{GORAN}, even sizably increasing the value of $v_{24}$.
In turn, the factor $s$ also increases, and the effect of NROs on
 low-energy physics can be significantly enhanced.  
We keep away from such a possibility by assuming $\lambda_{24}$ to be
 of ${\cal O}(1)$.
Thus, the effect of $ Op^{5_H}$ and $Op^{24_H}$ is that of producing
 small shifts in massive and massless parameters in the Higgs sector by
 ratios of ${\cal O}(s)$, and the vacuum structure of the minimal model
 can be regarded as a good approximation also when NROs in the Higgs
 sector are nonvanishing. 
We neglect these effects, which introduce only small flavour-blind
 shifts in sFVs induced by Yukawa couplings, and in the same spirit, we  
 neglect also the effect of NROs in the purely Higgs-boson sector of
 the seesaw of type~II~(see also \S\ref{sec:ValidityEffPic}).

Before closing this section, we would like to emphasize that the
 evolution equations for the VEVs in Eq.~(\ref{eq:VEVS24evol}) remain
 valid also when NROs in the Higgs sector are nonvanishing. 
They obviously apply in this case to the modified VEVs and must be 
 written in terms of the modified $\gamma_{24}$ and
 $\tilde{\gamma}_{24}$.
They remain valid, regardless of the value of $\lambda_{24}$, i.e., of 
 whether the VEVs modifications are small or large. 
As shown in Appendix~\ref{sec:New}, these RGEs hold quite
 model-independently.
Thus, the approximation made here plays no role on whether the picture
 of effective couplings can be maintained at the quantum level.

\subsection{NROs in the SUSY-conserving part of the K\"ahler potential}
\label{sec:KalerPNROS}
%
One of the lamented problems when dealing with NROs is that they spoil
 the usual requirement of minimality of the K\"ahler potential. 
Indeed, NROs such as 
\begin{eqnarray}
 Op^{{\cal K}_1}_{\bar{5}_M}  & = & 
 \sum_{m=1}^k 
  \bar5_M \,C_{0,m}^{{\cal K},\bar{5}_M}
  \left(\!\frac{24_H^\ast}{M_{\rm cut}}\!\right)^{\!\!m} \!\bar5_M^* ,
\label{eq:KpotNROsREAB}
\\
 Op^{{\cal K}_2}_{\bar{5}_M}  & = & 
 \hspace*{-0.4truecm}
 \sum_{
\scriptsize{\begin{array}{l}n+m=2\\[-1.5ex]n,m\ne0\end{array}}}^k\!\! 
 \hspace*{-0.2truecm}
  \bar5_M \,C_{n,m}^{{\cal K},\bar{5}_M}
  \left(\!\frac{24_H^T}{M_{\rm cut}}\!\right)^{\!\!n} \!\!
  \left(\!\frac{24_H^\ast}{M_{\rm cut}}\!\right)^{\!\!m}\! \bar5_M^* ,
\label{eq:KpotNROsNONREAB}
\end{eqnarray}
 their Hermitian conjugates,  $(Op^{{\cal K}_1}_{\bar{5}_M})^\ast$, 
 $(Op^{{\cal K}_2}_{\bar{5}_M})^\ast$, as well as NROs
 as\footnote{{}For the coupling $f_2$, we omit both {\it SU}(5) and
  flavour indices.}
\begin{equation}
 (1/M_{\rm cut})5_H^* 10_M f_1 \bar5_M,  \qquad
 (f_2/M_{\rm cut}^2) \bar5_M^* 10_M^* 10_M \bar5_M,
\label{eq:kpotNROextra}
\end{equation}
 do precisely that.
In addition, even if accidentally vanishing at the tree level, they are 
 induced at the quantum level by interactions in the superpotential.

In a basis with canonical kinetic terms at the renormalizable level, in
 which the renormalizable operator $\bar5_M \bar5_M^*$ has a coupling
 equal to the unit matrix, the coefficients of NROs in
  $Op^{{\cal K}_1,\bar{5}_M}$, $Op^{{\cal K}_2,\bar{5}_M}$, and $f_1$, 
  $f_2$ are, in general, flavour-dependent. 
Indeed, by postulating a nontrivial flavour structure of the
 coefficients in $Op^{5}$, $Op^{10}$, etc. in the superpotential, we
 have already assumed that the unknown dynamics generating NROs depends
 on flavour.

Nonrenormalizable operators of the first type, with only one antichiral
 or only one chiral superfield, can be reabsorbed by field
 redefinitions: 
\begin{equation}
 \bar{5}_M  \ \to \ 
 \bar{5}_M - \sum_{n=0}^k  \left(C_{0,m}^{{\cal K},\bar{5}_M}\right)^\ast
  \left(\!\frac{24_H}{M_{\rm cut}}\!\right)^{\!\!m}\!\bar5_M+\cdots.
\label{eq:fieldREDEF}
\end{equation}
Once this is done, however, it is impossible to reabsorb those of the
 second type through supersymmetric field redefinitions, which must
 preserve chirality.   
Thus, NROs of the second type, in general, cannot be avoided. 
The deviation from minimality that they produce is, however, at most of
 ${\cal O}(s^2)$.

We do not try to control in this framework the loop effects of such
 NROs, of ${\cal O}(s^2\times s_{\rm loop})$, where $s_{\rm loop}$ is
 the usual loop-suppression factor: 
\begin{equation}
  s_{\rm loop} = \frac{1}{16 \pi^2},
\label{eq:suppressFACTORS}
\end{equation}
 accidentally of the same order of $s$, but, possibly, smaller in models
 in which $M_{\rm GUT}$ and $M_{\rm cut}$ differ by less than the usual two
 orders of magnitude. 
This is because, as will be discussed in
 \S\ref{sec:EffPicLOOPlevel}, the picture of effective couplings at
 the quantum level can be retained if we restrict ourselves to an
 accuracy of ${\cal O}(s\times s_{\rm loop})$ for the calculation of
 sFVs.
Thus, NROs of dimension six can contribute to sFVs, but only at the tree
 level, whereas those of dimension five contribute also at the one-loop
 level.

Among the surviving dimension-six NROs, with more than one chiral and
 more than one antichiral superfield, those consisting only of light
 fields are completely irrelevant.
The operator $(f_2/M_{\rm cut}^2) \bar5_M^* 10_M^* 10_M \bar5_M$, in
 Eq.~(\ref{eq:kpotNROextra}), for example, violates baryon number.
Its contribution to the proton-decay rate, however, can be safely 
 neglected since it is much smaller than those from the dimension-five
 NRO in $Op^{\rm PD}$~(see Eq.~(\ref{eq:supNROforPD})) and the 
 effective dimension-six operators induced by the exchange of 
 {\it SU}(5) gauge bosons, which have a suppression factor 
 $1/M_{\rm GUT}^2$ instead of $1/M_{\rm cut}^2$.

Other dimension-six NROs containing a pair of one chiral and one
 antichiral dynamical field $24_H$  can only affect GUT-scale physics, 
 at least,{{{}}} at the tree level, and are, therefore, irrelevant to our
 discussion.
The same NROs in which the two fields $24_H$ are replaced by the
 scalar~VEV $v_{24}$ can certainly be removed by {\it SU}(5)-breaking
 field redefinitions. 
Whether this is true also in the case in which one or both fields
 $24_H$ acquire the~VEV $F_{24}$ will be discussed in
 \S\ref{sec:UNIVERSALITY}~(see in particular
 Eqs.~(\ref{eq:dim6reduced})-(\ref{eq:KpotALMOSTmin})).

\section{Parameter counting for effective Yukawa couplings}
\label{sec:NROsBC-YUKAWA} 
%
The high-scale boundary values of the {\it SU}(5)-breaking effective 
 Yukawa couplings must be selected in a large space of free parameters.
{}For a correct counting of the number of those that are physical, it is
 convenient to express the matrices of effective Yukawa couplings
 in terms of their diagonalized forms and rotation matrices.
Like the CKM matrix in the SM, only the matrices of mismatch between
 different rotation matrices are of physical relevance. 
Possible parametrizations for them will be discussed and the role that
 they play for sFVs will be highlighted.

{}For simplicity, we neglect NROs in the seesaw sector, as their effect,
 compared with the renormalizable ones with couplings of ${\cal O}(1)$, is
 expected to be much smaller than the effect they have in the 
 $\bar{5}_M$ and $10_M$ sectors. 
The generalization to the case in which they are nonvanishing is 
 straightforward and does not add any insight to this discussion. 
Moreover, when considering seesaw couplings, we concentrate on the
 seesaw of type~I.
Those of types~II and~III can be dealt in a similar way.

We start by reviewing in more detail the case of the MSSM(5) model,
 already outlined in \S\ref{sec:MatterSEC} and
 Appendix~\ref{sec:emb-SMfields}.

\subsection{Limit $s\to 0$}
\label{sec:NROsBC-YUKAWAmin}
%
In the case of vanishing NROs, suitable rotations of the fields $10_M$ 
 and $\bar{5}_M$ allow to reduce the set of Yukawa couplings $Y^5$, 
 $Y^{10}$, and $Y_N^{\rm I}$ to their physical parameters:
\begin{eqnarray}
  Y^5  \     &\to &  \widehat{Y}^5,                   
\nonumber\\
  Y^{10}      &\to & K_{\rm CKM}^T \,\widehat{Y}^{10} P_{10} K_{\rm CKM}, 
\nonumber\\ 
  Y_N^{\rm I} \ &\to &  e^{i\phi_{\rm I}} Y_\nu^{\rm I} P_{\rm I}.
\label{eq:reducedYUKAWA}
\end{eqnarray}
In addition to the three real and positive elements in each of the
 matrices $\widehat{Y}^5$ and $\widehat{Y}^{10}$, the four independent
 parameters of $K_{\rm CKM}$, and the fifteen elements in
 $Y_{\nu}^{\rm I}$, together with the three eigenvalues $\hat{M}_{N}$, 
 all present also in the MSSM with a seesaw of type~I, there are the
 five physical phases, $P_{10}$, $P_{\rm I}$ and $\phi_{\rm I}$.

Once we decompose in their SM representations the two Yukawa operators
 of the minimal model, and that for the seesaw of type~I, and we denote
 as $Y^5_i$~($i={\scriptstyle D,E,DU,LQ}$), $Y^{10}_j$
 ($j={\scriptstyle U,UEQQ}$), and
 $Y^{\rm I}_h$~($h={\scriptstyle N,ND}$) the couplings of the resulting
 operators (they are the couplings of
 Eqs.~(\ref{eq:5effSMdecomp}),~(\ref{eq:10effSMdecomp}), 
 and~(\ref{eq:effYsswIdef}) in the limit of vanishing NROs), these
 additional phases can be moved to the Yukawa operators involving only
 colored Higgs triplets. 
Rotations that break {\it SU}(5) are obviously needed to achieve
 this~(see Eqs.~(\ref{eq:SMfieldsID}) and~(\ref{eq:YUKAWAsswRotat})).
We move then to a basis in which the Yukawa couplings for the
 $\bar{5}_M$ and $10_M$ sectors can be parametrized as
\begin{eqnarray}
 Y_D^5                         \to&
  \widehat{Y}^5,                &\hspace*{1.5truecm}  
 Y_{DU}^5   \ \to \ \widehat{Y}^5 K_{\rm CKM}^\dagger P_{10}^\dagger,   
\nonumber \\[1.1ex]     
 Y_E^5                         \to&
 \widehat{Y}^5,                &\hspace*{1.5truecm} 
 Y_{LQ}^5   \ \to\  e^{-i\phi_{\rm I}}\,\widehat{Y}^5P_{\rm I}^\dagger, 
\nonumber \\[1.1ex]     
 Y_U^{10}                       \to&
 \widehat{Y}^{10}K_{\rm CKM},     &\hspace*{1.5truecm}
 Y_{UE}^{10} \ \to \ e^{i\phi_{\rm I}}\, 
                   \widehat{Y}^{10} K_{\rm CKM} P_{\rm I},
\nonumber \\[1.1ex]     
                           &   &\hspace*{1.5truecm}
 Y_{QQ}^{10} \ \to\  K_{\rm CKM}^T \widehat{Y}^{10} P_{10} K_{\rm CKM},  
\label{eq:YUKminSU5br}
\end{eqnarray}
 and those in the seesaw sector of type~I (after diagonalization of the 
 Majorana mass $M_N$) as
\begin{eqnarray}
\, Y_N^{\rm I}               \to&
   Y_{\nu}^{\rm I}, \phantom{K_{\rm CKM}} \,
                           &\hspace*{1.5truecm}
\  Y_{ND}^{\rm I}          \ \to \ 
   e^{i \phi_{\rm I}} Y_{\nu}^{\rm I} P_{\rm I}. 
                                       \phantom{K_{\rm CKM}}\, 
\quad\quad
\label{eq:sswYUkminSU5br}
\end{eqnarray}

The determination of the boundary conditions at $M_{\rm cut}$ is, in this
 case, quite straightforward, if we ignore the charged lepton couplings. 
We obtain the low-energy values of $\widehat{Y}_D$, $\widehat{Y}_U$, and
 $K_{\rm CKM}$ from experiments, and we renormalize them upwards making 
 use of the RGEs listed in Appendix~\ref{sec:RGEsSU5}.

At $M_{\rm ssw}$, the seesaw degrees of freedom are switched on: RHNs, in
 the case illustrated here. Unfortunately, their coupling
 $Y_{\nu}^{\rm I}$ is not fully known~(see Eq.~(\ref{eq:CASASrel})).
The arbitrariness by which $Y_{\nu}^{\rm I}$ is plagued opens up new
 directions in the parameter space of the problem, which can be
 surveyed by scanning over values of the unknown quantities, at least, in
 principle.

At the GUT threshold, the superheavy fields are introduced.
We move to a basis in which the Yukawa couplings involving only light
 fields, which we call now $Y_E^5$, $Y_D^5$, $Y_U^{10}$, and
 $Y_N^{\rm I}$, are as on the left columns of
 Eqs.~(\ref{eq:YUKminSU5br}) and~(\ref{eq:sswYUkminSU5br}), with $Y_E^5$
 identified to $Y_D^5$, and $Y_{\nu}^{\rm I}$ a fifteen parameter matrix.
The couplings involving superheavy Higgs fields, $Y_{DU}^5$, $Y_{LQ}^5$,
 $Y_{UE}^{10}$, $Y_{QQ}^{10}$ and $Y_{ND}^{\rm I}$ are then as on the
 right columns of the same equations, in which we have fixed the
 arbitrary phases $P_{10}$, $P_{\rm I}$, and $\phi_{\rm I}$.
All fields must then be rotated from this basis to one 
 {\it SU}(5)-symmetric, that is, one in which they can all be 
 accommodated in the two {\it SU}(5) multiplets $\bar{5}_M$ and $10_M$, 
 with couplings for the Yukawa operators given by only three Yukawa 
 couplings, $Y^5$, $Y^{10}$, and $Y_N^{\rm I}$. 
The basis in which these couplings are those of
 Eq.~(\ref{eq:reducedYUKAWA}) is reached with rotations opposite to
 those in Eq.~(\ref{eq:SMfieldsID}), in which $\phi_l$ and $P_l$ are
 $\phi_{\rm I}$ and $P_{\rm I}$, respectively. 
The resulting Yukawa couplings can then be finally evolved upwards
 through RGEs, whose solutions at $M_{\rm cut}$ provide the high-scale
 boundary conditions for our problem.

\subsection{Nonvanishing NROs}
\label{sec:NROsBC-YUKAWAgen}
%
When NROs are present, the rotation matrices needed for the
 diagonalization of different effective Yukawa couplings induced by the
 same class of NROs are no longer common, and the elimination of
 unphysical degrees of freedom is more involved.

The effective Yukawa couplings $\mbox{\boldmath{$Y$}}^{5}_i$
 and $\mbox{\boldmath{$Y$}}^{10}_j$ are diagonalized as
\begin{eqnarray}
  \mbox{\boldmath{$Y$}}^{5}_i\  =&   
 (V_{5\,i}^\dagger)^T\,       \widehat{\mbox{\boldmath{$Y$}}}_i^{5}
  V_{10\,i}^\dagger,               & \quad (i={\scriptstyle D,E,DU,LQ})
\nonumber\\[1.1ex]
  \mbox{\boldmath{$Y$}}^{10}_j  =& \ 
(W_{10\,j}^{\prime\dagger})^T\,\widehat{\mbox{\boldmath{$Y$}}}_j^{10}
 W_{10\,j}^\dagger,                & \quad (j= {\scriptstyle U,UE,QQ})  
\label{eq:Y5effRotations}
\end{eqnarray}
 with the elements of 
 $\widehat{\mbox{\boldmath{$Y$}}}_{D}^{5}$,  
 $\widehat{\mbox{\boldmath{$Y$}}}_{E}^{5}$, and 
 $\widehat{\mbox{\boldmath{$Y$}}}_{U}^{10}$ giving rise to the correct
 fermion spectrum, i.e., ${\rm diag}(y_d,y_s,y_b)$,
 ${\rm diag}(y_e,y_\mu,y_\tau)$, and ${\rm diag}(y_u,y_c,y_t)$, 
 respectively.
We have left the labels $5$ and $10$ in the diagonalization matrices 
 $V_{5\,i}$, $V_{10\,j}$, and $W_{10\,j}$ as a reminder of the case
 without NROs.
The matrices $W_{10\,j}^\prime$ were introduced because the couplings
 $\mbox{\boldmath{$Y$}}^{10}_{U}$ and
 $\mbox{\boldmath{$Y$}}^{10}_{UE}$ are, in general, not symmetric.
{}For $j={\scriptstyle QQ}$, it is $W_{10\,QQ}^\prime = W_{10\,QQ}$.

Two of the rotation matrices in Eq.~(\ref{eq:Y5effRotations}) can be
 absorbed by {\it SU}(5)-symmetric field redefinitions of $10_M$ and
 $\bar5_M$ similar to those made in the case of vanishing NROs.
We choose to eliminate $V_{10\,D}^\dagger$ and $V_{5\,E}^\dagger$.
We also rotate away the phases $P_{10}^{(1)}$ and $P_{10}^{(2)}$ that
 appear in the parametrization of the product of diagonalization
 matrices from which emerges now the CKM matrix: 
 $U_{\rm CKM} = W_{10\,U}^\dagger V_{10\,D} = 
 P_{10}^{(1)} K_{\rm CKM} P_{10}^{(2)} e^{i \phi_{10}}$.
We obtain
\begin{eqnarray}
 \mbox{\boldmath{$Y$}}^{5}_i  \ \to&  (\Delta V_{5\,i}^\dagger)^T\, 
  \widehat{\mbox{\boldmath{$Y$}}}_i^{5}\,  \Delta V_{10\,i}^\dagger, 
                                  & \quad (i={\scriptstyle D,E,DU,LQ})
\nonumber\\[1.1ex]
 \mbox{\boldmath{$Y$}}^{10}_j    \to&  
  \left[(\Delta W_{10\,j}^{\prime\dagger})K_{\rm CKM}\right]^T\,
  \widehat{\mbox{\boldmath{$Y$}}}_j^{10} P_{10}
  \left[\Delta W_{10\,j}^\dagger K_{\rm CKM}\right],
                                  & \quad (j= {\scriptstyle U,UE,QQ})  
\label{eq:YeffRotations}
\end{eqnarray}
 where the ten matrices of diagonalization mismatch, besides
 $K_{\rm CKM}$, are
\begin{eqnarray}
\Delta V_{5\,i}^\dagger  \ \, = &\  P_{10}^{(2)\dagger}
                                 V_{5\,i}^\dagger V_{5\,E}
                                 P_{10}^{(2)},
                            & \quad\quad (i={\scriptstyle D,DU,LQ})
\nonumber\\[1.1ex]
\Delta V_{10\,i}^\dagger   \, = &\ P_{10}^{(2)}
                                V_{10\,i}^\dagger V_{10\,D} 
                                P_{10}^{(2)\dagger}\, ,
                            & \quad\quad (i={\scriptstyle E,DU,LQ})  
\nonumber\\[1.5ex]
\Delta W_{10\,j}^{\prime\dagger}= &\ P_{10}^{(1)\dagger}
                                W_{10\,j}^{\prime \dagger} W_{10\,U}
                                P_{10}^{(1)} ,
                            & \quad\quad (j={\scriptstyle U,UE})
\nonumber\\[1.1ex]
\Delta W_{10\,j}^{\dagger}    = &\ P_{10}^{(1)\dagger}
                                W_{10\,j}^{\dagger} W_{10\,U}
                                P_{10}^{(1)} .
                            & \quad\quad (j={\scriptstyle UE,QQ})  
\label{eq:ROTATidentif}
\end{eqnarray}
Note that 
 $\Delta V_{5\,E}^\dagger$,
 $\Delta V_{10\,D}^\dagger$, and
 $\Delta W_{10\,U}^\dagger$ are not included in this list, because they
 are trivially equal to the unit matrix. 
In addition, since $W^{\prime}_{10\,QQ}= W_{10\,QQ}$, it is
 $\Delta W^{\prime\dagger}_{10\,QQ}=\Delta W^{\dagger}_{10\,QQ}$.
In the limit of vanishing NROs, all these mismatch matrices reduce to
 the unit matrix.

All ten of them seem necessary in order to parametrize the effective
 Yukawa couplings $\mbox{\boldmath{$Y$}}_{i}^{5}$ and
 $\mbox{\boldmath{$Y$}}_{j}^{10}$, together with 
 $\widehat{\mbox{\boldmath{$Y$}}}_{D}^{5}$, 
 $\widehat{\mbox{\boldmath{$Y$}}}_{E}^{5}$, 
 $\widehat{\mbox{\boldmath{$Y$}}}_{U}^{10}$, and $K_{\rm CKM}$, 
 and the diagonal matrices 
 $\widehat{\mbox{\boldmath{$Y$}}}_{DU}^{5}$,  
 $\widehat{\mbox{\boldmath{$Y$}}}_{LQ}^{5}$,  
 $\widehat{\mbox{\boldmath{$Y$}}}_{UE}^{10}$, and 
 $\widehat{\mbox{\boldmath{$Y$}}}_{QQ}^{10}$, also unknown, 
 at least at ${\cal O}(s)$.
As shown in Appendix~\ref{sec:emb-SMfields}, in a basis in which the Majorana 
 mass $M_N$ is diagonal, the Yukawa couplings for the seesaw sector are
 parametrized as
\begin{equation}
 \mbox{\boldmath{$Y$}}_N^{\rm I}      \ = \
 \mbox{\boldmath{$Y$}}_{ND}^{\rm I} \ \to \ 
 e^{i\phi_{\rm I}}\, Y_\nu^{\rm I}\, P_{\rm I},
\label{eq:sswYeffRotSU5sym}
\end{equation}
 where $Y_\nu^{\rm I}$ is the coupling of the minimal case.

It is easy to see, however, that in a basis reached through
 {\it SU}(5)-breaking rotations, in which 
 $\mbox{\boldmath{$Y$}}_{D}^{5}$,
 $\mbox{\boldmath{$Y$}}_{E}^{5}$,
 $\mbox{\boldmath{$Y$}}_{U}^{10}$, and
 $ \mbox{\boldmath{$Y$}}_{N}^{\rm I}$ match the corresponding couplings
 of the MSSM with a seesaw sector of type~I:
\begin{eqnarray}
  \mbox{\boldmath{$Y$}}^{5}_{D}   &\to& 
   \widehat{\mbox{\boldmath{$Y$}}}_{D}^{5}, 
\nonumber\\
  \mbox{\boldmath{$Y$}}^{5}_{E}   &\to& 
   \widehat{\mbox{\boldmath{$Y$}}}_{E}^{5},
\nonumber\\
  \mbox{\boldmath{$Y$}}^{10}_{U}  &\to&
   \widehat{\mbox{\boldmath{$Y$}}}_{U}^{10} K_{\rm CKM} ,
\nonumber\\
 \mbox{\boldmath{$Y$}}_N^{\rm I}  &\to&  Y_\nu^{\rm I} ,
\label{eq:YeffRotSU5br}
\end{eqnarray}
 fewer matrices are needed to parametrize the remaining couplings:
\begin{eqnarray}
\mbox{\boldmath{$Y$}}^{5}_{DU}    & \to &  
\left[\Delta V_{5\,DU}^\dagger \Delta V_{5\,D} \right]^T 
\! \widehat{\mbox{\boldmath{$Y$}}}_{DU}^{5}
\left[\Delta V_{10\,DU}^\dagger K_{\rm CKM}^\dagger
      \Delta W_{10\,U}^{\prime}P_{10}^{\dagger}\right],
\nonumber\\
\mbox{\boldmath{$Y$}}^{5}_{LQ}    & \to & 
 e^{-i\phi_{\rm I}}
\left[\Delta V_{5\,LQ}^\dagger  P_{\rm I}^\dagger\right]^T
 \,\widehat{\mbox{\boldmath{$Y$}}}_{LQ}^{5}\,
\left[ \Delta V_{10\,LQ}^\dagger \right],
\nonumber\\
\mbox{\boldmath{$Y$}}^{10}_{UE}   & \to & 
 e^{i\phi_{\rm I}}
\left[\Delta W_{10\,UE}^{\prime\dagger} 
      \Delta W_{10\,U}^{\prime}\right]^T
\!\widehat{\mbox{\boldmath{$Y$}}}_{UE}^{10}
\left[\Delta W_{10\,UE}^\dagger K_{\rm CKM} \Delta V_{10\,E}
      P_{\rm I} \right],
\nonumber\\
\mbox{\boldmath{$Y$}}^{10}_{QQ}   & \to & 
\left[\Delta W_{10\,QQ}^\dagger  K_{\rm CKM}\right]^T
\! \widehat{\mbox{\boldmath{$Y$}}}_{QQ}^{10} P_{10}
\left[\Delta W_{10\,QQ}^\dagger  K_{\rm CKM} \right].
\label{eq:HCYeffRotSU5br}
\end{eqnarray}
Note that the mismatch matrix $\Delta V_{5\,D}$ was not eliminated,
 but only shifted to the seesaw sector:
\begin{equation}
\mbox{\boldmath{$Y$}}_{ND}^{\rm I} 
\ \to \
   e^{i\phi_{\rm I}} Y_\nu^{\rm I}\,
   P_{\rm I}\Delta V_{5\,D}
\label{eq:sswYeffRotSU5br}.
\end{equation}
It turns out to be directly responsible for breaking the correlation
 between the seesaw-induced sQFVs and sLFVs of the minimal model.
The shift described here is not peculiar of the seesaw of type~I, but it
 is common also to the seesaw of types~II and~III, which have at least
 one operator coupling the field $D^c$ to a seesaw-mediator field.

In this basis, eight matrices in addition to $K_{\rm CKM}$ are
 sufficient to parametrize all Yukawa couplings, compared with the ten
 needed in the basis of
 Eqs.~(\ref{eq:YeffRotations}) and~(\ref{eq:sswYeffRotSU5sym}).
The remaining two matrices are shifted from the Yukawa sector to the
 interactions of the fields $X$ and $\bar{X}$ in the gauge sector, 
 which do not affect the physics studied here.\footnote{They do affect
  the physics in which the gauge bosons $X$ and $\bar{X}$ play a role,
  as for example, proton decay, which can proceed through dimension-six 
  operators induced integrating out these gauge bosons. See remark at
  the end of \S\ref{sec:PDconstBC}.}
Since the RGEs have a covariant form under flavour rotations and their
 results do not depend on the basis chosen, the 
 dependence on these two matrices present when working in the previous
 basis (obtained without {\it SU}(5)-breaking rotations) will have to 
 drop out of the RGE results.

In this analysis, we consistently use the basis in
 Eqs.~(\ref{eq:YeffRotations}) and~(\ref{eq:sswYeffRotSU5sym}).
This is because, in the basis of
 Eqs.~(\ref{eq:YeffRotSU5br}),~(\ref{eq:HCYeffRotSU5br}),
 and~(\ref{eq:sswYeffRotSU5br}), the ${\cal O}(s)$- and
 ${\cal O}(s^2)$-constraints stop having the simple form they had in
 bases obtained without {\it SU}(5)-breaking rotations, in which they 
 were derived, and depend, in general, on the mismatch matrices 
 eliminated to obtain 
 $\mbox{\boldmath{$Y$}}^{5}_{D}$,
 $\mbox{\boldmath{$Y$}}^{5}_{E}$,
 $\mbox{\boldmath{$Y$}}^{10}_{U}$, and
 $\mbox{\boldmath{$Y$}}_N^{\rm I}$
 in the basis of Eq.~(\ref{eq:YeffRotSU5br}).

The determination of the boundary conditions of the effective Yukawa
 couplings at $M_{\rm cut}$ differs from that for the couplings of the 
 minimal model in the following way. 
To begin with,
 $\mbox{\boldmath{$Y$}}^{5}_{E}$, $\mbox{\boldmath{$Y$}}^{5}_{D}$,
 $\mbox{\boldmath{$Y$}}^{10}_{U}$ and $K_{\rm KCM}$ are taken as
 low-energy inputs to be evolved up to $M_{\rm GUT}$, with the seesaw
 threshold dealt as in the minimal case, by introducing the coupling
 $Y_\nu^{\rm I}$. 
At $M_{\rm GUT}$, the resulting couplings are matched to 
 $\mbox{\boldmath{$Y$}}^{5}_{E}$, $\mbox{\boldmath{$Y$}}^{5}_{D}$,
 $\mbox{\boldmath{$Y$}}^{10}_{U}$ of Eq.~(\ref{eq:YeffRotations}), by 
 choosing three arbitrary unitary matrices for 
 $\Delta V_{5\,D}^\dagger$,  $\Delta V_{10\,E}^\dagger$, and
 $\Delta W_{10\,U}^{\prime \dagger}$.
The seesaw coupling $\mbox{\boldmath{$Y$}}_{N}^{\rm I}$ is obtained from
 $Y_\nu^{\rm I}$ attaching to it three arbitrary phases.  
Similarly, the couplings $\mbox{\boldmath{$Y$}}^{5}_{DU}$,
 $\mbox{\boldmath{$Y$}}^{5}_{LQ}$, $\mbox{\boldmath{$Y$}}^{10}_{UE}$,
 and $\mbox{\boldmath{$Y$}}^{10}_{QQ}$, parametrized by seven mismatch
 matrices and four diagonal matrices of couplings, are arbitrary.
The only restriction in their selection is that, together with 
 $\mbox{\boldmath{$Y$}}^{5}_{E}$, $\mbox{\boldmath{$Y$}}^{5}_{D}$ and
 $\mbox{\boldmath{$Y$}}^{10}_{U}$, they satisfy the ${\cal O}(s)$- and
 ${\cal O}(s^2)$-constraints of Eqs.~(\ref{eq:5Constraint})
 and~(\ref{eq:10Constraint}). 
Since NROs were omitted in the seesaw sector, the coupling
 $\mbox{\boldmath{$Y$}}_{ND}^{\rm I}$ coincides with
 $\mbox{\boldmath{$Y$}}_{N}^{\rm I}$.
(If included, it would differ from it at ${\cal O}(s)$, as shown by 
  Eq.~(\ref{eq:sswIConstraint}).)  
All these couplings are then evolved up to $M_{\rm cut}$.

Here, the boundary values for the soft terms are picked up and the full 
 set of RGEs is evolved downwards. 
At the GUT threshold, the two theories, the MSSU(5) model with NROs and
 the MSSM, both with a seesaw sector, are matched again by performing
 {\it SU}(5)-breaking rotations that bring 
 $\mbox{\boldmath{$Y$}}^{5}_{E}$, $\mbox{\boldmath{$Y$}}^{5}_{D}$,
 $\mbox{\boldmath{$Y$}}^{10}_{U}$ back to the form they have in 
 Eq.~(\ref{eq:YeffRotSU5br}).
In particular, the field $D^c$ is rotated as
\begin{equation}
 D^c \ \to \ 
 \left[ \Delta V_{5\,D}^\dagger \right]^\dagger D^c 
     \ = \ 
 \bigl(P_{\rm I}^\dagger \Delta_D\bigr) D^c .
\label{eq:DcROTAT}
\end{equation}
The following evolution downwards is that of the MSSM with one of the 
 three types of seesaw. 
At the electroweak scale, further rotations must be performed to extract
 the low-energy physical parameters. 
These rotations bring the Yukawa couplings to have, for example, the 
 form on the right-hand side of Eq.~(\ref{eq:YeffRotSU5br}).
Strictly speaking, the {\it SU}(5)-breaking rotations at $M_{\rm GUT}$,
 including that of Eq.~(\ref{eq:DcROTAT}), could be avoided, having
 only the final rotations at the electroweak scale.
By doing this, the typical GUT degrees of freedom that are unphysical
 below $M_{\rm GUT}$ are not eliminated and they are formally included
 in the RGEs below this scale.
Since the RGEs are covariant under unitary transformations, however,
 unphysical parameters do not play any role in the running, and the 
 RGE solutions do not depend on them.
The dependence on the matrix $\Delta_D$ is then recovered at
 $M_{\rm weak}$, when the low-energy rotations are performed.

The procedure outlined here is conceptually clear, but the large number
 of unknown matrices to be specified makes the problem very difficult to
 handle.
It is possible, however, that there exist some limiting cases, in which 
 the number of parameters needed to specify these unitary matrices is 
 smaller than the usual three for angles, plus six, for phases.
{}For example, some of the phases may be rotated away.

Before tackling this issue, we pause for a moment to compare our case
 with that of Ref.~\citen{BGOO}, where only one NRO is
 introduced.\footnote{All NROs of dimension five in $Op^5$, $Op^{10}$,
  and $Op^{N_{\rm I}}$ are discussed in Ref.~\citen{BGOO}. 
 Only one of them, however, the operator with coefficient $C_{1,0}^5$ 
  in $Op^5$, is actually considered when counting the number of 
  arbitrary but physical parameters introduced by NROs.}
In the {\it SU}(5)-symmetric basis, the effective Yukawa couplings are 
 in this case
\begin{eqnarray}
\mbox{\boldmath{${\cal Y}$}}^{5}_i    \      =& \
 Y^5 + C_{1,0}^{5}\,s\, (I_{\bar{5}_M})_i,         & \qquad 
 (i={\scriptstyle D,E,DU,LQ})
\nonumber\\[1.1ex]
\mbox{\boldmath{${\cal Y}$}}^{10}_j          =& \
  Y^{10}_j = Y^{10},                           & \qquad
 (j={\scriptstyle U,UE,QQ})
\label{eq:OKYuk}
\end{eqnarray}
 denoted with calligraphic symbols to distinguish them from the general
 ones.
The couplings $\mbox{\boldmath{${\cal Y}$}}^{10}_j$ coincide with those
 of the minimal model and are all symmetric matrices.
Thus, all matrices $W_{10\,j}^{\prime\dagger}$ and $W_{10\,j}^{\dagger}$
 coincide with the rotation matrix $W_{10}^\dagger$ of the minimal case,
 and the mismatch matrices  $\Delta W_{10\,j}^{\prime\,\dagger}$ and 
 $\Delta W_{10\,j}^{\dagger}$ are trivial:
\begin{equation} 
\Delta W_{10\,j}^{\prime\,\dagger}= \Delta W_{10\,j}^{\dagger}=
  {\mathop{\bf 1}}. \quad\quad (j={\scriptstyle U,UE,QQ})
\label{eq:DELTAW}
\end{equation}
Moreover, as the values of $(I_{\bar{5}_M})_i$ in
 Eq.~(\ref{eq:effY5hypercharges}) show, only two of the effective
 couplings
 $\mbox{\boldmath{${\cal Y}$}}^{5}_i$ $(i={\scriptstyle D,E,DU,LQ})$ are
 independent: 
\begin{equation}
 \mbox{\boldmath{${\cal Y}$}}^{5}_{D}   \ = \
 \mbox{\boldmath{${\cal Y}$}}^{5}_{DU}, \qquad
 \mbox{\boldmath{${\cal Y}$}}^{5}_{E}   \ = \
 \mbox{\boldmath{${\cal Y}$}}^{5}_{LQ}.
\label{eq:OKequalY5}
\end{equation}
The first of these two equalities implies that 
 $V_{5\,D}^\dagger  = V_{5\,DU}^\dagger$ and 
 $V_{10\,D}^\dagger = V_{10\,DU}^\dagger$, the second that
 $V_{5\,E}^\dagger  = V_{5\,LQ}^\dagger$ and
 $V_{10\,E}^\dagger = V_{10\,LQ}^\dagger$.
Therefore, not all mismatch matrices are independent or  nontrivial.
Indeed, the following relations hold
\begin{eqnarray}
 \Delta V_{5\,D}^{\dagger} \ \,   = &\ \Delta V_{5\,DU}^{\dagger},
                                &  \hspace*{1truecm} 
 \Delta V_{10\,DU}^{\dagger}       = {\mathop{\bf 1}},
\nonumber\\[1.001ex]
 \Delta V_{10\,LQ}^{\dagger}       = &\ \Delta V_{10\,E}^{\dagger},
                                &  \hspace*{1truecm}
 \Delta V_{5\,LQ}^{\dagger} \      = {\mathop{\bf 1}}.
\label{eq:DeltaV}
\end{eqnarray}
By plugging these and the relations in Eq.~(\ref{eq:DELTAW}) in
 Eq.~(\ref{eq:HCYeffRotSU5br}), it is easy to see that, in this case, 
 only two mismatch matrices are needed to parametrize the effective 
 Yukawa couplings: $\Delta V_{5\,D}^\dagger$ and 
 $\Delta V_{10\,E}^\dagger$, multiplied by some phase matrices.
In addition, no arbitrariness exists for the boundary conditions of 
 $\widehat{\mbox{\boldmath{${\cal Y}$}}}_{DU}^{5}$, 
 $\widehat{\mbox{\boldmath{${\cal Y}$}}}_{LQ}^{5}$,
 $\widehat{\mbox{\boldmath{${\cal Y}$}}}_{UE}^{10}$, and  
 $\widehat{\mbox{\boldmath{${\cal Y}$}}}_{QQ}^{10}$, as these couplings
  are linked to
 $\widehat{\mbox{\boldmath{${\cal Y}$}}}_{D}^{5}$,
 $\widehat{\mbox{\boldmath{${\cal Y}$}}}_{E}^{5}$, and 
 $\widehat{Y}^{10}$, which are fixed by low-energy data.

The apparent simplicity of this case should not be deceiving, as the 
 number of parameters that it involves is large.
To show this, we parametrize $\Delta V_{5\,D}^\dagger$ and
 $\Delta V_{10\,E}^\dagger$ as usual:
\begin{eqnarray}
 \Delta V_{5\,D}^\dagger\, & = & 
   P_D^{(1)} \,K_D^\dagger \,P_D^{(2)} e^{i\phi_D}, 
\nonumber\\
 \Delta V_{10\,E}^\dagger   & = & 
   P_E^{(1)} \,K_E^\dagger \,P_E^{(2)} e^{i\phi_E} ,
\label{eq:PARAMETRIZEmismatch}
\end{eqnarray}
 with CKM-like matrices $K_E$ and $K_D$, and we express the effective
 Yukawa couplings $\mbox{\boldmath{${\cal Y}$}}_i^5$ and
 $\mbox{\boldmath{${\cal Y}$}}_j^{10}$ after 
 {\it SU}(5)-breaking rotations as
\begin{eqnarray}
 \mbox{\boldmath{${\cal Y}$}}_D^5                \to&
   \widehat{\mbox{\boldmath{${\cal Y}$}}}_D^5,   &\qquad \quad
 \mbox{\boldmath{${\cal Y}$}}_{DU}^5             \ \to \ 
   \widehat{\mbox{\boldmath{${\cal Y}$}}}_D^5\,
   K_{\rm CKM}^\dagger P_{10}^\dagger,   
\nonumber\\[1.2ex]
 \mbox{\boldmath{${\cal Y}$}}_E^5                \to&
   \widehat{\mbox{\boldmath{${\cal Y}$}}}_E^5,   &\qquad \quad   
 \mbox{\boldmath{${\cal Y}$}}_{LQ}^5             \ \to 
 \  e^{-i(\phi_{\rm I}-\phi_{E})}
   \,\widehat{\mbox{\boldmath{${\cal Y}$}}}_E^5
   \,\Delta_E{^\dagger},
 \qquad
\nonumber\\[1.2ex]
 \mbox{\boldmath{${\cal Y}$}}_U^{10}             \to& \ 
   \widehat{Y}^{10}\,K_{\rm CKM},                 &\qquad  \quad 
 \mbox{\boldmath{${\cal Y}$}}_{UE}^{10}         \ \to \
 \  e^{i(\phi_{\rm I}-\phi_{E})}
   \,\widehat{Y}^{10}\,
   K_{\rm CKM} \,\Delta_E,
\nonumber\\[1.3ex]
     &                                        &\qquad \quad
 \mbox{\boldmath{${\cal Y}$}}_{QQ}^{10}         \ \to\
   K_{\rm CKM}^{T} P_{10}
   \,\widehat{Y}^{10} \, K_{\rm CKM}.  
\label{eq:effYUKparFIN} 
\end{eqnarray}
%
The seesaw couplings obtain the forms
\begin{eqnarray}
 \mbox{\boldmath{${\cal Y}$}}_N^{\rm I}          \to & \
   Y_{\nu}^{\rm I}, \quad \,                    &\qquad \qquad \quad
 \mbox{\boldmath{${\cal Y}$}}_{ND}^{\rm I}      \ \to \ 
   e^{i(\phi_{\rm I}-\phi_D)}\,
   Y_{\nu}^{\rm I} \,\Delta_D .  \qquad \qquad  
\label{eq:NYUKparFIN}
\end{eqnarray}
where the matrices $\Delta_D$ and $\Delta_E^\dagger$ are
\begin{equation}
\Delta_E^\dagger =
   \bigl( P_{\rm I}^{\dagger} P_{E}^{(1)} \bigr)
    \,K_E^\dagger\, P_E^{(2)}  ,
\qquad \,
\Delta_D =
   \bigl( P_{\rm I} P_{D}^{(2)\dagger} \bigr)
    \,K_D   P_D^{(1)\dagger}.  \qquad
\label{eq:DeltaMatr} 
\end{equation}

When compared with Eqs.~(\ref{eq:YUKminSU5br})
 and~(\ref{eq:sswYUkminSU5br}), these equations show that by introducing
 only one NRO, the number of physical parameters increases by two
 CKM-like matrices, $K_E$ and $K_D$, three $P$-type phase matrices, with
 two phases each, and one overall phase, for a total of nine phases and
 six mixing angles. 
If we add also the three eigenvalues that distinguish now 
 $\widehat{\mbox{\boldmath{${\cal Y}$}}}_D^5$ from 
 $\widehat{\mbox{\boldmath{${\cal Y}$}}}_E^5$,  the number of new
 physical parameters sums up to eighteen, which is the number of
 parameters of the complex matrix $C_{1,0}^{5}$.
That is, all the new parameters due to the NRO introduced in this 
 case are physical. 
This is, however, not true in general, when more NROs are present.

\subsection{Mismatch-matrices approximation}
\label{sec:NROsBC-YUKAWAappr}
%
An approximated parametrization of the mismatch matrices, with a
 consequent reduction of the number of physical parameters, can be
 obtained when 
 i) $\tan \beta$ is large, and ii) $s$ does not exceed the value
 of $\sim 10^{-2}$.
This approximated parametrization is valid irrespective of the number
 of NROs introduced.

In the basis of
 Eqs.~(\ref{eq:YeffRotations}) and~(\ref{eq:sswYeffRotSU5sym}), it is
 easy to obtain from the ${\cal O}(s)$-constraint in
 Eq.~(\ref{eq:5Constraint}), with $i={\scriptstyle D}$ and
 $i^\prime={\scriptstyle E}$, that~\cite{BGOO}
\begin{equation}
 \left\vert (\Delta V_{5\,D}^\dagger)_{(h,3)}  \right\vert
  \lesssim \frac{s}{y_b}, 
\quad\quad
 \left\vert (\Delta V_{10\,E}^\dagger)_{(k,3)} \right\vert
  \lesssim \frac{s}{y_\tau}. 
\quad \quad\quad 
 (h,\,k\ne 3)
\label{eq:5Dand10Einequalities}
\end{equation}
{}For the derivation of these inequalities, we have made the approximation
 $(\widehat{\mbox{\boldmath{$Y$}}}_{E}^{5})_{ii} =
  (\widehat{\mbox{\boldmath{$Y$}}}_{D}^{5})_{ii}$, since the differences
 between these Yukawa couplings do not affect our estimate.
We recall that all ${\cal O}(s)$- and ${\cal O}(s^2)$-constraints were
 derived in an {\it SU}(5)-symmetric basis.
Since for $s\sim 10^{-2}$, it is $s/y_b, s/y_\tau \sim 1/\tan\beta$, we
 conclude that, for relatively large values of $\tan\beta$, the elements
 (1,3) and (2,3) of the matrices $\Delta V_{5\,D}^\dagger$ and
 $\Delta V_{10\,E}^\dagger$ are small, whereas the elements in the
 upper-left $2\times 2$ sub-block are unconstrained.

Thus, in this limit, the matrix $\Delta V_{5\,D}^\dagger$ has one mixing
 angle only and four phases and we express it as
\begin{equation}
 \Delta V_{5\,D}^\dagger \ = \
  P_D^{(1)}\big\vert_2 \,K_D^T\big\vert_2\,P_D^{(2)} \,e^{i\phi_D}, 
\label{eq:newPARAMmismTWO}
\end{equation}
 where $P_D^{(2)}$ is one of the usual two-phase $P$-matrices,
 $K_D\big\vert_2$ and $P_D^{(1)}\big\vert_2$ are, respectively, a
 $3\times 3$ orthogonal matrix and a one-phase diagonal matrix with
 $\det P_D^{(1)}\big\vert_2=1$, of type
\begin{equation}
  K_D\big\vert_2        =
\left(\begin{array}{c|c}
  \begin{array}{rr}
     \cos \theta & - \sin \theta \\ \sin\theta & \cos \theta 
  \end{array}         &  0_2                            \\[1.6ex]
  \hline\\[-1.6ex] 
   0_2^T              &   1                             \\ 
\end{array}\right),
\quad\quad
  P_D^{(1)}\big\vert_2  =
\left(\begin{array}{c|c}
  \begin{array}{ll}
      e^{i\phi_D^{(1)}}  &  0  \\ 0 & e^{-i \phi_D^{(1)}}   
  \end{array}         & 0_2                             \\[1.5ex]
   \hline\\[-1.6ex]
   0_2^T              &  1                              \\ 
\end{array}\right).
\label{eq:PARAMETROkadaRes}
\end{equation}
The symbol $0_2$ denotes here a two-component null vector. 
Clearly, we could have also chosen to parametrize
 $\Delta V_{5\,D}^\dagger$ as
\begin{equation}
\Delta V_{5\,D}^\dagger\ = \
 P_D^{(1)\,\prime} \,K_D^T\big\vert_2 \,
 P_D^{(2)\,\prime}\big\vert_2 \,e^{i\phi_D}. 
\label{eq:otherPARAMmismTWO}
\end{equation}
The two parametrizations can be reduced to each other because a
 two-phase $P$-matrix can always be decomposed as 
 $P_D^{(2)} =P_D^{(8)}\, P_D^{(2)\,\prime}\big\vert_2$, with
 $P_D^{(2)\,\prime}\big\vert_2$ of the type shown in
 Eq.~(\ref{eq:PARAMETROkadaRes}), and
 $P_D^{(8)} =
 {\rm diag}(e^{i\phi_D^{(8)}},e^{i\phi_D^{(8)}},e^{-2i\phi_D^{(8)}})$, 
 which obviously commute with $K_D^T\big\vert_2$.

The matrices $\Delta_D^\dagger$ and $\Delta_E^\dagger$, then, can also
 be expressed in either one of these two forms.
An explicit parametrization of $\Delta_D$ is
\begin{equation}
  \Delta_D            =
\left(\begin{array}{c|c}
\!\!\left(\!
 \begin{array}{ll}
  \cos \theta_D  \,e^{\,i\rho_D}               &\!
  - \sin \theta_D\,e^{\,i\sigma_D}             \\[1.0ex]
  \sin\theta_D   \,e^{\,-i\sigma_D}    &\!
 \phantom{-} 
    \cos \theta_D\,e^{\,-i\rho_D}
 \end{array}   
\!\!\!\right) e^{\,-\frac{i}{2}\tau_D}\!
                      &  0_2             \\[2.5ex]
\hline                                   \\[-1.7ex] 
   0_2^T              &  e^{\,i\tau_D}   \\ 
\end{array}\right).
\label{eq:DeltaDparam}
\end{equation}

It is easy to see that the same type of approximated parametrization is
 also possible for all the mismatch matrices in
 Eq.~(\ref{eq:ROTATidentif}).
{}For example, with the {\it SU}(5)-symmetric rotation 
 $\bar{5}_M \to \Delta V_{5 \,LQ} \,\bar{5}_M$, it is possible to remove
 $(\Delta V_{5 \,LQ}^\dagger)^T$ in the expression for 
 $\mbox{\boldmath{$Y$}}^{5}_{LQ}$ in Eq.~(\ref{eq:YeffRotations}),
 reducing it to the same form that $\mbox{\boldmath{$Y$}}^{5}_{E}$ had
 when we derived the inequalities of
 Eq.~(\ref{eq:5Dand10Einequalities}).
The role that $\Delta V_{5\,D}^\dagger$ had then is now played by
 $\Delta V_{5\,D}^\dagger\Delta V_{5\,LQ}$.
Thus, the ${\cal O}(s)$-constraint in Eq.~(\ref{eq:5Constraint}) with
 $i =\scriptstyle{D}$ and $i^\prime =\scriptstyle{LQ}$, gives
\begin{equation}
 \left\vert (\Delta V_{5\,D}^\dagger \Delta V_{5\,LQ})_{(3,h)}  
 \right\vert \
 \lesssim \frac{s}{y_b},
\quad\quad
 \left\vert (\Delta V_{10\,LQ}^\dagger)_{(k,3)} \right\vert
 \lesssim \frac{s}{y_\tau},
\quad \quad \quad 
 (h,\,k\ne 3) 
\label{eq:5LQand10LQinequalities}
\end{equation}
 which allows us to conclude that also $\Delta V_{5\,LQ}$ and
 $\Delta V_{10\,LQ}$ can be parametrized in the same way.
We have used here
 $(\widehat{\mbox{\boldmath{$Y$}}}_{LQ}^{5})_{ii} =
  (\widehat{\mbox{\boldmath{$Y$}}}_{D}^{5})_{ii}$.
As in the corresponding approximation made for 
 Eq.~(\ref{eq:5Dand10Einequalities}), the differences between these
 elements are irrelevant to this estimate.  
Thus, the same procedure can be iterated for the other mismatch matrices
 in Eq.~(\ref{eq:ROTATidentif}), showing therefore that in the limit of
 $s$ not larger than $10^{-2}$, all the $\Delta V_{5\,i}^\dagger$ and 
 $\Delta V_{10\,i}^\dagger$ are of the same type, if $\tan \beta$ is
 large.
However, no clear hierarchy exists among the different elements of these
 matrices when $\tan \beta$ is not really large, even for values such as
 $\tan\beta \lesssim 10$. 
The same parametrization is valid also for the mismatch matrices
 $\Delta W_{10\,j}^\dagger$ and  $\Delta W_{10\,j}^{\prime\dagger}$, but
 in this case, irrespective of the value of $\tan \beta$.

It should be stressed that the approximated form of the mismatch 
 matrices is indeed an approximation.
The elements (1,3) and (2,3) in these matrices are not vanishing 
 but small: all elements in these matrices
 are, in general, modified by small first--third and second--third{{}} 
 generation mixing angles that are bounded from above by
 $s/y_b$, $s/y_\tau$, or $s/y_t$. 
Moreover, they are not even absolutely{{{}}} small, for example when
 compared with the corresponding elements of the CKM matrix, but they 
 are small with respect to the element of the MNS matrix.
As will be discussed in \S\ref{sec:Results}, however, the fact
 that the elements (1,3) and (2,3) in these matrices are small in the
 sense discussed here does not imply that the
 first--third and second--third generation flavour transitions in the
 scalar sector are not affected by NROs.

When applied to the case in which only one NRO is introduced, this
 approximation reduces the number of arbitrary mixing angles to be
 specified at $M_{\rm GUT}$ from six to two~\cite{BGOO}.
The number of arbitrary phases in addition to the five already present
 in the minimal model, is five, one of which is an overall phase.

We would like to emphasize here that the case of only one NRO, that of
 dimension five with coefficient $C_{1,0}^{5}$ in $Op^{5}\vert_5$ is
 somewhat special.
If the second of the two NROs in $Op^{5}|_5$ with coefficient
 $C_{0,1}^{5}$ is introduced, the number of physical mismatch matrices
 increases to four.
This can be seen by taking 
 $\Delta W_{10\,j}^{\prime\,\dagger}= \Delta W_{10\,j}^{\dagger}=
  {\mathop{\bf 1}}$, for all $j$, in
 Eqs.~(\ref{eq:YeffRotSU5br}) and~(\ref{eq:HCYeffRotSU5br}), and by
 using the ${\cal O}(s^2)$-constraint of Eq.~(\ref{eq:5Constraint}). 
Even when the approximation discussed here can be applied, there are
 still four arbitrary mixing angles to be inputted at $M_{\rm GUT}$,
 together with a fairly large number of phases.

We do not attempt to count them in this case, nor in the general one
 with ten mismatch matrices. 
Given how large the number of independent parameters is,
 these exercises are not particularly significant, at least at this
 stage. 
Indeed, it remains to be seen whether the constraints induced by the 
 suppression of the proton-decay rate affect the form of the mismatch
 matrices.
If these constraints do not reduce the number of arbitrary physical
 parameters, it is difficult to imagine the feasibility of a
 phenomenological study of sFVs, which attempts to include all of them.
Nevertheless, these matrices are not all at the same level when it comes
 to the modifications that they bring to the seesaw-induced sFVs. 
Thus, one reasonable simplification may be that of selecting specific
 directions of the very large parameter space opened up by the various
 mismatch matrices where these modifications are largest.

\section{Proton-decay constraints on mismatch matrices}
\label{sec:PDconstBC}
%
As is well known, even if their mass $M_{H^C}$ is of
 ${\cal O}(M_{\rm GUT})$, colored Higgs fields give a very large
 contribution to the {\it SU}(5)-breaking dimension-five superpotential
 operators:
\begin{equation}
 W^{\rm PD}_{H^C} = \frac{1}{M_{H^C}} \left[
 \left(C^{\rm PD}_{\rm LLLL}\right)^{hijk}\!\!
                    \left(Q_h Q_i\right)\!\left(L_j Q_k\right)
\ + \ 
 \left(C^{\rm PD}_{\rm RRRR}\right)^{hijk}\!\!
                    \left(U^c_h E^c_i \right)\!
                    \left(D^c_j U^c_k \right) 
                                      \right],
\label{eq:PDcolorHIGGS}
\end{equation}
 with the first ones, $(Q_h Q_i)(L_j Q_k)$, known as the LLLL-operators,
 and the second ones, $(U^c_h E^c_i )(D^c_j U^c_k )$, as the RRRR-operators.
The corresponding Wilson coefficients, before and after the
 {\it SU}(5)-breaking flavour rotations of Eq.~(\ref{eq:SMfieldsID}) (or 
 in the basis of Eq.~(\ref{eq:YUKminSU5br})), are
\begin{eqnarray}
\left(C^{\rm PD}_{\rm LLLL}\right)^{hijk}        
& = &
  \displaystyle{\frac{1}{2}}  Y^{10}_{(h,i)} Y^{5}_{(j,k)}              
\ = \ 
  \displaystyle{ \frac{1}{2} }\!
  \left[ K^T_{\rm CKM}\widehat{Y}^{10} K_{\rm CKM} 
  \right]_{(h,i)}\! 
  \left[ \widehat{Y}^{5}
  \right]_{(j,k)} ,
\nonumber\\
\left(C^{\rm PD}_{\rm RRRR}\right)^{hijk}        
& = & 
\phantom{\displaystyle{\frac{1}{2}}}  Y^{10}_{(h,i)} Y^{5}_{(j,k)}              
\ = \
  \left[ \widehat{Y}^{10} K_{\rm CKM}       
  \right]_{(h,i)}\! 
  \left[ \widehat{Y}^{5}  K_{\rm CKM}^\dagger
  \right]_{(j,k)} .
\label{eq:WCPDmin}
\end{eqnarray}
Here, and throughout this section, we neglect overall phases and 
 two-phase matrices. 
Thus, despite being proportional to small Yukawa couplings (some of
 the flavour indices are of first generation), these operators are only
 suppressed by one power of $1/M_{H^C}$, inducing, in general, a decay 
 of the proton unacceptably rapid~\cite{ExperimentPD,TheoryPD,MP}. 
This is in contrast to the dimension-six operators mediated by GUT gauge
 bosons, also inducing proton decay, which are suppressed by two powers
 of superheavy masses.

Nonrenormalizable operators can be effective in suppressing the
 proton-decay rate because they can change the value of the Yukawa
 couplings involved in this calculation, without however altering the
 couplings that reproduce a correct fermion spectrum~\cite{ZURAB,GORAN}. 
Once NROs are introduced, the dimension-five superpotential operators
 $W^{\rm PD}_{H^C}$ remains formally unchanged, but the Wilson
 coefficients $C^{\rm PD}_{\rm LLLL}$ and $C^{\rm PD}_{\rm RRRR}$ are
 now replaced by boldface ones, 
 $\mbox{\boldmath{$C$}}^{\rm PD}_{\rm LLLL}$ and
 $\mbox{\boldmath{$C$}}^{\rm PD}_{\rm RRRR}$,
 expressed in terms of effective Yukawa couplings:\footnote{Despite
  the language of effective couplings used, the material presented here
  is independent of the way NROs are treated at the quantum level.} 
\begin{eqnarray}
  \left(\mbox{\boldmath{$C$}}^{\rm PD}_{\rm LLLL}\right)^{hijk}
& =  &
  \displaystyle{ \frac{1}{2} }\!
  \left(\mbox{\boldmath{$Y$}}^{10}_{QQ} \right)_{(h,i)}
  \left(\mbox{\boldmath{$Y$}}^{5}_{LQ}  \right)_{(j,k)} ,
\nonumber\\
  \left(\mbox{\boldmath{$C$}}^{\rm PD}_{\rm RRRR}\right)^{hijk}
& =  &
  \phantom{ \displaystyle{ \frac{1}{2} } \!}
  \left(\mbox{\boldmath{$Y$}}^{10}_{UE}\right)_{(h,i)}
  \left(\mbox{\boldmath{$Y$}}^{5}_{DU} \right)_{(j,k)} .
\label{eq:WCPDgeneric}
\end{eqnarray}
The additional flavour structure induced by the mismatch matrices
 present in the effective couplings can be tuned to suppress them.
How many NROs are needed for this suppression, i.e., at which
 order in $s$ the two series of operators $Op^{5}$ and $Op^{10}$ can be
 truncated, is a question that requires a dedicated calculation, clearly
 beyond the scope of this paper. 
It is conceivable that the precision sufficient for the evaluation of
 sFVs, i.e., ${\cal O}(s)$ in the effective Yukawa couplings, may not be 
 adequate to slow down the decay of the proton.

The tuning of the above Wilson coefficients can have an important
 feedback for the determination of sFVs.
We can show this for specific values of
 $\mbox{\boldmath{$Y$}}^{5}_{LQ}$,
 $\mbox{\boldmath{$Y$}}^{5}_{DU}$,
 $\mbox{\boldmath{$Y$}}^{10}_{QQ}$, and
 $\mbox{\boldmath{$Y$}}^{10}_{UE}$ that were found in
 Ref.~\citen{DESYpeople} to provide acceptable proton-decay rates.
In one of two cases studied in this reference (the second one), just
 below the GUT threshold, in the MSSM, these are 
\begin{eqnarray}
 \mbox{\boldmath{$Y$}}^{10}_{QQ}           & = &
 \widehat{\mbox{\boldmath{$Y$}}}^{10}_{QQ} \ = \
  {\rm diag}(0,0,y_t),
\nonumber\\
 \mbox{\boldmath{$Y$}}^{10}_{UE}           & = &
 \widehat{\mbox{\boldmath{$Y$}}}^{10}_{UE} \ = \
  {\rm diag}(0,0,y_t),
\nonumber\\
 \mbox{\boldmath{$Y$}}^{5}_{DU}            & = &
 \widehat{\mbox{\boldmath{$Y$}}}^{5}_{DU}  \ = \ 
  {\rm diag}(y_d -y_e, y_s -y_\mu, y_b),
\nonumber\\
 \mbox{\boldmath{$Y$}}^{5}_{LQ}            & = & 
 \widehat{\mbox{\boldmath{$Y$}}}^{5}_{LQ}  \ = \  
  {\rm diag}(0,0,y_\tau), 
\label{eq:DESYansatz}
\end{eqnarray}
 in a basis in which the effective Yukawa couplings
 $\mbox{\boldmath{$Y$}}^{5}_{D}$,
 $\mbox{\boldmath{$Y$}}^{5}_{E}$,
 $\mbox{\boldmath{$Y$}}^{10}_{U}$
 are\footnote{We thank D. Emmanuel-Costa for confirming that this 
 is indeed the basis used in Ref.~\citen{DESYpeople}.}
\begin{eqnarray}
 \mbox{\boldmath{$Y$}}^{5}_{D}             & = &
 \widehat{\mbox{\boldmath{$Y$}}}_{D}^{5}   \ = \
  {\rm diag}(y_d,y_s, y_b),
\nonumber\\
 \mbox{\boldmath{$Y$}}^{5}_{E}             & = &
 \widehat{\mbox{\boldmath{$Y$}}}_{E}^{5}   \ = \
  {\rm diag}(y_e,y_\mu, y_\tau),
\nonumber\\
 \mbox{\boldmath{$Y$}}^{10}_{U}            & = &
  K_{\rm CKM}^T\,\widehat{\mbox{\boldmath{$Y$}}}_{U}^{10}K_{\rm CKM}
                                           \ = \
  K_{\rm CKM}^T\,{\rm diag}(y_u,y_c,y_t)K_{\rm CKM}.
\label{eq:ECWbasis}
\end{eqnarray}
(The seesaw couplings do not play any role in this discussion.)

Since they must be color antisymmetrized, the three $Q$'s in the
 operators $QQQL$, two of which have the same {\it SU}(2) index, cannot
 belong to the same generation.
Hence, the special forms of $\mbox{\boldmath{$Y$}}^{10}_{QQ}$ and 
 $\mbox{\boldmath{$Y$}}^{5}_{LQ}$  give a vanishing contribution to the
 coefficient $\mbox{\boldmath{$C$}}^{\rm PD}_{\rm LLLL}$.
The contribution to $\mbox{\boldmath{$C$}}^{\rm PD}_{\rm RRRR}$, in 
 contrast, is nonvanishing. 
According to the analysis of Ref.~\citen{DESYpeople}, this ansatz leads
 to a decay rate of the proton smaller than the existing experimental
 constraints for $\tan \beta \lesssim12$, when the mass of the lightest
 stop squark, $m_{\widetilde{t}_1}$, is $\sim 400\,{\rm GeV}$, and the
 mass for the colored Higgs triplet is of ${\cal O}(M_{\rm GUT})$.

We recall, however, that the calculation in this reference is based on
 the inclusion of only dimension-five NROs in the classes $Op^{5}$ and
 $Op^{10}$.
Thus, a further reduction of the proton-decay rate, which relaxes the
 bound on $\tan\beta$, can be achieved with a simple modification of the
 above ansatz, without having to increase the value of
 $m_{\widetilde{t}_1}$.
For example, by adding NROs of higher dimensions in $Op^{5}$, which are
 practically irrelevant to the evaluation of sFVs, it is possible to
 tune the value of the element (1,1) of 
 $\widehat{\mbox{\boldmath{$Y$}}}^{5}_{DU}$ in
 Eq.~(\ref{eq:DESYansatz}) to be vanishing. 
According to Ref.~\citen{DESYpeople}, the simultaneous vanishing of 
 the elements (1,1) of 
 $\widehat{\mbox{\boldmath{$Y$}}}^{5}_{DU}$ and 
 $\widehat{\mbox{\boldmath{$Y$}}}^{10}_{UE}$ is a sufficient condition
 for the contribution to proton decay from the RRRR operators to 
 vanish as well~(see the first ansatz used by the authors of this 
 reference).
Clearly, it is enough to simply suppress the element (1,1) of 
 $\widehat{\mbox{\boldmath{$Y$}}}^{5}_{DU}$ and therefore to suppress
 this contribution, without having to make it vanish altogether. 
This is somewhat irrelevant for the following discussion and, 
 hereafter, whenever the effective couplings in
 Eq.~(\ref{eq:DESYansatz}) are mentioned, it is assumed that the
 element (1,1) of 
 $\widehat{\mbox{\boldmath{$Y$}}}^{5}_{DU}$ is exactly vanishing.

The basis in Eqs.~(\ref{eq:DESYansatz}) and~(\ref{eq:ECWbasis}) can be
 interpreted as a basis obtained without {\it SU}(5)-breaking rotations,
 as it satisfies the ${\cal O}(s)$- and ${\cal O}(s^2)$-constraints of
 Eqs.~(\ref{eq:5Constraint}) and~(\ref{eq:10Constraint}). 
Thus, a comparison with the effective couplings in 
 Eq.~(\ref{eq:YeffRotations}), allows us to conclude that all the
 mismatch matrices are trivial, or the inverse of $K_{\rm CKM}$ (which
 for the purposes of this discussion is nearly trivial) and all
 additional phases are vanishing. 
This choice, of course, selects one particular point in the large space
 of parameters opened up by the NROs. 
In this case, the predictions for the seesaw-induced sFVs do not differ
 from those of the minimal model, with vanishing NROs.

On the other hand, the light-field Yukawa couplings in
 Eq.~(\ref{eq:ECWbasis}) can also be interpreted as obtained through
 {\it SU}(5)-breaking rotations.
Then, the effective Yukawa couplings associated with the triplet Higgs
 bosons
 $\mbox{\boldmath{$Y$}}^{5}_{DU}$,
 $\mbox{\boldmath{$Y$}}^{5}_{LQ}$,
 $\mbox{\boldmath{$Y$}}^{10}_{UE}$,
 $\mbox{\boldmath{$Y$}}^{10}_{QQ}$,
 have the complicated expressions:
\begin{eqnarray}
\mbox{\boldmath{$Y$}}^{5}_{DU}    & \to & 
 \left[\Delta V_{5\,DU}^\dagger \Delta V_{5\,D} 
 \right]^T \!
\widehat{\mbox{\boldmath{$Y$}}}_{DU}^{5}
 \left[\Delta V_{10\,DU}^\dagger K_{\rm CKM}^\dagger 
       \Delta W_{10\,U}^\prime   K_{\rm CKM} 
\right],
\nonumber\\
\mbox{\boldmath{$Y$}}^{5}_{LQ}    & \to & 
 \left[\Delta V_{5\,LQ}^\dagger 
 \right]^T \,
\widehat{\mbox{\boldmath{$Y$}}}_{LQ}^{5}\,
 \left[ \Delta V_{10\,LQ}^\dagger 
 \right],
\nonumber\\
\mbox{\boldmath{$Y$}}^{10}_{UE}   & \to & 
 \left[\Delta W_{10\,UE}^{\prime\dagger}
       \Delta W_{10\,U}  K_{\rm CKM}
 \right]^T \!
\widehat{\mbox{\boldmath{$Y$}}}_{UE}^{10}
 \left[\Delta W_{10\,UE}^\dagger K_{\rm CKM} \Delta V_{10\,E}
 \right],
\nonumber\\
\mbox{\boldmath{$Y$}}^{10}_{QQ}   & \to & 
 \left[\Delta W_{10\,QQ}^\dagger  K_{\rm CKM}
 \right]^T \!
\widehat{\mbox{\boldmath{$Y$}}}_{QQ}^{10}
 \left[\Delta W_{10\,QQ}^\dagger  K_{\rm CKM} 
 \right],
\label{eq:HCYeffRotDESY}
\end{eqnarray}
 where we have neglected phases, as this discussion has only
 demonstrative purposes.
The identification of these couplings with those of
 Eq.~(\ref{eq:DESYansatz}) now requires:
\begin{eqnarray}
 \Delta V_{5\,DU} \        & = &  \Delta V_{5\,D},
\nonumber \\[1.1ex]
 \Delta V_{10\,DU}         & = &  K_{\rm CKM}^\dagger 
                                   \Delta W_{10\,U}^\prime K_{\rm CKM}, 
\nonumber \\[1.1ex]
 \Delta V_{5\,LQ}\  \      & = &  \Delta V_{10\,LQ}
                            \ = \ {\mathop{\bf 1}},
\nonumber \\[1.1ex]
 \Delta W_{10\,UE}^\prime  & = &  \Delta W_{10\,U} K_{\rm CKM},
\nonumber \\[1.1ex]
 \Delta W_{10\,UE}         & = &  K_{\rm CKM} \Delta V_{10\,E},
\nonumber\\
 \Delta W_{10\,QQ}         & = &  K_{\rm CKM},
\label{eq:DESYmismMATR}
\end{eqnarray}
 which fixes three of the mismatch matrices.
It leaves the others still undetermined, although with some relations
 among them, due to the existence of the ${\cal O}(s)$- and
 ${\cal O}(s^2)$-constraints for effective couplings. 
Since the simple form for these constraints given in
 Eqs.~(\ref{eq:5Constraint}) and~(\ref{eq:10Constraint}) applies to
 couplings in bases obtained without {\it SU}(5)-breaking rotations, we need 
 to obtain the original set of effective couplings in the
 {\it SU}(5)-conserving basis from which those in Eqs.~(\ref{eq:ECWbasis})
 and~(\ref{eq:HCYeffRotDESY}) emerge.
To this end, it is sufficient to insert in Eq.~(\ref{eq:YeffRotations})
 the diagonal form of the effective Yukawa couplings in
 Eqs.~(\ref{eq:DESYansatz}) and~(\ref{eq:ECWbasis}) and the mismatch
 matrices in Eq.~(\ref{eq:DESYmismMATR}).
This results in the following effective Yukawa couplings:
\begin{eqnarray}
 \mbox{\boldmath{$Y$}}^{5}_{D}  \ \,             & \to  & 
 \Delta V_{5\,D}^{\dagger\,^T}\
 \widehat{\mbox{\boldmath{$Y$}}}_{DU}^{5},
\nonumber\\
 \mbox{\boldmath{$Y$}}^{5}_{E}  \ \,             & \to  & 
 \widehat{\mbox{\boldmath{$Y$}}}_{DU}^{5}\,
 \Delta V_{10\,E}^\dagger,
\nonumber\\
 \mbox{\boldmath{$Y$}}^{5}_{LQ}                  & \to  &  
 \widehat{\mbox{\boldmath{$Y$}}}_{LQ}^{5},
\nonumber\\
 \mbox{\boldmath{$Y$}}^{5}_{DU}                  & \to  & 
 \Delta V_{5\,D}^{\dagger\,^T}\
 \widehat{\mbox{\boldmath{$Y$}}}_{DU}^{5} \,
 \Delta V_{10\,DU}^\dagger,
\label{eq:5secCOUPL}
\end{eqnarray}
 for the $\bar{5}_M$ sector, and
\begin{eqnarray}
 \mbox{\boldmath{$Y$}}^{10}_{U} \                & \to  & 
 \Delta V_{10\,DU}^{\dagger\,^T} K_{\rm CKM}^T\ 
 \widehat{\mbox{\boldmath{$Y$}}}_{U}^{10}\,
  K_{\rm CKM} ,
\nonumber\\
 \mbox{\boldmath{$Y$}}^{10}_{QQ}                 & \to  &  
 \widehat{\mbox{\boldmath{$Y$}}}_{QQ}^{10},
\nonumber\\
 \mbox{\boldmath{$Y$}}^{10}_{UE}                 & \to  & 
 \Delta V_{10\,DU}^{\dagger\,^T} \
 \widehat{\mbox{\boldmath{$Y$}}}_{U}^{10}\,
 \Delta V_{10\,E}^{\dagger}, 
\label{eq:10secCOUPL}
\end{eqnarray}
 for the $10_M$ sector.

A considerable simplification can be made at this point by applying
 the approximation of \S\ref{sec:NROsBC-YUKAWAappr} to the mismatch
 matrices.
Having established that the limitation on $\tan \beta$ obtained when
 using the ansatz of Eqs.~(\ref{eq:ECWbasis}) and~(\ref{eq:DESYansatz})
 can be evaded with some modifications, we can make use of this
 approximation, which is valid only for large $\tan \beta$, at least for
 some of the mismatch matrices.  
It becomes then a simple (although tedious) exercise to work out the
 expressions of the ${\cal O}(s)$- and ${\cal O}(s^2)$-constraints. 
By neglecting the first generation Yukawa coupling $y_u$ and the CKM
 mixing angles, we obtain
\begin{eqnarray}
\left\vert
  \displaystyle{\frac{1}{2}}\left(\Delta V_{10\,DU}\right)_{(1,2)}y_c 
\right\vert \hspace*{3truecm}
 &\leq & {\cal O}(s^2), 
\nonumber\\
\left\vert
  \left(\Delta V_{5\,D}\right)_{(1,2)}\left\{
  \left[1-\left(\Delta V_{10\,DU}\right)_{(2,2)}\right]y_s + 
  \left(\Delta V_{10\,DU}\right)_{(2,2)} y_\mu \right\}
\right\vert
 &\leq & {\cal O}(s^2). 
\label{eq:nontrivDELTAV}
\end{eqnarray}
{}For the typical values that the Yukawa couplings in these expressions
 have at $M_{\rm GUT}$, i.e., $y_c \sim 10^{-3}$ and 
 $y_s \sim  y_\mu/3 \sim 10^{-4} \tan \beta$, the absolute value of 
 $(\Delta V_{5\,D})_{(1,2)}$ cannot exceed 0.1 for $\tan \beta > 10$. 
Thus, the mismatch matrix $\Delta V_{5\,D}$, which coincides with 
 $\Delta_D$ of Eqs.~(\ref{eq:DcROTAT}) and~(\ref{eq:sswYeffRotSU5br})
 in the limit of vanishing phases, is approximated by the unit matrix, 
 at least as much as $K_{\rm CKM}$ is. 
As a consequence, the predictions for sFVs induced by the seesaw
 couplings do not deviate substantially from those of the MSSU(5) model
 with vanishing NROs.

The situation may be different if $\tan \beta$ is small.
In such a case, none of the three mixing angles of the mismatch matrix
 $\Delta V_{5\,D}$ needs to be small and the seesaw-induced sFVs may
 have patterns different from those observed in the MSSU(5) model
 without NROs.

Moreover, we would like to emphasize that it is not possible to claim
 at this point that the results obtained with the ansatz of
 Eq.~(\ref{eq:DESYansatz}) are generic.
This specific ansatz is claimed in Ref.~\citen{DESYpeople} to be only a 
 sufficient one to suppress the proton-decay rate, but not necessary.

We have also observed that the inclusion of NROs with dimension higher
 than five tends to facilitate the suppression of the proton-decay rate.
In contrast, the use of only one NRO made in  Ref.~\citen{BGOO}, i.e., 
 that with coefficient $C_{1,0}^{5}$ in $Op^{5}$, while adequate to 
 obtain the correct fermion spectrum, may not be sufficient for this 
 suppression.
The two coefficients corresponding to the LLLL- and RRRR-operators are
 now
\begin{eqnarray}
\left(\mbox{\boldmath{${\cal C}$}}^{\rm PD}_{\rm LLLL}\right)^{hijk}
& = &
 \displaystyle{ \frac{1}{2} }\!
 \left[K^T_{\rm CKM}\widehat{Y}^{10} K_{\rm CKM}
 \right]_{(h,i)}
 \left[\widehat{\mbox{\boldmath{${\cal Y}$}}}^{5}_E \,\Delta V_{10\,E}
 \right]_{(j,k)} ,
\nonumber\\
\left(\mbox{\boldmath{${\cal C}$}}^{\rm PD}_{\rm RRRR}\right)^{hijk}  
& = &
 \phantom{ \displaystyle{ \frac{1}{2} }\!}
 \left[\widehat{Y}^{10} K_{\rm CKM} \,\Delta V_{10\,E}^\dagger
 \right]_{(h,i)}
 \left[\widehat{\mbox{\boldmath{${\cal Y}$}}}^{5}_D K_{\rm CKM}^\dagger
 \right]_{(j,k)}  .
\label{eq:WCPDOkadaNRO}
\end{eqnarray}
Although a check should actually be made, the fact that only one
 mismatch matrix is present, $\Delta V_{10\,E}$, may make it impossible 
 to neutralize the effects of the dangerous products
 $\widehat{Y}^{10} K_{\rm CKM}$ and
 $K^T_{\rm CKM}\widehat{Y}^{10} K_{\rm CKM}$.

Presumably, one way to suppress the decay rate of the proton without 
 affecting the sFVs analysis is to tune NROs belonging to the class
 $Op^{\rm PD}$~\cite{BMYtalks} in Eq.~(\ref{eq:supNROforPD}) to cancel
 those from the effective operators in
 $\mbox{\boldmath{$W$}}^{\rm PD}_{H^C}$.  
Naively thinking, such a cancellation seems, indeed, possible. 
The dimension-five NRO in $Op^{\rm PD}$, in fact, has a stronger
 suppression than those in $\mbox{\boldmath{$W$}}^{\rm PD}_{H^c}$, as
 $1/M_{H^C} \sim 1/M_{\rm GUT}$ is larger than
 $1/M_{\rm cut}$, but its coefficients can be of ${\cal O}(1)$,
 whereas  $C^{\rm PD}_{\rm LLLL}$ and $C^{\rm PD}_{\rm RRRR}$ (as well as
 $\mbox{\boldmath{$C$}}^{\rm PD}_{\rm LLLL}$ and 
 $\mbox{\boldmath{$C$}}^{\rm PD}_{\rm RRRR}$) are suppressed by small
 Yukawa couplings and CKM mixing elements.
An explicit check of such a possibility should be made, and the
 results will be presented elsewhere.

We would like to close this section with the following observation. 
If they are not reduced to be trivial by the tuning required to suppress
 the Higgs-triplet-{{{}}}induced decay rate of the proton, some of the 
 mismatch matrices may appear in the gauge-boson-induced operators of
 dimension six, which also induce proton decay. 
Their presence may alter the relative size of the widths of the
 different decay modes of the proton, with respect to those obtained in
 the MSSU(5) model by the same dimension-six operators.

\section{Effective couplings at the tree level in the SUSY-breaking
         sector}
\label{sec:SOFTNROS}
%
An important role is played here by the auxiliary VEV $F_{24}$, 
 overlooked in the literature devoted to the evaluation of sFVs within
 nrMSSU(5) models.
The fact that $F_{24} \ll  F_{\sf X}$, the largest SUSY-breaking
 auxiliary VEV of ${\cal O}(\widetilde{m} M_{\rm cut})${{{}}} may
 perhaps induce us to think that the effect of the much smaller 
 SUSY-breaking VEV $F_{24}$, of
 ${\cal O}(\widetilde{m}M_{\rm GUT})$, is negligible.
In reality, the hierarchy between $F_{\sf X}$ and $F_{24}$,
\begin{equation}
  \frac{F_{24}}{F_{\sf X}}  = {\cal O}(s),
\label{eq:F24FXhyerarchy}
\end{equation}
 is of the same order as the expansion parameter in the series defining
 effective couplings. 
Thus, $F_{24}$ turns out to be essential for the identification of the
 correct expression of the soft effective couplings in terms of the
 original parameters of the model~(discussed in this section), in the
 determination of the boundary conditions for the soft effective
 couplings~(to be discussed in \S\S\ref{sec:NROsBC-SOFT}
 and \ref{sec:UNIVERSALITY}), and in guaranteeing that the soft
 effective couplings evolve like the soft couplings of an MSSU(5) model
 with the {\it SU}(5) symmetry broken at ${\cal O}(s)$ everywhere except in
 the gauge sector~(to be shown in \S\ref{sec:EffPicLOOPlevel} and
 Appendix~\ref{sec:proof}).

\subsection{Effective trilinear couplings} 
\label{sec:trilSOFTsecNROS}
%
One may naively think that effective trilinear couplings can be simply 
 read off from soft operators such as
\begin{equation}
 {\widetilde O}p^{5} = 
 \sum_{n+m=0}^k 
\sqrt{2} \, \widetilde{\bar{5}}_M \, \widetilde{C}_{n,m}^{5} 
   \left(\!\frac{24_H^T}{M_{\rm cut}}\!\right)^{\!\!n}
 \!\widetilde{10}_M 
   \left(\!\frac{24_H}{M_{\rm cut}}\!\right)^{\!\!m}
 \!\bar5_H ,
\label{eq:NRO5tilde}
\end{equation}
 when the field $24_H^{24}$ is replaced by its~VEV $v_{24}$.
These operators are obtained from the superpotential ones:
\begin{equation}
  Op^{5}({\sf{X}})  = 
 \sum_{n+m=0}^k
   \left(\!\frac{{\sf{X}}}{M_{\rm cut}}\!\right)        
   \sqrt{2} \,  \bar{5}_M \,  a_{n,m}^{5} C_{n,m}^{5} 
   \left(\!\frac{24_H^T}{M_{\rm cut}}\!\right)^{\!n}  \! {10}_M 
   \left(\!\frac{24_H}{M_{\rm cut}}\!\right)^{\!m}  \! \bar5_H ,
\label{eq:NRO5tildeX}
\end{equation}
 where ${\sf{X}}$ is the field whose auxiliary component, $F_{\sf{X}}$,
 breaks SUSY.
Thus, the coefficients $\widetilde{C}_{n,m}^{5}$ are given by
\begin{equation}
 \widetilde{C}_{n,m}^{5}  = f_{\sf{X}} \, a_{n,m}^{5} \,C_{n,m}^{5}
                    \equiv A_{n,m}^{5} \,C_{n,m}^{5},  
\label{eq:widetildeC5nmdef}
\end{equation} 
 where $f_{\sf{X}}$ is 
\begin{equation}
 f_{\sf{X}} = \frac{F_{\sf{X}}}{M_{\rm cut}}.
\label{eq:fxDEF}
\end{equation}

In addition, since the~VEV $F_{24}$ is, in general, nonvanishing,
 trilinear couplings obtain{{{}}} contributions also simply from $Op^{5}$
 itself.
As a consequence, the coefficient $\widetilde{C}_{n,m}^{5}$ gets shifted
 by the quantity $\left(n+m\right)f_{24} C_{n,m}^{5}$, where $f_{24}$ is
 also of ${\cal O}(\widetilde{m})$~(see Eq.~(\ref{eq:f24DEF})).

As in the case of effective Yukawa couplings, effective renormalizable
 trilinear terms are generated when the field $24_H$ is replaced by
 its~VEVs:
\begin{equation}
-\widetilde{D}^c \mbox{\boldmath{$A$}}^{5}_{D\phantom u}
 \widetilde{Q} H_d,                                        \quad\,
-\widetilde{E}^c(\mbox{\boldmath{$A$}}^{5}_{E\phantom q})^T
 \widetilde{L} H_d,                                        \quad\,
-\widetilde{D}^c \mbox{\boldmath{$A$}}^{5}_{DU}
 \widetilde{U}^c H_D^C,                                    \quad\,
-\widetilde{L}\, \mbox{\boldmath{$A$}}^{5}_{LQ}
 \widetilde{Q} H_D^C,  
\label{eq:tilde5effSMdecomp}
\end{equation}
 with couplings $\mbox{\boldmath{$A$}}^{5}_i$
 $(i={\scriptstyle D,E,DU,LQ})$ given by
\begin{eqnarray}
 \mbox{\boldmath{$A$}}^{5}_i  \                       & = &  \ 
 \mbox{\boldmath{$A$}}^{5}_{i,F_{\sf X}} +
 \mbox{\boldmath{$A$}}^{5}_{i,F_{24}}                      =    
  \displaystyle{\sum_{n+m=0}^k}                             
 \mbox{\boldmath{$A$}}^{5}_i\vert_{n+m}                   =
  \displaystyle{\sum_{n+m=0}^k}                             
 \mbox{\boldmath{$A$}}^{5}_{i,F_{\sf X}}\vert_{n+m} +
 \mbox{\boldmath{$A$}}^{5}_{i,F_{24}}\vert_{n+m}  
\nonumber\\                                          & = & 
  \displaystyle{\sum_{n+m=0}^k} \!
   \left[
  A_{n,m}^{5}+\!\left(n\!+\!m\right)f_{24}\right]C_{n,m}^{5} \, s^{(n+m)}  
   \left(\left(I_{\bar{5}_M}\right)_i\right)^n
   \left(\left(I_{\bar{5}_H}\right)_i\right)^m ,
\label{eq:effA5def}
\end{eqnarray}
 where
 $\mbox{\boldmath{$A$}}^{5}_{i,F_{\sf{X}}}$ and
 $\mbox{\boldmath{$A$}}^{5}_{i,F_{24}}$ are the parts generated by the
 auxiliary component of ${\sf{X}}$ and by the auxiliary component of
 the field $24_H$, respectively.  
The hypercharge factors~$(I_{\bar{5}_M})_i$ and~$(I_{\bar{5}_H})_i$ are
 those of Eq.~(\ref{eq:effY5hypercharges}), and 
 $\widetilde{C}_{0,0}^{5}=A_{0,0}^{5}\, Y^5$ is the coupling $A^5$ in
 Eq.~(\ref{eq:VminSU5terms}).

Similar considerations apply also to the $10_M$ and the seesaw sectors,
 with effective trilinear parameters 
 $\mbox{\boldmath{$A$}}^{10}_U$,
 $\mbox{\boldmath{$A$}}^{10}_{QQ}$,
 $\mbox{\boldmath{$A$}}^{10}_{UE}$, and
 $\mbox{\boldmath{$A$}}^{\rm I}_{N}$,
 $\mbox{\boldmath{$A$}}^{\rm I}_{ND}$, to be defined in analogy to the
 various $\mbox{\boldmath{$A$}}^{5}_i$.
In an analogous way, we can also define the various effective trilinear
 couplings{{}}
 of the Higgs sector and all effective bilinear couplings.

There are constraints on the couplings 
 $\mbox{\boldmath{$A$}}^{5}_i$, $\mbox{\boldmath{$A$}}^{10}_j$ and
 $\mbox{\boldmath{$A$}}^{\rm I}_h$, completely similar to the
 ${\cal O}(s)$- and ${\cal O}(s^2)$-constraints on the effective
 Yukawa couplings.
{}For example, the couplings $\mbox{\boldmath{$A$}}^{5}_i$ must satisfy
\begin{equation}
\begin{array}{cll}
\left\vert \left(
      \mbox{\boldmath{$A$}}^{5}_i  -\mbox{\boldmath{$A$}}^{5}_{i'}
           \right)_{(h,k)} \right\vert     &\leq&
 {\cal O}(s\widetilde{m}), \quad  (i,i^\prime={\scriptstyle D,E,DU,LQ})
\\[1.5ex]
\left\vert \left(
      \mbox{\boldmath{$A$}}^{5}_{D} - \mbox{\boldmath{$A$}}^{5}_{E} 
     +\mbox{\boldmath{$A$}}^{5}_{LQ}- \mbox{\boldmath{$A$}}^{5}_{DU}
           \right)_{(h,k)} \right\vert     &\leq&
 {\cal O}(s^2\widetilde{m}).
\label{eq:A5ConstraintGeneral}
\end{array}
\end{equation}
In \S\ref{sec:NROsBC-TrilSOFT}, we shall elaborate more on the form 
 that the constraints on effective trilinear couplings can have at
 $M_{\rm cut}$, if we impose a certain type of universality for the soft
 massive parameters.

\subsection{Effective sfermion masses}
\label{sec:massSOFTsecNROS}
%
The soft mass for the field  $\bar{5}_M$ arises in the MSSU(5) model
 from the usual K\"ahler potential operator:
\begin{equation}
 Op^{{\cal K}_0}_{\bar{5}_M}({\sf{X}}^\ast\!\!,{\sf{X}}) \, = \, 
 \left(\!\frac{{\sf{X}}^\ast}{M_{\rm cut}}\!\right)
       \bar{5}_M \,\vert d^0_{0,0} \vert^2 \,  \bar{5}_M^\ast 
 \left(\!\frac{{\sf{X}}}{M_{\rm cut}}\!\right) ,
\label{eq:usualSFTM}
\end{equation}
 $\widetilde{m}_{\bar{5}_M}^2 =|f_{\sf{X}}|^2 \vert d^0_{0,0}\vert^2$.
To this, contributions from the class of operators:
\begin{eqnarray}
 Op^{{\cal K}_1}_{\bar{5}_M}({\sf{X}}^\ast\!\!,{\sf{X}})  & = & 
 \left(\!\frac{{\sf{X}}^\ast}{M_{\rm cut}}\!\right)\!
 \Bigl[ \sum_{m=1}^k                     
        \bar{5}_M
      \,\left(d^{0}_{0,m}\right)^{2}\, C_{0,m}^{{\cal K},\bar{5}_M}\! 
        \left(\!\frac{24_H^\ast}{M_{\rm cut}}\!\right)^{\!\!m} \!\! 
        \bar{5}_M^\ast   
 \Bigr] \left(\!\frac{{\sf{X}}}{M_{\rm cut}}\!\right), 
\label{eq:SFTM0F24-1}
\\
 Op^{{\cal K}_2}_{\bar{5}_M}({\sf{X}}^\ast\!\!,{\sf{X}})  & = &
 \left(\!\frac{{\sf{X}}^\ast}{M_{\rm cut}}\!\right)\!
\Bigl[\!\!
 \sum_{\scriptsize{\begin{array}{l}n+m=2\\[-1.5ex]n,m\ne0
                     \end{array}}}^k \!\!\!\!
        \bar{5}_M
      \,\left(d^{0}_{n,m}\right)^{2}\, C_{n,m}^{{\cal K},\bar{5}_M}\! 
        \left(\!\frac{24_H^T}{M_{\rm cut}}\!\right)^{\!\!n}\!\!
        \left(\!\frac{24_H^\ast}{M_{\rm cut}}\!\right)^{\!\!m} \!\!
        \bar{5}_M^\ast  
\Bigr] \left(\!\frac{{\sf{X}}}{M_{\rm cut}}\!\right), 
\nonumber\\ &&
\label{eq:SFTM0F24-2}
\end{eqnarray}
 and their Hermitian conjugates, 
 $(Op^{{\cal K}_1}_{\bar{5}_M}({\sf{X}}^\ast\!\!,{\sf{X}}))^\ast$ and 
 $(Op^{{\cal K}_2}_{\bar{5}_M}({\sf{X}}^\ast\!\!,{\sf{X}}))^\ast$,
 must be added.
The couplings $C_{n,m}^{{\cal K},\bar{5}_M}$ are matrices in flavour
 space,{{{}}} as the dynamics that generates these NROs, as well as those in
 the Yukawa sector, is not flavour blind.

Since $F_{24} \ne 0$, contributions also come from
\begin{equation}
 Op^{{\cal K}_1}_{\bar{5}_M}({\sf X}) \, = 
 \Bigl[ \sum_{m=1}^k 
        \bar{5}_M \,\left(d^{1}_{0,m}\right)^2  C_{0,m}^{{\cal K},\bar{5}_M}\! 
        \left(\!\frac{24_H^\ast}{M_{\rm cut}}\!\right)^{\!\!m}\!\!    
        \bar{5}_M^\ast 
 \Bigr] \left(\!\frac{{\sf{X}}}{M_{\rm cut}}\!\right) ,   
\label{eq:SFTM1F24-1}
\end{equation}
\begin{equation}
 Op^{{\cal K}_2}_{\bar{5}_M}({\sf X})  \, = 
 \Bigl[\!\!
 \sum_{\scriptsize{\begin{array}{l} n+m=2\\[-1.5ex]n,m\ne0
                     \end{array}}}^k \!\!\!\!
        \bar{5}_M \,\left(d^{1}_{n,m}\right)^2 C_{n,m}^{{\cal K},\bar{5}_M}\! 
        \left(\!\frac{24_H^T}{M_{\rm cut}}\!\right)^{\!\!n}\!\!
        \left(\!\frac{24_H^\ast}{M_{\rm cut}}\!\right)^{\!\!m}\!\!    
        \bar{5}_M^\ast 
 \Bigr] \left(\!\frac{{\sf{X}}}{M_{\rm cut}}\!\right),    
\label{eq:SFTM1F24-2}
\end{equation}
 and from 
 $(Op^{{\cal K}_1}_{\bar{5}_M}({\sf X}))^\ast$ and
 $(Op^{{\cal K}_2}_{\bar{5}_M}({\sf X}))^\ast$, when one field
 $24_H$ gets the $F_{24}$~VEV, all others, the~VEV $v_{24}$.
The upper index in the coefficients $d^{1}_{0,m}$ and $d^{1}_{n,m}$ is a
 reminder of this.
Contributions are also due to the very same operator
 $Op^{{\cal K}_2}_{\bar{5}_M}$ of Eq.~(\ref{eq:KpotNROsNONREAB}).
They are obtained when one $24_H^T$ field and one $24_H^\ast$ field are
 replaced by the~VEV $F_{24}$, whereas all others acquire
 the~VEV $v_{24}$.

Once all the fields $24_H$ are replaced by their~VEVs, these
 operators induce a splitting between the soft mass for
 $\widetilde{D}^c$ and $\widetilde{L}$:
\begin{equation}
  \widetilde{\bar{5}}_M \,\widetilde{m}_{\bar{5}_M}^2
  \widetilde{\bar{5}}_M^\ast 
\ \to \ 
  \widetilde D^c \, \widetilde{\mbox{\boldmath{$m$}}}_{D^c}^{2}
  \widetilde D^{c\,\ast} +
  \widetilde L^\ast\,(\widetilde{\mbox{\boldmath{$m$}}}_{L}^{2})^{\ast}
  \widetilde L .
\label{eq:NROm25eff}
\end{equation}
The two effective mass parameters
 $\widetilde{\mbox{\boldmath{$m$}}}_{f}^{2}$ $(f={\scriptstyle D^c,L})$
 are both equal to $\widetilde{m}_{\bar{5}_M}^2$ in the lowest order in 
 the $s$-expansion, but already at ${\cal O}(s)$,{{{}}} they obtain{{{}}} different 
 contributions proportional to $ C_{0,1}^{{\cal K},\bar{5}_M}$, and therefore, 
 in general,{{{}}} flavour violating.

Considerations analogous to those made here hold also for the effective 
 masses in the $10_M$ sector.  
As for the effective soft masses in the the seesaw sectors, that for
 the RHNs $N^c$ coincide with the coefficient of the renormalizable 
 mass term, $\widetilde{m}_{N^c}^2$ up to an overall factor
 $\prod_j {\rm Tr} ({24_H}/{M_{\rm cut}}\!)^{k_j}$, because the RHNs are 
 {\it SU}(5) singlets.
The soft mass for the fields $15_H$ and $\bar{15}_H$, in the seesaw of 
 type~II, is also the coefficient of the renormalizable mass term,
 because we neglect NROs in the Higgs sector.
In the case of the field $24_M$ in the seesaw of type~III, in contrast,
 the various effective masses differ from $\widetilde{m}_{24_M}^2$, the 
 coefficient of the renormalizable mass term~(see
 Eq.~(\ref{eq:SU5sswSOFTPOT})).

\section{Universal boundary conditions with sFVs}
\label{sec:NROsBC-SOFT}
%
In the MSSU(5) model, as always when trying to highlight sFVs induced 
 at the quantum level, the ansatz of universality and proportionality
 of the boundary conditions for the  soft SUSY-breaking parameters is
 adopted. 
Thus, by taking again the seesaw of type~I as a case study, all soft
 masses squared  in Eqs.~(\ref{eq:VminSU5terms})
 and~(\ref{eq:SU5sswSOFTPOT}) are assumed at $M_{\rm cut}$ to be 
 flavour-{{{}}}independent and all equal to the coupling $\widetilde{m}_0^2$. 
In the same equations, the bilinear couplings $B_5$, $B_{24}$, $B_N$, 
 and the trilinear couplings $A^5$, $A^{10}$, $A_{N}^{\rm I}$  are such
 that
\begin{equation}
\frac{B_5(M_{\rm cut})}{M_5(M_{\rm cut})}     =
\frac{B_{24}(M_{\rm cut})}{M_{24}(M_{\rm cut})} =
\frac{B_N(M_{\rm cut})}{M_N(M_{\rm cut})}     =  B_0 ,
\label{eq:BinMSSU5}
\end{equation}
 and
\begin{equation}
\frac{A^5(M_{\rm cut})}{Y^5(M_{\rm cut})}     =
\frac{A^{10}(M_{\rm cut})}{Y^{10}(M_{\rm cut})} =
\frac{A_N^{\rm I}(M_{\rm cut})}{Y_N^{\rm I}(M_{\rm cut})} = A_0 ,
\label{eq:AinMSSU5}
\end{equation}
 with flavour- and field-independent massive couplings $B_0$ and $A_0$. 
Once the boundary conditions for the superpotential massive parameters  
 $M_5$, $M_{24}$, and $M_{N}$, are determined at the cutoff scale, and 
 those for the Yukawa couplings $Y^5$, $Y^{10}$, and $Y_N^{\rm I}$, are
 fixed in the way outlined in the previous section, those for the
 bilinear couplings $B_5$, $B_{24}$, and $B_{N}$, and for the trilinear
 ones, $A^5$, $A^{10}$, and $A_N^{\rm I}$ are also fixed.

The situation is more complicated when NROs are present.

\subsection{Effective trilinear couplings}
\label{sec:NROsBC-TrilSOFT}
%
We consider here specifically the couplings
 $\mbox{\boldmath{$A$}}^{5}_i$, but the discussion holds for all
 effective trilinear couplings.
Note that, for a target precision of ${\cal O}(s\times s_{\rm loop})$
 in the evaluation of sQFVs and sLFVs, the relevant terms in the
 expansions of $\mbox{\boldmath{$A$}}^{5}_i$ are up to $k=2$.
That is, the tree-level contribution of dimension-six NROs is needed.

A comparison of $\mbox{\boldmath{$A$}}^{5}_i$ in Eq.~(\ref{eq:effA5def})
 with $\mbox{\boldmath{$Y$}}^{5}_i$ in Eq.~(\ref{eq:effY5def}) shows
 that the effective trilinear couplings, in general, are not aligned
 with the effective Yukawa couplings at $M_{\rm cut}$.
Clearly, a simple extension of the conventional notion of universality
 used in the absence of NROs, which assigns the same proportionality
 constant to different $n$-linear couplings ($n\le 3$) in operators of
 the same dimensionality:
\begin{equation}
 A_{1,0}^{5} =A_{0,1}^{5} \equiv A_1, 
 \hspace*{1truecm}
 A_{2,0}^{5} =A_{0,2}^{5} =A_{1,1}^{5}=A_2,
 \hspace*{1truecm}
 \cdots,
\label{eq:usualUNIVERS}
\end{equation}
 is not sufficient to guarantee the proportionality of
 $\mbox{\boldmath{$A$}}^{5}_i(M_{\rm cut})$ and
 $\mbox{\boldmath{$Y$}}^{5}_i(M_{\rm cut})$,
 as it has only the effect of replacing $A_{n,m}^{5}$ with 
 $A_{n+m}^{5}$ in Eq.~(\ref{eq:effA5def}).

A more restrictive concept of universality, in which all the
 proportionality constants, $A_0$, $A_1$, $A_2$, $\cdots$, are equal
 can solve the problem if the auxiliary~VEV $F_{24}=0$ vanishes
 at $M_{\rm cut}$, which implies
\begin{equation}
 f_{24} = 
 B_0 - A_0 = 0  
\label{eq:VANISHINGf24}
\end{equation}
 (see Eq.~(\ref{eq:f24DEF})).
When this condition is realized, the equality of the proportionality
 constant of all trilinear and $n$-linear couplings,
\begin{equation}
 B_0 = A_0 =A_1 = A_2 =\cdots,
\label{eq:StrongUNIV1}
\end{equation}
 leads to
\begin{equation}
       \mbox{\boldmath{$A$}}^{5}_i (M_{\rm cut})  =  A_0
       \mbox{\boldmath{$Y$}}^{5}_i(M_{\rm cut}).
\label{eq:effA5align}
\end{equation}
If such a notion of universality is adopted, the tuning parameter of the
 scalar sector, $\Delta$, discussed in \S\ref{sec:HiggsSECandVACUA},
 vanishes identically.
It should also be noted that, despite a vanishing boundary
 condition at $M_{\rm cut}$, $f_{24}$ is generated at different scales
 by radiative corrections, since it evolves according to 
 Eq.~(\ref{eq:f24evol}).

Alternatively, if $f_{24}\ne 0$, it is possible to guarantee the
 alignment of $\mbox{\boldmath{$A$}}^{5}_i(M_{\rm cut})$ and
 $\mbox{\boldmath{$Y$}}^{5}_i(M_{\rm cut})$ by requiring that, for every
 value of $n+m$, the squared bracket in
 Eq.~(\ref{eq:effA5def}), with $A_{n,m}^{5}$ replaced by
 $A_{n+m}^{5}$, is equal to $A_0$:
\begin{equation}
 B_0 - f_{24}    =
 A_0             =
 A_1 + f_{24}    =
 A_2 +2f_{24}    =
 \cdots          = 
 A_n +nf_{24}.
\label{eq:StrongUNIVgen}
\end{equation}
Two parameters are needed to characterize this condition, which we
 rewrite as:
\begin{eqnarray}
    B_0     & = & A_0^\prime -2 f_{24},
\nonumber\\
    A_0     & = & A_0^\prime -3 f_{24},
\nonumber\\
    A_1     & = & A_0^\prime -4 f_{24},
\nonumber\\
    A_2     & = & A_0^\prime -5 f_{24},
\nonumber\\
   \cdots,  &   &                      
\nonumber\\
    A_n     & = & A_0^\prime -(n+3)f_{24}. 
\label{eq:StrongUNIVgen1}
\end{eqnarray}
The two parameters are $f_{24}$ and $A_0$, or $f_{24}$ and
 $A_0^\prime$, and the relation of Eq.~(\ref{eq:effA5align}) still holds.

The condition in Eq.~(\ref{eq:StrongUNIV1}), characterized by only 
 one parameter, $A_0$, coincides with this in the limit $f_{24}\to 0$.
It is easy at this point to see that there is yet another one-parameter
 condition, which also guarantees the alignment of 
 $\mbox{\boldmath{$A$}}^{5}_i(M_{\rm cut})$ and
 $\mbox{\boldmath{$Y$}}^{5}_i(M_{\rm cut})$.
This is obtained by setting $A_0^\prime =0$.
In this case, the parameter that fixes the soft couplings in question
 is $f_{24}$:
\begin{equation}
 B_0 = - 2f_{24},   \quad
 A_0 = - 3f_{24},   \quad
 A_1 = - 4f_{24},   \quad
 \cdots,          \quad  
 A_n = - (3+n)f_{24}. 
\label{eq:StrongUNIV2}
\end{equation}
The expressions for $\mbox{\boldmath{$A$}}^{5}_i(M_{\rm cut})$ are still
 those in Eq.~(\ref{eq:effA5align}), with $A_0=-3 f_{24}$.
It will be shown in \S\ref{sec:UNIVERSALITY} that the two conditions
 in Eqs.~(\ref{eq:StrongUNIV1}) and~(\ref{eq:StrongUNIV2}) correspond
 to very precise patterns of mediation of SUSY breaking.
We anticipate that, in order to realize the condition in
 Eq.~(\ref{eq:StrongUNIV1}), the SUSY-breaking field ${\sf X}$ can only
 couple to the superpotential, whereas to realize the condition in
 Eq.~(\ref{eq:StrongUNIV2}), ${\sf X}$ couples only to the K\"ahler
 potential, in a certain basis. 
The more general condition in Eq.~(\ref{eq:StrongUNIVgen1}) is an 
 interpolation of the other two.

If neither of these conditions is satisfied, and the usual universality
 for soft couplings is used, in which
 $B_0$, $A_0$, $A_1$, $A_2, \cdots$ are all independent parameters,
 the proportionality between
 $\mbox{\boldmath{$A$}}^{5}_i(M_{\rm cut})$ and
 $\mbox{\boldmath{$Y$}}^{5}_i(M_{\rm cut})$ is, in general, broken at
 ${\cal O}(s)$.
Thus, sFVs of this order exist at the tree level.
We can nevertheless make the effort of assessing the level of
 arbitrariness introduced in the problem, as we have done in the case of
 the effective Yukawa couplings.
The hope is that, given the drastic reduction of parameters that the
 usual notion of universality still entails, the constraints acting on
 the effective trilinear couplings
 $\mbox{\boldmath{$A$}}^{5}_i(M_{\rm cut})$ are sufficiently numerous to
 eliminate a large part of this arbitrariness.

The situation is, indeed, better than it may look at first glance{{{}}}. 
Since the soft bilinear, trilinear, $n$-linear couplings of the original
 NROs are still aligned with the corresponding superpotential
 couplings, the two sets of effective couplings
 $\mbox{\boldmath{$Y$}}^{5}_i(M_{\rm cut})$,
 $\mbox{\boldmath{$Y$}}^{10}_j(M_{\rm cut}),\cdots$, and  
 $\mbox{\boldmath{$A$}}^{5}_i(M_{\rm cut})$,
 $\mbox{\boldmath{$A$}}^{10}_j(M_{\rm cut}),\cdots$ are not completely
 independent. 
This is because the matrices that constitute the input parameters in
 the set of effective Yukawa couplings
 $\mbox{\boldmath{$Y$}}^{5}_i(M_{\rm cut})$,
 $\mbox{\boldmath{$Y$}}^{10}_j(M_{\rm cut}),\cdots$ and in that of
 effective trilinear couplings are the same up to the coefficients
 $A_0$, $A_1$, $A_2$,~$\cdots$, whereas the number
 of output parameters, namely, the two sets of effective Yukawa and 
 trilinear couplings, is now doubled.
Additional constraints with respect to those listed in
 \S\S\ref{sec:effYUK} and \ref{sec:trilSOFTsecNROS} are, therefore,
 expected.
Besides the ${\cal O}(s \widetilde{m})$- and
 ${\cal O}(s^2 \widetilde{m})$-constraints, there exist also 
 ${\cal O}(s^3 \widetilde{m})$- and
 ${\cal O}(s^3 \widetilde{m}^2)$-constraints, whereas no
 ${\cal O}(s^3)$-constraint exists for the effective Yukawa couplings.

To be specific, in the limit $s \to 0$, we have the three
 ${\cal O}(s)$-constraints for the effective couplings
 $\mbox{\boldmath{$Y$}}^{5}_i(M_{\rm cut})$ of
 Eq.~(\ref{eq:5Constraint}), and the
 four ${\cal O}(s\widetilde{m})$-constraints for the soft parameters
 $\mbox{\boldmath{$A$}}^{5}_i(M_{\rm cut})$, which say trivially that in
 this limit the soft trilinear couplings are aligned with the Yukawa
 couplings. 
The proportionality constant is $A_0$.

At the next order, there exist five constraints: the 
 ${\cal O}(s^2)$-constraint on the effective couplings
 $\mbox{\boldmath{$Y$}}^{5}_i(M_{\rm cut})$ of
 Eq.~(\ref{eq:5Constraint}), and the following four
 ${\cal O}(s^2\widetilde{m})$-constraints on  
 $\mbox{\boldmath{$A$}}^{5}_i(M_{\rm cut})$:
\begin{eqnarray}
&&
\left\vert \Bigl(\mbox{\boldmath{$A$}}^{5}_i(M_{\rm cut})  
 -(A_1\!+\!f_{24}) \mbox{\boldmath{$Y$}}^{5}_{i}(M_{\rm cut})  
\right.
\nonumber\\
&&
\hspace*{1truecm}
\left.
 -(A_0\!-\!A_1\!-\!f_{24})
           \Bigl(\displaystyle{\frac25}
                 \mbox{\boldmath{$Y$}}^{5}_{E}
                 +\displaystyle{\frac35}
                 \mbox{\boldmath{$Y$}}^{5}_{DU}
           \Bigr)(M_{\rm cut})  
           \Bigr)_{(h,k)} \right\vert     
 \leq {\cal O}(s^2\widetilde{m}), 
\label{eq:s2A5Constraint}
\end{eqnarray}
 for $i={\scriptstyle D,E,DU,LQ}$, expressed in terms of $A_0$ and
 $A_1+f_{24}$.
Thus, the ${\cal O}(s\widetilde{m})$ of the matrices
 $\mbox{\boldmath{$A$}}^{5}_i(M_{\rm cut})$ is fixed by these
 parameters and the effective Yukawa couplings.
This means that if only NROs of dimension five are introduced, the
 matrices $\mbox{\boldmath{$A$}}^{5}_i(M_{\rm cut})$, and therefore, all
 misalignments between
 $\mbox{\boldmath{$A$}}^{5}_i(M_{\rm cut})  $ and
 $\mbox{\boldmath{$Y$}}^{5}_i(M_{\rm cut})  $, are known exactly once the 
 couplings $\mbox{\boldmath{$Y$}}^{5}_i(M_{\rm cut})$ are fixed.

Similar considerations hold also for the couplings
 $\mbox{\boldmath{$A$}}^{10}_j(M_{\rm cut})$
 ($j={\scriptstyle U,UE,QQ}$), which satisfy the following 
 ${\cal O}(s^2\widetilde{m})$-constraints:
\begin{eqnarray}
&&
\left\vert 
\Bigl(\mbox{\boldmath{$A$}}^{10}_j(M_{\rm cut})  
  -(A_1\!+\!f_{24}) \mbox{\boldmath{$Y$}}^{10}_{j}(M_{\rm cut})  
\right.
\nonumber\\
&&
\hspace*{1truecm}
\left.
  +(A_1\!-\!A_0\!+\!f_{24})
    \Bigl(\displaystyle{\frac{2}{5}}
    \left(\mbox{\boldmath{$Y$}}^{10}_{U}\right)^S
       +\!\displaystyle{\frac{3}{5}}
          \mbox{\boldmath{$Y$}}^{10}_{QQ}
    \Bigr)(M_{\rm cut})  
\Bigr)_{(h,k)}  \right\vert     
 \leq {\cal O}(s^2\tilde m). 
\label{eq:s2A10Constraint}
\end{eqnarray}

The contribution of ${\cal O}(s^2\widetilde{m})$ to the matrices
 $\mbox{\boldmath{$A$}}^{5}_i(M_{\rm cut})$ and
 $\mbox{\boldmath{$A$}}^{10}_j(M_{\rm cut})$ is not
 known, and it is only restricted to satisfy higher-order constraints.
In the case of $\mbox{\boldmath{$A$}}^{5}_i(M_{\rm cut})$, these are
\begin{eqnarray}
&&
\left\vert \Bigl(
    (\mbox{\boldmath{$A$}}^{5}_D   \!-\!\mbox{\boldmath{$A$}}^{5}_E
\!+\!\mbox{\boldmath{$A$}}^{5}_{LQ}\!-\!\mbox{\boldmath{$A$}}^{5}_{DU}
    )(M_{\rm cut})  
\right.
\nonumber\\
&&
\hspace*{1truecm}
\left.
 -\left(A_2\!+\!2 f_{24}\right)(
     \mbox{\boldmath{$Y$}}^{5}_D   \!-\!\mbox{\boldmath{$Y$}}^{5}_E
\!+\!\mbox{\boldmath{$Y$}}^{5}_{LQ}\!-\!\mbox{\boldmath{$Y$}}^{5}_{DU}
                             )(M_{\rm cut})  
           \Bigr)_{(h,k)} \right\vert
 \leq {\cal O}(s^3\widetilde{m}), 
\label{eq:s3A5Constr1}
\end{eqnarray}
 a ${\cal O}(s^3\widetilde{m})$-constraint, and a constraint at
 ${\cal O}(s^3\widetilde{m}^2)$:
\begin{eqnarray}
&&
\left\vert \Bigl(
  6\left(A_2-\!A_0+\!2f_{24}\right)
   \left(
   \mbox{\boldmath{$A$}}^{5}_E  -\!\mbox{\boldmath{$A$}}^{5}_{DU}
   \right)(M_{\rm cut})  
\right.
\nonumber\\
&&
\hspace*{1truecm}
  -\left(A_2-\!A_1+f_{24}\right)
   \left(
  2\mbox{\boldmath{$A$}}^{5}_E +\!3\mbox{\boldmath{$A$}}^{5}_{DU}
   \right)(M_{\rm cut})  
\phantom{\Bigr)_{(h,k)}}
\nonumber\\
&&
\hspace*{1truecm}
  -6 \left(A_1+\!f_{24}\right)\left(A_2-\!A_0+\!2f_{24}\right)
   \left(
   \mbox{\boldmath{$Y$}}^{5}_{E}-\!\mbox{\boldmath{$Y$}}^{5}_{DU}
   \right)(M_{\rm cut})   
\nonumber\\
&&
\hspace*{1truecm}
\left.
  +A_0\left(A_2-\!A_1+f_{24}\right)
   \left(
  2\mbox{\boldmath{$Y$}}^{5}_{E}+\!3\mbox{\boldmath{$Y$}}^{5}_{DU}
   \right)(M_{\rm cut})  
           \Bigr)_{(h,k)} \right\vert     
 \leq {\cal O}(s^3\widetilde{m}^2). 
\label{eq:s3A5Constr2}
\end{eqnarray}
It should be recalled here that these constraints as well as the
 expressions for the effective trilinear couplings themselves are 
 obtained in {\it SU}(5)-symmetric bases.
As mentioned at the end of \S\ref{sec:KalerPNROS}, 
 {\it SU}(5)-breaking field redefinitions must be performed to 
 eliminate from the K\"ahler potential NROs of dimension six in which 
 the fields $24_H$ acquire~VEVs.
These dimension-six operators will be discussed explicitly in 
 \S\ref{sec:UNIVERSALITY}. 
In the basis in which such operators are removed, the 
 ${\cal O}(s^3\widetilde{m})$- and
 ${\cal O}(s^3\widetilde{m}^2)$-constraints cannot be used.

This situation is clearly very different from that in which not even
 the usual universality for soft couplings is required. 
In this case, the two sets of couplings
 $\mbox{\boldmath{$Y$}}^{5}_i(M_{\rm cut})$,
 $\mbox{\boldmath{$Y$}}^{10}_j(M_{\rm cut}),\cdots$, and 
 $\mbox{\boldmath{$A$}}^{5}_i(M_{\rm cut})$,
 $\mbox{\boldmath{$A$}}^{10}_j(M_{\rm cut}),\cdots$ are completely
 independent. 
The counting of output parameters (the effective couplings) and 
 input parameters (the original couplings of the renormalizable and
 nonrenormalizable operators) is the same for the two sets.
The constraints for the effective trilinear couplings
 $\mbox{\boldmath{$A$}}^{5}_i(M_{\rm cut})$,
 $\mbox{\boldmath{$A$}}^{10}_j(M_{\rm cut}),\cdots$ are then formally
 equal to those for
 $\mbox{\boldmath{$Y$}}^{5}_i(M_{\rm cut})$,
 $\mbox{\boldmath{$Y$}}^{10}_j(M_{\rm cut})\cdots$~(see
 \S\ref{sec:trilSOFTsecNROS}). 
It is clearly very difficult to deal with a situation like this.
The only hope is that, by expressing these effective couplings in the
 {\it SU}(5)-symmetric basis of Eq.~(\ref{eq:YeffRotations}) as
\begin{eqnarray}
   \mbox{\boldmath{$A$}}^{5}_i
                                       \    \to &  
  (\Delta\widetilde V_{5\,i}^\dagger)^T\, 
   \,\widehat{\!\mbox{\boldmath{$A$}}}_i^{5}\, 
   \Delta\widetilde V_{10\,i}^\dagger ,             & 
  \quad (i={\scriptstyle D,E,DU,LQ})
\nonumber\\
 \,\mbox{\boldmath{$A$}}^{10}_j              \to &  
  \left[(\Delta\widetilde W_{10\,j}^{\prime\dagger})
               \widetilde{K}_{\rm CKM}\right]^T
     \widehat{\mbox{\boldmath{$A$}}}_j^{10}  P_{10} 
  \left[\Delta\widetilde W_{10\,j}^\dagger
              \widetilde{K}_{\rm CKM}\right],     & 
  \quad (j= {\scriptstyle U,UE,QQ})  
\label{eq:AeffRotations}
\end{eqnarray}
 it is possible to find some limit in which the number of parameters
 needed to specify the mismatch matrices in these equations can be
 reduced considerably, as in the case of the effective Yukawa couplings. 
Note that, unlike $\Delta V_{5\,E}$, $\Delta V_{10\,D}$ and
 $\Delta W_{10\,U}$~(see Eq.~(\ref{eq:ROTATidentif})), the mismatch
 matrices $\Delta \widetilde V_{5\,E}$, $\Delta \widetilde V_{10\,D}$
 and $\Delta \widetilde W_{10\,U}$ do not coincide with the unit matrix.
These and all mismatch matrices in these equations, including
 $\widetilde{K}_{\rm CKM}$, have a tilde to distinguish them from the
 mismatch matrices for the effective Yukawa couplings.
They do reduce to those matrices in the limit $s\to 0$.

\subsection{Effective sfermion masses}
\label{sec:NROsBC-MassSOFT}
%
As in the case of trilinear soft couplings, a simple generalization 
 of the concept of universality that assigns the same
 flavour-independent coupling to operators with the same dimensionality:
\begin{equation}
  d^0_{0,0} = d^0_0 ,                         
\qquad
  d^{0}_{n,m} = d^{0}_{n+m} , 
\qquad
  d^{1}_{n,m} = d^{1}_{n+m} , 
\label{eq:usualUNIVERS2}
\end{equation}
 with $n=0,1,\cdots$, and $m=1,2,\cdots$, is not sufficient to guarantee
 the flavour independence of the various contributions to the effective 
 masses, at ${\cal O}(s)$ in the expansion for 
 $\widetilde{\mbox{\boldmath{$m$}}}_{f}^2$ ($f={\scriptstyle D^c,L}$).
These contributions appear in the right--right sector of the down-squark
 mass matrix and in the left--left sector of the charged slepton mass
 matrix, precisely those affected by the large seesaw couplings. 
Moreover, being of ${\cal O}(s)$, they are numerically of the same size
 as the contributions induced through RGEs by the couplings of the
 various seesaw mediators, which are{{}} of ${\cal O}(s_{\rm loop})$.

{}Further restrictions on the parameters in
 Eq.~(\ref{eq:usualUNIVERS2}), such as
\begin{equation}
  d^0_0  =  d^{0}_{n+m}  =  d^{1}_{n+m}= \cdots = \delta, 
\label{eq:KaehlerstrUNIVERS}
\end{equation}
 could simplify considerably the form of the SUSY-breaking part of the
 K\"ahler potential, leading, for example, to the K\"ahler potential
 sector: 
\begin{equation}
\left[ 1 + \delta f_{\sf X} \theta^2 + 
           \delta^\ast f_{\sf X}^\ast \bar{\theta}^2 + 
     \vert \delta\vert^2 \vert f_{\sf X}
     \vert^2 {\theta}^2 \bar{\theta}^2  
\right]
 \Bigl\{
   \bar5_M \bar5_M^*
 +\Bigl[
     Op^{{\cal K}_1}_{\bar{5}_M} 
 +\! Op^{{\cal K}_2}_{\bar{5}_M}  + {\rm H.c.}
  \Bigr]
 \Bigr\} ,
\label{eq:KaehlerstrUNIVERS2}
\end{equation}
 where we have replaced ${\sf X}/M_{\rm cut}$ with
 $f_{\sf X}\,\theta^2$.

Clearly, the conditions under which these tree-level sFVs vanish are the
 conditions under which field redefinitions can remove simultaneously
 NROs in the SUSY-conserving and SUSY-breaking parts of the K\"ahler
 potential.
As mentioned earlier, it is sufficient to be able to remove from 
 Eq.~(\ref{eq:KaehlerstrUNIVERS2}) those NROs, which give 
 contributions up to ${\cal O}(s^2)$, once the field $24_H$ is 
 replaced by its~VEVs~(see next section).

\section{Universal boundary conditions free from sFVs}
\label{sec:UNIVERSALITY}
%
To eliminate sFVs at the tree level, an extension of the
 conventional notion of universality to the case with nonvanishing NROs
 must require at least that the field ${\sf X}$, which communicates
 SUSY breaking, couples to operators in the superpotential and K\"ahler 
 potential in a generation-independent way, and without distinguishing
 different fields.\footnote{Such a definition excludes the mechanism of
  gauge mediation of SUSY breaking. 
  It can, however, be generalized to include it, by restricting the
  above assumption to flavour independence only.
  In our setup, the mediation of SUSY breaking must happen above
  $M_{\rm GUT}$, at the Planck scale, or quite close to it, if any of
  the effects discussed in this paper is to be detected. 
  Thus, a component of gauge mediation, if present, has to coexist with
  that of gravity mediation.
  The latter is inevitably present at such scales and, in general, 
  dominant since it does not suffer from the loop suppression that 
  plagues the former.
  In the following, we restrict our discussion to the case of pure
  gravity mediation of SUSY breaking.}
In particular, within each of the two potentials, it couples to
 operators of the{{{}}} same dimensionality in the same way.

We refer to the concept of universality outlined above as weak or
 unrestricted universality, as opposite to that in which the different
 couplings of ${\sf X}$ to the various operators in the superpotential
 and K\"ahler potential are, in addition, restricted by special
 conditions.
We call this second type of universality restricted or strong
 universality.  
Examples of such conditions were outlined already in
 \S\S\ref{sec:NROsBC-TrilSOFT} and \ref{sec:NROsBC-MassSOFT}.
Here, we reexamine  in a systematic way the problem of which conditions 
 are to be imposed to the superpotential and K\"ahler potential 
 in order eliminate sFVs at the tree level.

In the case of the superpotential, a weak universality implies
\begin{eqnarray}
 W  &=&
   M_{ij}  \left(1 + b f_X \theta^2\right)\phi_i\phi_j 
\nonumber\\
&&
\hspace*{-0.5truecm}
  +Y_{ijk} \left(1 + a f_X \theta^2\right)\phi_i\phi_j\phi_k
  +\displaystyle{\frac{1}{M_{\rm cut}}}
    C_{ijkl}\left(1 +a_1 f_X\theta^2\right)\phi_i\phi_j\phi_k\phi_l
  +\cdots,
\label{eq:Wuniv}
\end{eqnarray}
 where the notation is simplified with respect to that of the previous
 sections, as indices denote both field types and generations. 
The fields $\phi_i$ in this superpotential are {\it SU}(5) multiplets, 
 $M$ is $M_5$ or $M_{24}$, or a seesaw massive parameter, $Y$ any
 trilinear coupling, and $C$ any NRO coupling.
(Here, we limit ourselves to NROs of dimension five, for simplicity.)
The parameter $f_X$ is the usual ratio $F_X/M_{\rm cut}$ that signals
 the breaking of SUSY, and the quantities $b$, $a$, $a_1$, $\dots$ are
 just numbers.
A comparison with the notation of \S\ref{sec:NROsBC-TrilSOFT} 
 allows the identifications: $B_0 = b f_{\sf X}$,  $A_0 = a f_{\sf X}$, 
 $A_1 = a_1 f_{\sf X}$, $\cdots$, and $f_{24} = (b-a)f_{\sf X}$.

We postpone the listing of the general form of what we call the weakly 
 universal K\"ahler potential in a noncanonical form, to draw 
 immediately some observations about the behaviour of the superpotential
 under field redefinitions.
These are required to reduce the K\"ahler potential to be minimal.
In general, field redefinitions look like
\begin{equation}
 \phi_i \to
  \left(c_{ij} +\tilde{c}_{ij} f_X \theta^2\right)\phi_j
 +\frac{1}{M_{\rm cut}}
  \left((d_5)_{ijk}+(\tilde{d}_5)_{ijk}f_X \theta^2\right)\phi_j\phi_k
 + \cdots ,
\label{eq:FRedefGENERAL}
\end{equation}
 where the second parenthesis and the following ones, hidden in the
 dots, are present only when NROs are nonvanishing.
(The field ${\sf X}$, treated here as a spurion, is not included in the 
 counting of the dimensionality of the various operators.)
It is obvious that if the mechanism of mediation of SUSY breaking is
 even only weakly universal, the couplings $\tilde{c}_{ij}$ and
 $(\tilde{d}_5)_{ijk}$ are not arbitrary.  
That is, they cannot depend on field type and generations.
Therefore, Eq.~(\ref{eq:FRedefGENERAL}) should be reduced to 
\begin{equation}
 \phi_i         \to
  c_{ij}\left(1 +\delta_K f_X \theta^2\right)\phi_j
 +\frac{1}{M_{\rm cut}}
  \left(d_5\right)_{ijk}
        \left(1+\delta_N f_X \theta^2\right)\phi_j\phi_k +
  \cdots ,
\label{eq:FRedefMinUNIV}
\end{equation}
 where $\delta_K$ and $\delta_N$ are just numbers.

With these field redefinitions, the above superpotential is modified 
 as
\begin{eqnarray}
 W          &\to&
   M_{lm} c_{li} c_{mj}
   \left[1 + \left(b+2 \delta_K\right) f_X \theta^2\right]
                              \phi_i\phi_j 
\nonumber\\[1.5ex] &   &
\hspace*{-0.5truecm} 
+\ Y_{lmn} c_{li} c_{mj} c_{nk} 
   \left[1 + \left(a+3 \delta_K\right) f_X \theta^2\right]
                              \phi_i\phi_j\phi_k
\nonumber\\[1.1ex] &   &
\hspace*{-0.5truecm} 
   +\displaystyle{\frac{1}{M_{\rm cut}}}
 2 M_{lm} c_{li} \left(d_5\right)_{mjk}
   \left[1 + \left(b+ \delta_K+\delta_N\right) f_X \theta^2\right]
                             \phi_i\phi_j\phi_k
\nonumber\\ &   &
\hspace*{-0.5truecm} 
   +\displaystyle{\frac{1}{M_{\rm cut}}}
   C_{lmno} c_{li} c_{mj} c_{nk} c_{oh}
   \left[1 +\left(a_1 +4 \delta_K\right) f_X\theta^2\right]
                                \phi_i\phi_j\phi_k\phi_h
\nonumber\\ &   &
\hspace*{-0.5truecm} 
   +\displaystyle{\frac{1}{M_{\rm cut}}}
 3 Y_{lmn} c_{li} c_{mj} \left(d_5\right)_{nkh}
   \left[1 + \left(a+ 2\delta_K+\delta_N\right) f_X \theta^2\right]
                              \phi_i\phi_j\phi_k\phi_h
\nonumber\\ &   &
\hspace*{-0.5truecm} 
   +\displaystyle{\frac{1}{M_{\rm cut}^2}}
   M_{lm} \left(d_5\right)_{lij} \left(d_5\right)_{mkh}
   \left[1 + \left(b+ 2\delta_N\right) f_X \theta^2\right]
                              \phi_i\phi_j\phi_k\phi_h
+ \cdots.
\label{eq:Wrotated}
\end{eqnarray}
The minimal assumption of Eq.~(\ref{eq:FRedefMinUNIV}), however, is
 not sufficient to preserve the alignment of bilinear, trilinear,
 $n$-linear soft terms with the corresponding SUSY-conserving terms,
 i.e., to preserve weak universality, unless
\begin{equation}
 \delta_N  =  a  -b + 2\delta_K  
           =  a_1-a + 2\delta_K 
           = \cdots .  
\label{eq:STABLEunivCONDIT}
\end{equation}
Together with the definition of $f_{24}$, this implies
\begin{equation}
  a_1 -a = a -b  = \cdots = -f_{24}/f_{\sf X},  
\label{eq:STABLEunivCONDITa}
\end{equation}
 which, although derived in a different way, coincides remarkably
 with the condition of Eq.~(\ref{eq:StrongUNIVgen}).

While the stability conditions in Eq.~(\ref{eq:STABLEunivCONDITa})
 relate only SUSY-breaking parameters in the superpotential, those in
 Eq.~(\ref{eq:STABLEunivCONDIT}) link SUSY-breaking parameters in the
 superpotential and in the K\"ahler potential.
These latter conditions  indicate that field redefinitions can remove
 SUSY-breaking operators in the K\"ahler potential without disturbing
 the alignment of SUSY-conserving and SUSY-breaking parameters in the
 superpotential only for specific couplings in both potentials.

Note that even SUSY-conserving field redefinitions, with
 $\delta_K=\delta_N=0$, in general, spoil weak universality, except
 for the particular choice of SUSY-breaking parameters in the
 superpotential: 
\begin{equation}
 b=a=a_1=a_2=\cdots, \qquad f_{24}=0.
\label{eq:couplEQUAL}
\end{equation}
This is the first notion of strong universality introduced in the
 previous section~(see Eq.~(\ref{eq:StrongUNIV1})).
It is encoded in the following particularly simple form for the
 superpotential:
\begin{equation}
 W              =
 \left(1 + a f_X \theta^2\right)
 \left[
     M_{ij}  \phi_i\phi_j 
    +Y_{ijk}\, \phi_i\phi_j\phi_k
    +\displaystyle{\frac{1}{M_{\rm cut}}}
     C_{ijkl}\,\phi_i\phi_j\phi_k\phi_l +\cdots \right].
\label{eq:WunivSTR1}
\end{equation}
This type of strong universality is, indeed, stable only if
 $\delta_K=\delta_N=0$, that is, if ${\sf X}$ does not couple to the
 K\"ahler potential, which can then be minimized by 
 SUSY-conserving field redefinitions.
They modify the superpotential by changing the definition of the
 parameters $M$, $Y$, $C, \cdots$, but leave unchanged the overall
 parenthesis $(1+a f_{\sf X}\theta^2)$, and therefore, also the alignment
 of effective trilinear couplings and effective Yukawa couplings. 
The soft masses squared are vanishing at $M_{\rm cut}$ and are then 
 generated radiatively, driven by gaugino masses and trilinear
 couplings.

It would be interesting at this point to check how the superpotential 
 of Eq.~(\ref{eq:WunivSTR1}) gets modified when  ${\sf X}$ couples also 
 to the K\"ahler potential and field redefinitions have nonvanishing
 $\delta_K$ and $\delta_N$.

We postpone momentarily this issue and we consider the other, very
 restrictive notion of universality for bilinear, trilinear, and
 $n$-linear soft parameters of Eq.~(\ref{eq:StrongUNIV2}), determined
 also by one parameter only, $f_{24}$.
This can now be expressed as
\begin{equation}
 b:a:a_1: \cdots = 2: 3:4: \cdots, \qquad
 a f_{\sf X} =  -3 f_{24}.
\label{eq:couplRATIOS}
\end{equation}
In this case, the conditions in Eq.~(\ref{eq:STABLEunivCONDIT})
 guarantee that, under field redefinitions,
 $a f_{\sf X}$, $b f_{\sf X}$, $a_1 f_{\sf X}$ are shifted in such a way
 as to leave the ratios $b:a:a_1\cdots$ unchanged.
{}From Eq.~(\ref{eq:Wrotated}), it is easy to see that $a f_{\sf X}$
 becomes
\begin{equation}
  a f_{\sf X} = -3 f_{24}\to -3 \left(f_{24} -\delta_K f_{\sf X} \right).  
\label{eq:aSHIFT}
\end{equation} 
The superpotential, which has the form
\begin{eqnarray}
 W         &=&
   M_{ij}  \left(1 -2 f_{24} \theta^2\right)\phi_i\phi_j 
\nonumber\\ &&
\hspace*{-0.45truecm}
  +Y_{ijk} \left(1 -3 f_{24} \theta^2\right)\phi_i\phi_j\phi_k
  +\displaystyle{\frac{1}{M_{\rm cut}}}
   C_{ijkl}\left(1 -4 f_{24} \theta^2\right)\phi_i\phi_j\phi_k\phi_l
  +\cdots
\label{eq:WunivSTR2}
\end{eqnarray}
 before field redefinitions, remains of the same type, except for a
 modification of the parameters $M$, $Y$, $C, \cdots$ and for the shift 
 of $f_{24}$:
\begin{equation}
  f_{24} \to  \left(f_{24} -\delta_K f_{\sf X} \right). 
\label{eq:f24SHIFT}
\end{equation} 
The alignment
 $\mbox{\boldmath{$A$}}^{5}_i = A_0 \mbox{\boldmath{$Y$}}^{5}_i$ of 
 Eq.~(\ref{eq:effA5align}) is satisfied for $A_0 =-3 f_{24}$, with
 different values of $f_{24}$ before and after field redefinitions.
Note that, if $\delta_K f_{\sf X}= f_{24}$, it is possible to find a
 basis in which the shifted value of $f_{24}$, and therefore, all the
 parameters $b$, $a$, $a_1$, $\cdots$, are vanishing.
In this basis, ${\sf X}$ couples only to the K\"ahler potential.
Bilinear, trilinear and $n$-linear soft terms satisfying the condition
 of Eq.~(\ref{eq:couplRATIOS}) are generated as a result of field
 redefinitions.

This fact illustrates why the superpotential in
 Eq.~(\ref{eq:WunivSTR1}) tends to be destabilized, unless $X$ is
 decoupled from the K\"ahler potential. 
It is simply due to the fact that SUSY-breaking field redefinitions
 generate the parameter $f_{24}$, even if this vanishes in one basis.

As the previous one, the more general notion of universality of 
 Eq.~(\ref{eq:StrongUNIVgen1}) seems also stable under the field 
 redefinitions of Eq.~(\ref{eq:FRedefMinUNIV}).
Depending on the two parameters $A_0^\prime$ and $f_{24}$, this general 
 notion of universality is nothing else but an interpolation of the 
 two notions considered above. 
It can now be expressed as
\begin{equation}
 b-a^\prime:a-a^\prime:a_1-a^\prime \cdots = 2: 3:4 \cdots,
 \qquad
 (a-a^\prime)f_{\sf X} = -3 f_{24},
\label{eq:couplRATIOS2}
\end{equation}
 with $a^\prime = A_0^\prime/f_{\sf X}$.
By rewriting the parameters $b$, $a$, $a_1$, $\cdots$ as 
 $a^\prime +(b-a^\prime)$, 
 $a^\prime +(a-a^\prime)$,
 $a^\prime +(a_1-a^\prime)$, $\cdots$,
 we obtain the following form for Eq.~(\ref{eq:STABLEunivCONDIT}):
\begin {equation}
 \delta_N         =
 \left[\left(a-a^\prime\right) -\left(b-a^\prime\right)\right] +
  2 \delta_K      = 
 \left[\left(a_1-a^\prime\right) -\left(a-a^\prime\right)\right] +
  2 \delta_K .
\label{eq:STABLEunivCONDIT2}
\end{equation}
These conditions guarantee that the ratios
 $b-a^\prime:a-a^\prime:a_1-a^\prime\cdots$ are not modified by field
 redefinitions, which shift $(a-a^\prime)f_{\sf X}$ as follows:
\begin{equation}
  (a-a^\prime)f_{\sf X}
  =  -3 f_{24}
 \to -3\left(f_{24}-\delta_K f_{\sf X}\right). 
\label{eq:f24SHIFT2}
\end{equation}
Thus, the superpotential, which has the interesting form
\begin{eqnarray}
 W              &=&
  \left(1 +a^\prime f_{\sf X} \theta^2\right)
\left[
   M_{ij}  \left(1 -2 f_{24} \theta^2\right)\phi_i\phi_j 
  +Y_{ijk} \left(1 -3 f_{24} \theta^2\right)\phi_i\phi_j\phi_k
\phantom{\displaystyle{\frac{1}{M_{\rm cut}}}}
\right.
\nonumber\\     & & 
\left.
\phantom{
  \left(1 +a^\prime f_{\sf X} \theta^2\right)}
\hspace*{-0.3truecm}
  +\displaystyle{\frac{1}{M_{\rm cut}}}
   C_{ijkl}\left(1 -4 f_{24} \theta^2\right)\phi_i\phi_j\phi_k\phi_l
  +\cdots
\right]
\label{eq:WunivSTR3}
\end{eqnarray}
 remains formally unchanged by field redefinitions, except for the shift
 of $f_{24}$ of Eq.~(\ref{eq:f24SHIFT2}), and a modification of the
 parameters $M$, $Y$, $C,\cdots$.
The alignment
 $\mbox{\boldmath{$A$}}^{5}_i = A_0 \mbox{\boldmath{$Y$}}^{5}_i$ of 
 Eq.~(\ref{eq:effA5align}) is again satisfied for
 $A_0 =A_0^\prime -3 f_{24}$, with different values of $f_{24}$ before and
 after field redefinitions.

It is to the form in Eq.~(\ref{eq:WunivSTR3}) that the superpotential
 in Eq.~(\ref{eq:WunivSTR1})
 is brought by field redefinitions with $\delta_K f_{\sf X}=-f_{24}$. 
Vice versa, field redefinitions with $\delta_K f_{\sf X}=f_{24}$ bring
 the superpotential in Eq.~(\ref{eq:WunivSTR3}) to the form in
 Eq.~(\ref{eq:WunivSTR1}).

We have discovered so far that a weakly universal superpotential with
 NROs as in Eq.~(\ref{eq:Wuniv}) is unstable under field redefinitions,
 even SUSY-conserving ones, if NROs exist also in the K\"ahler
 potential.
There would be no problem at all if such a superpotential could be
 somehow realized in a basis in which the K\"ahler potential is already
 canonical. 
In general, however, this is not the case and the superpotential must be
 strongly universal to prevent field redefinitions from spoiling even 
 its weak universality.
Any of the two forms in Eq.~(\ref{eq:WunivSTR1}) or~(\ref{eq:WunivSTR3})
 is equally suitable.

Note that in the MSSM limit, i.e., if only light superfields
 are present, NROs are negligible in both the superpotential and 
 the K\"ahler potential, and the need for a restricted notion of
 universality to avoid sFVs at the tree level does not exist.

A weakly{{{}}} universal K\"ahler potential, to which
 Eq.~(\ref{eq:usualUNIVERS2}) already alludes, is  
\begin{eqnarray}
 K({\sf X}^\ast,{\sf X})          & = & 
 \left[
 1 - \alpha_4 f_{\sf X} \theta^2 
   - \alpha_4^\ast f_{\sf X}^\ast \bar{\theta}^2 
   + \left(\left\vert\alpha_4\right\vert^2 - d^2 \right)
       \left\vert f_{\sf X}\right\vert^2 \theta^2 \bar{\theta}^2
 \right] \phi^\ast_i \phi_i
\nonumber\\[1.8ex]                 &   &
\hspace*{-0.5truecm}
-
\Bigl\{
 \frac{\left(d_5\right)_{ijk}}{M_{\rm cut}}
 \left[
 1 +\alpha_5^0 f_{\sf X} \theta^2 
   +\beta_5^{0\ast} f_{\sf X}^\ast \bar{\theta}^2
   +\gamma_5^0
    \left\vert f_{\sf X}\right\vert^2 \theta^2\bar{\theta}^2 
 \right] \phi^\ast_i \phi_j \phi_k  +{\rm H.c.}
\Bigr\}
\nonumber\\[1.8ex]                 &   &
\hspace*{-0.5truecm}
-\ \frac{\left(d_6\right)_{ijkh}}{M_{\rm cut}^2}
 \left[
 1 +\alpha_6^0 f_{\sf X} \theta^2 
   +\alpha_6^{0\ast} f_{\sf X}^\ast \bar{\theta}^2 
   +\gamma_6^0
    \left\vert f_{\sf X}\right\vert^2 \theta^2\bar{\theta}^2 
 \right] \phi^\ast_i \phi_j^\ast \phi_k \phi_h
\nonumber\\[1.8ex]                 &   &
\hspace*{-0.5truecm}
-
\Bigl\{
 \frac{\left(d_6^\prime\right)_{ijkh}}{M_{\rm cut}^2}
 \left[
 1 + \alpha_6^\prime f_{\sf X} \theta^2 
   + \beta_6^{\prime\ast} f_{\sf X}^\ast \bar{\theta}^2 
   +\gamma_6^\prime
    \left\vert f_{\sf X}\right\vert^2 \theta^2\bar{\theta}^2 
 \right] \phi^\ast_i \phi_j \phi_k \phi_h
   + {\rm H.c.}
\Bigr\} , 
\nonumber\\                         &   &
\label{eq:KaelerWUNIV}
\end{eqnarray}
 where, without loss of generality, we have already assumed the
 canonical form for the SUSY-conserving dimension-four operators.
The parameter  $d^2$ is assumed to be real and positive, since, as will
 be shown later, it determines the soft mass squared of the fields
 $\phi_i$.
The couplings $(d_5)_{ijk}$ and $(d_6^\prime)_{ijkl}$ are symmetric
 under permutations of the last two and the last three indices,
 respectively, and $d_6$ satisfies the following relations:
 $\left(d_6\right)_{ijkh}= \left(d_6\right)_{jikh}= 
  \left(d_6\right)_{ijhk}= \left(d_6^{\ast}\right)_{khij}$.
{}For simplicity, we also limit this discussion to operators up to
 dimension six.
As will be shown in \S\ref{sec:EffPicLOOPlevel}, within the precision
 of our calculation, these are sufficient for the determination of the
 boundary conditions for effective soft masses.

As already discovered in the case of weakly universal superpotentials,
 this K\"ahler potential is also unstable under field redefinitions.
It is a simple exercise to show this. 
As in the case of the superpotential, the stability of the weak
 universality implies conditions among its various parameters and the
 field-redefinition parameters, which we do not report here explicitly.
It is straightforward, although tedious, to show that for a 
 K\"ahler potential and a superpotential in the basis specified by 
 Eqs.~(\ref{eq:KaelerWUNIV}) and~(\ref{eq:WunivSTR1}), constraints on
 the various parameters in $K({\sf X}^\ast,{\sf X})$ are induced by the
 requirement that NROs of dimension five can be eliminated.
These constraints, together with the conditions of stability of the 
 weak universality of both potentials, allow us to rewrite
 $K({\sf X}^\ast,{\sf X})$ in the following factorizable form:
\begin{equation}
 K({\sf X}^\ast,{\sf X}) = 
\Big[ 
 1 - \delta_K      f_{\sf X} \theta^2 
   - \delta_K^\ast f_{\sf X}^\ast \bar{\theta}^2 
   + \left(\left\vert \delta_K\vert^2 - d ^2 \right)
           \right\vert f_{\sf X} \vert^2 \theta^2 \bar{\theta}^2
\Big] K ,
\label{eq:KAELERstrongUNIV}
\end{equation} 
 where $\delta_K = \alpha_4$, and $K$ is the
 SUSY-conserving part of the K\"ahler potential:
\begin{eqnarray}
 K          &=&
  \phi^\ast_i \phi_i
 -\left[\frac{\left(d_5\right)_{ijk}}{M_{\rm cut}}
              \phi^\ast_i \phi_j \phi_k  + {\rm H.c.}  \right] 
 -\frac{(d_6)_{ijkh}}{M_{\rm cut}^2}
              \phi^\ast_i \phi_j^\ast \phi_k \phi_h
\nonumber\\[1.1ex] & & 
\hspace*{1truecm}
 -\left[\frac{(d_6^\prime)_{ijkh}}{M_{\rm cut}^2}
              \phi^\ast_i \phi_j \phi_k \phi_h + {\rm H.c.} \right] . 
\label{eq:KAELERnoX}
\end{eqnarray}
The form in Eq.~(\ref{eq:KAELERstrongUNIV}) is one example of a K\"ahler
 potential with strong universality.
The superpotential is still that of Eq.~(\ref{eq:WunivSTR1}), and
 $f_{24}$ vanishes.\footnote{If in the basis of
  Eq.~(\ref{eq:KaelerWUNIV}) for $K({\sf X}^\ast,{\sf X})$ the
  superpotential is that of Eq.~(\ref{eq:WunivSTR3}), the same procedure
  followed in the previous case yields a slightly more complicated form
  for the K\"ahler potential than that in
  Eq.~(\ref{eq:KAELERstrongUNIV}), but equivalent to it.
 {}Field redefinitions, indeed, bring it to the factorizable form of
  Eq.~(\ref{eq:KAELERstrongUNIV}), with
  $\delta_K f_{\sf X} = \alpha_4 f_{\sf X} -f_{24}$, while the
  superpotential is brought to the form in Eq.~(\ref{eq:WunivSTR1}).
 Thus, irrespective of whether the superpotential is in the form of 
  Eq.~(\ref{eq:WunivSTR1}) or~(\ref{eq:WunivSTR3}) in the basis of 
  Eq.~(\ref{eq:KaelerWUNIV}) for the K\"ahler potential, the final form
  of the superpotential in the basis in which the K\"ahler potential is
  factorized as in Eq.~(\ref{eq:KAELERstrongUNIV}) is always that of
  Eq.~(\ref{eq:WunivSTR1}), i.e., also a factorized one.}
Note that the guess made in \S\ref{sec:NROsBC-MassSOFT} (see
 Eq.~(\ref{eq:KaehlerstrUNIVERS2})) for a K\"ahler potential, which
 would not lead to sFVs at the tree{{{}}} level, is in the right
 direction but not completely correct.
The expression in Eq.~(\ref{eq:KaehlerstrUNIVERS2}) is, indeed, very
 similar to that in Eq.~(\ref{eq:KAELERstrongUNIV}), when $\delta_K$
 is identified with $-\delta$, but, in reality, less general, as it has
 one less parameter.

It is interesting to see that when $d^2=\vert\delta_K\vert^2$,
 the scalar potential obtained from
 Eqs.~(\ref{eq:WunivSTR1}) and~(\ref{eq:KAELERstrongUNIV})
 coincides with that obtained in a supergravity model in which the 
 superpotential and the K\"ahler potential for the visible sector are 
 assumed to be sequestered and have the same expression as 
 those in Eqs.~(\ref{eq:WunivSTR1}) and (\ref{eq:KAELERstrongUNIV}) with 
 ${\sf X}\to0$.

Thanks to the factorization of Eq.~(\ref{eq:KAELERstrongUNIV}), the
 effort in bringing $K({\sf X}^\ast,{\sf X})$ to a form as close as
 possible to the canonical one reduces to that of minimizing $K$.
The elimination in $K$ of the dimension-five NROs, which is possible to 
 achieve, automatically eliminates also all the corresponding
 SUSY-breaking NROs of the same dimensionality.
In the language of the previous sections, since it is possible to
 redefine away the dimension-five NROs in
 $Op^{{\cal K}_1}_{\bar{5}_M}$, {{{}}} all the dimension-five NROs in 
 $Op^{{\cal K}_1}_{\bar{5}_M}({\sf X})$,
 $Op^{{\cal K}_1}_{\bar{5}_M}({\sf X}^\ast)$, and
 $Op^{{\cal K}_1}_{\bar{5}_M}({\sf X}^\ast\!,{\sf X})$ can also{{{}}} be
 eliminated.
This is particularly important as it enables the elimination of the most
 dangerous arbitrary sFVs in the boundary values for soft masses,
 i.e., those that can be of the same size as the sFVs induced
 radiatively by the neutrino seesaw couplings.

In fact, the situation is even better than that just described.
Both terms in square brackets in Eq.~(\ref{eq:KAELERnoX}) can be removed
 through field redefinitions, as they both have all fields except one
 with the same chirality, therefore reducing $K({\sf X}^\ast,{\sf X})$
 to
$$
 K({\sf X}^\ast,{\sf X}) = 
\Big[ 
 1 - \delta_K      f_{\sf X} \theta^2 
   - \delta_K^\ast f_{\sf X}^\ast \bar{\theta}^2 
   + \left(\left\vert \delta_K\vert^2- d^2\right)
           \right\vert f_{\sf X} \vert^2 \theta^2 \bar{\theta}^2
\Big]
$$
\begin{equation}
\hspace*{5truecm}
\times\left\{
  \phi^\ast_i \phi_i
 -\frac{(d_6)_{ijkh}}{M_{\rm cut}^2}
              \phi^\ast_i \phi_j^\ast \phi_k \phi_h
\right\}.
\label{eq:KAELERstrongUNIV2}
\end{equation} 
While these field redefinitions are SUSY conserving, SUSY-breaking ones
 are needed to bring $K({\sf X}^\ast,{\sf X})$ to the even simpler form:
$$
 K({\sf X}^\ast,{\sf X}) = 
\Big[ 1 - d^2 \vert f_{\sf X}\vert^2
                         \theta^2 \bar{\theta}^2
\Big]  \phi^\ast_i \phi_i
\hspace*{8truecm}
$$
\begin{equation}
+\Big[ 
 1 + \delta_K      f_{\sf X} \theta^2 
   + \delta_K^\ast f_{\sf X}^\ast \bar{\theta}^2 
   + \left(\left\vert \delta_K\vert^2 - d^2\right)
           \right\vert f_{\sf X} \vert^2 \theta^2 \bar{\theta}^2
\Big]
\frac{(-d_6)_{ijkh}}{M_{\rm cut}^2}
              \phi^\ast_i \phi_j^\ast \phi_k \phi_h ,
\label{eq:KAELERdim4and6}
\end{equation} 
 in which the form of the dimension-four operators is canonical.
Through them, $f_{24}$ is generated again, with the value
 $f_{24} = -\delta_K f_{\sf X}$, and the superpotential is brought from
 the unstable form of Eq.~(\ref{eq:WunivSTR1}) to the more general form
 of Eq.~(\ref{eq:WunivSTR3}).
It is clear at this point why the guess of
 Eq.~(\ref{eq:KaehlerstrUNIVERS2}) for the K\"ahler potential is not
 correct. 
These SUSY-breaking field redefinitions would reduce the soft masses to
 be vanishing.

{}For all practical purposes, we can neglect all dimension-six NROs,
 except when they contain two $24_H$ fields, with opposite chirality, as
 these fields can acquire~VEVs, thus contributing in general in a
 flavour-{{{}}}dependent way to soft masses.
If these two fields are $\phi_j^\ast \phi_k$, then the term multiplied
 by the square bracket in the second line of
 Eq.~(\ref{eq:KAELERdim4and6}) becomes
\begin{equation}
s^2
(-d_6)_{ijjh}
\left(1 + f_{24} \theta^2 + f_{24}^\ast \bar{\theta}^2 + 
 \vert f_{24} \vert^2 \theta^2 \bar{\theta}^2 \right) 
              \phi^\ast_i \phi_h ,
\label{eq:dim6reduced}
\end{equation}
 and the index $j$ labels the components of the field $24_H$.
Group theoretical factors are suppressed in this expression and
 are assumed to be reabsorbed by $d_6$.
The second line of Eq.~(\ref{eq:KAELERdim4and6}) is
$$
\Big[ 1
+\Bigl(\delta_K     f_{\sf X}     +\!f_{24}   \Bigr)     \theta^2 
+\Bigl(\delta_K^\ast f_{\sf X}^\ast +\!f_{24}^\ast\Bigr)\bar{\theta}^2 
+\Bigl(\left\vert \delta_K f_{\sf X} +\!f_{24}\right\vert^2
        -d^2 \vert f_{\sf X} \vert^2         \Bigr)\theta^2\bar{\theta}^2
\Big]
$$
\begin{equation}
 \hspace*{1truecm} \times s^2 (-d_6)_{ijjh} \phi^\ast_i \phi_h , 
\label{eq:dim6}
\end{equation}
 and recalling that $f_{24}=-\delta_K f_{\sf X}$, 
 the K\"ahler potential of Eq.~(\ref{eq:KAELERdim4and6}) reduces to
\begin{equation}
 K({\sf X}^\ast,{\sf X}) \simeq 
\Big[ 1 - d^2 \vert f_{\sf X}\vert^2
                         \theta^2 \bar{\theta}^2
\Big]
\left(\delta_{ih} - s^2 (d_6)_{ijjh}\right) \phi^\ast_i \phi_h ,
\label{eq:KpotALMOSTmin} 
\end{equation}
 up to negligible NROs of dimension six.
There is something very special about this cancellation, which rests on
 the simultaneous presence of strong universality in both potentials.
Had it been  $\delta_K f_{\sf X}+f_{24}\ne 0$, {\it SU}(5)-breaking field
 redefinitions could have removed the two terms proportional to
 $\theta^2$ and $\bar{\theta}^2$, but not the term proportional to
 $\theta^2 \bar{\theta}^2$, which multiplies $d_6$.
These field redefinitions would have also shifted $f_{24}$ of a quantity
 of ${\cal O}(s^2)$ proportional to $d_6$.
Hence, misalignments would have been introduced in the superpotential, 
 correlated to the residual sFVs of ${\cal O}(s^2)$ in the K\"ahler
 potential.

A further SUSY-conserving, but {\it SU}(5)-breaking field redefinition,
 yields
\begin{equation}
 K({\sf X}^\ast,{\sf X}) \simeq 
\Big[ 1 - d^2 \vert f_{\sf X}\vert^2
                         \theta^2 \bar{\theta}^2
\Big]
  \phi^\ast_i \phi_i .
\label{eq:KpotMIN} 
\end{equation}
Note that these are also flavour-violating field redefinitions of 
 ${\cal O}(s^2)$. 
Since they are SUSY-conserving, their effect is only that of modifying
 the parameters $M$, $Y$, $C,\cdots$ in the superpotential, that is,
 of modifying simultaneously supersymmetric parameters and corresponding 
 soft parameters, without introducing spurious
 misalignments at the tree level.\footnote{Possible misalignments may be
 introduced at the loop level, which, however, would be of 
 ${\cal O}(s^2 \times s_{\rm loop})$ and, therefore, completely negligible
 for our purposes.}
The superpotential is still formally that of Eq.~(\ref{eq:WunivSTR3}),
 as if these field redefinitions had not taken place. 
They can, therefore, be totally ignored if one needs to use this
 superpotential in an {\it SU}(5)-symmetric basis.
Three free parameters, together with the gaugino mass, are needed in
 this case to specify all soft terms:
\begin{equation}
 a^\prime, \quad f_{24}, \quad d^2,
\label{eq:STRUNIVone} 
\end{equation}
 with $d^2$ real and positive. 
Had we taken $d^2 =\vert \alpha_4 \vert^2$ in
 Eq.~(\ref{eq:KaelerWUNIV}), Eq.~(\ref{eq:KpotMIN}) would now be
\begin{equation}
 K({\sf X}^\ast,{\sf X}) \simeq 
\Big[ 1 - \vert f_{24}\vert^2 \theta^2 \bar{\theta}^2 
\Big]  \phi^\ast_i \phi_i ,
\label{eq:KpotMINtwo} 
\end{equation}
 and only the parameters $ a^\prime$ and $f_{24}$ would have to be
 specified in addition to the gaugino mass.

In summary, for a SUSY-breaking mediator coupled to operators of both
 potentials, the condition to avoid sFVs at the tree level is that of
 having, in the same basis, the superpotential of
 Eq.~(\ref{eq:WunivSTR1}) and the K\"ahler potential of
 Eq.~(\ref{eq:KAELERstrongUNIV}), or equivalently, in the form of
 Eqs.~(\ref{eq:WunivSTR3}) and~(\ref{eq:KpotMIN}), respectively.

As mentioned earlier, it is possible that the SUSY-breaking mediator
 couples only to the superpotential.
The K\"ahler potential, now SUSY conserving, is in this case that of
 Eq.~(\ref{eq:KAELERnoX}).
This can be easily reduced with SUSY-conserving field redefinitions to
 have the form of the potential in curly brackets in
 Eq.~(\ref{eq:KAELERstrongUNIV2}).
This possibility was brought to attention when discussing the
 superpotential in Eq.~(\ref{eq:WunivSTR1}).
Since in this basis $f_{24}=0$, the K\"ahler potential reduces directly
 to 
\begin{equation}
 K   \simeq \ 
 \Big[ \delta_{ih} - s^2 (d_6)_{ijjh}
 \Big] \phi^\ast_i \phi_h ,
\label{eq:KnoscaleALMOSTmin} 
\end{equation}
 when the fields $\phi_j^\ast \phi_j$, identified with 
 $24_H^\ast 24_H$, acquire their scalar~VEV.
Again, an {\it SU}(5)-breaking field redefinition can remove the second 
 term in parentheses.

Note that, if the condition of universality for the superpotential is 
 relaxed to be weak instead of strong in the basis of
 Eq.~(\ref{eq:KaelerWUNIV}) for the K\"ahler potential, it is still 
 possible to tune the K\"ahler potential to have the form in 
 Eq.~(\ref{eq:KAELERstrongUNIV}).
An attempt to minimize it in this case yields
\begin{equation}
 K({\sf X}^\ast,{\sf X}) \simeq  \
\Big[ \delta_{ih} + 
  \Bigl( \vert d\vert^2 \vert f_{\sf X} \vert^2  \delta_{ih}
  - s^2 (d_6)_{ijjh}    \vert f_{24}+\delta_K f_{\sf X}\vert^2
  \Bigr)\theta^2 \bar{\theta}^2
\Big] \phi^\ast_i \phi_h 
\label{eq:KpotNOmin6} 
\end{equation}
 up to negligible dimension-six NROs, but destroys also the weak  
 universality of the superpotential.
Thus, situations in which the superpotential is weakly universal and
 the K\"ahler potential is canonical, up to terms of ${\cal O}(s^2)$, 
 which are arbitrary, and, in general, flavour- and {\it CP}-violating,
 can only be achieved at the expense of some severe tuning.  
Highly tuned is also a situation in which the superpotential is
 weakly universal and the K\"ahler potential is already canonical, 
 as in Eq.~(\ref{eq:KpotMIN}).

\section{Emerging scenarios}
\label{sec:EmerScen}
%
The analysis of the previous sections shows clearly that the
 seesaw-induced predictions for sFVs in an MSSU(5) setting can vary
 considerably. 
There are of course the interesting differences induced by the
 various types of seesaw mechanism.
If large enough, they may help distinguish which of the three types of
 seesaw mediators gives rise to the low-energy dimension-five neutrino
 operator of Eq.~(\ref{eq:nu}).

There are also the worrisome modifications that the seesaw-induced
 sFVs may undergo, depending on the type of NROs introduced and the role
 that they play. 
The arbitrariness of NROs is the key ingredient exploited to solve the
 two major problems of the MSSU(5) model, the unacceptable predictions
 for the values of fermion masses and for the proton-decay rate. 
How to use it to correct the fermion spectrum is straightforward and
 well known, whereas the recipe to be followed to suppress the decay
 rate of the proton is not simple and presumably not unique. 
Indeed, we have argued in \S\ref{sec:PDconstBC} that
 baryon-number-violating NROs may also produce a sufficient suppression
 of this rate,
 in addition to that due to the freedom in the effective Yukawa
 couplings of  matter fields to colored Higgs
 triplets~\cite{ZURAB,GORAN,DESYpeople}.
Thus, it is not yet clear whether the constraints from proton decay 
 imply that the  mismatch matrices
 $\Delta V_{5\,i}$,
 $\Delta V_{10\,i}$ ($i={\scriptstyle D,E,DU,LQ}$), and  
 $\Delta W_{10\,j}$, 
 $\Delta W_{10\,j}^\prime$ ($j= {\scriptstyle U,UE,QQ}$)
 discussed in \S\ref{sec:NROsBC-YUKAWA} are trivial or not. 
Waiting to get a definite answer on this issue, we assume here that
 these matrices are nontrivial, i.e., that they are substantially
 different from the unit matrix.

As for the arbitrariness that NROs can induce in the boundary conditions
 for soft parameters, the results obtained in
 \S\ref{sec:UNIVERSALITY} can be summarized as follows.
In general, this arbitrariness causes particularly dangerous tree-level 
 sFVs already at ${\cal O}(s)$, which completely obscure the effect of the
 large seesaw Yukawa couplings. 
There exist, however, special field- and flavour-independent couplings of
 the SUSY-breaking mediator field ${\sf X}$ to the superpotential and
 the K\"ahler potential, or strongly universal couplings, which are not 
 destabilized by field redefinitions, and which allow us to eliminate
 such tree-level sFVs.
In addition to the gaugino mass, $M_{1/2}$, only three
 parameters are sufficient to specify all effective soft couplings:
\begin{equation}
\widetilde{m}_0^2, \quad A_0^\prime, \quad f_{24}, 
\label{eq:STRUNIVpar}
\end{equation}
 the first two of which are expressed in \S\ref{sec:UNIVERSALITY} as 
 $d^2 \vert f_{\sf X}\vert^2 $ and $a^\prime f_{\sf X}$. 
At $M_{\rm cut}$, all effective soft masses squared are equal to 
 $\widetilde{m}_0^2$:
\begin{equation}
 \widetilde{\mbox{\boldmath{$m$}}}_{i}^2(M_{\rm cut}) =  
 \widetilde{m}_0^2 \, {\mathop{\bf 1}},
\label{eq:StrongStrongBC1}
\end{equation}
 for any field $i$.
The effective trilinear couplings are aligned with the effective Yukawa
 couplings, as for example in
\begin{equation}
 \mbox{\boldmath{$A$}}^{5}_i(M_{\rm cut}) =
 A_0 \mbox{\boldmath{$Y$}}^{5}_i  \qquad (i={\scriptstyle D,E,DU,LQ})
\label{eq:StrongStrongBC2}
\end{equation}
 with
\begin{equation}
    A_0 = A_0^\prime - 3 f_{24}. 
\label{eq:STRUNIVA0}
\end{equation}
All $B$ parameters are expressed in terms of the corresponding 
 superpotential massive parameter as\footnote{We have not given
  explicitly definitions of effective bilinear couplings in terms of the
  original renormalizable and nonrenormalizable operators. 
  They are completely analogous to those for the trilinear effective
  couplings.}
\begin{equation}
 \mbox{\boldmath{$B$}}_i (M_{\rm cut}) =  B_0  M_i , 
\label{eq:StrongStrongBC3}
\end{equation}
 with
\begin{equation}
 B_0 = A_0^\prime - 2 f_{24}
     = A_0 + f_{24}.
\label{eq:STRUNIVB0}
\end{equation}
That is, we recover for effective couplings the type of universality
 usually advocated for models without NROs.
Indeed, the choice of three parameters at $M_{\rm cut}$ can be the much
 more familiar
\begin{equation}
\widetilde{m}_0^2, \quad A_0, \quad B_0, 
\label{eq:STRUNIVparALTERN}
\end{equation}
 with $f_{24}$ fixed in terms of $B_0$ and $A_0$: $f_{24} = B_0 - A_0$. 
The choice $d^2 = \vert\alpha_4\vert^2$ in Eq.~(\ref{eq:STRUNIVone})
 corresponds to fixing $B_0$ among the above parameters to be
 $(B_0 - A_0)^2 =\widetilde{m}_0^2$, reproducing the minimal supergravity 
 result~\cite{BFS}.
Note that all the bilinear terms discussed above are for the
 superheavy fields, and the MSSM $B$-term can be arbitrary even in this
 case, depending on the fine tuning for the $\mu$ term, as shown in
 \S\ref{sec:HiggsSECandVACUA}.

The case discussed so far is a somewhat special case, which rests
 uniquely on the requirement of simultaneous strong universality in the
 superpotential and in the K\"ahler potential.
With this choice of universality, the arbitrariness induced by NROs
 remains confined to the Yukawa sector. 
An estimate of how the predictions of the seesaw-induced sFVs are
 modified by the mismatch matrices will be given later in this section.

Beyond this special case, the situation becomes considerably more 
 complicated. 
{}Following \S\ref{sec:UNIVERSALITY}, we know that by relaxing the
 condition of universality for the superpotential to be
 weak instead of strong in the basis of Eq.~(\ref{eq:KaelerWUNIV})
 for the K\"ahler potential, we can tune the K\"ahler potential to be 
 as in Eq.~(\ref{eq:KAELERstrongUNIV}), which leads to the form in 
 Eq.~(\ref{eq:KpotNOmin6}), by field redefinitions.
That is, effective soft masses squared are universal at ${\cal O}(1)$, 
 but arbitrary at ${\cal O}(s^2)$.
In addition, the field redefinitions through which such form of the
 K\"ahler potential is obtained, spoil also the weak universality of the
 superpotential at ${\cal O}(s)$. 
At this order, the effective trilinear couplings, not aligned anymore to
 the effective Yukawa couplings, are arbitrary $3\times 3$ matrices.
It is possible to parametrize them as shown in
 Eq.~(\ref{eq:AeffRotations}), in terms of diagonal matrices and new
 unitary matrices of diagonalization mismatch, 
 $\Delta \widetilde V_{5\,i}$,
 $\Delta \widetilde V_{10\,i}$ ($i={\scriptstyle D,E,DU,LQ}$),  
 $\Delta \widetilde W_{10\,j}$, 
 $\Delta \widetilde W_{10\,j}^\prime$ ($j= {\scriptstyle U,UE,QQ}$), and
 $\widetilde{K}_{\rm CKM}$.
The number of free parameters, with which one has to deal,
 is, however, overwhelmingly large.

As explained in \S\ref{sec:UNIVERSALITY}, it is also possible to
 envisage another tuned situation, in which the superpotential has weak
 universality in the basis of Eq.~(\ref{eq:KpotMIN}) for the K\"ahler
 potential.
The effective soft masses squared are in this case universal.
Superpotential couplings and corresponding soft couplings are in general
 aligned, but the alignment is lost at  ${\cal O}(s)$ for effective
 Yukawa couplings and effective trilinear couplings.
This is the scenario analyzed in Ref.~\citen{BGOO}.
It is considerably simpler than the one discussed earlier, because,
 although not aligned to the effective Yukawa couplings, the effective
 trilinear couplings depend on them in precise ways, at least, at
 ${\cal O}(s)$.
In fact, the constraint in Eq.~(\ref{eq:s2A5Constraint}) shows that the
 misalignment between, for example, 
 $\mbox{\boldmath{$A$}}^{5}_i$ and the corresponding
 $\mbox{\boldmath{$Y$}}^{5}_i$ is given in terms of various effective
 Yukawa couplings and a new parameter, $A_1$, in addition to $A_0$ and
 $f_{24}$:
\begin{equation}
 \mbox{\boldmath{$A$}}^{5}_i(M_{\rm cut})    =
\left\{   (A_1+\!f_{24})
    +\frac{A_0-\!A_1-\!f_{24}}{5}
\left[\frac{
    2 \mbox{\boldmath{$Y$}}^5_{E}
 +\!3 \mbox{\boldmath{$Y$}}^5_{DU}}
           {\mbox{\boldmath{$Y$}}^{5}_{i}}
\right] (M_{\rm cut})
\right\}   {\mbox{\boldmath{$Y$}}^{5}_{i}}(M_{\rm cut}).
\label{eq:A5atORDs}
\end{equation}
Thus, at this order, all effective soft couplings can be specified in
 terms of
\begin{equation}
\widetilde{m}_0^2,  \quad A_0, \quad A_1, \quad f_{24},
\label{eq:tunedCASEpar}
\end{equation}
 and the gaugino mass, $M_{1/2}$. 
The remaining arbitrariness in the expressions of the various effective
 trilinear couplings is, at this order, the arbitrariness of the
 effective Yukawa couplings, which are not fixed by their 
 ${\cal O}(s^2)$-constraints given in 
 Eq.~(\ref{eq:5Constraint}). 
Constraints at ${\cal O}(s^2)$ among the various effective trilinear 
 couplings do exist.
These are the ${\cal O}(s^3\widetilde{m})$- and the
 ${\cal O}(s^3\widetilde{m}^2)$-constraints given explicitly for the
 couplings 
 $\mbox{\boldmath{$A$}}^{5}_i$ ($i={\scriptstyle D,E,DU,LQ}$), in
 Eqs.~(\ref{eq:s3A5Constr1}) and~(\ref{eq:s3A5Constr2}).
They link effective trilinear couplings and effective Yukawa couplings
 at this order, making use also of the parameters $A_0$, $A_1$, $f_{24}$,
 and an additional one, $A_2$.
Although insufficient to fix the effective trilinear couplings
 completely, they seem nevertheless useful to limit their arbitrariness.

\section{Picture of effective couplings at the quantum level}
\label{sec:EffPicLOOPlevel}
%
The picture of effective couplings outlined so far may be valid at the
 quantum level under two obvious conditions: 
 i) the dynamical part of the field $24_H$ that appears in NROs can be
 neglected when considering low-energy physics;
ii) NROs that do not contain the field $24_H$ are also irrelevant for
 low energy, except perhaps for proton decay and
 cosmology~\cite{ADDGHR}.
These two conditions are not valid in general, but only for a limited
 accuracy of our calculation of sFVs. 
It can be easily shown that, within this accuracy, the effective Yukawa
 couplings evolve in the same way as the couplings of an MSSU(5) model
 with the {\it SU}(5) symmetry broken at ${\cal O}(s)$ everywhere except
 that in the gauge sector. 
A key ingredient for this proof is the running of the VEVs of the
 adjoint Higgs field $24_H$, discussed also in
 \S\ref{sec:HiggsSECandVACUA} and Appendix~\ref{sec:New}.

\begin{wrapfigure}{r}{6.6cm}
\centerline{\includegraphics[width=6.6cm]{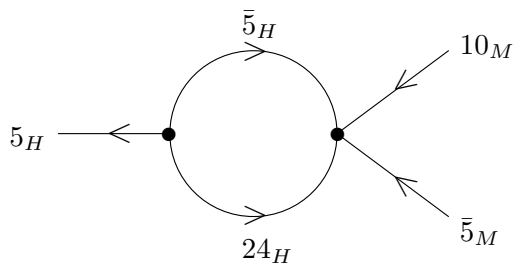}}
\caption{NRO induced by
         $Op^{5}$ and $\lambda_5 5_H 24_H \bar{5}_H$ in
         $W^{\rm MSSU(5)}_H$.}
\label{fig:NROloop}
\end{wrapfigure}
%
\subsection{ Effective-picture validity and allowed accuracy}
\label{sec:ValidityEffPic}
%
It is easy to see that the conditions i) and ii) listed above are
 in general violated. 
We show it here in the case of effective Yukawa couplings, but the
 same discussion can be applied to the holomorphic SUSY-breaking
 terms. 
It is shown in Fig.~\ref{fig:NROloop} how two operators in the
 superpotential, say, for definiteness, the NRO in $Op^{5}\vert_5$ with
 coefficient $C_{0,1}^{5}$ and the renormalizable one
 $\lambda_5 5_H 24_{H} \bar{5}_H$, induce in the K\"ahler potential the
 NRO
\begin{equation}
 x \lambda_5\, 5_H^\ast 10_M \,C_{0,1}^{5\,T}\,\bar{5}_M.  
  \quad \quad 
 \bigl(x \propto \frac{s_{\rm loop}}{M_{\rm cut}}\bigr) 
\label{eq:LOOPlambda5NRO}
\end{equation}
The field redefinition needed to reabsorb it,
\begin{equation}
 5_H \ \to \ 5_H - x \lambda_5 10_M \,C_{0,1}^{5\,T}\,\bar{5}_M, 
\label{eq:shift}
\end{equation}
 shifts also the various terms in the superpotential to induce
 NROs such as 
\begin{equation}
 -x \lambda_5 \,10_M \,Y^{10}\,10_M 10_M\,C_{0,1}^{5\,T}\,\bar{5}_M ,
\quad\quad
 -x \lambda_5 \, N^c \,Y^{\rm I}_N\,\bar{5}_M 10_M\,C_{0,1}^{5\,T}
                               \,\bar{5}_M ,
\label{eq:LOOPinducPDoper}
\end{equation}
 which do not contain the field $24_H$ and are, therefore, not included
 in the linear combinations that define the various
 $\mbox{\boldmath{$Y$}}^5_i$, as well as
\begin{equation}
 -x \lambda_5 \left(  M_5 \bar{5}_M \,C_{0,1}^{5}\,10_M \bar{5}_H   +   
    \lambda_5\,\bar{5}_M \,C_{0,1}^{5}\,10_M 24_H \bar{5}_H \right). 
\label{eq:24dynamEFFECT}
\end{equation}
Thus, at the one-loop level, the dynamical part of the field $24_H$
 can give rise to corrections to some of the coefficients included
 in the effective couplings $\mbox{\boldmath{$Y$}}^5_i$, although not
 to the complete linear combinations that define them.
In our specific case, for example, it induces a correction to the
 coefficient of the renormalizable operator $Op^{5}\vert_4$ of
 ${\cal O}(s\times s_{\rm loop})$, since $M_5$ is of the same order of
 $v_{24}$, and one to the coefficients of $Op^{5}\vert_5$ itself, of
 ${\cal O}(s_{\rm loop})$ and, therefore, of
 ${\cal O}(s\times s_{\rm loop})$ to the effective couplings
 $\mbox{\boldmath{$Y$}}^5_i$.

\begin{wrapfigure}{l}{6.6cm}
\centerline{\includegraphics[width=6.6cm]{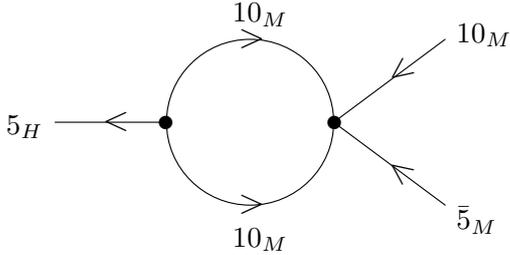}}
\caption{NRO induced by
         $Op^{\rm PD}$ and $ 10_M 10_M 5_H$ in
         $W^{\rm MSSU(5)}_M$. }
\label{fig:PDNROloop}
\end{wrapfigure}
Similarly, as shown in Fig.~\ref{fig:PDNROloop}, $Op^{5}\vert_4$ and
 the dimension-five operator in the class $Op^{\rm PD}$ in
 Eq.~(\ref{eq:supNROforPD}), also give rise at the one-loop level to
 the K\"ahler potential NRO~$5_H^\ast 10_M \bar{5}_M$ and, therefore,
 to superpotential NROs such as those listed above. 

The suppression factor is also in this case 
 $ x \propto {s_{\rm loop}}/{M_{\rm cut}}$ if the dimension-five NRO in
 $Op^{\rm PD}$ exists at the tree level and the various components of
 its coefficient  $C_0^{\rm PD}$ are of ${\cal O}(1)$. 
Strictly speaking, some of the flavour components of this coefficient
 must be sufficiently suppressed to avoid conflict with the experimental
 lower bound on the lifetime of the proton.
Others, however, may happen to be unsuppressed.

The same apparently problematic results are also induced, for example,
 by the tree-level NRO $N^c \bar{5}_M 10_M \bar{5}_M$ and the seesaw
 operator $-N^c Y_N^{\rm I} \bar{5}_M 5_H $.

A closer inspection shows, however, that the only corrections of 
 ${\cal O}(s\times s_{\rm loop})$ are those to terms of 
 $\mbox{\boldmath{$Y$}}^5_{DU}$ and 
 $\mbox{\boldmath{$Y$}}^5_{LQ}$, i.e., effective couplings of
 Yukawa interactions involving the superheavy Higgs triplet $H_D^C$.
Through RGEs, these can, therefore, induce sFVs at
 ${\cal O}(s\times s_{\rm loop}^2)$.
In contrast, because the combination
 $\mu_2= M_5-(1/2)\sqrt{6/5} \lambda_5 v_{24}$ has already been tuned to
 be of ${\cal O}(M_{\rm weak})$ (see \S\ref{sec:HiggsSECandVACUA}),
 the corrections that the two operators in 
 Eq.~(\ref{eq:24dynamEFFECT}) bring to the effective couplings of the
 Yukawa interactions that involve light Higgs fields,
 $\mbox{\boldmath{$Y$}}^5_D$ and 
 $\mbox{\boldmath{$Y$}}^5_E$, are negligibly small, of 
 ${\cal O}((M_{\rm weak}/M_{\rm cut})\times s_{\rm loop})$.

Thus, the picture of effective couplings can be retained at the quantum 
 level if we limit the precision of our calculation of sFVs, to be at
 most of ${\cal O}(s\times s_{\rm loop})$, but it cannot be used for 
 other quantities, such as the proton-decay Wilson coefficients, which
 depend at the tree level on the Yukawa couplings of the superheavy
 fields.
It will be shown in the next subsection, and proven explicitly in 
 Appendix~\ref{sec:proof}, that, up to a precision of 
 ${\cal O}(s\times s_{\rm loop})$ for sFVs, the effective couplings behave
 as couplings of renormalizable operators, i.e., they obey RGEs of
 renormalizable operators.

This limitation in precision is certainly not a problem, since it does
 not seem possible to probe sFVs beyond
 ${\cal O}(s\times s_{\rm loop}) =10^{-4}$.
If a higher sensitivity could, hypothetically, be achieved, the picture
 of effective coupling would have to be replaced by a different 
 treatment of the NROs, or, possibly, would require an enlargement of 
 basis, to contain also effective couplings for the NROs themselves.

The precision ${\cal O}(s\times s_{\rm loop})$ is that induced at the
 one-loop level by NROs of dimension five, that is, $Op^{5}\vert_5$ and
 $Op^{10}\vert_5$ in the Yukawa sector, with a nondynamical $24_H$. 
Thus, our calculation of low-energy quantities is not sensitive to
 quantum contributions from higher-dimension NROs, as well as to
 resummation effects obtained by solving the RGEs for effective
 couplings.
We nevertheless retain this picture, because we find it physically
 very clear.   
It leads to straightforward matching conditions at $M_{\rm GUT}$ and to
 a transparent interpretation of the boundary conditions at the cutoff
 scale.

There is one effect of this order that we do not include. 
As mentioned already, we neglect NROs in the Higgs sector.  
Barring small values of the coupling $\lambda_{24}$, the effect of these
 NROs is that of producing small shifts in the mass of the
 superheavy fields, such as the colored Higgs triplets.  
Shifts in the mass of these fields result in a modification of the
 scale at which these fields decouple from light matter.
On the product of effective couplings that enter in the evaluation of
 sLFVs:
\begin{equation}
 \frac1{16\pi^2}
 \mbox{\boldmath{$Y_N^{{\rm I}\,\dagger}$}}
 \mbox{\boldmath{$Y_N^{\rm I}$}} 
 \ln\left(\frac{M_{H^C}}{M_{\rm cut}}\right)
\sim 
 s_{\rm loop} 
 \left\{Y_N^{{\rm I}\,\dagger}Y_N^{\rm I} 
  -\frac{1}{2}\,s
 \left[ Y_N^{{\rm I}\,\dagger}   C_1^{N_{\rm I}} +   
        C_1^{N_{\rm I}\,\dagger} Y_N^{\rm I}
 \right] \cdots
 \right\}   ,  
\label{LoopEffect}
\end{equation}
 the effect of the shift $M_{H^C}\to M_{H^C}(1+{\cal O}(s))$ 
 is 
\begin{equation}
 \frac1{16\pi^2}
 \mbox{\boldmath{$Y_N^{{\rm I}\,\dagger}$}}
 \mbox{\boldmath{$Y_N^{\rm I}$}}
 \ln\left(\frac{M_{H^C}(1+{\cal O}(s))}{M_{H^C}}\right) 
\sim
 s \times s_{\rm loop} 
 \left\{ Y_N^{{\rm I}\,\dagger} Y_N^{\rm I} + \cdots
 \right\}.
\end{equation}
This term is of the same order of the term with a square bracket in 
 Eq.~(\ref{LoopEffect}). 
Differently from that term, however, this is only a flavour-blind shift
 of the leading term $s_{\rm loop}Y_N^{{\rm I}\dagger}Y_N^{\rm I}$, 
 which does not introduce an additional flavour structure.  
A simple treatment of the GUT threshold in which all superheavy masses
 are decoupled at the same scale also automatically disregards such 
 effects.

As an alternative to the effective-coupling picture, it is possible to use
 the procedure adopted by the authors of Ref.~\citen{BGOO}, in which
 each coupling of the various NROs is treated separately instead of
 being collected in a few effective couplings. 
We refer the reader to their work for details. 
As in our case, their treatment is also valid at most at
 ${\cal O}(s\times s_{\rm loop})$, and our analysis of boundary conditions
 for the NROs is also valid in their case.

\subsection{ RGE treatment of effective couplings}
\label{sec:NROinRGE}
%
The evolution of effective couplings is analogous to that of couplings
 of renormalizable operators, even above $M_{\rm GUT}$ when the field 
 $24_H$ is still active, provided the terms that break this picture are
 very small.
As shown in the previous subsection, these terms do exist, but can be
 neglected for a target accuracy  of ${\cal O}(s \times s_{\rm loop})$
 in the evaluation of sFVs.

The proof follows exactly the steps usually made when deriving the RGEs
 in renormalizable SUSY models. 
Since the superpotential is not renormalized, only the K\"ahler potential
 is subject to loop corrections, which determine the anomalous
 dimensions for the various fields.
As a result of these corrections, the K\"ahler potential loses its
 minimality. 
Its canonical (or nearly canonical) form can be recovered through field
 redefinitions, which affect also the superpotential.
The parameters in the superpotential and K\"ahler potential, expressed
 in terms of the redefined fields, are modified with respect to the 
 original ones. 
These modifications give rise to the RGEs for the model.

A detailed proof in this case of effective couplings is given in
 Appendix~\ref{sec:proof}.
Here, we concentrate only on some final steps of this proof, which allow
 a comparison of our analysis to others existing in the literature.

{}For simplicity, let us consider a generic superpotential term 
 $\mbox{\boldmath{$Y$}}(24_H)\Phi_1\Phi_2\Phi_3$, with 
\begin{equation}
\mbox{\boldmath{$Y$}}(24_H)=\sum_{n=0} C_n 24_H^n,
\label{eq:exTobeYEC}
\end{equation}
 and suppressed {\it SU}(5) indices. 
The quantity $\boldsymbol{Y}(\langle{24_H}\rangle)$ will become the effective 
 coupling for the operator $\Phi_1\Phi_2\Phi_3$, once the field $24_H$ 
 acquires its VEVs.

Through the procedure outlined above, we obtain the following 
 RGEs for the original couplings:
\begin{equation}
\dot C_{n}                        = 
\left(\mbox{\boldmath{$\gamma $}}_1 + 
      \mbox{\boldmath{$\gamma $}}_2 + 
      \mbox{\boldmath{$\gamma $}}_3 +
      n \mbox{\boldmath{$\gamma $}}_{24_H}
\right)       C_{n}, 
\label{eq:CnRGEgen}
\end{equation}
 in terms of effective couplings, on which the anomalous dimensions
 $\mbox{\boldmath{$\gamma$}}_i$ depend. 
We have neglected here the fact that $C_n$ and the various 
 $\mbox{\boldmath{$\gamma$}}_i$ in general do not commute. 
Since different $C_n$'s have different RGEs, with the ratio
 ${{C_n}/{C_0}}$ evolving as
 $n \mbox{\boldmath{$\gamma $}}_{24_H} {C_{n}}/{C_0}$, it seems that
 the anomalous dimension $\mbox{\boldmath{$\gamma $}}_{24_H}$ destroys
 the picture of effective couplings at the loop level.
The largest effect of $\gamma_{24}$ on the effective Yukawa couplings is
 of ${\cal O}(s\times s_{\rm loop})$. 
Through the Yukawa couplings, the effective soft masses squared are
 also affected, but the largest flavour-dependent effect is of
 ${\cal O}(s\times s_{\rm loop}^2)$, and can be neglected. 
There is, however, a larger effect induced by the anomalous dimension
 $\gamma_{24}$ on the effective soft masses squared, of 
 ${\cal O}(s\times s_{\rm loop})$.
This is due to the wave-function renormalization on the $n$-th term in
 the expansion of these masses in powers of the field $24_H$, analogous
 to the expansion for the Yukawa couplings. 
This effect is flavour-independent, and therefore, irrelevant for all
 practical purposes.

Similarly, for the coefficients of 
 $\mbox{\boldmath{$A$}}_{F_{\sf X}}(24_H)$,
\begin{equation}
\mbox{\boldmath{$A$}}_{F_{\sf X}}(24_H)=\sum_{n=0} \widetilde{C}_n 24_H^n,
\label{eq:exTobeAECFX}
\end{equation}
the part of the effective trilinear coupling corresponding to
 $\mbox{\boldmath{$Y$}}(24_H)$ (the only one considered in the
 literature), we obtain
\begin{equation}
\dot{\widetilde{C}}_{n}                        = 
\left(\mbox{\boldmath{$\gamma $}}_1 + 
      \mbox{\boldmath{$\gamma $}}_2 + 
      \mbox{\boldmath{$\gamma $}}_3 +
      n \mbox{\boldmath{$\gamma $}}_{24_H}
\right)       \widetilde{C}_{n} 
+
\left(\widetilde{\mbox{\boldmath{$\gamma $}}}_1 + 
      \widetilde{\mbox{\boldmath{$\gamma $}}}_2 + 
      \widetilde{\mbox{\boldmath{$\gamma $}}}_3 +
      n \widetilde{\mbox{\boldmath{$\gamma $}}}_{24_H}
\right)       {C}_{n} .
\label{eq:tildeCnRGEgen}
\end{equation}
The picture of effective couplings seems now to be destroyed by both 
 $\mbox{\boldmath{$\gamma $}}_{24_H}$ and
 $\widetilde{\mbox{\boldmath{$\gamma $}}}_{24_H}$.
Even assuming universal boundary conditions for all the
 $\widetilde{C}_n$ at the cutoff scale,
 $\widetilde{C}_n = A_n C_n$, with $A_n = A_0$, differences such as 
\begin{equation} 
 \Delta A = A_1-A_0 
\label{eq:DeltaA}
\end{equation}
 are generated at the one-loop level, and a misalignment of
 ${\cal O}(s\times s_{\rm loop})$ between the effective Yukawa couplings
 and trilinear couplings is introduced by
 $\mbox{\boldmath{$\gamma $}}_{24_H}$ and
 $\widetilde{\mbox{\boldmath{$\gamma $}}}_{24_H}$, i.e., precisely 
 of the order of the effects that we would like to retain in our
 calculation.
This is in addition to the usual misalignment induced also in the MSSM
 at ${\cal O}(s_{\rm loop})$  by the corresponding RGEs, expressed by the
 fact that the ratio $({\widetilde{C}_0}/{C_0})$ evolves as
 $(\tilde{\gamma}_1 + \tilde{\gamma}_2 + \tilde{\gamma}_3)$.
The former is smaller by a factor of ${\cal O}(s)$, but it is not
 governed by the CKM elements as the latter one.

All these effects are fictitious. 
To begin with, the running of the VEV $v_{24}$ cancels exactly the
 terms $n\gamma_{24}$ in the RGEs for $C_n$, giving for the effective 
 couplings:
\begin{equation}
\dot{\mbox{\boldmath{$Y$}}}\vert_{n}=
\dot{(C_{n} v_{24}^n)} = 
\left(\mbox{\boldmath{$\gamma $}}_1 + 
      \mbox{\boldmath{$\gamma $}}_2 + 
      \mbox{\boldmath{$\gamma $}}_3 \right)   
 \mbox{\boldmath{$Y$}}\vert_{n}. 
\label{eq:v24CnRGEgen}
\end{equation}
It cancels also the term $n\gamma_{24}$ in the RGE for $\widetilde{C}_n$,
 giving for
 $\mbox{\boldmath{$A$}}_{F_{\sf X}}\vert_{n}
   =(\widetilde{C}_{n} v_{24}^n)$:
\begin{equation}
\dot{\mbox{\boldmath{$A$}}}_{F_{\sf X}}\vert_{n} =
\left(\mbox{\boldmath{$\gamma $}}_1 + 
      \mbox{\boldmath{$\gamma $}}_2 + 
      \mbox{\boldmath{$\gamma $}}_3 
\right)       \mbox{\boldmath{$A$}}_{F_{\sf X}}\vert_{n}
+
\left(\widetilde{\mbox{\boldmath{$\gamma $}}}_1 + 
      \widetilde{\mbox{\boldmath{$\gamma $}}}_2 + 
      \widetilde{\mbox{\boldmath{$\gamma $}}}_3 
   + n \widetilde{\mbox{\boldmath{$\gamma $}}}_{24_H}
\right)   \mbox{\boldmath{$Y$}}\vert_{n}.
\label{eq:v24tildeCnRGEgen}
\end{equation}
The term $n \widetilde{\mbox{\boldmath{$\gamma $}}}_{24_H}$ in this
 equation is also fictitious.
It is cancelled by the running of the VEV $F_{24}$ in
\begin{equation}
\mbox{\boldmath{$A$}}_{F_{24}}(24_H)=\sum_{n=0} n f_{24}{C}_n 24_H^n , 
\label{eq:exTobeAECF24}
\end{equation}
 the other contribution to effective trilinear terms, neglected in
 the literature, whose $n$th term evolves as
\begin{equation}
\dot{\mbox{\boldmath{$A$}}}_{F_{24}}\vert_{n} =
\dot{(n f_{24} C_{n} v_{24}^n)} = 
\left(\mbox{\boldmath{$\gamma $}}_1 + 
      \mbox{\boldmath{$\gamma $}}_2 + 
      \mbox{\boldmath{$\gamma $}}_3 
\right)  \mbox{\boldmath{$A$}}_{F_{24}}\vert_{n} 
 - n \widetilde{\mbox{\boldmath{$\gamma $}}}_{24_H}
     \mbox{\boldmath{$Y$}}\vert_{n}.
\label{eq:f24CnRGEgen}
\end{equation}
Thus, the correct equation for the complete effective trilinear 
 coupling,
 $\mbox{\boldmath{$A$}}\vert_{n} =
 \mbox{\boldmath{$A$}}_{F_{\sf X}}\vert_{n} +
 \mbox{\boldmath{$A$}}_{F_{24}}\vert_{n} $, is recovered.

The picture of effective couplings is used at the quantum level also in
 Ref.~\citen{NROsFLsAHCH}.
No evolution of the VEVs of the field $24_H$ is considered there
 and the correct RGEs are obtained only as an approximation, simply by
 neglecting the terms $n\gamma_{24_H}$ and $n \tilde{\gamma}_{24_H}$ in
 Eqs.~(\ref{eq:CnRGEgen}) and~(\ref{eq:tildeCnRGEgen}).
The misalignment that these terms produce at
 ${\cal O}(s\times s_{\rm loop})$ between the effective trilinear
 couplings and effective Yukawa couplings is within the precision of our
 calculation and we cannot neglect it.

The authors of Ref.~\citen{BGOO} use a different procedure for their 
 analysis. 
We can, nevertheless, try to discuss it using our language of effective
 couplings.
Their RGEs between $M_{\rm cut}$ and $M_{\rm GUT}$ for the parameters
 corresponding to our  $C_n$ and $\widetilde{C}_n$ coincide with those
 in our Eqs.~(\ref{eq:CnRGEgen}) and~(\ref{eq:tildeCnRGEgen}).
These authors therefore claim that a term $\Delta A$, as in 
 Eq.~(\ref{eq:DeltaA}), is generated. 
As explained, this should be cancelled by the running of the VEVs
 in the effective couplings, which they ignore, together with the
 running of the parameters in the Higgs sector.
Thus, such a cancellation is not incorporated in their analysis. 
In particular, because the running of $F_{24}$ is not included, they do
 not realize that their vanishing value of $F_{24}$ at $M_{\rm GUT}$,
 which we deduce from their matching condition at this scale, is
 incompatible with an equally vanishing value at $M_{\rm cut}$, which
 they seem to adopt. 
Thus, their effective trilinear couplings at $M_{\rm cut}$ contain 
 already tree-level sFVs, which depend on the angle $\theta_D$ of the 
 mismatch matrix that we call $\Delta_D$ in Eqs.~(\ref{eq:NYUKparFIN})
 and~(\ref{eq:DeltaMatr}) (see Eq.~(\ref{eq:s2A5Constraint})).
We believe that the large dependence on this angle, which they find in
 the rate for $\mu\to e\gamma$ (see their Fig.~9), is due to sLFVs in
 their boundary values.

\section{Low-energy sFVs from seesaw mechanism and NROs}
\label{sec:Results}
%
The simplest scenario to deal with, and the one to which we restrict 
 ourselves in this section, is the first discussed in
 \S\ref{sec:EmerScen}.
This emerges essentially by imposing that the universal couplings of the
 SUSY-breaking field ${\sf X}$ to the superpotential and K\"ahler
 potential are stable under the field redefinitions of
 Eq.~(\ref{eq:FRedefMinUNIV}), and that the NROs of dimension five can
 be eliminated from the K\"ahler  potential. 
It requires the minimum number of soft parameters, guarantees
 universality of the effective soft masses squared and the alignment
 of trilinear effective couplings with effective Yukawa couplings.

The modifications induced in this scenario by NROs in the sFVs of the
 MSSU(5) model, through the effective Yukawa couplings, are twofold.
On one side, NROs contribute to sFVs through RGEs and alter the 
 pattern of sFVs of the MSSU(5) model at
 ${\cal O}(s\times s_{\rm loop})$.
The largest effect, however, is that of the mismatch matrices.

The mismatch in diagonalization of the various effective Yukawa
 couplings generated by the same class of operators, for example, 
 $\mbox{\boldmath{$Y$}}^{5}_{D}$, $\mbox{\boldmath{$Y$}}^{5}_{E}$, 
 $\mbox{\boldmath{$Y$}}^{5}_{LQ}$, and 
 $\mbox{\boldmath{$Y$}}^{5}_{DU}$, in the case of 
 $Op^{\bar{5}_M}$, affects also the various sectors of the sfermion mass
 matrices, thereby spoiling the correlations between sQFVs and sLFVs.  
It should be emphasized that these rotations can only modify 
 existing sFVs, but cannot generate them if they do not already exist. 
(For example, the term 
 $\widetilde{D}^c \widetilde{\mbox{\boldmath{$m$}}}_{D^c}^2
 \widetilde{D}^{c\,\ast}$ cannot be affected by mismatch matrices if 
 $\widetilde{\mbox{\boldmath{$m$}}}_{D^c}^2$ is proportional to the 
 unit matrix.)
Moreover, since their entries, or at least, their largest entries, can 
 be of ${\cal O}(1)$, these mismatch matrices alter the patterns of the
 MSSU(5)-model sFVs, producing modifications that can be of the same
 order of the sFVs themselves.

Note that there is only one mismatch matrix affecting in this indirect
 way the seesaw-induced sFVs in
 $\widetilde{\mbox{\boldmath{$m$}}}_{D^c}^2$, driven by
 $\mbox{\boldmath{$Y$}}^{\rm I}_{ND}$.
This is the matrix $\Delta_D$ defined in terms of $\Delta V_{5\,D}$
 and a diagonal matrix of phases in Eq.~(\ref{eq:sswYeffRotSU5br}).
{}For the fields in the $\bar{5}_M$ sector, there are also sFVs 
 induced by $\mbox{\boldmath{$Y$}}^{5}_{DU}$ and
 $\mbox{\boldmath{$Y$}}^{5}_{LQ}$, which involve several mismatch
 matrices, and for which the level of arbitrariness is large. 
These are, however, considerably smaller than those induced by the
 seesaw couplings.

Similarly, other mismatch matrices affect sFVs for fields in the $10_M$
 representation, which are driven by 
 $\mbox{\boldmath{$Y$}}^{10}_U$,
 $\mbox{\boldmath{$Y$}}^{10}_{QQ}$,
 $\mbox{\boldmath{$Y$}}^{10}_{UE}$. 
Thus, this sector is also plagued by a large degree of
 uncertainty/arbitrariness.    
This is true in particular for the right--right sector of the charged
 slepton mass matrix, which, in the MSSU(5) model, exhibits sFVs induced
 by the Yukawa couplings of the top quark~\cite{GUTeLFV}.

We leave the analysis of all possible sFVs for future work.
Here, we concentrate on the modifications induced by NROs of the
 MSSU(5)-model correlations between the seesaw-induced sQFVs and sLFVs,
 for each of the three types of seesaw mechanism. 
Explicit formulae for such correlations in the case of the seesaw of
 types~II and~III do not exist in the literature and we derive them in 
 the following, before going to the case of nrMSSU(5) models. 
Clearly, these predictions in the MSSU(5) model are also predictions 
 for those points in the parameter space opened up by NROs in 
 which the mismatch matrices are approximately the unit matrix.
Such points are obtained in situations like that described in
 \S\ref{sec:PDconstBC}, in which the approximate triviality of these
 matrices is induced by the proton-decay constraints.

We start by recalling the well-known relation between sLFVs and 
 sQFVs in the right--right down sector, for the seesaw of type~I:
\begin{equation}
\widetilde{m}^{2}_{D^c\,(i,j)}      =  
 \left(\!1+\frac{t_{\rm ssw}}{t_{\rm GUT}}\!\right)^{\!\!-1} \!\!\!
 \times\,
\widetilde{m}^{2\,\ast}_{L\,(i,j)},
 \quad \quad (i\ne j)
\label{eq:MoroiREL}
\end{equation}
 with $t_{\rm GUT}$ and $t_{\rm ssw}$ given by
\begin{equation}
 t_{\rm GUT}   = \frac{1}{16 \pi^2}
    \ln \left(\frac{M_{\rm GUT}}{M_{\rm cut}}\right), 
\quad \quad \quad 
 t_{\rm ssw}=  \frac{1}{16 \pi^2}
    \ln \left(\frac{M_{\rm ssw}}{M_{\rm GUT}}\right) .
\label{eq:tDEFs} 
\end{equation}
This relation can be easily obtained in an approximated way by
 integrating the RGEs for the off-diagonal elements of
 $\widetilde{m}^{2\,\ast}_{L}$ and $\widetilde{m}^{2}_{D^c}$ induced by
 the interaction  $- N^c Y_{N}^{\rm I} \bar{5}_M 5_H$  of
 Eq.~(\ref{eq:YUKAWAsswRotat}), neglecting contributions from all
 other Yukawa interactions and the scale dependence of the
 coefficients of these RGEs.
We obtain
\begin{eqnarray}
\widetilde{m}^{2\,\ast}_{L\,(i,j)} \ 
&=& 
 \left(t_{\rm GUT}\!+\!t_{\rm ssw}\right)    \times
\left. 
 {\cal F}\bigl(Y_{N}^{{\rm I}\,T},
         \widetilde{m}_{\bar{5}_M}^2,
         \widetilde{m}_{N^c}^{2\,\ast},
         \widetilde{m}_{5_H}^2, 
         A_{N}^{{\rm I}\,T}\bigr)_{(i,j)}  
\right\vert_{Q=M_{\rm cut}} , \qquad
\nonumber\\[1.01ex]
\widetilde{m}^{2}_{D^c\,(i,j)}               
&=& 
\quad \ \left(t_{\rm GUT}\right) \quad \times                   
\left.
 {\cal F}\bigl(Y_{N}^{{\rm I}\,T},
         \widetilde{m}_{\bar{5}_M}^2,
         \widetilde{m}_{N^c}^{2\,\ast},
         \widetilde{m}_{5_H}^2,
         A_{N}^{{\rm I}\,T}\bigr)_{(i,j)}   
\right\vert_{Q=M_{\rm cut}} , \qquad
\label{eq:onestepINTEGRATION}
\end{eqnarray}
 where the function ${\cal F}$, defined in Eq.~(\ref{eq:calFuncDEF}), 
 acquires the familiar expression
\begin{equation}
 \left. 
  {\cal F}\left(Y_{N}^{{\rm I}\,T},
          \widetilde{m}_{\bar{5}_M}^2,\widetilde{m}_{N^c}^{2\,\ast},
          \widetilde{m}_{5_H}^2, A_{N}^{{\rm I}\,T}\right)_{(i,j)}  
  \right\vert_{Q=M_{\rm cut}}  
 \!\!\!\!\! = \ 
\left.
  2 \left(3 \widetilde{m}^2_0 + A^2_0 \right)
  \left(Y^{{\rm I}\,T} _N Y^{{\rm I}\,\ast}_N\right)_{(i,j)}  \right\vert_{Q=M_{\rm cut}}  , 
\label{eq:onestepFUNCTION}
\end{equation}
 under the assumption of universality of soft masses.

As the relation in Eq.~(\ref{eq:MoroiREL}) shows, in the case of the 
 seesaw of type~I, sLFVs are larger than sQFVs in absolute value. 
This is because the Higgs triplets decouple at $M_{\rm GUT}$, a couple 
 orders of magnitude above $M_{\rm ssw}$.

In the case of the seesaw of type~II, the full {\it SU}(5) interaction 
 $\bar{5}_MY_{N}^{\rm II} 15_H\bar{5}_M$ remains active down to $M_{\rm ssw}$  
(see the decomposition of this term in Eq.~(\ref{eq:YUKAWAsswRotat})).
Thus, sLFVs and sQFVs are in this case expected to be of the same
 order up to subleading {\it SU}(5)-breaking effects in the RG flows 
 below $M_{\rm GUT}$.
As for the type~III, because two of the interactions inducing sLFVs and
 one of those inducing sQFVs survive between $M_{\rm GUT}$ and
 $M_{\rm ssw}$ (see the third decomposition in 
 Eq.~(\ref{eq:YUKAWAsswRotat})), the relation between sLFVs and sQFVs
 depends on group-theoretical factors.

An explicit calculation is particularly simple within the same 
 approximation used for the seesaw of type~I.
Indeed, relations similar to that shown in Eq.~(\ref{eq:MoroiREL})
 are obtained, where $t_{\rm ssw}$ has to be replaced by the product 
 of $t_{\rm ssw}$ and the functions $r_{\rm II}(t_{\rm ssw},t_{\rm GUT})$ and  
 $r_{\rm III}(t_{\rm ssw},t_{\rm GUT})$, respectively.
Including now the seesaw of type~I, we define a function
 $r_{\rm ssw}(t_{\rm ssw},t_{\rm GUT})$, which,{{{}}} in the three 
 cases, has the values
 \begin{equation}
  r_{\rm I}
  = 1, \quad\quad
  r_{\rm II}
 = 0, \quad\quad
  r_{\rm III}
 = -\left(24+10\frac{t_{\rm ssw}}{t_{\rm GUT}}\right)^{-1}. 
 \label{eq:rSSW} 
 \end{equation} 
The relation in Eq.~(\ref{eq:MoroiREL}) then becomes 
\begin{equation}
\widetilde{m}^{2}_{D^c\,(i,j)}      =  
 R_{\rm ssw} 
 \times\,
\widetilde{m}^{2\,\ast}_{L\,(i,j)},
 \quad \quad (i\ne j)
\label{eq:MoroiRELssw}
\end{equation}
 with 
\begin{equation}
 R_{\rm ssw} =
  \left[1+r_{\rm ssw}(t_{\rm ssw},t_{\rm GUT})
  \frac{t_{\rm ssw}}{t_{\rm GUT}}\right]^{\!\!-1} .
\label{eq:Rseesaw}
\end{equation}
Since the ratio $t_{\rm ssw}/t_{\rm GUT}$ is assumed here to be
 $ \lesssim 1$,  $r_{\rm III}$ is at the percent level, and 
 sLFVs and sQFVs differ also at the same order, as in the seesaw of 
 type~II.
We summarize the situation in {{{}}}Table~\ref{tab:seesaw}, where we list
 the {\it SU}(5) Yukawa interactions for the multiplets in which the seesaw
 mediators are embedded, their SM decompositions (omitting 
 signs and numerical coefficients) and the 
 size of sLFVs versus sQFVs, defined as the ratio 
 $\vert \widetilde{m}^{2}_{L\,(i,j)}/\widetilde{m}^{2}_{D^c\,(i,j)}\vert$ 
 with $i\neq j$ for the three types of seesaw mechanism.
We assume in this table an exact integration of the relevant RGEs, 
 which explain the value $\sim 1$ for this ratio in the case of 
 the seesaw of type~II. 
(The value $r_{\rm II}=0$ was obtained in a particular approximation.)

\begin{table}[pt]
\begin{center}                      
\caption{{\it SU}(5) Yukawa interactions for the seesaw multiplets 
         required by the three different types of seesaw mechanism, and
         their SM decompositions.
         Signs and numerical coefficients reported in the text are
         omitted here.
         The expected sizes of the induced sLFVs versus sQFVs,
         as defined in the text, are also given.}
\begin{tabular}{lccc}               
\hline          &&&\\[-1.9ex]       
           & type~I           & type~II            & type~III~(and~I) 
\\[1.01ex]\hline\hline&&&\\[-1.9ex]     
mediator    & $N^c$            & $15_H$            & $24_M$ 
\\[1.01ex]
interaction & $N^c\bar5_M5_H$  &\quad \quad $\bar5_M15_H\bar5_M$  
                               & $5_H24_M\bar5_M$ 
\\[1.01ex]\hline\hline&&&\\[-1.9ex]
 \multirow{2}{25mm}{only sLFVs}
  &  $N^cLH_u$  &  $LTL$ &  $H_uW_ML,\,H_uB_ML$ \\ 
  & -  &  -  & $H_U^CX_ML$
\\[1.05ex]\hline&&&\\[-1.9ex]
sLFVs\,\&\,sQFVs&    -           & $D^cLQ_{15}$    &    -  
\\[1.01ex]\hline&&&\\[-1.9ex]
 \multirow{2}{25mm}{only sQFVs}
  &  - & $D^cSD^c$ & $H_u\bar X_M D^c$ \\ 
  & \quad\quad $N^cD^cH_U^C$ & - & \quad $H_U^CG_M D^c,\,H_U^CB_M D^c$ 
\\[1.01ex]\hline&&&\\[-1.9ex]
$\vert$sLFVs/sQFVs$\vert$ \quad & $>1$ & $\sim1$ & $\sim1$ 
\\[1.01ex] \hline&&&                   
\end{tabular}  
\label{tab:seesaw}
\end{center}                           
\end{table}
%

In the nrMSSU(5) models, the parameters
 $\widetilde{m}_{D^c}^{2}$ and $\widetilde{m}_{L}^{2}$ are replaced by 
 $\widetilde{\mbox{\boldmath{$m$}}}_{D^c}^2$ and
 $\widetilde{\mbox{\boldmath{$m$}}}_{L}^2$.
A one-step integration of the corresponding RGEs for the
 off-diagonal elements of these matrices yields in the case of the
 seesaw of type~I:
\begin{eqnarray}
\widetilde{\mbox{\boldmath{$m$}}}^{2\,\ast}_{L\,(i,j)} \
&=& 
 \left(t_{\rm GUT}\! +\! t_{\rm ssw}\right)  \times
\left. 
{\cal F}\bigl(\mbox{\boldmath{$Y$}}_{N}^{{\rm I}\,T},
              \widetilde{\mbox{\boldmath{$m$}}}_{L}^{2\,\ast},
              \widetilde{m}_{N^c}^{2\,\ast},
              \widetilde{m}_{H_u}^2,
              \mbox{\boldmath{$A$}}_{N}^{{\rm I}\,T}\bigr)_{(i,j)}
\right\vert_{Q=M_{\rm cut}} ,   \qquad
\nonumber\\[1.01ex]
\widetilde{\mbox{\boldmath{$m$}}}^{2}_{D^c\,(i,j)} 
&=& 
 \quad \left(t_{\rm GUT}\right) \quad \times
\left.
{\cal F}\bigl(\mbox{\boldmath{$Y$}}_{ND}^{{\rm I}\,T},
              \widetilde{\mbox{\boldmath{$m$}}}_{D^c}^2,
              \widetilde{m}_{N^c}^{2\,\ast},
              \widetilde{m}_{H_U^C}^2,
              \mbox{\boldmath{$A$}}_{ND}^{{\rm I}\,T}\bigr)_{(i,j)}
\right\vert_{Q=M_{\rm cut}} . \qquad
\label{eq:onestepINTnro}
\end{eqnarray}
Thus, the elements $\widetilde{\mbox{\boldmath{$m$}}}_{L\,(i,j)}^{2}$
 are, to a good approximation,
{{{}}}
\begin{eqnarray}
 \widetilde{\mbox{\boldmath{$m$}}}_{L\,(i,j)}^{2\,\ast} \
 &=& 
 \left(t_{\rm GUT}\! +\! t_{\rm ssw}\right) \times 
\left.
  2 \left(3 \widetilde{m}^2_0 + A^2_0 \right)
 (\mbox{\boldmath{$Y$}}^{{\rm I}\,T}_{N}
  \mbox{\boldmath{$Y$}}^{{\rm I}\,\ast}_{N})_{(i,j)}  \right\vert_{Q=M_{\rm cut}}  ,
\nonumber\\[1.01ex]
 \widetilde{\mbox{\boldmath{$m$}}}_{D^c\,(i,j)}^{2}
 &=& 
 \ \quad \left(t_{\rm GUT}\right) \,\quad \times
\left.
  2 \left(3 \widetilde{m}^2_0 + A^2_0 \right)
 (\mbox{\boldmath{$Y$}}^{{\rm I}\,T}_{ND}
  \mbox{\boldmath{$Y$}}^{{\rm I}\,\ast}_{ND})_{(i,j)}  \right\vert_{Q=M_{\rm cut}}  .
\label{eq:sLFVs}
\end{eqnarray}
Since we have neglected NROs in the seesaw sectors,
 in {\it SU}(5)-symmetric bases, it is 
 $\mbox{\boldmath{$Y$}}_N^{\rm I} =
  \mbox{\boldmath{$Y$}}_{ND}^{\rm I} =Y_N^{\rm I}$. 
After the {\it SU}(5)-breaking rotations
 $D^c \to P_{\rm I}^\dagger \Delta_D D^c$ and 
 $L\to e^{-i \phi_{\rm I}} P_{\rm I}^\dagger L$, and generalizing our 
 expressions to include all three seesaw types, we obtain
\begin{equation}
\widetilde{\mbox{\boldmath{$m$}}}^2_{D^c\,(i,j)}      =  
 R_{\rm ssw}\times\,
\left(
 \Delta_D^{T}\ \widetilde{\mbox{\boldmath{$m$}}}^{2\,\ast}_L\,
 \Delta_D^{\ast}
\right)_{(i,j)} ,
\label{eq:modMoroiREL}
\end{equation}
 where the presence of the mismatch matrices clearly breaks the
 simple correlation between the seesaw-induced sQFVs and sLFVs
 present in the MSSU(5) model.
This fact was already strongly emphasized in Ref.~\citen{BGOO}. 
While the authors of this paper, however, stressed this effect for the
 case of first--second generation transitions, it is easy to see that
 the correlations of the minimal model are also lost in the case of
 second--third and first--third generation transitions~\cite{ParkEt}.
Indeed, for the flavour transitions $(i,3)$, with $i=1,2$, we have
\begin{eqnarray}
\widetilde{\mbox{\boldmath{$m$}}}^2_{D^c\,(i,3)}    
&=&
 R_{\rm ssw} \times 
\left\{\, \Delta_{D\,(1,i)}     \,  \Delta^\ast_{D\,(2,3)}\,
          \widetilde{\mbox{\boldmath{$m$}}}^{2\,\ast}_{L\,(1,2)}
    +     \Delta_{D\,(2,i)}     \,  \Delta^\ast_{D\,(1,3)}\,
          \widetilde{\mbox{\boldmath{$m$}}}^{2\,\ast}_{L\,(2,1)} \right.
\nonumber\\
& &
\phantom{
 R_{\rm ssw}\times}
    +     \Delta_{D\,(1,i)}     \,  \Delta^\ast_{D\,(3,3)}\,
          \widetilde{\mbox{\boldmath{$m$}}}^{2\,\ast}_{L\,(1,3)}
    +     \Delta_{D\,(3,i)}     \,  \Delta^\ast_{D\,(1,3)}\,
          \widetilde{\mbox{\boldmath{$m$}}}^{2\,\ast}_{L\,(3,1)}
\nonumber\\
& & 
\phantom{
 R_{\rm ssw}\times }
    +     \Delta_{D\,(2,i)}     \,  \Delta^\ast_{D\,(3,3)}\,
          \widetilde{\mbox{\boldmath{$m$}}}^{2\,\ast}_{L\,(2,3)}
    +     \Delta_{D\,(3,i)}     \,  \Delta^\ast_{D\,(2,3)}\,
          \widetilde{\mbox{\boldmath{$m$}}}^{2\,\ast}_{L\,(3,2)}
\nonumber\\
& & 
\phantom{
 R_{\rm ssw}\times }
    +     \Delta_{D\,(i,i)}     \,\, \Delta^\ast_{D\,(i,3)}\, 
\!\!\left[\widetilde{\mbox{\boldmath{$m$}}}^{2\,\ast}_{L\,(i,i)}\, 
    -     \widetilde{\mbox{\boldmath{$m$}}}^{2\,\ast}_{L\,(j,j)} \right]
\nonumber\\
& & 
\phantom{
 R_{\rm ssw}\times }
    +     \Delta_{D\,(3,i)}     \,   \Delta^\ast_{D\,(3,3)}\, 
\!\!\left[\widetilde{\mbox{\boldmath{$m$}}}^{2\,\ast}_{L\,(3,3)}\! 
    -     \widetilde{\mbox{\boldmath{$m$}}}^{2\,\ast}_{L\,(j,j)} \right]
\Bigr\},
\label{eq:flTRANSwithNRO}
\end{eqnarray}
 with $j=(1,2)$ and $j\ne i$, where the unitarity of the matrix
 $(\Delta_D)$ was used.
If no hierarchies exist among the elements of the unitary matrix
 $\Delta_D$, all terms in this equation contribute in the same way to
 the off-diagonal elements,
 $\widetilde{\mbox{\boldmath{$m$}}}^{2}_{D^c\,(i,3)}$, except perhaps 
 those in the last two lines, where the diagonal elements of 
 $\widetilde{\mbox{\boldmath{$m$}}}_{L}^{2}$ tend to cancel each other
 under the assumption of universal boundary conditions for the soft
 terms.
An equally complicated expression is obtained for 
 $\widetilde{\mbox{\boldmath{$m$}}}^2_{D^c\,(1,2)}$ in terms of all
 elements in the matrix 
 $\widetilde{\mbox{\boldmath{$m$}}}^2_L$ and of the mismatch matrix
 $\Delta_D$, which we do not report here.
We list instead the expression that both 
 $\widetilde{\mbox{\boldmath{$m$}}}^2_{D^c\,(1,2)}$ and
 $\widetilde{\mbox{\boldmath{$m$}}}^2_{D^c\,(i,3)}$ get when the
 approximation of \S\ref{sec:NROsBC-YUKAWAappr} is used: 
\begin{eqnarray}
\widetilde{\mbox{\boldmath{$m$}}}^2_{D^c\,(1,2)}               & = &
 R_{\rm ssw} \times 
\Bigl\{\, \Delta_{D\,(1,1)}     \,
          \Delta^\ast_{D\,(2,2)}\,
 \widetilde{\mbox{\boldmath{$m$}}}_{L\,(1,2)}^{2\,\ast} +
          \Delta_{D\,(2,1)}     \, 
          \Delta^\ast_{D\,(1,2)}\,
 \widetilde{\mbox{\boldmath{$m$}}}_{L\,(2,1)}^{2\,\ast}
\nonumber\\                                                &   & 
\phantom{
 R_{\rm ssw} \times  }
    +     \Delta_{D\,(1,1)}     \,
          \Delta^\ast_{D\,(1,2)}\, 
\!\!\left[\widetilde{\mbox{\boldmath{$m$}}}^{2\,\ast}_{L\,(1,1)} 
    -     \widetilde{\mbox{\boldmath{$m$}}}^{2\,\ast}_{L\,(3,3)}
\right]
\nonumber\\                                                &   & 
\phantom{
 R_{\rm ssw} \times  }
    +     \Delta_{D\,(2,1)}     \,
          \Delta^\ast_{D\,(2,2)}\, 
\!\!\left[\widetilde{\mbox{\boldmath{$m$}}}^{2\,\ast}_{L\,(2,2)}\! 
    -     \widetilde{\mbox{\boldmath{$m$}}}^{2\,\ast}_{L\,(3,3)}
\right]
\Bigr\},
\\[1.1ex]
\widetilde{\mbox{\boldmath{$m$}}}^2_{D^c\,(i,3)}              & = &
 R_{\rm ssw} \times 
\Bigl\{\, \Delta_{D\,(1,i)}     \,
 \widetilde{\mbox{\boldmath{$m$}}}_{L\,(1,3)}^{2\,\ast} +
          \Delta_{D\,(2,i)}\,
 \widetilde{\mbox{\boldmath{$m$}}}_{L\,(2,3)}^{2\,\ast} 
\Bigr\} \,\Delta^\ast_{D\,(3,3)}.
\label{eq:flTRANSwithNROapp}
\end{eqnarray}

In the above relations, the prediction of the MSSU(5) model is recovered
 for 
 $(\Delta_D)_{(1,2)}\sim 0$, but it is strongly violated if
 $(\Delta_D)_{(1,2)}\sim (\Delta_D)_{(1,1)}$. 
Even in such a case, there is a relation between elements of
 $\widetilde{\mbox{\boldmath{$m$}}}^{2}_{D^c}$ and
 $\widetilde{\mbox{\boldmath{$m$}}}^{2}_{L}$ that remains unchanged by
 the rotation matrix $\Delta_D$:
\begin{equation}
\left[
  \left\vert \widetilde{\mbox{\boldmath{$m$}}}^{2}_{D^c\,(1,3)}
  \right\vert^2   +\!     
  \left\vert \widetilde{\mbox{\boldmath{$m$}}}^{2}_{D^c\,(2,3)}
  \right\vert^2
\right]^{1/2}        =  \ 
 R_{\rm ssw} \times 
\left[
  \left\vert \widetilde{\mbox{\boldmath{$m$}}}^{2\,\ast}_{L\,(1,3)}
  \right\vert^2 +\! 
  \left\vert \widetilde{\mbox{\boldmath{$m$}}}^{2\,\ast}_{L\,(2,3)}
  \right\vert^2
\right]^{1/2} \!\!\!\!\!,  
\label{eq:modifRELAT}
\end{equation}
 reported also in Ref.~\citen{ParkEt}.

\section{Summary}
\label{sec:summary}
%
We have analyzed the issue of flavour and {\it CP} violation induced by
 seesaw Yukawa couplings in the low-energy soft parameters of the
 MSSU(5) model with NROs and the seesaw mechanism, or nrMSSU(5) models. 
Although our way of dealing with these NROs at the quantum level is 
 closer to the procedure proposed in Ref.~\citen{NROsFLsAHCH}, we have
 largely built on the analysis presented in Ref.~\citen{BGOO}, the first
 to deal in this context with NROs to correct the erroneous predictions
 of the MSSU(5) model for fermion masses.
The addition of NROs is a minimal ultraviolet completion sufficient to
 rescue the MSSU(5) model also from the other fatal problem by which it
 is plagued, i.e., that of predicting a rate for the decay of the
 proton that is too fast.

We have gone beyond the analysis of Ref.~\citen{BGOO} in several
 directions. 
We have not tried to restrict the number and type of NROs introduced,
 complying with the idea that, barring accidentally vanishing couplings,
 it is difficult to confine NROs to a certain sector of the model, or
 impose that they are of a certain type.
Their physical relevance, however, decreases when their dimensionality
 increases and, depending on the accuracy required for the problem under
 consideration, NROs of certain dimensions can be completely negligible.
For our calculation of sFVs at a precision of
 ${\cal O}(s\times s_{\rm loop})$, it is sufficient to include NROs of
 dimension five at the quantum level and the tree-level boundary 
 values of NROs of dimension six. 
Despite this, we have kept our discussion as general as possible.

Since their impact is of little consequence for sFVs, we have also
 disregarded the effects of NROs in the Higgs sector.
This approximation has allowed us to neglect possible modifications of
 the vacuum structure, which is, therefore, that of the MSSU(5) model,
 although a generalization to the case with nonvanishing NROs in this
 sector is conceptually straightforward.
We have studied this vacuum in detail, emphasizing the scaling behaviour
 of the scalar and auxiliary~VEVs $v_{24}$ and $F_{24}$ of the
 field $24_H$.
We have also shown what an important role these two~VEVs play for
 the determination of the boundary conditions of various soft parameters
 and for their evolution.

We have introduced the concept of effective couplings, which get
 contributions from the couplings of renormalizable operators and those
 of nonrenormalizable ones when the field $24_H$ acquires~VEVs.  
This concept can be applied to the case of other GUT groups (it was,
 indeed, introduced in Ref.~\citen{NROsFLsAHCH}, for a SUSY~{\it SO}(10)
 model) and can also be used when the superheavy Higgs fields 
 acquiring VEVs are not in the adjoint representation.

By making use of these couplings, we have analyzed in full generality the
 amount of arbitrariness introduced by NROs in the Yukawa sector, which,
 in general, spoils the correlation between the seesaw-induced sQFVs and
 sLFVs typical of the MSSU(5) model.
This arbitrariness can be parametrized by  several diagonal matrices of 
 effective Yukawa couplings and several matrices of diagonalization
 mismatch, each of them expressed in terms of mixing angles and phases.
The parameter space opened up by the introduction of NROs in this sector
 is, indeed, quite large. 
It is precisely this arbitrariness that allows us to decrease the decay
 rate of the proton.

We have pointed out that the suppression of this rate through NROs in
 the Yukawa sector may have an important feedback for the evaluation of
 sFVs. 
We have shown that the correlation between the seesaw-induced sQFVs and
 sLFVs typical of the MSSU(5) model remains practically unchanged in a
 particular point of the above parameter space in which the decay rate
 of the proton is suppressed.
We have also pointed out that different NROs, in addition to those in
 the Yukawa sector, could yield the necessary suppression of the
 proton-decay rate and argued, as a consequence, that the suppression of
 this rate by NROs does not necessarily guarantee that the pattern of
 sFVs in nrMSSU(5) models remains that of the MSSU(5) one.

The concept of effective couplings has allowed us to formulate in a
 general way the problem of possible sFVs at the tree level, i.e.,
 in the boundary conditions of soft parameters, which NROs, in general,
 introduce. 
It has provided a tool to address the problem of finding conditions 
 to be imposed on the couplings of the SUSY-breaking mediator ${\sf X}$
 to the superpotential and K\"ahler potential to avoid such tree-level
 sFVs (see \S\S\ref{sec:NROsBC-TrilSOFT} 
 and \ref{sec:NROsBC-MassSOFT}).
This is because the usual requirement of universality of these
 couplings, that is, of blindness to flavour and field type, is not
 sufficient in the  context of models with NROs.
We have called this blindness weak universality.

We have shown that the needed special conditions can actually be
 obtained in a general way in models with NROs, by imposing that this
 universality of couplings of ${\sf X}$ to the superpotential and
 K\"ahler potential is stable under  the field redefinitions of
 Eq.~(\ref{eq:FRedefMinUNIV}), and that the dimension-five operators in
 the K\"ahler potential can be removed. 
This has led us to identify a special type of universal couplings of
 ${\sf X}$ to the superpotential and K\"ahler potential, or
 strongly universal couplings, which amounts to having, in the same
 basis, the factorizable form of Eqs.~(\ref{eq:WunivSTR1})
 and~(\ref{eq:KAELERstrongUNIV}) for these two potentials.
Clearly, in models with vanishing NROs, such as the MSSM, the two
 notions of universality coincide.

When this type of strong universality is advocated, the soft
 SUSY-breaking terms can be described in terms of the usual four 
 parameters, $\widetilde{m}_0^2$, $A_0$, $B_0$ and $M_{1/2}$, and the 
 only arbitrariness induced by NROs in the predictions of the 
 seesaw-induced sFVs is that of the Yukawa sector. 
If we impose $(B_0 -A_0)^2 = \widetilde{m}_0^2$, the situation is like
 that emerging by taking the flat limit of models embedded in minimal
 supergravity.
In both cases, an ambiguity appears at the GUT scale in the
 determination of the MSSM $B$ parameter, depending on how the tuning
 for this and the MSSM $\mu$ parameters are made.

We have found that it is possible to extend the picture of effective
 couplings to the quantum level, obviously not in general, but within
 certain limits.
These limits restrict the accuracy that can be obtained for the
 evaluation of sFVs in nrMSSU(5) models treated in this way.
The  maximal accuracy that we can achieve is of
 ${\cal O}(s\times s_{\rm loop})$.
Nevertheless, this is more than adequate for our study of sFVs. 
We have shown that within this accuracy, the nrMSSU(5) models can be
 treated like renormalizable ones, as the effective couplings do obey
 RGEs typical of renormalizable models. 
Clearly, this is not true for the original couplings of the various NROs.
The evolution of the~VEVs of the field $24_H$, however, corrects
 for the difference in the running of these original couplings, which 
 enter in the definition of the effective couplings, and the running of
 the effective couplings themselves. 
As shown in Appendix~\ref{sec:New}, this fact is completely general, and
 it applies to
 {\it i)} other GUT groups, {\it ii)} the case in which the fields that
 acquire VEVs are in representations different than the adjoint, and
 {\it iii)} the case in which NROs are included in the Higgs sector.

Within the accuracy of our analysis, the one-loop results obtained with
 this method are not different from those obtained with the more
 conventional method of Ref.~\citen{BGOO}, when the running of the
 two~VEVs $v_{24}$ and $F_{24}$ is kept into account.
The effect of this running is missed in the existing
 literature~\cite{BGOO,ParkEt} 
(see for example the discussion at the end of \S\ref{sec:NROinRGE}).

We have insisted on this effective picture as it gives,
 in our opinion, a very clear physical interpretation of the parameters
 of these models and includes the running of the VEVs in a 
 natural way.

We have given complete lists of RGEs for the nrMSSU(5) models, together
 with those for the MSSU(5) model, and for the MSSM, for all the three
 types of seesaw mechanism. 
(This last set of evolution equations are needed at scales below
 $M_{\rm GUT}$.) 
Some of these RGEs are presented here for the first time.
The others provide a check of RGEs existing in the literature, which we
 correct when possible.

{}Finally, by making use of approximated solutions of the above RGEs, we
 have sketched how the predictions for the seesaw-induced sFVs can be
 modified by the presence of NROs in the Yukawa sector, in the scenario
 of strongly universal couplings of ${\sf X}$ to the K\"ahler potential
 and superpotential.
Modifications are induced not only in first--second generation sFVs, but
 also in those involving the third generation.

\section*{Acknowledgements}
The authors are grateful to S.~Mishima for collaboration at an early
 stage of this work.
They also thank T.~Goto, J.~Hisano, S.~Martin, Y.~Nir, and
 S.~Teraguchi, for critically reading parts of this manuscript, as well
 as S.~Bertolini, T.~Goto, Y.~Okada, K.-I.~Okumura, G.~Senjanovic,
 T.~Shindou, and in particular, Y.~Kawamura,{{{}}} for discussions.  
The hospitality, at various stages of this project, of KIAS, Korea, of
 Nagoya University, Japan, and of the Galileo Galilei Institute, Italy
 (where F.~B. was financially supported by the INFN, sezione di Firenze) is
 acknowledged by F.~B., that of National Central University, Taiwan, by
 T.~Y..
Last, but not least, F.~B. expresses her gratitude to K.~Hagiwara
 for his unrelenting support and encouragement throughout the completion
 of this work.
The work of F.~B. is partially supported by the Excellency Research
 Project of National Taiwan University, Taiwan:
 ``Mass generation, heavy flavours, neutrinos at the particle physics
 frontier'', grant Nos.{{{}}}~97R0066-60 and 98R0066-60, that of T.~Y.
 by the Japanese Society for the Promotion of Science.

\appendix
%
\section{Embeddings in GUT Representations: \\Normalizations and
              Group Factors}
\label{sec:emb-GROUPfact} 
%
Lepton and Higgs doublets of {\it SU}(2) are contained in 5plets of
 {\it SU}(5); antidoublets in $\bar{5}$plets, with the antidoublet of,
 say, $L$ given by  $\epsilon L$. 
We recall that $\epsilon$ is the $2\times2$ matrix proportional to the
 Pauli matrix $\sigma_2$ ($\epsilon = i \sigma_2$), with elements
 $\epsilon_{11}=\epsilon_{22}=0$, $\epsilon_{12}=-\epsilon_{21}=1$.
Thus, the SM decompositions of $\bar{5}_M$, $5_H$ and $\bar{5}_H$ are
\begin{equation}
(\bar{5}_M)_A = \left(
  \begin{array}{c} D^c_a     \\(\epsilon L)_\alpha \end{array} 
                \right),
\hspace*{0.5truecm}
(5_H)_A = \left(
  \begin{array}{c} (H^C_U)_a \\ (H_u)_\alpha \end{array} 
          \right),
\hspace*{0.5truecm}
(\bar{5}_H)_A = \left(
  \begin{array}{c} (H^C_D)_a \\(\epsilon H_d)_\alpha \end{array} 
                \right),
\label{eq:5reprSU5}
\end{equation}
 where $A$, like all {\it SU}(5) indices denoted with upper-case latin
 letters, is decomposed into {\it SU}(3) and {\it SU}(2) indices,
 denoted respectively by lower-case latin letters and lower-case Greek
 ones. 
The 10plet of {\it SU}(5) is usually expressed as a $5\times 5$
 antisymmetric matrix:
\begin{equation}
 (10_M)_{AB}  = \frac{1}{\sqrt{2}}
\left(
 \begin{array}{cc}
\epsilon_{abc} U^c_c &  -Q^T_{a\beta}                 \\[1.0001ex]
    Q_{\alpha b}     & -\epsilon_{\alpha\beta} E^c  
 \end{array}
\right),
\label{eq:10reprSU5}
\end{equation}
 with $\epsilon_{abc}$ antisymmetric in the exchange of two contiguous
 indices, and $\epsilon_{123}=1$. 
The  multiplication rule 
 $10_M^\ast 10_M = (10_M^\ast)_{AB} (10_M)_{AB}={\rm Tr}(10_M^\dagger 10_M)$,
 together with the coefficient $1/\sqrt{2}$, guarantees that the kinetic
 terms for $10_M$ are correctly normalized.

The {\it SU}(2) doublets $Q$, $L$, $H_u$ and $H_d$ in the previous two 
 equations are, as usual,
\begin{equation}
Q = \left(\begin{array}{c} U \\ D \end{array} \right),
\hspace*{0.8truecm}
L = \left(\begin{array}{c} N \\ E \end{array} \right),
\hspace*{0.8truecm}
H_d = \left(\begin{array}{c} H_d^0 \\ H_d^- \end{array} \right),
\hspace*{0.8truecm}
H_u = \left(\begin{array}{c} H_u^+ \\ H_u^0 \end{array} \right).
\label{eq:SU2doublets}
\end{equation}

The {\it SU}(5) multiplication of a $5$ and a $\bar{5}$ multiplet, such
 as in $W_{\rm RHN}$ in Eq.~(\ref{eq:SU5sswSUPERP}), is understood to be
 trivially $\bar{5}_M 5_M \equiv (\bar{5}_M)_A (5_M)_A$. 
The first two terms of $W^{\rm MSSU(5)}_{\rm M}$ in
 Eq.~(\ref{eq:WminSU5m}) are
\begin{eqnarray}
 \bar{5}_M 10_M \bar{5}_H \ 
&\equiv& 
 (\bar{5}_M)_A(10_M)_{AB} (\bar{5}_H)_B, 
\nonumber\\
 10_M  10_M 5_H 
&\equiv& 
 \epsilon_{ABCDE} (10_M)_{AB} (10_M)_{CD} (5_H)_E.
\label{eq:5multiplRULES}
\end{eqnarray}
The tensor $\epsilon_{ABCDE}$ is fully antisymmetric, with
 $\epsilon_{12345}=1$ and 
 $\epsilon_{abc\alpha\beta} =\epsilon_{abc}\epsilon_{\alpha\beta}$.
These multiplication rules and the embedding of doublets and
 antidoublets in $5$ and $\bar{5}$ multiplets specified above are
 consistent with the usual multiplication rule for {\it SU}(2) doublets, 
 $L H_d \equiv L_{\alpha} \epsilon_{\alpha \beta} (H_d)_{\beta}$,
 and the {\it SU}(3) multiplication,
 $D^c U^c H^C_D \equiv \epsilon_{abc} D^c_a U^c_b (H^C_D)_c $.

Given all the above definitions, it is easy to see that the SM
 decomposition of the two terms in $W^{\rm MSSU(5)}_{\rm M}$ is
\begin{eqnarray}
\sqrt{2} \, \bar{5}_M Y^5 10_M \bar{5}_H 
 &=&  \!\!
- D^{c}  Y^5     Q    H_d 
- E^{c}  Y^{5\,T}  L    H_d 
- L      Y^5    Q    H^C_D      
- D^{c}  Y^5     U^{c}  H^C_D ,
\nonumber\\
-\frac{1}{4} 10_M Y^{10} 10_M 5_H         
 &=& \
             U^{c} Y^{10} Q    H_u  
+\frac{1}{2} Q    Y^{10} Q    H_U^C 
+            U^{c} Y^{10} E^{c} H^C_U ;
\label{eq:minSU5mattDECOMP}
\end{eqnarray}
 that for the Yukawa interaction in $W_{\rm RHN}$ can be read off
 from Eq.~(\ref{eq:YUKAWAsswRotat}).

The Higgs field $24_H$ is accommodated into the traceless $5\times 5$
 matrix: 
\begin{equation}
 (24_H)_{AB} = 
\left( \begin{array}{cc} 
             (G_H)_{ab}         & (X_H)_{a \beta}       \\[1.0001ex]
 (\epsilon \bar{X}_H)_{\alpha b}  & (\epsilon W_H)_{\alpha \beta} 
       \end{array}  \right)
 + 
  B_H \sqrt{\frac{6}{5}}
\left( \begin{array}{cc}
 \frac{1}{3}\delta_{ab}  &  0                    \\[1.0001ex]   
     0                   & -\frac{1}{2} \delta_{\alpha \beta} 
       \end{array}  \right) , 
\label{eq:24reprSU5}
\end{equation}
 where the subscript index $H$ distinguishes its components from those
 of the gauge field in the adjoint representation $24$, $G$, $X$,
 $\bar{X}$, $W$, and $B$.
We define $(24_H)_{AB}\equiv \sqrt2\,24_H^i T^i_{AB}$~($i=1,...,24$),
 and similarly,
 $(G_H)_{ab} \equiv \sqrt2\,G_H^i  T^i_{ab}$~($i= 1,...,8$) and 
 $(W_H)_{\alpha \beta} \equiv
 \sqrt2\,W_H^i T^{i+20}_{\alpha \beta}$~($i=1,2,3$).
The symbols $T^i$ denote, as usual, {\it SU}(5) generators, with 
 ${\rm Tr} T^i T^j = (1/2) \delta_{ij}$.
The field $X_H$ is an {\it SU}(2) doublet and an {\it SU}(3) antitriplet
 and has hypercharge $5/6$; $\bar{X}_H$, with hypercharge $-5/6$, is an
 antidoublet of {\it SU}(2) and a triplet of {\it SU}(3). 
The field $B_H$, or $24_H^{24}$, is a SM singlet, and
 $\sqrt2 T^{24}_{AB}$ is the quantity
 $\sqrt{6/5}\,{\rm diag}(1/3,1/3,1/3,-1/2,-1/2)$ that multiplies it in
 Eq.~(\ref{eq:24reprSU5}). 
It is the field $B_H$ that acquires the {\it SU}(5)-breaking~VEV
 $\langle B_H \rangle$:
\begin{equation}
 B_H \to \ \left \langle B_H \right \rangle + B_{H}^\prime.
\label{eq:BHshift}
\end{equation}
In the limit of exact SUSY,{{{}}} this~VEV coincides with $v_{24}$ of
 Eq.~(\ref{eq:v24noSOFT}).

With these ingredients, it is easy to obtain the SM decomposition 
 of the terms in
 $W^{\rm MSSU(5)}_{\rm H}$~(Eq.~(\ref{eq:WminSU5H})).
The first two terms give rise to
\begin{equation}
 \mu_3 H_U^C H_D^C  + \mu_2 H_u H_d, 
\label{eq:mu2mu3noSOFTredef}
\end{equation}
 with $\mu_2$ and $\mu_3$ defined in 
 Eqs.~(\ref{eq:mu2noSOFT}) and~(\ref{eq:mu3noSOFT}), and to
$$
 \lambda_5 
\left\{ H_U^C 
 \left[G_H +\sqrt{\frac{6}{5}}\frac{1}{3}B_{H}^\prime \right] H_D^C 
+\left[H_U^C X_H H_d + H_u \bar{X}_H H_D^C \right]
\right.
\hspace*{1truecm}
$$
\begin{equation}
\hspace*{3truecm}
\left.
+H_u \left[W_H -\sqrt{\frac{6}{5}}\frac{1}{2}B_{H}^\prime\right]
 H_d \right\} .
\label{eq:524bar5DECOMP}
\end{equation}
{}Finally, for the products of $24_H$, we recall that
\begin{eqnarray}
 (24_H)^2 &=& {\rm Tr}(24_H \,24_H), 
\nonumber\\
 (24_H)^3 &=& \frac{1}{2}{\rm Tr}(\{24_H,24_H\}24_H) = 
            {\rm Tr}(24_H \,24_H\, 24_H). 
\label{eq:prods24}
\end{eqnarray}
The last relation is not general, but applies to the specific case 
 considered here. 
Thus, for the last two terms in $W^{\rm MSSU(5)}_{\rm H}$, we obtain the
 decompositions
\begin{eqnarray}
{24_H}^2  &= &
 \quad 
 {B_H}^2 + \sum_i (G_H^i)^2 + \sum_i (W_H^i)^2+
  \left(X_H \bar{X}_H - \bar{X}_H X_H \right),
\hspace*{3truecm}
\nonumber\\
{24_H}^3  &= &  
 -\frac{1}{\sqrt{30}}{B_H}^3
 +3\sqrt{\frac{6}{5}}B_H 
\nonumber\\
          &  & \times        
 \left[\frac{1}{3}\sum_i (G_H^i)^2 -\frac{1}{2}\sum_i (W_H^i)^2
      +\frac{1}{3}X_H \bar{X}_H +\frac{1}{2}\bar{X}_H X_H 
   \right]+\cdots.
\label{eq:decWIHIGGS}
\end{eqnarray}
Once the field $B_H$ acquires its~VEV, the superpotential mass 
 terms for the fields $B_H^\prime$, $G_H^i$, and $W_H^i$ become, in the 
 supersymmetric limit: 
\begin{equation}
-\frac{1}{2} M_{24}\left(B_H^{\prime}\right)^2
+\frac{1}{2}\left(5 M_{24}\right)\sum_i (G_H^i)^2
-\frac{1}{2}\left(5 M_{24}\right)\sum_i (W_H^i)^2,
\label{eq:24Hmasses}
\end{equation}
 whereas the Goldstone bosons $X_H$ and $\bar{X}_H$ remain massless.

The matter fields $24_M$ are as $24_H$, with the subscript index $_H$
 replaced by $_M$ everywhere. 
Three such fields are introduced for the implementation of the seesaw 
 of type~III, labelled by a flavour index, which is suppressed in this
 appendix. 
The decomposition of the mass term $24_M M_{24_M} 24_M$, appearing in 
 $W_{\rm 24M}$ of Eq.~(\ref{eq:SU5sswSUPERP}), follows that of
 $24_H^2$.
The product $24_M 24_M 24_H$ in the last term of $W_{\rm 24M}$, in 
 contrast, is not as simple as the product $(24_H)^3$ discussed above.
This is because there are, in reality, two ways to obtain a singlet out
 of the product of three adjoint representations of {\it SU}(5). 
Indeed, the product of two $24$'s is 
\begin{equation}
 24\times 24 = 1 + 24_S + 24_A + \cdots,
\end{equation}
 where $24_S$ and $24_A$ correspond, respectively, to a symmetric and 
 an antisymmetric product.
In the case of $(24_H)^3$, the antisymmetric part vanishes.
In contrast, since more than one field $24_M$ exists, the
 antisymmetric product of two different $24_M$'s does not vanish, and
 $24_M 24_M 24_H$ gets contributions from both the symmetric and 
 antisymmetric products:
\begin{equation}
 (24_M^2 24_H) = 
 \frac{1}{2}{\rm Tr}(\{24_M,24_M\}\,24_H)+
 \frac{1}{2}{\rm Tr}([24_M,24_M]\,24_H).
\end{equation}
Thus, two distinct interaction terms with different Yukawa couplings
 $Y_{24_M}^S$ and $Y_{24_M}^A$ exist.
The decomposition of the term  $5_H 24_M Y_{N}^{\rm III} \bar{5}_M$ in 
 $W_{\rm 24M}$ is given in Eq.~(\ref{eq:YUKAWAsswRotat}), where the
 terms $H_u W_M H_d$ and $H_u B_M H_d$ are understood as
 $H_u^T \epsilon W_M \epsilon H_d$ and
 $H_u^T B_M \epsilon H_d$, respectively.

The $15_H$ and $\bar{15}_H$ representations are expressed 
 as $5\times 5$ symmetric matrices: 
\begin{eqnarray}
 (15_H)_{AB}  &=& 
\left(\! \begin{array}{cc}
  S_{ab}      & 
\displaystyle{\frac{1}{\sqrt{2}}}   (Q^T_{15})_{a \beta} \\
\displaystyle{\frac{1}{\sqrt{2}}}   (Q_{15})_{\alpha b} 
             &                       T_{\alpha \beta}         
         \end{array}
\!\right),
\nonumber\\[1.1ex]
 (\bar{15}_H)_{AB} &=& 
\left(\! \begin{array}{cc}
 \bar{S}_{ab}      & 
\displaystyle{\frac{1}{\sqrt{2}}} 
                      (\bar{Q}^T_{15}\epsilon^T)_{a \beta} \\
\displaystyle{\frac{1}{\sqrt{2}}} 
                    (\epsilon\bar{Q}_{15})_{\alpha b}
                  & (\epsilon \bar{T}\epsilon^T)_{\alpha\beta}
         \end{array}
\!\right).
\label{eq:15Handbar15HreprSU5}
\end{eqnarray}
In these matrices, $S$ and $\bar{S}$ are singlets of {\it SU}(2) and a
 6plet and $\bar6$plet of {\it SU}(3), respectively; $Q_{15}$ has the
 same {\it SU}(3) and {\it SU}(2) quantum numbers of $Q$ in
 Eq.~(\ref{eq:SU2doublets}), $\bar{Q}_{15}$ the opposite ones: 
\begin{equation}
 Q_{15} = \left( \begin{array}{c} U_{15} \\ D_{15} 
                 \end{array} \right), 
\hspace*{1truecm}
 \bar{Q}_{15} = \left( \begin{array}{c} -\bar{D}_{15}\\ \ \bar{U}_{15} 
                 \end{array} \right), 
\label{eq:Q15repres}
\end{equation}
 $T$ and $\bar{T}$ are singlets of {\it SU}(3) and a triplet and an
 antitriplet of {\it SU}(2), respectively. 
Explicitly,
\begin{equation}
 T =      \left(\!  \begin{array}{cc}
                                   T^{++} & 
\displaystyle{\frac{1}{\sqrt{2}}}  T^+    \\
\displaystyle{\frac{1}{\sqrt{2}}}  T^+    &  
                                   T^0 
                    \end{array} \!\right),
\hspace*{1truecm}
\bar{T} = \left(\!  \begin{array}{cc}
                                  \bar{T}^0  & 
\displaystyle{\frac{1}{\sqrt{2}}} \bar{T}^-  \\
\displaystyle{\frac{1}{\sqrt{2}}} \bar{T}^-  & 
                                  \bar{T}^{--} 
                    \end{array} \!\right).
\label{eq:Tmatrices}
\end{equation}
The factor $1/\sqrt{2}$ in the off-diagonal elements of $15_H$ and
 $\bar{15}_H$ is introduced in order to obtain the correct
 normalization of the kinetic terms for $15_H$ and $\bar{15}_H$.
We recall that 
\begin{eqnarray}
         15_H^\ast 15_H 
&=&     (15_H^\ast)_{AB}(15_H)_{AB},
\nonumber\\
    \bar{15}_H^\ast\bar{15}_H 
&=&(\bar{15}_H^\ast)_{AB}(\bar{15}_H)_{BA}, 
\nonumber\\
         15_H \bar{15}_H
&=&     (15_H)_{AB} (\bar{15}_H)_{BA}, 
\nonumber\\
     \bar{5}_M 15_H \bar{5}_M 
&=& (\bar{5}_M)_A (15_H)_{AB} (\bar{5}_M)_B,
\nonumber\\
      15_H 24_H \bar{15}_H 
&=&  (15_H)_{AB} (24_H)_{BC} (\bar{15}_H)_{CA}.  
\label{eq:15multiplRULES}
\end{eqnarray}
Similar rules holding for $5_H \bar{15}_H 5_H$ and
 $\bar{5}_H 15_H \bar{5}_H$.

Thus, the bilinear combination of the fields $15_H$ and $\bar{15}_H$ 
 in the mass term $ M_{15} 15_H \bar{15}_H $ in $W_{\rm 15H}$ in 
 Eq.~(\ref{eq:SU5sswSUPERP}) is
\begin{equation}
 S \bar{S} +  Q_{15} \bar{Q}_{15} + T \bar{T} ,
\label{eq:Mass15}
\end{equation}
 where $T \bar{T}= T \epsilon \bar{T} \epsilon^T$. 
The SM decomposition of the first interaction in $W_{\rm 15H}$ can be
 read off from Eq.~(\ref{eq:SU5sswSUPERP}).
The remaining terms proportional to $\lambda_D$ and $\lambda_U$ in
 $W_{\rm 15H}$ are
\begin{eqnarray}
    \frac{1}{\sqrt{2}} \lambda_D \bar{5}_H {15}_H \bar{5}_H  
&=& \frac{1}{\sqrt{2}} \lambda_D                            \left\{
      H_D^C S H_D^C  + \sqrt{2} H_D^C Q_{15}H_d  + H_d T H_d   \right\},
\nonumber\\
    \frac{1}{\sqrt{2}} \lambda_U 5_H \bar{15}_H 5_H   
&=& \frac{1}{\sqrt{2}} \lambda_U                            \left\{
      H_U^C \bar{S}H_U^C -\sqrt{2}H_D^C \bar{Q}_{15}H_u +H_u \bar{T} H_u
                                                            \right\},
\label{eq:LambdaUandDterms}
\end{eqnarray}
 where the {\it SU}(2) multiplications of $H_d T H_d$ (as well as of 
 $L T L$ in Eq.~(\ref{eq:YUKAWAsswRotat})) and $H_u\bar{T}H_u$ are
 understood to be $ H_d^T \epsilon^T T \epsilon H_d$ and  
 $ H_u^T \epsilon \bar{T} \epsilon^T H_u$, respectively.

As for the proton-decay NROs in Eq.~(\ref{eq:supNROforPD}), the  
 product of matter fields is understood as 
$$
 [(10_M     )_h (10_M)_i] [(\bar{5}_M)_j (10_M)_k]=
 \hspace*{8truecm}
$$
\begin{equation}
 \hspace*{2truecm}
 \epsilon_{ABCDE}
 [((10_M)_{AB})_h ((10_M)_{CD})_i] [((\bar{5}_M)_X)_j ((10_M)_{XE})_k],
\label{eq:Dim5opPDEXPAND}
\end{equation}
 with all {\it SU}(5) indices $A,B,C,D,X$ summed over $\{1,2,3,4,5\}$. 
The fields $24_H$ in this NRO are understood to be inserted anywhere
 along the string of matter fields. 
That is, $Op^{\rm PD}$  shares the same type of complication that
 $Op^{10}$ has.

The NROs in $Op^{10}$ are, indeed, 
\begin{eqnarray}
Op^{10} 
& = & -\frac{1}{4}
 \epsilon_{ABCDE} \,
 (10_M)_{A^\prime B^\prime}
 (10_M)_{C^\prime D^\prime} (5_H)_{E^\prime} 
 \!\sum_{\sum n_i=0}^k \!
 \left(\!\frac{1}{M_{\rm cut}}\!\right)^{\!(\sum n_i)\!}
 \hspace*{2truecm}  
\nonumber\\[1.001ex]
&   & 
 \times \,C^{10}_{n_1,n_2,n_3,n_4,n_5}\,
  (24_H)^{n_1}_{A^\prime A} 
\,(24_H)^{n_2}_{B^\prime B}
\,(24_H)^{n_3}_{C^\prime C} 
\,(24_H)^{n_4}_{D^\prime D}
\,(24_H)^{n_5}_{E^\prime E}, 
\label{eq:OP10multiplRULE}
\end{eqnarray}
 where it is assumed that
\begin{equation}
 (24_H)^{n}_{A B} = \delta_{A B} 
 \quad\quad   {\rm for} \ n=0.
\label{eq:24powerZERO}
\end{equation}
Note that when $k=1$, i.e., in the case of NROs of dimension
 five, the five terms in the above sum are not all independent: they
 reduce to the two terms of $Op^{10}\vert_5$ in Eq.~(\ref{eq:NROdim5}). 
The number of independent terms, however, increases rapidly with the
 number of dimensions.
There are five, when $k=2$:  
\begin{eqnarray}
Op^{10}\vert_6
 & = & 
 -\displaystyle{\frac{1}{4 M_{\rm cut}}}
\left[
 10_M\,C_{2;a}^{10}\,(24_H)^2\, 10_M 5_H          \,+\,
 10_M\,C_{2;b}^{10}\,          10_M \,5_H \,(24_H)^2 
\right.
\nonumber\\     
 &   & \hspace*{1.2truecm}
 +10_M\,C_{1,1;a}^{10}\,(24_H)\,10_M\,(24_H)\,5_H \,+\, 
  10_M\,C_{1,1;b}^{10}\,(24_H)\,10_M\,5_H\,(24_H) 
\nonumber\\[1ex]     
 &   &  
\left.\hspace*{1.2truecm}
 +10_M\,C_{1,1;c}^{10}\,(24_H^T) 10_M\,(24_H)\,5_H 
\right],  
\label{eq:10NROdim6}
\end{eqnarray}
 with coefficients that are linear combinations of the original 
 $C_{n_1,n_2,n_3,n_4,n_5}^{10}$'s.

The contractions of {\it SU}(5) indices for the NROs in $Op^{5}\vert_5$
 and $Op^{10}\vert_5$ are
\begin{eqnarray}
\bar{5}_M 24_H^T 10_M \bar5_H 
&=& 
 (\bar{5}_M)_{A} (24_H)_{BA} (10_M)_{BC} (\bar5_H)_{C},
\nonumber\\
 \bar{5}_M 10_M 24_H \bar5_H
&=& 
 (\bar{5}_M)_{A} (10_M)_{AB} (24_H)_{BC} (\bar5_H)_{C},
\nonumber\\
 10_M 10_M 24_H 5_H
&=& 
 \epsilon_{ABCDE}
  (10_M)_{AB} (10_M)_{CD^\prime}(24_H)_{D^\prime D}  (5_H)_{E},
\nonumber\\
10_M 10_M 5_H 24_H
&=& 
 \epsilon_{ABCDE} 
  (10_M)_{AB} (10_M)_{CD} (5_H)_{E^\prime}(24_H)_{E^\prime E}.
\label{eq:NROs5RULES}
\end{eqnarray}
Those for the NROs in $Op^{10}\vert_6$ are at this point a simple
 exercise.

\section{Flavour Rotations and Identification of SM Fields}
\label{sec:emb-SMfields}
%
We recall that two different unitary matrices are needed to diagonalize
 a generic matrix $Y_G$, whereas only one is needed to diagonalize a
 symmetric one $Y_S$:
\begin{equation}
 \widehat{Y}_G = U_G^T Y_G V_G , \hspace*{1truecm}
 \widehat{Y}_S = W_S^T Y_S W_S ,  
\label{eq:diagMatrices}
\end{equation} 
 and that any unitary matrix can be parametrized as 
\begin{equation}
 U = e^{i \phi_U} P_U^{(1)} K_U P_U^{(2)} .
\label{eq:unitary}
\end{equation}
Here, $K_U$ and $P_U^{(i)}$ ($i=1,2$), like all $K$- and $P$-matrices
 in this paper, are respectively a unitary matrix with three mixing
 angles and one phase (the CKM matrix is of this type) and diagonal
 phase matrices with two nonvanishing phases and with
 $\det P_U^{(i)}=1$; $\phi_U$ is an overall phase.

Thus, $Y_G$ and $Y_S$ can be recast as 
\begin{eqnarray}
 Y_G  & = &  e^{i \phi_G}
 \left(K_{U_G} P_{U_G}^{(2)}\right)^T  \widehat{Y}_G \, P_G 
 \left(K_{V_G} P_{V_G}^{(2)}  \right),
\nonumber\\
 Y_S  & = &  e^{i \phi_S}
 \left(K_S P_S^{(2)}\right)^T \ \widehat{Y}_S \, P_S
 \left(K_S P_S^{(2)} \right) . 
\label{eq:diagGandS}
\end{eqnarray}
This is nothing but a reshuffling of the parameters of $Y_G$ and $Y_S$.
The eighteen independent parameters of $Y_G$ are distributed among the 
 three real and positive eigenvalues collected in $\widehat{Y}_G$, six
 mixing angles in $K_{U_G}$ and $K_{V_G}$, and nine phases: two in
 each of the matrices $P_{U_G}^{(2)}$, $P_{V_G}^{(2)}$, $P_G$, one in
 each of the two matrices $K_{U_G}$ and $K_{V_G}$ and one in the
 overall factor $e^{i \phi_G}$. 
Similarly, the twelve parameters of $Y_S$ are distributed among three
 real and positive eigenvalues, three mixing angles in $K_S$, and six
 phases, four in $P_S$ and $P_S^{(2)}$, one in $K_S$, and one in the
 overall factor.

If the Yukawa matrices $Y^5$ and $Y^{10}$ in $W^{\rm MSSU(5)}_{\rm M}$ in
 Eq.~(\ref{eq:WminSU5m}) are diagonalized as
\begin{equation}
 \widehat{Y}^5 =  V_5^T Y^5 V_{10},
\hspace*{1truecm}
 \widehat{Y}^{10} = W_{10}^T Y^{10} W_{10},
\label{eq:Y5Y10diagon}
\end{equation}
 the following flavour rotations of the complete $SU(5)$ multiplets
 $\bar{5}_M$, $10_M$: 
\begin{equation}
  \bar{5}_M \rightarrow V_{5} \bar{5}_M, 
\hspace*{1truecm}
  10_M      \rightarrow V_{10} 10_M, 
\label{eq:10Mbar5Mrotations}
\end{equation}
 reduce the first interaction term of $W^{\rm MSSU(5)}_{\rm M}$ to be
 diagonal. 
By expressing the unitary matrix $W_{10}^\dagger V_{10}$ as 
\begin{equation}
 W_{10}^\dagger V_{10}
    \  = \  P_{10}^{(1)} K_{10}      P_{10}^{(2)} e^{i \phi_{10}}
    \ \equiv \  U_{\rm CKM},
\label{eq:rewrite}
\end{equation}
 where $K_{10}$ will be identified with the high-scale CKM matrix, 
 and by further redefining $\bar{5}_M$ and $10_M$ as 
\begin{equation}
  \bar{5}_M  \rightarrow  e^{i \phi_{10}} P_{10}^{(2)} \bar{5}_M ,
\hspace*{1truecm}
      10_M   \rightarrow  e^{-i\phi_{10}} P_{10}^{(2)\,\dagger} 10_M ,
\label{eq:moreRotat}
\end{equation}
 we obtain the form of $W^{\rm MSSU(5)}_{\rm M}$ given in 
 Eq.~(\ref{eq:WminSU5mRot}).

The choice of the {\it SU}(5)-breaking rotation of
 Eq.~(\ref{eq:SMfieldsID}) is specifically made to have the CKM matrix
 in the up-quark sector of the MSSM, as $K_{10}$ is in the $10_M$-sector
 of the MSSU(5) model, whereas the rotation
\begin{equation}
  \bar{5}_M = \{D^c, e^{-i\phi_l} P_l^\dagger L\},
  \hspace*{1truecm}
  10_M  = \{K^\dagger_{10} Q,
            K^\dagger_{10} P_{10}^\dagger U^c, e^{i\phi_l}P_l E^c\},
\label{eq:SMfieldsIDnew}
\end{equation}
 has the effect of shifting the CKM matrix in the down-squark sector
 of the MSSM:
\begin{equation}
  W^{\prime\,{\rm MSSM}}  = 
  U^c  \,\widehat{Y}_U                               \,Q H_u 
- D^c  \left(\widehat{Y}_D K_{\rm CKM}^\dagger \right) Q H_d 
- E^c  \,\widehat{Y}_D                               \,L H_d . 
\label{eq:WMSSMprimeNEW}
\end{equation}
The flavour bases in this equation and in Eq.~(\ref{eq:WMSSMprime}) are
 the most commonly used in the MSSM. 
The basis in which $W^{\prime\,{\rm MSSM}}$ has the form
\begin{equation}
  W^{\prime\,{\rm MSSM}} =  
  U^c  \left(K_{\rm CKM}^T \widehat{Y}_U K_{\rm CKM}  \right) Q H_u 
- D^c  \,\widehat{Y}_D                                 \,Q H_d 
- E^c  \,\widehat{Y}_D                                 \,L H_d , 
\label{eq:WMSSMprimeNEWNEW}
\end{equation}
 is used in Ref.~\citen{DESYpeople} (see \S\ref{sec:PDconstBC}). 
Except for the removal of the phases in $P_{10}$, this special form does
 not require {\it SU}(5)-breaking rotations to match the MSSU(5) model
 with the MSSM at $M_{\rm GUT}$.

Note that we could have also chosen to parametrize $Y^5$ and $Y^{10}$ in
 $W^{\rm MSSU(5)}_M$ in such a way to have the matrix $K_{10}$ in the
 $\bar{5}_M$ sector:
\begin{equation}  
 W^{\rm MSSU(5)}_{\rm M} =  
  \sqrt{2} \, \bar{5}_M
  \left(\widehat{Y}^5 K_{10}^\dagger P_{10}^{(1)}\right) 
                   10_M \bar{5}_H  
 -\frac{1}{4} 10_M \, \widehat{Y}^{10}\, 10_M 5_H .
\label{eq:WminSU5mRotNEW}
\end{equation}
{}For this, it is sufficient to rotate the field $10_M$ with the matrix 
 $W_{10}$ instead than with $V_{10}$ as done above, and $\bar{5}_M$ as
 in Eq.~(\ref{eq:moreRotat}). 
Starting from this superpotential, which basis for the up- and
 down-quark sectors for the MSSM superpotential should be used is again
 a matter of choice.

Of the two complex matrices in $W_{\rm RHN}$ in
 Eq.~(\ref{eq:SU5sswSUPERP}), $Y^{\rm I}_N$ is a generic complex matrix,
 $M_N$, a symmetric one, and both{{{}}} are diagonalized as 
\begin{equation}
 \widehat{Y}_N^{\rm I}  =
   U_N^{{\rm I}\,T}   Y_N^{\rm I}  V_N^{\rm I},
\hspace*{1.0truecm}
 \widehat{M}_N          =
   W_N^{{\rm I}\,T}   M_N   W_N^{\rm I}. 
\label{eq:YandMsswRotat}
\end{equation}
Thus, for the diagonalization of the mass term in $W_{\rm RHN}$, it is
 sufficient to flavour rotate $N^c$: 
\begin{equation}
  N^c   \rightarrow W_N^{\rm I}  \, N^c.
\label{eq:IsswMediatRot}
\end{equation}
This and the two rotations of $\bar{5}_M$ transform the Yukawa term into
\begin{equation}
 e^{i \phi_{10}}    N^c 
 \left( U_N^{{\rm I}\,\dagger} W_N^{\rm I} \right)^T 
  \widehat{Y}_N^{\rm I}
 \left( V_N^{{\rm I}\,\dagger} V_5 \right)
 P_{10}^{(2)} \    \bar{5}_M 5_H ,
\label{eq:IsswYrotONE}
\end{equation}
 which can be recast in the form
\begin{equation}
 e^{i \phi_{\rm I}}    N^c 
 \left( P_{\rm I}^a K_{U_N}^{{\rm I}\,T} \right) 
  \widehat{Y}_N^{\rm I}
 \left( P_{\rm I}^b K_{V_N}^{\rm I} \right) P_{\rm I} \
                       \bar{5}_M 5_H ,
\label{eq:IsswYrotated}
\end{equation}
 once the two unitary matrices $( U_N^{{\rm I}\,\dagger} W_N^{\rm I} )$
 and $( V_N^{{\rm I}\,\dagger} V_5 ) P_{10}^{(2)}$ are parametrized as
 in Eq.~(\ref{eq:unitary}).

That is, no reduction of the original eighteen parameters of
 $Y_N^{\rm I}$ is possible.   
However, below $M_{\rm GUT}$, the freedom in identifying the {\it SU}(2)
 doublet and singlet leptonic components of $\bar{5}_M$ and $10_M$ (see
 Eq.~(\ref{eq:SMfieldsID})) allows the elimination of three phases in
 the purely leptonic part of this interaction term, say,
 $e^{i \phi_{\rm I}} P_{\rm I}$, and the identification
\begin{equation}
  Y_\nu^{\rm I} = 
  P_I^a K_{U_N}^{I\,T}\,\widehat{Y}_N^{\rm I}\,
  P_{\rm I}^b K_{V_N}^{\rm I}.
\label{eq:YnuSSWIcoupl}
\end{equation}

The situation for the seesaw of type~III mirrors exactly that for the
 seesaw of type~I.
The matrix $M_{\rm 24M}$ is diagonalized as $M_N$, reducing the Yukawa
 interaction to
\begin{equation}
 e^{i \phi_{\rm III}} \,  5_H \, 24_M  
 \left( P_{\rm III}^a K_{U_N}^{{\rm III}\,T} \right) 
  \widehat{Y}_N^{\rm III}
 \left( P_{\rm III}^b K_{V_N}^{\rm III} \right) P_{\rm III}\, 
                       \bar{5}_M ,
\label{eq:IIIsswYrotated}
\end{equation}
 and the connection between $Y_\nu^{\rm III}$ and
 $\widehat{Y}_N^{\rm III}$ is as in Eq.~(\ref{eq:YnuSSWIcoupl}), with
 the replacement ${\rm I}\to {\rm III}$.

{}For the seesaw of type~II, the symmetric matrix $Y_N^{\rm II}$ is
 diagonalized as $Y_S$ in Eq.~(\ref{eq:diagMatrices}), and no
 diagonalization is needed for $M_{15}$, which is just a number. 
Thus, the first Yukawa interaction in $W_{\rm 15H}$ reduces to 
\begin{equation}
 e^{i 2\phi_{\rm II}} \frac{1}{\sqrt{2}} \bar{5}_M P_{\rm II} 
\left(K_{W_N}^{{\rm II}}P_{\rm II}^a \right)^T\widehat{Y}_N^{\rm II}
\left(K_{W_N}^{{\rm II}}P_{\rm II}^a \right)    P_{\rm II} 
   15_H  \bar{5}_M .
\label{eq:IIsswYrotated}
\end{equation}
As before, the phase $\phi_{\rm II}$ and the two in $P_{\rm II}$ can
 be eliminated in the purely leptonic part of this interaction.
The Yukawa coupling of this part has, therefore, only nine physical 
 parameters, those in the symmetric matrix $Y_\nu^{\rm II}$, which
 can be expressed as
\begin{equation}
 Y_\nu^{\rm II} = 
 \left(K_{W_N}^{{\rm II}} P_{\rm II}^a \right) \widehat{Y}_N^{\rm II}
 \left(K_{W_N}^{{\rm II}} P_{\rm II}^a \right)^T.  
\label{eq:YnuSSWIIcoupl}
\end{equation}

\section{Evolution of the~VEVs of the Field $24_H$}
\label{sec:New}
%
We show here that the RGEs in Eq.~(\ref{eq:VEVS24evol}) hold in a
 model-independent way for the leading components of the scalar and
 auxiliary VEVs of any chiral superfield $\phi_j$, with anomalous
 dimension $\gamma_j$, provided the scalar VEV of $\phi_j$ {\it i)}
 exists in the SUSY limit, i.e., it is determined by superpotential
 parameters, and {\it ii)} is much larger than the SUSY-breaking scale
 $\widetilde{m}$.
We call $v_j$ the VEV of $\phi_j$ in the SUSY limit, and $M$ the 
 superpotential large scale that determines it.

As usual, we expand the superpotential and K\"ahler potential in
 powers of the SUSY-breaking VEV $F_{\sf X}$ over the cutoff scale, 
 $F_{\sf X}/M_{\rm cut} =f_{\sf X}$, which we denote generically by
 $\widetilde{m}$:
\begin{eqnarray}
 W(\widetilde{m})
&=&
 W +\widetilde{m} \theta^2 \widetilde W, 
\nonumber\\
 K(\widetilde{m})
&=&
 K +[\widetilde{m}\theta^2 \widetilde K+ {\rm H.c.}]
   +{\cal O}(\widetilde{m}^2). 
\label{eq:genericWKpot}
\end{eqnarray}
The scalar potential is then given by 
\begin{equation}
 V= -F^i K_i^j F_j
    + [(W^j - \widetilde{m} \tilde K^j)F_j
    + \widetilde{m} \widetilde W + {\rm H.c.}]
   +{\cal O}(M^2 \widetilde{m}^2), 
\label{eq:generalVpot}
\end{equation}
 where we use covariant and contravariant indices, which have been 
 avoided elsewhere in the main text. 
The symbols $F_j$~($F^i$) denote the auxiliary components of the
 chiral~(antichiral) multiplets $\phi_j$~($\phi^i$).
In contrast, a subscript~(superscript) index in $V$, $W$, $\widetilde{W}$,
 $K$, and $\widetilde{K}$ denotes the derivative of these functions with
 respect to $\phi^j$~($\phi_i$).

{}From the derivative of the scalar potential with respect to $F_j$, we
 find 
\begin{equation}
 F^i =  (W^j - \widetilde{m} \widetilde K^j)(K^{-1})^i_j,   
\label{eq:FvevGeneral}
\end{equation}
 which allows to recast the scalar potential in the form
\begin{equation}
 V= (W^j - \widetilde{m}\widetilde K^j)(K^{-1})^i_j
    (W_i - \widetilde{m}\widetilde K_i)
   + [\widetilde{m} \widetilde W + {\rm H.c.}]
   +{\cal O}(M^2 \widetilde{m}^2).
\label{eq:generalVpotNEW}
\end{equation} 
Note that if it is possible to reduce the K\"ahler potential to its
 canonical form, with $\widetilde K=0$ and $K^i_j=\delta^i_j$, then the
 scalar potential reduces to that of Eq.~(\ref{eq:VBHcomplete}) in the
 specific case of the MSSU(5) model.
The condition of minimality of the K\"ahler potential is, however, not
 required in the following.

In the SUSY limit $\widetilde{m} \to 0$, the auxiliary component $F^j$
 must vanish and, therefore, 
\begin{equation}
 W^j = 0,
\label{eq:scVEVleading}
\end{equation}
 which determines the scalar VEV $v_j$ independently of the K\"ahler
 potential, in general, of ${\cal O}(M)$.

{}For nonvanishing  $\widetilde{m}$, this scalar VEV so obtained is
 shifted by terms of ${\cal O}(\widetilde{m})$, and an auxiliary VEV is
 also generated. 
At ${\cal O}(M^2 \widetilde{m})$, the vacuum condition is
\begin{equation}
V^k \,=\, W^{jk}(K^{-1})^i_j (W_i - \widetilde{m}\widetilde K_i)
   + \widetilde{m} \widetilde W^k +{\cal O}(M \widetilde{m}^2)
    \, =\, 0,
\end{equation} 
 where we have used the fact that $W_i$, and therefore, also
 $(W_i- \widetilde{m} \widetilde K_i)$, are at most of 
 ${\cal O}(M \widetilde{m})$, from which we obtain the leading,
 nonvanishing contribution to the  auxiliary VEV, of
 ${\cal O}(M \widetilde{m})$:
\begin{equation}
 F_j = (K^{-1})^i_j (W_i - \widetilde{m}\widetilde K_i) 
     = -\widetilde{m} \frac{\widetilde W^k}{W^{jk}}
       +{\cal O}(\widetilde{m}^2).   
\end{equation}
Despite the apparent dependence on the K\"ahler potential shown by the
 first equality in this equation, $F_j$ is in reality determined by the
 superpotential only.

Through the same analysis, it is easy to see that the higher order terms
 $\delta v_j$, $\delta F_j$ and $\delta^2 v_j$ of the scalar and
 auxiliary VEVs of $\phi_j$ expanded in $\widetilde{m}/M$ do depend on
 the K\"ahler potential.
Thus, their evolution equations are sensitive to the vertex corrections
 that this potential obtains at the quantum level and are, therefore,
 quite different from those for $v_j$ and $F_j$.

Having established that $v_j$ and $F_j$ are independent of the K\"ahler
 potential, it is then easy to obtain their evolution equations
 independently of their specific expression.
Redefinitions of the field $\phi_i$ act as
\begin{equation}
 \phi_j    \to
 \exp\left[\left(\gamma_j+\tilde{\gamma}_j\theta^2
           \right)\,t\right]\,
 \left(\phi_j\right)^r,
\label{eq:24Hredefin}
\end{equation}
 with the label $r$ distinguishing the redefined field.
Since these redefinitions are just a renaming, we can equate the~VEVs
 of the left-hand side and right-hand side in this equation, obtaining
\begin{equation}
\frac{1}{t} 
\left\{ \left<\left(\phi_j\right)^r\right>
       -\left<\phi_j\right> 
\right\}  = - 
\left\{        \mbox{\boldmath{$\gamma$}}_j v_j + 
\theta^2 \left[\mbox{\boldmath{$\gamma$}}_j F_j +
    \widetilde{\mbox{\boldmath{$\gamma$}}}_j v_j
       \right]
\right\} +{\cal O}(t),
\label{eq:vevsEVOL}
\end{equation}
 which reproduces Eq.~(\ref{eq:VEVS24evol}) in the limit $t\to 0$.

\section{Renormalization of Effective Couplings}
\label{sec:proof}  
%
We prove in this appendix that the effective couplings evolve all the
 way up to the cutoff scale following RGEs that are formally those of an 
 MSSU(5) model broken at ${\cal O}(s)$, but with gauge interactions
 respecting the {\it SU}(5) symmetry. 
We prove this specifically for the $\bar{5}_M$ sector with a type~I
 seesaw.
We rewrite the superpotential class of operators $Op^{5}$ as 
\begin{equation}
 Op^{5} \ = \ 
 \sqrt{2}\, \left(\bar{5}_M\right)_{A}
 \left(\mbox{\boldmath{$Y$}}^5(24_H\right))_{ABCD}\,
 \left(10_M\right)_{BC} 
 \left(\bar{5}_H\right)_D , 
\label{eq:Op5rewrite}
\end{equation}
 where the effective coupling $\mbox{\boldmath{$Y$}}^5(24_H)$ is
 defined as
\begin{equation}
 \left(\mbox{\boldmath{$Y$}}^{5}(24_H)\right)_{ABCD}   \, = \, 
 \sum_{n+m=0}^k 
 C_{n,m}^{5}
 \left(\displaystyle{\frac{24_H^T}{M_{\rm cut}}}\right)^{\!\!n}_{\!\!AB}  
 \left(\displaystyle{\frac{24_H}{M_{\rm cut}}} \right)^{\!\!m}_{\!\!CD},  
\label{eq:effYUK524def}
\end{equation}
 with $(24_H)^0_{AB}$ defined in Eq.~(\ref{eq:24powerZERO}).
It will be shown later how the effective couplings
 $\mbox{\boldmath{$Y$}}^{5}_i$ ($i=\scriptstyle{D,E,DU,LQ}$) used in 
 \S\ref{sec:effYUK} are related to this.  
Similarly, we rewrite $Op^{5}({\sf{X}})$ in Eq.~(\ref{eq:NRO5tildeX}),
 which gives rise to ${\widetilde O}p^{5}$, as
\begin{equation}
 Op^{5}({\sf{X}})  \ = \ 
 \sqrt{2}\, \left(\bar{5}_M\right)_A 
 \left(\widetilde{\mbox{\boldmath{$Y$}}}^5\!
        ({\sf{X}},24_H)\right)_{ABCD}\,
 \left(10_M\right)_{BC} 
 \left(\bar{5}_H\right)_D , 
\label{eq:tildeOp5rewrite}
\end{equation}
 with $(\widetilde{\mbox{\boldmath{$Y$}}}^5\!({\sf{X}},24_H))_{ABCD}$
 defined as 
\begin{equation}
\left(\widetilde{\mbox{\boldmath{$Y$}}}^5\!
 ({\sf{X}},24_H)\right)_{ABCD}                           \, = \, 
 \left(\!\frac{{\sf{X}}}{M_{\rm cut}}\!\right)
 \sum_{n+m=0}^k 
 a_{n,m}^{5} C_{n,m}^{5}
 \left(\displaystyle{\frac{24_H^T}{M_{\rm cut}}}\right)^{\!\!n}_{\!\!AB} 
 \left(\displaystyle{\frac{24_H}{M_{\rm cut}}}\right)^{\!\!m}_{\!\!CD} .  
\label{eq:effTildeYUK524def}
\end{equation}

We expand $\mbox{\boldmath{$Y$}}^{5}(24_H)$ around the~VEV of $24_H$: 
 $\langle 24_H \rangle=\sqrt2\,\langle 24_H^{24}\,\rangle T^{24}+24_H'$,
 with $T^{24}$ the 24th generator of {\it SU}(5), and $24_H'$ denoting 
 the quantum fluctuations of the field, including the Nambu-Goldstone
 modes:
\begin{equation}
 \mbox{\boldmath{$Y$}}^{5}(24_H)                        \ = \
 \mbox{\boldmath{$Y$}}^{5}(\langle 24_H \rangle )
 +{\mbox{\boldmath{$Y$}}^{5}}'(\langle 24_H \rangle )24_H'
 +{\cal O}({24_H'}^2) ,
\end{equation}
 where we have suppressed here all {\it SU}(5) indices.
(These indices should be handled with care in the second term, where
 the prime indicates the derivative with respect to the field $24_H$.)
We shall reinstate them back when it is important to show the
 {\it SU}(5) structure in quantities containing these couplings.

As discussed in \S\ref{sec:HiggsSECandVACUA},
 $\langle 24_H^{24}\rangle$ is decomposed in the~VEVs of the scalar and
 auxiliary components, $\langle B_H\rangle$ and
 $\langle F_{B_H}\rangle$, as in Eq.~(\ref{eq:V24decomp}).
At the leading order in $\widetilde{m}/M_{\rm GUT}$, which is sufficient
 for this discussion, it is $\langle B_H\rangle=v_{24}$ and
 $\langle F_{B_H}\rangle=F_{24}$.
Thus, the lowest order in this expansion is 
\begin{equation}
 \mbox{\boldmath{$Y$}}^5(\langle 24_H \rangle) \equiv  
 \mbox{\boldmath{$Y$}}^5 + \theta^2  
 \mbox{\boldmath{$A$}}^{5}_{F_{24}}                 =  
 \sum_{n+m=0}^k \, C_{n,m}^{5}
 \left(\!\displaystyle{\frac{\langle 24_H^{24}\rangle}{M_{\rm cut}}}\!
 \right)^{\!\!n+m} \!\! 
 \left(\!\!\sqrt2 T^{24}_{\bar{5}_M}\!\right)^{\!\!n} \!
 \left(\!\!\sqrt2 T^{24}_{\bar{5}_H}\!\right)^{\!\!m},
\label{eq:effYUK5SU5symm}
\end{equation}
 where the lower labels in the generator $T^{24}$ denote whether it is
 acting on the representation $\bar{5}_M$ or $\bar{5}_H$.
Note the use of the symbol $\mbox{\boldmath{$Y$}}^5$ for the term in
 which all~VEVs are scalar, and of
 $\mbox{\boldmath{$A$}}^{5}_{F_{24}}$ for the term in which one of them
 is the auxiliary~VEV.
It is justified by the fact that $\mbox{\boldmath{$Y$}}^5$ and  
 $\mbox{\boldmath{$A$}}^{5}_{F_{24}}$ reduce to the couplings in
 Eq.~(\ref{eq:effY5def}) and to the couplings
 $\mbox{\boldmath{$A$}}^{5}_{i,{F_{24}}}$ discussed in 
 \S\ref{sec:trilSOFTsecNROS}, when the generators $T^{24}$ in this
 equation are replaced by their eigenvalues.
Contrary to that of \S\ref{sec:NROps}, the formulation given here
 allows us to maintain an {\it SU}(5)-symmetric structure.

Similarly, the lowest order in the expansion of 
 $\widetilde{\mbox{\boldmath{$Y$}}}^5\!({\sf{X}},24_H)$, with ${\sf{X}}$
 also replaced by the~VEV of its auxiliary component, is
\begin{equation}
\widetilde{\mbox{\boldmath{$Y$}}}^5\!
(\langle {\sf{X}}\rangle ,\langle 24_H \rangle)        \, \equiv \,
 \theta^2 \mbox{\boldmath{$A$}}^{5}\vert_{F_{\sf{X}}}  \, = \, 
 \theta^2 \!\! \sum_{n+m=0}^k \!
  \widetilde{C}_{n,m}^{5}\!
 \left(\!\displaystyle{\frac{\langle 24_H^{24}\rangle}{M_{\rm cut}}}\!
 \right)^{\!\!n+m} \!\! 
 \left(\!\sqrt2 T^{24}_{\bar{5}_M}\right)^{\!\!n} \!
 \left(\!\sqrt2 T^{24}_{\bar{5}_H}\right)^{\!\!m} ,
\label{eq:effTildeYUK5SU5symm}
\end{equation}
 and the complete effective coupling $\mbox{\boldmath{$A$}}^5$ is
\begin{equation}
 \mbox{\boldmath{$A$}}^5           = 
 \mbox{\boldmath{$A$}}^{5}_{F_{\sf{X}}} +
 \mbox{\boldmath{$A$}}^{5}_{F_{24}}. 
\label{eq:effA5SU5symm}
\end{equation}

We are now in a position to calculate the one-loop corrections to the
 K\"ahler potential coming from the above effective couplings.
{}For example, the corrections to $\bar{5}_M \bar{5}_M^\ast$ due to the
 effective coupling $\mbox{\boldmath{$Y$}}^{5}(24_H)$, with the fields
 $10_M$ and $\bar5_H$ exchanged in the loop, are 
\begin{eqnarray}
 \delta{\cal K} \ \supset  - 4t  \bar{5}_M  \Bigl[ 
&&
 \mbox{\boldmath{$Y$}}^{5}(24_H)                       
 \mbox{\boldmath{$Y$}}^{5\,\dagger}(24_H)                   +   
 \widetilde{\mbox{\boldmath{$Y$}}}^{5}\!({\sf{X}},24_H)
            \mbox{\boldmath{$Y$}}^{5\,\dagger}(24_H)  
\nonumber\\
&&\hspace{-0.45truecm}+  
 \mbox{\boldmath{$Y$}}^{5}(24_H)                       
 \widetilde{\mbox{\boldmath{$Y$}}}^{5\,\dagger}\!({\sf{X}},24_H) + 
 \widetilde{\mbox{\boldmath{$Y$}}}^{5}\!({\sf{X}},24_H)       
 \widetilde{\mbox{\boldmath{$Y$}}}^{5\,\dagger}\!({\sf{X}},24_H)
  \Bigr] \bar{5}_M^\ast ,\quad 
\label{eq:KP5ast5corr}
\end{eqnarray}
 where $t$ is the usual factor $t=\ln(Q/Q_0)/(16\pi^2)$.
The saturation of {\it SU}(5) indices not shown in this equation, is,
 for example for the first term, as follows
\begin{equation} 
 \left(\mbox{\boldmath{$Y$}}^{5}(24_H)\right)_{A'B'C'D'}\,
 \frac{1}{2}
 \left(\delta_{B'B}\delta_{C'C}-\delta_{B'C}\delta_{C'B}\right)
         \delta_{D'D}\,
 \left(\mbox{\boldmath{$Y$}}^{5\,\dagger}(24_H)\right)_{ABCD}.
\label{eq:KP5ast5corrGTF}
\end{equation}
In this expression, the indices $A$ and $A'$ are to be contracted with
 those of $\bar{5}_M$ and $\bar{5}_M^\ast$, and the factors
 $(1/2)(\delta_{B'B}\delta_{C'C}-\delta_{B'C}\delta_{C'B})$ and
 $\delta_{D'D}$ come, respectively, from the propagators of the fields
 $10_M$ and $\bar5_H$ exchanged in the loop.
{}For vanishing NROs, it is 
 $\mbox{\boldmath{$Y$}}^5(24_H)_{ABCD}=Y^5\delta_{AB}\delta_{CD}$, and
 the above product reduces to 
\begin{equation} 
 2\delta_{A'A}Y^{5}Y^{5\,\dagger} .
\label{eq:KP5ast5corrGTFinmsSU(5)}
\end{equation}
The other terms in Eq.~(\ref{eq:KP5ast5corr}) for $\delta{\cal K}$ have
 the same {\it SU}(5) structure.

The form of $\delta{\cal K}$ simplifies considerably if we neglect the
 quantum fluctuation $24_H'$.
In this limit, it is easy to see that the corrections to 
 $\bar{5}_M \bar{5}_M^\ast$  listed above, indeed, decompose in the
 corrections to the terms $D^{c} D^{c\,\ast}$ and $L^\ast L$.
We show this explicitly for the first term of
 Eq.~(\ref{eq:KP5ast5corr}).

When the dynamical part of the field $24_H$ is neglected, the coupling
 $\mbox{\boldmath{$Y$}}^{\!5}\!(24_H\!)$ reduces to that of
 Eq.~(\ref{eq:effYUK5SU5symm}), and the {\it SU}(5) indices $ABCD$ are 
 now carried by the generators $T^{24}$:
\begin{equation}
 \left(\!\sqrt2 T^{24}_{\bar{5}_M}\right)^{\!n}_{AB} 
 \left(\!\sqrt2 T^{24}_{\bar{5}_H}\right)^{\!m}_{CD}    \ = \
 \left(\!\sqrt2 T^{24}_{\bar{5}_M}\right)^{\!n}_{AA} 
 \left(\!\sqrt2 T^{24}_{\bar{5}_H}\right)^{\!m}_{DD} 
 \delta_{AB} \,\delta_{CD}.    
\label{eq:effY5Su5symmINDEX}
\end{equation}
By plugging this expression in Eq.~(\ref{eq:KP5ast5corrGTF}), we obtain
\begin{eqnarray}
&&
 \sum_{n'\!+\!m'=0}^k  \sum_{n\!+\!m=0}^k 
 {C}_{n',m'}^{(5)} {C}_{n,m}^{(5)\,\dagger}
\left(\!\displaystyle{\frac{\langle 24_H^{24}\rangle}{M_{\rm cut}}}\!
\right)^{\!\!n'\!+\!m'} \!\!
\left(\!\displaystyle{\frac{\langle 24_H^{24}\rangle}{M_{\rm cut}}}\!
\right)^{\!\!n\!+\!m} 
 \displaystyle{\frac{1}{2}}
 \sum_D
\left(\delta_{A'A}\delta_{DD}-\delta_{A'D}\delta_{AD}\right)
\nonumber\\
&&
 \hspace*{3.5truecm} \times
\left(\!\sqrt2 T^{24}_{\bar{5}_M}\!\right)^{\!n'}_{A'A'}\! 
\left(\!\sqrt2 T^{24}_{\bar{5}_H}\!\right)^{\!m'}_{DD} 
\left(\!\sqrt2 T^{24}_{\bar{5}_M}\!\right)^{\!n}_{AA}\! 
\left(\!\sqrt2 T^{24}_{\bar{5}_H}\!\right)^{\!m}_{DD} \! .
\label{eq:KP5at5interm1}
\end{eqnarray}
We focus on the first Kronecker $\delta$ in parenthesis. 
The second one, due to the antisymmetry of field $10_M$, taken with its
 minus sign, gives the same result of the first and cancels the factor
 $1/2$ in front of the parenthesis. 
The original {\it SU}(5) indices appearing in
 Eq.~(\ref{eq:KP5ast5corrGTF}) are now fixed as follows: $A'=B'=B=A$
 and $C'=D'=D=C$.
The index $A$, equal to $A'$, fixes the component of the external fields
 $\bar{5}_M$ and $\bar{5}_M^\ast$ and, therefore, whether the K\"ahler
 potential correction in Eq.~(\ref{eq:KP5ast5corrGTF}) is a correction
 to the term $D^{c}D^{c\,\ast}$ or $L^\ast L$.
Similarly, the value of the summed index $D$ fixes the component of the
 field $\bar{5}_H$ running in the loop. 
This is a component of $H_D^C$ for $D\supset \{1,2,3\}$ and a component
 of $H_d$ for $D\supset \{4,5\}$.
The fact that the indices $B$ and $C$ are fixed indicates that the
 component of the field $10_M$ in the other propagator of the loop is
 also automatically fixed by the symmetry. 
That is, the following choice for $A$ and $D$: $A\supset \{4,5\}$ and
 $D\supset \{4,5\}$, but with $A\ne D$, fixes the external fields to be
 $L^\ast$ and $L$ and the fields exchanged in the loop $H_d$ and $E^c$.
By choosing $D$ in the set $\{1,2,3\}$, the fields exchanged in the loop
 are $H_D^C$ and $Q$. 
Thus, for each of these choices, the external fields $\bar{5}_M$ and
 $\bar{5}_M^\ast$ decompose in their SM components, and the previous
 product becomes
\begin{equation}
 \sum_{n'\!+\!m'=0}^k \,
 \sum_{n\!+\!m=0}^k 
   {C}_{n',m'}^{(5)}\,
   {C}_{n,m}^{(5)\,\dagger}
  \left(\!\displaystyle{\frac{\langle 24_H^{24}\rangle}{M_{\rm cut}}}\!
  \right)^{\!\!n'\!+\!m'} \!\!
  \left(\!\displaystyle{\frac{\langle 24_H^{24}\rangle}{M_{\rm cut}}}\!
  \right)^{\!\!n\!+\!m} \!
 \sum_i
  \left(\!I_{\bar{5}_M}\!\right)^{\!n'}_i\! 
  \left(\!I_{\bar{5}_H}\!\right)^{\!m'}_i
  \left(\!I_{\bar{5}_M}\!\right)^{\!n}_i\! 
  \left(\!I_{\bar{5}_H}\!\right)^{\!m}_i ,
\label{eq:KP5at5interm2}
\end{equation}
 with the sum over $D$ now replaced by the sum over 
 $i= \scriptstyle{D,E,DU,LQ}$. 
Thus, in the limit in which the dynamical part of the field $24_H$ is
 neglected, the right-hand side of Eq.~(\ref{eq:KP5ast5corr}) decomposes
 into corrections to the term $D^{c}D^{c\,\ast}$ and $L^\ast L$.
Those to the term $D^{c}D^{c\,\ast}$, in particular, are
\begin{eqnarray}
 \delta{\cal K} 
&\supset &    - t  D^{c} \sum_i c_{(D^c,i)}^\ast
\Bigl[ 
 \mbox{\boldmath{$Y$}}_i^{5}(24_H)                   
 \mbox{\boldmath{$Y$}}_i^{5\,\dagger}(24_H)                 
+\widetilde{\mbox{\boldmath{$Y$}}}_i^{5} ({\sf{X}},24_H)
            \mbox{\boldmath{$Y$}}_i^{5\,\dagger} (24_H)  
\nonumber\\
&&
 \hspace*{2truecm}
+\mbox{\boldmath{$Y$}}_i^{5}(24_H)                   
 \widetilde{\mbox{\boldmath{$Y$}}}_i^{5\,\dagger} ({\sf{X}},24_H)
+\widetilde{\mbox{\boldmath{$Y$}}}_i^{5} ({\sf{X}},24_H)   
 \widetilde{\mbox{\boldmath{$Y$}}}_i^{5\,\dagger} ({\sf{X}},24_H)
 \Bigr] D^{c\,\ast} \!, \qquad
\label{eq:KPDastDcorr}
\end{eqnarray}
 with the group factors $c_{(D^c,i)}^\ast$ and $c_{(L,i)}$, in general,
 different for distinct indices $i$. 
Their sum over $i$, however, reproduces the coefficient $4$ in
 Eq.~(\ref{eq:KP5ast5corr}),
 $\sum_i c_{(D^c,i)}^\ast = \sum_i c_{(L,i)} =4$.

\begin{figure}[t]
\centerline{\includegraphics[width=11.3cm]{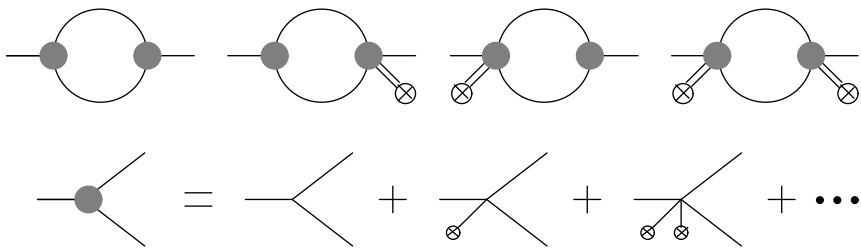}}
\caption{One-loop diagrams contributing corrections to the
         K\"ahler potential and to effective vertices: the black
         circle denotes the effective coupling as shown in the
         lower part of the figure. Here, the single and double
         lines ending in the symbol $\otimes$ indicates the~VEV
         of $24_H$ and ${\sf{X}}$, respectively.}
\label{fig:Diagrams}
\end{figure}

After the above discussion, it is easy to see that the one-loop
 diagrams expressed in terms of effective coupling correctly sum up all
 the one-loop diagrams expressed in terms of the original couplings, as
 schematically illustrated in Fig.~\ref{fig:Diagrams}.
In this figure, the four upper diagrams correspond, in order, to the
 four terms in Eq.~(\ref{eq:KP5ast5corr}).

The part of the corrections in which $24_H'$ is neglected can be 
 reabsorbed by field redefinitions, as the chiral field $24_H$ is 
 replaced by its~VEVs (numbers) and an {\it SU}(5) generator.
This is not true for the contributions to these corrections in which a
 chiral field $24_H$ and an antichiral one acquire the auxiliary~VEV.
These terms will end up contributing to the effective soft masses
 squared, as usual.
This procedure, repeated for all fields, defines all anomalous
 dimensions in terms of the effective couplings. 
Thus, suppressing {\it SU}(5) indices, we can express
 $\mbox{\boldmath{$\gamma $}}_{\bar{5}_M}$ as
\begin{equation}
  \mbox{\boldmath{$\gamma$}}_{\bar{5}_M} =  \frac{1}{2}  
\Bigl\{\bigl[{\rm gauge}\ {\rm contr}\bigr] +
 4\mbox{\boldmath{$Y$}}^{5\,\ast}\mbox{\boldmath{$Y$}}^{5\,T} +
 \cdots
\Bigr\}
 + \mbox{\boldmath{$\zeta$}}_{\gamma_{\bar{5}_M}} , 
\label{eq:EFFanomDimDEF}
\end{equation}
 where in the dots are included the contributions from all other
 effective Yukawa couplings, whereas all remaining quantities are
 collected in $\mbox{\boldmath{$\zeta$}}_{\gamma_{\bar{5}_M}}$.
Note that, in the limit of vanishing NROs, the product of effective
 couplings in this anomalous dimension reduces to the expression in
 Eq.~(\ref{eq:KP5ast5corrGTFinmsSU(5)}) and we recover the 
 numerical coefficient in the anomalous dimension $\gamma_{\bar{5}_M}$ 
 of Eq.~(\ref{eq:I-anomDIMfromGUT}). 
The corresponding quantity
 $\tilde{\mbox{\boldmath{$\gamma $}}}_{\bar{5}_M}$ is
\begin{equation}
  \widetilde{\mbox{\boldmath{$\gamma$}}}_{\bar{5}_M} =   
\Bigl\{-\bigl[{\rm gaugino}\ {\rm contr}\bigr] +
 4 \mbox{\boldmath{$Y$}}^{5\,\ast}
   \mbox{\boldmath{$A$}}^{5\,T} + \cdots
\Bigr\}
 + \widetilde{\mbox{\boldmath{$\zeta$}}}_{\gamma_{\bar{5}_M}} ,
\label{eq:EFFanomDimTildeDEF}
\end{equation}
 with 
 $\widetilde{\mbox{\boldmath{$\zeta$}}}_{\gamma_{\bar{5}_M}}$ playing the
 same role that 
 $\mbox{\boldmath{$\zeta$}}_{\gamma_{\bar{5}_M}}$
 has for 
 $\mbox{\boldmath{$\gamma$}}_{\bar{5}_M}$.
Note that it is truly the complete effective coupling 
 $\mbox{\boldmath{$A$}}^5$ that appears here, to which both
 $\mbox{\boldmath{$Y$}}^5\!(24_H)$ and
 $\widetilde{\mbox{\boldmath{$Y$}}}^5\!({\sf{X}},24_H)$ contribute. 
The minus sign in front of the $[{\rm gaugino\ contr.}]$ is just a 
 symbolic reminder of the fact that this term has an opposite sign to the
 term  $[{\rm gauge\ contr.}]$ in the definition of 
 $\mbox{\boldmath{$\gamma$}}_{\bar{5}_M}$.

We recall that the contributions expressed in terms of effective
 couplings in $\mbox{\boldmath{$\gamma$}}_{\bar{5}_M}$ and 
 $\widetilde{\mbox{\boldmath{$\gamma$}}}_{\bar{5}_M}$ are just numbers
 (with generators) and do not consist of any dynamical field.
This is not true for the quantities
 $\mbox{\boldmath{$\zeta $}}_{\gamma_{\bar{5}_M}}$ and
 $ \widetilde{\mbox{\boldmath{$\zeta$}}}_{\gamma_{\bar{5}_M}}$, which
 collect all the contributions that cannot be expressed in terms of
 effective couplings, in particular, those discussed in
 \S\ref{sec:ValidityEffPic}. 
As already mentioned there, these quantities do break the picture of
 effective couplings.
They are not, in general, subleading with respect to the other terms in
 $\mbox{\boldmath{$\gamma$}}_{\bar{5}_M}$ and
 $\widetilde{\mbox{\boldmath{$\gamma$}}}_{\bar{5}_M}$, but they give
 rise to subleading contributions to sFVs.
Thus, we neglect them in the following discussion.

{}Field redefinitions through the just defined anomalous dimension, such
 as 
 $\bar{5}_M \to \exp{
       [(\mbox{\boldmath{$\gamma$}}_{\bar{5}_M} \!+\!
  \tilde{\mbox{\boldmath{$\gamma$}}}_{\bar{5}_M} \theta^2)\, t ]\,
                    } \bar{5}_M $, 
 and similar ones for all the other fields, allow us to recover the
 minimality of the K\"ahler potential, up to soft masses squared (and
 the contributions collected in 
 $\mbox{\boldmath{$\zeta $}}_{\gamma_{\bar{5}_M}}\!\!$ and
 $ \widetilde{\mbox{\boldmath{$\zeta$}}}_{\gamma_{\bar{5}_M}}\!$), by 
 shifting the couplings in the superpotential.
It is then very easy to express the one-loop leading logarithm
 modifications  of the original couplings in the superpotential, such as
 $C_{n,m}^{5}$, in terms of these anomalous dimensions.
To simplify further our notation, we absorb also the group theoretical
 factors into the original superpotential coefficients:
\begin{equation}
 \mbox{\boldmath{$Y$}}^5_{ABCD}      \ = \ 
  \sum_{n=0}^k 
   \left(C_{n}^{5}\right)_{ABCD}
  \left(\!\displaystyle{\frac{\langle B_H \rangle }{M_{\rm cut}}}\!
  \right)^{\!\!n}.
\label{eq:effC5}
\end{equation}
Similarly, 
 $\mbox{\boldmath{$A$}}^5_{F_{\sf{X}}}$ and
 $\mbox{\boldmath{$A$}}^5_{F_{24}}$ now have the simplified expressions
\begin{eqnarray}
 \left(\mbox{\boldmath{$A$}}^5_{F_{\sf{X}}}\right)_{ABCD} 
   & = & 
  \displaystyle{\sum_{n=0}^k} 
   \left(\widetilde{C}_{n}^{5}\right)_{ABCD}
  \left(\!\displaystyle{\frac{\langle B_H \rangle }{M_{\rm cut}}}\!
  \right)^{\!\!n},
\nonumber\\
 \left(\mbox{\boldmath{$A$}}^5_{F_{24}}\right)_{ABCD}
   & = & 
  \displaystyle{\sum_{n=0}^k} 
    n f_{24}
   \left(C_{n}^{5}\right)_{ABCD}
  \left(\!\displaystyle{\frac{\langle B_H \rangle }{M_{\rm cut}}}\!
  \right)^{\!\!n}.
\label{eq:effAFXAF24}
\end{eqnarray}

By keeping into account the evolution of the~VEVs of the field
 $24_H$, it is easy to obtain the RGEs for the effective Yukawa and
 trilinear couplings:
\begin{eqnarray}
\dot{\left(\mbox{\boldmath{$Y$}}^{5} \big\vert_{n}\right)} &=&
 \left\{\mbox{\boldmath{$\gamma $}}_{\bar{5}_M}^T \,
        \mbox{\boldmath{$Y$}}^{5} \big\vert_{n}
      + \mbox{\boldmath{$Y$}}^{5} \big\vert_{n}
        \mbox{\boldmath{$\gamma $}}_{10_M} 
      + \mbox{\boldmath{$\gamma $}}_{\bar{5}_H}
        \mbox{\boldmath{$Y$}}^{5} \big\vert_{n}  \right\} , 
\nonumber \\
\dot{\left(\mbox{\boldmath{$A$}}^{5}\big\vert_{n}\right) } &=&
\left\{\mbox{\boldmath{$\gamma $}}_{\bar{5}_M}^T 
       \mbox{\boldmath{$A$}}^{5}\big\vert_{n}         
   +   \mbox{\boldmath{$A$}}^{5}\big\vert_{n}
       \mbox{\boldmath{$\gamma $}}_{10_M}           
   +   \mbox{\boldmath{$\gamma $}}_{\bar{5}_H} 
       \mbox{\boldmath{$A$}}^{5}\big\vert_{n}   \right\}  
\nonumber\\                                               &&                    
 \qquad +
 \left\{
 \widetilde{\mbox{\boldmath{$\gamma $}}}_{\bar{5}_M}^T 
           \mbox{\boldmath{$Y$}}^{5} \big\vert_{n}             
+          \mbox{\boldmath{$Y$}}^{5} \big\vert_{n}
 \widetilde{\mbox{\boldmath{$\gamma $}}}_{10_M}                 
+\widetilde{\mbox{\boldmath{$\gamma $}}}_{\bar{5}_H}        
           \mbox{\boldmath{$Y$}}^{5} \big\vert_{n}   \right\} ,
\label{eq:RGEforEffY5A5}
\end{eqnarray}
 where we have suppressed the {\it SU}(5) indices for simplicity.
The one-loop evolution equation for the coefficients $C_{n}^{5}$ and
 $\widetilde{C}_{n}^{5}$ needed to derive them are discussed in detail in
 \S\ref{sec:NROinRGE}. 
The evolution equation for 
 $(\mbox{\boldmath{$A$}}^{5}_{F_{{\sf{X}}}})\!\vert_n$ and that for 
 $(\mbox{\boldmath{$A$}}^{5}_{F_{24}})\vert_n$ are formally like the
 equation for $\mbox{\boldmath{$A$}}^{5}\vert_{n}$, except for one 
 spurious term. 
This is 
 $ n \widetilde{\mbox{\boldmath{$\gamma $}}}_{24_H}$ in the RGE for
 $(\mbox{\boldmath{$A$}}^{5}_{F_{{\sf{X}}}})\!\vert_n$, and 
 $- n \widetilde{\mbox{\boldmath{$\gamma $}}}_{24_H}${{{}}} in that for
 $(\mbox{\boldmath{$A$}}^{5}_{F_{24}})\vert_n$.
Thus, the role played by $\mbox{\boldmath{$A$}}^{5}_{F_{24}}$ is very
 important for the correct evolution of
 $\mbox{\boldmath{$A$}}^{5} \big\vert_{n}$.
The RGEs for the effective couplings 
 $\mbox{\boldmath{$Y$}}^{5}$ and
 $\mbox{\boldmath{$A$}}^{5}$ are obtained from 
 $\mbox{\boldmath{$Y$}}^{5}\vert_{n}$ and
 $\mbox{\boldmath{$A$}}^{5}\vert_{n}$ with a summation over $n$.

Having proven already how the products of the {\it SU}(5)-symmetric
 effective couplings introduced in this section correctly decompose into
 products of the decomposed effective couplings of
 \S\ref{sec:NROps}, we can split the obtained RGEs for {\it SU}(5)
 multiplets into RGEs for their SM components.
{}For example, the RGEs for the four effective couplings
 $\mbox{\boldmath{$Y$}}^{5}_i$ are then
\begin{eqnarray}
\dot{\mbox{\boldmath{$Y$}}}^{5}_D  \     & = &
  \mbox{\boldmath{$\gamma$}}_{D^c}^T \,\mbox{\boldmath{$Y$}}^{5}_D  \
 +\mbox{\boldmath{$Y$}}^{5}_D          \mbox{\boldmath{$\gamma$}}_Q \
 +\mbox{\boldmath{$\gamma$}}_{H_d}     \mbox{\boldmath{$Y$}}^{5}_D,
\nonumber\\
\dot{\mbox{\boldmath{$Y$}}}^{5}_E  \     & = & 
  \mbox{\boldmath{$\gamma$}}_{L}^T \,\mbox{\boldmath{$Y$}}^{5}_E    \
 +\mbox{\boldmath{$Y$}}^{5}_{E}      \mbox{\boldmath{$\gamma$}}_{E^c}\ 
 +\mbox{\boldmath{$\gamma$}}_{H_d}   \mbox{\boldmath{$Y$}}^{5}_E,
\nonumber\\                            
\dot{\mbox{\boldmath{$Y$}}}^{5}_{DU}     & = &
  \mbox{\boldmath{$\gamma$}}_{D^c}^T\,\mbox{\boldmath{$Y$}}^{5}_{\!DU}
 +\mbox{\boldmath{$Y$}}^{5}_{\!DU}    \mbox{\boldmath{$\gamma$}}_{U^c} 
 +\mbox{\boldmath{$\gamma$}}_{H_D^C}  \mbox{\boldmath{$Y$}}^{5}_{\!DU},
\nonumber\\                              
\dot{\mbox{\boldmath{$Y$}}}^{5}_{LQ}     & = &
  \mbox{\boldmath{$\gamma$}}_{L}^T \,\mbox{\boldmath{$Y$}}^{5}_{\!LQ} 
 +\mbox{\boldmath{$Y$}}^{5}_{\!LQ}   \mbox{\boldmath{$\gamma$}}_{Q} 
 +\mbox{\boldmath{$\gamma$}}_{H_D^C} \mbox{\boldmath{$Y$}}^{5}_{\!LQ} ,
\label{eq:Y5NRORGE}
\end{eqnarray}
 where the decompositions 
$\gamma_{\bar5_M}\to
 \{\mbox{\boldmath{$\gamma$}}_{D^c},
   \!\mbox{\boldmath{$\gamma$}}_{L}\}$,
$\gamma_{10_M}\to
 \{\mbox{\boldmath{$\gamma$}}_{U^c},
   \!\mbox{\boldmath{$\gamma$}}_Q,
   \mbox{\boldmath{$\gamma$}}_{E^c}\}$, and
$\gamma_{\bar5_H}\to
 \{\mbox{\boldmath{$\gamma$}}_{H_D^C},
   \!\mbox{\boldmath{$\gamma$}}_{H_d}\}$
 are also made.
Among the decomposed anomalous dimensions, 
 $\mbox{\boldmath{$\gamma$}}_{L}$ and
 $\mbox{\boldmath{$\gamma$}}_{D^c}$, taken here as examples, are given
 by 
\begin{eqnarray}
\mbox{\boldmath{$\gamma$}}_{L}           & = &  
-2 \left(\displaystyle{\frac{12}{5}} g_5^2 \right)  {\mathop{\bf 1}}  +
     \mbox{\boldmath{$Y$}}^{5\,\ast}_E 
     \mbox{\boldmath{$Y$}}^{5\,T}_E              +
    3\mbox{\boldmath{$Y$}}^{5\,\ast}_{\!LQ}
     \mbox{\boldmath{$Y$}}^{5\,T}_{\!LQ}         +
     \mbox{\boldmath{$Y$}}_N^{{\rm I}\,\dagger}
     \mbox{\boldmath{$Y$}}_N^{\rm I} ,
\nonumber\\
\mbox{\boldmath{$\gamma$}}_{D^c}\!       & = &  
-2 \left(\displaystyle{\frac{12}{5}} g_5^2 \right)  {\mathop{\bf 1}}  +
    2\mbox{\boldmath{$Y$}}^{5\,\ast}_D 
     \mbox{\boldmath{$Y$}}^{5\,T}_D              +
    2\mbox{\boldmath{$Y$}}^{5\,\ast}_{\!DU}
     \mbox{\boldmath{$Y$}}^{5\,T}_{\!DU}         + 
     \mbox{\boldmath{$Y$}}_{ND}^{{\rm I}\,\dagger}
     \mbox{\boldmath{$Y$}}_{ND}^{{\rm I}} ,
\label{eq:anomdimLandDNRO}
\end{eqnarray}
 and we refer the reader to Appendix~\ref{sec:RGEGUTtoPL-NRO} for the 
 remaining ones.

Similarly, $\mbox{\boldmath{$A$}}^{5}_D$ among the four effective 
 trilinear couplings evolves according to
\begin{equation}
 \dot{\mbox{\boldmath{$A$}}}^{5}_D               \, = \,
\left(
  \mbox{\boldmath{$\gamma$}}_{D^c}^T                
  \mbox{\boldmath{$A$}}^{5}_D                     +
  \mbox{\boldmath{$A$}}^{5}_D
  \mbox{\boldmath{$\gamma$}}_{Q}                  +
  \mbox{\boldmath{$\gamma$}}_{H_d} 
  \mbox{\boldmath{$A$}}^{5}_D
\right) +
\left(
  \widetilde{\mbox{\boldmath{$\gamma $}}}_{D^c}^T 
             \mbox{\boldmath{$Y$}}^{5}_D          +
             \mbox{\boldmath{$Y$}}^{5}_D                
  \widetilde{\mbox{\boldmath{$\gamma $}}}_Q       +
  \widetilde{\mbox{\boldmath{$\gamma $}}}_{H_d} 
             \mbox{\boldmath{$Y$}}^{5}_D
\right),
\label{eq:A5NRORGE}
\end{equation}
 with the quantities 
 $ \widetilde{\mbox{\boldmath{$\gamma$}}}_i$ also decomposed:
 $ \widetilde{\mbox{\boldmath{$\gamma$}}}_{\bar5_M}\to 
 \{\widetilde{\mbox{\boldmath{$\gamma$}}}_{D^c},
   \widetilde{\mbox{\boldmath{$\gamma$}}}_{L}\}$,
 $ \widetilde{\mbox{\boldmath{$\gamma$}}}_{\bar5_H}\to \{
   \widetilde{\mbox{\boldmath{$\gamma$}}}_{H_D^C},
   \widetilde{\mbox{\boldmath{$\gamma$}}}_{H_d}\}$,
and 
 $ \widetilde{\mbox{\boldmath{$\gamma$}}}_{10_M}\to
 \{\widetilde{\mbox{\boldmath{$\gamma$}}}_{U^c},
   \widetilde{\mbox{\boldmath{$\gamma$}}}_{Q},
   \widetilde{\mbox{\boldmath{$\gamma$}}}_{E^c}\}$.

The RGEs listed above are those of a renormalizable model in which 
 only the gauge interactions respect the {\it SU}(5) symmetry, whereas
 all the other couplings break it at ${\cal O}(s)$.

It is simple to prove that also the effective soft masses squared obey
 RGEs analogous to those for soft masses squared in this broken
 MSSU(5) model.
A key role in this proof is played again by the fact that the wave
 function renormalization of the field $24_H$ is cancelled by the
 evolution of the $24_H$~VEVs.
In particular, the RGE for ${\widetilde{m}}_{\bar{5}_M}^2$ is now replaced
 by those for $\widetilde{\mbox{\boldmath{$m$}}}_{L}^{2}$ and
 $\widetilde{\mbox{\boldmath{$m$}}}_{D^c}^{2}$:
\begin{eqnarray}
 \dot{\widetilde{\mbox{\boldmath{$m$}}}}_{L}^{2}   & = & 
 -8 \left(\!\displaystyle{\frac{12}{5}} g_5^2 M_G^2 \!\right)
          \!{\mathop{\bf 1}}                       
+ \  {\cal F}(\mbox{\boldmath{$Y$}}_E^{5\,\ast},
             \widetilde{\mbox{\boldmath{$m$}}}_{L}^{2},
             \widetilde{\mbox{\boldmath{$m$}}}_{E^c}^2,
             \widetilde{\mbox{\boldmath{$m$}}}_{H_d}^2,
             \mbox{\boldmath{$A$}}_E^{5\,\ast})
\nonumber\\                                      &   &    
\phantom{ -8 \left(\!\displaystyle{\frac{12}{5}} g_5^2 M_G^2 \!\right)
          \!{\mathop{\bf 1}}}                   
+ 3  {\cal F}(\mbox{\boldmath{$Y$}}_{LQ}^{5\,\ast},
             \widetilde{\mbox{\boldmath{$m$}}}_{L}^{2},
             \widetilde{\mbox{\boldmath{$m$}}}_{Q}^{2\,\ast},
             \widetilde{\mbox{\boldmath{$m$}}}_{H_D^C}^2,
             \mbox{\boldmath{$A$}}_{LQ}^{5\,\ast})   
\nonumber\\                                      &   &
\phantom{ -8 \left(\!\displaystyle{\frac{12}{5}} g_5^2 M_G^2 \!\right)
          \!{\mathop{\bf 1}}}                   
+ \  {\cal F}(\mbox{\boldmath{$Y$}}_N^{{\rm I}\,\dagger},
             \widetilde{\mbox{\boldmath{$m$}}}_{L}^{2},
             \widetilde{\mbox{\boldmath{$m$}}}_{N^c}^{2},
             \widetilde{\mbox{\boldmath{$m$}}}_{H_u}^2, 
             \mbox{\boldmath{$A$}}_N^{{\rm I}\,\dagger}) ,  
\nonumber\\[1.0011ex]
 \dot{\widetilde{\mbox{\boldmath{$m$}}}}_{D^c}^{2}  & = & 
 -8 \left(\!\displaystyle{\frac{12}{5}} g_5^2 M_G^2 \!\right)
          \!{\mathop{\bf 1}}      
+  2 {\cal F}(\mbox{\boldmath{$Y$}}_D^{5},
             \widetilde{\mbox{\boldmath{$m$}}}_{D^c}^{2},
             \widetilde{\mbox{\boldmath{$m$}}}_{Q}^{2},
             \widetilde{\mbox{\boldmath{$m$}}}_{H_d}^2,
             \mbox{\boldmath{$A$}}_D^{5})   
\nonumber\\                                      &   &
\phantom{ -8 \left(\!\displaystyle{\frac{12}{5}} g_5^2 M_G^2 \!\right)
          \!{\mathop{\bf 1}}}                   
+  2 {\cal F}(\mbox{\boldmath{$Y$}}_{DU}^{5},
             \widetilde{\mbox{\boldmath{$m$}}}_{D^c}^{2},
             \widetilde{\mbox{\boldmath{$m$}}}_{U^c}^{2\,\ast},
             \widetilde{\mbox{\boldmath{$m$}}}_{H_D^C}^2,
             \mbox{\boldmath{$A$}}_{DU}^{5})   
\nonumber\\                                      &   &
\phantom{ -8 \left(\!\displaystyle{\frac{12}{5}} g_5^2 M_G^2 \!\right)
          \!{\mathop{\bf 1}}}                   
+ \  {\cal F}(\mbox{\boldmath{$Y$}}_{ND}^{{\rm I}\,T},
             \widetilde{\mbox{\boldmath{$m$}}}_{D^c}^{2},
             \widetilde{\mbox{\boldmath{$m$}}}_{N^c}^{2\,\ast},
             \widetilde{\mbox{\boldmath{$m$}}}_{H_U^C}^2, 
             \mbox{\boldmath{$A$}}_{ND}^{{\rm I}\,T}) ,  
\label{eq:RGEbar5MwithNRO}
\end{eqnarray}
 where  $\widetilde{m}_{\bar{5}_H}^2$ and $\widetilde{m}_{5_H}^2$
 were also replaced by 
 $\widetilde{\mbox{\boldmath{$m$}}}_{H_D^C}^2,
  \widetilde{\mbox{\boldmath{$m$}}}_{H_d}^2$ and
 $\widetilde{\mbox{\boldmath{$m$}}}_{H_U^C}^2,
  \widetilde{\mbox{\boldmath{$m$}}}_{H_u}^2$, respectively, 
 and $\widetilde{m}_{10_M}^2$ by  
 $\widetilde{\mbox{\boldmath{$m$}}}_{U^c}^2$,
 $\widetilde{\mbox{\boldmath{$m$}}}_{Q}^2$, and
 $\widetilde{\mbox{\boldmath{$m$}}}_{E^c}^2$.

A complete list of all other RGEs for effective couplings and explicit
 definitions of the various $\mbox{\boldmath{$\gamma$}}_{i}$ and 
 $\widetilde{\mbox{\boldmath{$\gamma$}}}_{i}$ can be found in
 Appendix~\ref{sec:RGEGUTtoPL-NRO}.

\section{MSSU(5) and nrMSSU(5) Models with Seesaw: RGEs}
\label{sec:RGEsSU5}  
%
We collect in this appendix the one-loop RGEs for the MSSU(5) models
 and for the nrMSSU(5) models with seesaw sectors, for
 different values of the scale $Q$ from $M_{\rm weak}$ to $M_{\rm cut}$.
We refer the reader to
 Refs.~\citen{RGES,RGESfalck,RGEs2loopYY,RGEs2loop,SLANSKY,CUI} for
 the ingredients useful to derive them.
The number of RGEs to be listed is very large.
Thus, we omit some of these equations, giving algorithms to obtain
 them from other equations, which we list, or from beta function
 coefficients and/or anomalous
 dimensions~\cite{RGEs2loopYY,RGEsfromWaveFncRen}. 
We illustrate in the following how to obtain the omitted RGEs.

\begin{list}{$\bullet$}{}
\item Gauge and gaugino RGEs: \hfill \\
It is well known that for a collection of gauge groups $G_i$
 with gauge couplings $g_i$, and beta function coefficients $b_i$,
 the gauge couplings and the gaugino masses $M_i$ satisfy the
 evolution equations:
\begin{eqnarray}
 \dot{g}_i \, &=&  \, b_i \, g_i^3 , 
\nonumber\\ 
 \dot{M}_i    &=&  2b_i\, g_i^2 M_i .
\label{eq:GaugGauginoRGEs}
\end{eqnarray} 
Thus, we omit these RGEs, giving only the beta function coefficient for
 each specific gauge group. 
\item Trilinear soft parameters: \hfill \\
We omit also the RGEs for all couplings of trilinear soft terms since
 they can be obtained from those of the corresponding Yukawa couplings
 as follows.
We assume here a Yukawa interaction $S Y_xT V$, where $S$ and $T$ are
 two generic fields with flavour, $V$ a flavourless one, i.e., a
 Higgs field.
Given the RGE for $Y_x$,
\begin{equation}
 \dot{Y}_x    = \gamma^T_S Y_x + Y_x \gamma_T + \gamma_V Y_x ,
\label{eq:GenericYRGE}
\end{equation}
 with $\gamma_S$, $\gamma_T$, and $\gamma_V$ the anomalous dimensions of
 the fields $S$, $T$, $V$, respectively, the RGE for the corresponding
 $A_x$ in the soft term $\widetilde{S} A_x \widetilde{T} V$ is 
\begin{equation}
\dot{A}_x    = 
\left(\gamma^T_S A_x + A_x \gamma_T + \gamma_V A_x \right) +
\left(\tilde{\gamma}^T_S Y_x + Y_x \tilde{\gamma}_T 
                                  +\tilde{\gamma}_V Y_x \right).
\label{eq:GenericARGE} 
\end{equation}
The quantities $\tilde{\gamma}_S$, $\tilde{\gamma}_T$, and
 $\tilde{\gamma}_V$ can be built from the corresponding anomalous
 dimensions ${\gamma}_S$, ${\gamma}_T$, and ${\gamma}_V$ with the
 replacements:
\begin{equation}
 g_i^2           \to  - 2 g_i^2 M_i,        \hspace*{1truecm}
 Y_x^\dagger Y_x \to    2 Y_x^\dagger A_x,  \hspace*{1truecm}
 Y_x^\ast Y_x^T  \to    2 Y_x^\ast A_x^T.  
\label{eq:replacements}
\end{equation}
In short,{{{}}}   
\begin{equation}
 \dot{A}_x = 
 \dot{Y}_x \vert_{Y_x \to A_x} 
+\dot{Y}_x \vert_{\gamma_j \to \tilde{\gamma}_j}. 
  \quad\quad (j=S,T,V)
\label{eq:TRILINEARrgeALGORITHM}
\end{equation}
\item Bilinear soft parameters: \hfill \\
A similar algorithm can be used to obtain the RGE of the coupling $B$ of
 a bilinear soft term $\widetilde{S} B \widetilde{T}$ from that of the
 coupling $M$ of the corresponding superpotential term $S M T$:
\begin{equation}
 \dot{M} = \gamma^T_S M + M \gamma_T .
\label{eq:GenericMRGE}
\end{equation}
The two fields $S$ and $T$ are assumed here to have flavour, but they
 may also be flavourless.
The RGE for $B$ is
\begin{equation}
 \dot{B} = \dot{M} \vert_{M \to B} + 
           \dot{M}   \vert_{\gamma_j \to \tilde{\gamma}_j}.
  \quad\quad (j=S,T)
\label{eq:BTERMSrgeALGORITHM}
\end{equation}
Thus, we omit also all the RGEs of bilinear soft couplings. 
\item Soft sfermion and Higgs masses: \hfill \\
It is also possible to deduce the RGEs for the soft masses of sfermion
 and Higgs fields from the expressions of their anomalous dimensions, 
 by following the algorithm given hereafter.

The anomalous dimension of any field, say, for example the generic field
 with flavour $T$ introduced above, obtains, in general, contributions from
 gauge interactions as well as Yukawa interactions (in our specific
 case, $SY_xTV$):
\begin{equation}
 \gamma_T = \gamma_{G,T} + \gamma_{Y,T},  
\label{eq:adGAMMAsplit}
\end{equation}
 with a similar splitting holding also for $\gamma_S$ and $\gamma_V$.
The two contributions $\gamma_{G,T}$ and $\gamma_{Y,T}$ are
\begin{equation}
 \gamma_{G,T} = \sum_i c_T^i g_i^2,  
\hspace{1truecm}
  \gamma_{Y,T}= c_T^x Y_x^\dagger Y_x,
\label{eq:TanDimPARTS}
\end{equation}
 where $c_T^i$ and $c_T^x$ are group theoretical factors. 
{}For the fields $S$ and $V$, the Yukawa-coupling-induced $\gamma_{Y,S}$
 and $\gamma_{Y,V}$ are
\begin{equation}
 \gamma_{Y,S} = c_S^x Y_x^\ast Y_x^T,
\hspace{1truecm}
 \gamma_{Y,V}= c_V^x {\rm Tr} Y_x^\dagger Y_x
             = c_V^x {\rm Tr} Y_x^\ast Y_x^T.
\label{eq:SVanDimYukPARTS}
\end{equation}

The RGE for the soft mass of the field $T$, $\widetilde{m}_T^2$, as 
 well as those for $\widetilde{m}_S^2$ and $\widetilde{m}_V^2$, 
 can also be split in two parts:
\begin{equation}
 \dot{\widetilde{m}}_T^2 = M_{G,T}^2 + M_{Y,T}^2.      
\label{eq:mRGEsplit}
\end{equation}
Note that $\widetilde{m}_T^2$ and $\widetilde{m}_S^2$ are hermitian
 matrices, whereas $\widetilde{m}_V^2$, like all soft masses of Higgs
 fields, is a real number.

The terms $M_{G,T}^2$, $M_{G,S}^2$, and $M_{G,V}^2$, are obtained from
 $\gamma_{G,T}$,  $\gamma_{G,S}$, and $\gamma_{G,V}$, respectively,{{{}}} with the
 replacements: 
\begin{equation}
  g_i^2 \to 4 g_i^2 M_i^2 .
\label{eq:anGAMMAGreplac}
\end{equation} 
The gaugino masses are assumed here to be real; otherwise, the correct
 replacements would be $4 g_i^2 \vert M_i\vert^2$ .

As for the terms $M_{Y,T}^2$, $M_{Y,S}^2$, and $M_{Y,V}^2$, we start
 assuming that the soft mass terms for the fields $T$, $S$ and $V$ are
 of types 
\begin{equation}
\widetilde{T}^\ast \widetilde{m}_T^2 \widetilde{T},
\qquad
\widetilde{S}^\ast \widetilde{m}_S^2 \widetilde{S},
\qquad
\widetilde{V}^\ast \widetilde{m}_V^2 \widetilde{V}.
\label{eq:TSVsoftmasses1}
\end{equation}
Then, $M_{Y,T}^2$, $M_{Y,S}^2$, and $M_{Y,V}^2$ are obtained from
 $\gamma_{Y,T}$, $\gamma_{Y,S}$, and $\gamma_{Y,V}$, respectively, as follows.
{}For the two fields with flavour, $T$ and $S$, we have
\begin{eqnarray}
 \gamma_{Y,T} =
c_T^x \,Y_x^\dagger Y_x & \ \to &  
c_T^x \,{\cal F}(Y_{x}^\dagger,
                 \widetilde{m}_{T}^2,
                 \widetilde{m}_{S}^{2\,\ast},
                 \widetilde{m}_{V}^2,
                 A_{x}^\dagger) 
              = M_{Y,T}^2,
\nonumber\\
 \gamma_{Y,S} = 
c_S^x \,Y_x^\ast Y_x^T \! & \to &  
c_S^x \,{\cal F}(Y_{x}^\ast,
                 \widetilde{m}_{S}^2,
                 \widetilde{m}_{T}^{2\,\ast},
                 \widetilde{m}_{V}^2,
                 A_{x}^\ast)
              = M_{Y,S}^2,
\label{eq:YukGAMMAreplacFL}
\end{eqnarray}
with the function ${\cal F}$ defined as
\begin{eqnarray}
{\cal F}(Y_h^\dagger,
         \widetilde{m}^2_j,
         \widetilde{m}^{2\,\ast}_k,
         \widetilde{m}^2_l,
         A_h^\dagger)
&=& 
   Y_h^\dagger Y_h \widetilde{m}^2_j
+ \widetilde{m}^2_j Y_h^\dagger Y_h
\nonumber \\ 
&&
+ 2\left(   Y_h^\dagger \widetilde{m}^{2\,\ast}_k Y_h +\!
            \widetilde{m}^2_l Y_h^\dagger Y_h         +\! 
            A_h^\dagger A_h              \right) ,\quad
\label{eq:calFuncDEF}
\end{eqnarray}
 where $Y_h$ is any of the Yukawa couplings in the superpotential,
 $\widetilde{m}^2_j$ is the mass squared for which the RGE in question
 is derived,
 $\widetilde{m}^2_k$ and $\widetilde{m}^2_l$ are the masses squared of
 the particles exchanged in the loops that induce the RGE, and $A_h$ is
 the soft counterpart of $Y_h$. 
As for the order in which $\widetilde{m}^2_k$ and $\widetilde{m}^2_l$
 must appear in ${\cal F}$, the following rules apply. 
{}For flavour-dependent interactions, and if $\widetilde{m}^2_j$ is the
 mass squared of a field with flavour, as in this case,
 then, in the second position, there must be the conjugate of the mass
 squared of the other field with flavour, in the third, that of the Higgs
 field.
Moreover, it goes without saying that, here and in the following, 
 when the Yukawa coupling on which ${\cal F}$ depends is 
 $Y_h^\ast$,  $Y_h^T$, or $Y_h$, instead of $Y_h^\dagger$, the
 sequences $Y_h^\dagger\cdots Y_h$ in the terms on the right-hand side
 of Eq.~(\ref{eq:calFuncDEF}) must be replaced by
 $Y_h^\ast\cdots Y_h^T$, $Y_h^T\cdots Y_h^\ast$, and
 $Y_h\cdots Y_h^\dagger$, respectively.

{}For the flavourless field $V$, both the anomalous dimension and
 the soft mass are real numbers, and the replacement of 
\begin{equation}
 \gamma_{Y,V} = 
 c_V^x \,{\rm Tr} Y_x^\dagger Y_x \ = \
 c_V^x \,{\rm Tr} Y_x^\ast Y_x^T   \ = \ 
 c_V^x \,{\rm Tr} Y_x^T Y_x^\ast   \ = \
 c_V^x \,{\rm Tr} Y_x Y_x^\dagger ,   
\label{eq:GAMMAtermsNOFL}
\end{equation}
 is any of the following terms: 
\begin{eqnarray}
   c_V^x \, {\rm Tr}
 {\cal F}(Y_{x}^\dagger,
          \widetilde{m}_{V}^2,
          \widetilde{m}_{S}^{2\,\ast},
          \widetilde{m}_{T}^2,
          A_{x}^\dagger)
&=&c_V^x \, {\rm Tr}
 {\cal F}(Y_{x}^\ast,
          \widetilde{m}_{V}^2,
          \widetilde{m}_{T}^{2\,\ast},
          \widetilde{m}_{S}^2,
          A_{x}^\ast) 
\nonumber\\
&=&c_V^x \, {\rm Tr}
 {\cal F}(Y_{x}^T,
          \widetilde{m}_{V}^2,
          \widetilde{m}_{S}^{2},
          \widetilde{m}_{T}^{2\,\ast},
          A_{x}^T)             
\nonumber\\
&=&c_V^x \, {\rm Tr}
 {\cal F}(Y_{x},
          \widetilde{m}_{V}^2,
          \widetilde{m}_{T}^{2},
          \widetilde{m}_{S}^{2\,\ast},
          A_{x})                 
\nonumber\\
&=& M_{Y,V}^2. 
\label{eq:YukGAMMAreplacNOFL}
\end{eqnarray}
In this case, the order in which the other two masses
 $\widetilde{m}_{T}^{2}$ and $\widetilde{m}_{S}^{2}$ appear in the 
 function ${\cal F}$, in particular, whether it is the second or the
 third mass the one chosen to be conjugated is not important,
 provided the Yukawa coupling and the corresponding trilinear coupling
 are modified according to Eq.~(\ref{eq:YukGAMMAreplacNOFL}). 
This is because of the presence of a trace in front of ${\cal F}$.

{}Note that if the Yukawa interaction is among flavourless fields,
 $Y_h$ is simply a number, the three masses in ${\cal F}$ are real
 numbers, and the order in which these three masses appear is totally
 irrelevant.

It is possible that the soft mass for the field $T$ is of type
 $\widetilde{T} \widetilde{m}_T^2 \widetilde{T}^\ast$, whereas those for
 $S$ and $V$ are as before:
\begin{equation}
\widetilde{T} \widetilde{m}_T^2 \widetilde{T}^\ast,
\qquad
\widetilde{S}^\ast \widetilde{m}_S^2 \widetilde{S},
\qquad
\widetilde{V}^\ast \widetilde{m}_V^2 \widetilde{V}.
\label{eq:TSVsoftmasses2}
\end{equation}
Since $\widetilde{m}_T^2$ is Hermitian, it is 
 $\widetilde{T} \widetilde{m}_T^2 \widetilde{T}^\ast = 
  \widetilde{T}^\ast \widetilde{m}_T^{2\,\ast} \widetilde{T}$. 
Thus, the RGEs for $\widetilde{m}_S^2$ and $\widetilde{m}_V^2$ are as
 those described before, with the replacement
 $\widetilde{m}_T^2 \leftrightarrow \widetilde{m}_T^{2\,\ast}$ in
 $M_{Y,S}^2$  and $M_{Y,V}^2$.
In the case of the RGE of $\widetilde{m}_T^2$ itself, the same
 replacement
 $\widetilde{m}_T^2 \leftrightarrow \widetilde{m}_T^{2\,\ast}$ has to be
 made.
In addition an overall conjugation of $M_{Y,T}^2$ is also needed.
In other words, the replacement of $\gamma_{Y,V}$ in
 Eq.~(\ref{eq:GAMMAtermsNOFL}) is any of the following terms:
\begin{eqnarray}
   c_V^x \, {\rm Tr}
 {\cal F}(Y_{x}^\dagger,
          \widetilde{m}_{V}^2,
          \widetilde{m}_{S}^{2\,\ast},
          \widetilde{m}_{T}^{2\,\ast},
          A_{x}^\dagger)              
&=&c_V^x \, {\rm Tr}
 {\cal F}(Y_{x}^\ast,
          \widetilde{m}_{V}^2,
          \widetilde{m}_{T}^{2},
          \widetilde{m}_{S}^2,
          A_{x}^\ast)
\nonumber\\
&=&c_V^x \, {\rm Tr}
 {\cal F}(Y_{x}^T,
          \widetilde{m}_{V}^2,
          \widetilde{m}_{S}^{2},
          \widetilde{m}_{T}^{2},
          A_{x}^T) 
\nonumber\\           
&=&c_V^x \, {\rm Tr}
 {\cal F}(Y_{x},
          \widetilde{m}_{V}^2,
          \widetilde{m}_{T}^{2\,\ast},
          \widetilde{m}_{S}^{2\,\ast},
          A_{x})             
\nonumber\\
&=&  M_{Y,V}^2   , 
\label{eq:YGAMreplacNOFLdiffMASS}
\end{eqnarray}
%
 whereas the replacements of the terms $\gamma_{Y,T}$ and $\gamma_{Y,S}$ 
 are
\begin{eqnarray}
\gamma_{Y,T} = \
c_T^x \,Y_x^\dagger Y_x & \ \to & \  
c_T^x \,{\cal F}(Y_{x}^T,
                 \widetilde{m}_{T}^2,
                 \widetilde{m}_{S}^{2},
                 \widetilde{m}_{V}^2,
                 A_{x}^T)
             = M_{Y,T}^2,
\nonumber\\
\gamma_{Y,S} = \
c_S^x \,Y_x^\ast Y_x^T \! & \to & \  
c_S^x \,{\cal F}(Y_{x}^\ast,
                 \widetilde{m}_{S}^2,\widetilde{m}_{T}^{2},
                 \widetilde{m}_{V}^2,A_{x}^\ast)
             = M_{Y,S}^2.
\label{eq:YGAMreplacFLdiffMASS}
\end{eqnarray}

In summary, a Yukawa coupling $Y$ (and the corresponding coupling $A$)
 can appear in the function ${\cal F}$ as $Y$, $Y^\ast$, $Y^\dagger$,
 $Y^T$.
The fact that the mass of a field with flavour has a conjugate or not in
 ${\cal F}$ depends on the type of interaction of this field,
 i.e., in which way it contributes to its anomalous dimension,
 and on the type of soft term this field has.

We illustrate the above sets of rules with two examples.
We start with the {\it SU}(2) doublet $Q$ of the MSSM, which has an 
 anomalous dimension: 
\begin{equation}
 \gamma_Q           =  
 -2\left(\frac{4}{3}g^2_3 +\frac{3}{4}g^2_2 +\frac{1}{60}g^2_1 
   \right) \!{\mathop{\bf 1}}
 +Y_U^\dagger Y_U +Y_D^\dagger Y_D , 
\label{eq:adQexample}
\end{equation}
 where $g_1$, $g_2$, and $g_3$ are the three MSSM gauge couplings.
The RGE for $\widetilde{m}^2_Q$ is 
\begin{equation}
 \dot{\widetilde{m}}^2_Q       =    M_{G,Q}^2
+ {\cal F}(Y_U^\dagger,
           \widetilde{m}_Q^2,
           \widetilde{m}_{U^c}^2,
           \widetilde{m}_{H_u}^2,
           A_U^\dagger)
+ {\cal F}(Y_D^\dagger,
           \widetilde{m}_Q^2,
           \widetilde{m}_{D^c}^2,
           \widetilde{m}_{H_d}^2,
           A_D^\dagger),
\label{eq:tildemQexample}
\end{equation}
 where $M_{G,Q}^2$ is given by  
\begin{equation}
 M_{G,Q}^2 =  
 -8\left(\frac{4}{3}g^2_3  M_3^2 
        +\frac{3}{4}g^2_2  M_2^2
        +\frac{1}{60}g^2_1 M_1^2  \right) \!{\mathop{\bf 1}}.
\label{eq:MgammaQ}
\end{equation}  
In the argument of the two functions ${\cal F}$, 
 $\widetilde{m}_{U^c}^{2}$ and $\widetilde{m}_{D^c}^{2}$ are not
 conjugated because we have defined the soft mass terms of the 
 {\it SU}(2) singlets $U^c$, $D^c$, and $E^c$ as
\begin{equation}
 \widetilde{U}^c \widetilde{m}_{U^c}^2 \widetilde{U}^{c\,\ast}, \quad
 \widetilde{D}^c \widetilde{m}_{D^c}^2 \widetilde{D}^{c\,\ast}, \quad
 \widetilde{E}^c \widetilde{m}_{E^c}^2 \widetilde{E}^{c\,\ast}, 
\label{eq:SOFTmassSINGL}
\end{equation}
 i.e., differently from those for the doublets, $Q$ and $L$,
 which are
\begin{equation}
 \widetilde{Q}^\ast \widetilde{m}_Q^2 \widetilde{Q}, \quad
 \widetilde{L}^\ast \widetilde{m}_L^2 \widetilde{L}.
\label{eq:SOFTmassDOUBL}
\end{equation}
With this definition, the mass parameters entering in the $6\times 6$
 sfermion mass matrices, conventionally written in the basis of the
 superpartners of left-{{}}handed fermions 
 ($\widetilde{u}_L=\widetilde{U}$, $\widetilde{d}_L=\widetilde{D}$,
  $\widetilde{e}_L=\widetilde{E}$) and right-handed fermions
 ($\widetilde{u}_R=\widetilde{U}^c$, $\widetilde{d}_R=\widetilde{D}^c$,
  $\widetilde{e}_R=\widetilde{E}^c$), are
  $\widetilde{m}^2_Q$ and $\widetilde{m}^2_{U^c}$ for up squarks,
  $\widetilde{m}^2_Q$ and $\widetilde{m}^2_{D^c}$ for down squarks, 
 and $\widetilde{m}^2_L$ and $\widetilde{m}^2_{E^c}$ for sleptons. 
In contrast, the choices 
 $\widetilde{\bar{5}}_M \widetilde{m}_{5_M}^2\,
                        \widetilde{\bar{5}}_M^\ast$, 
 $\widetilde{N^c} \widetilde{m}_{N^c}^2 \widetilde{N^c}^\ast$ and 
 $\widetilde{24}_M \widetilde{m}_{24_M}^2 \widetilde{24}_M^\ast$
 made in the soft SUSY-breaking potentials for the MSSU(5) model with a 
 seesaw of types~I and~III are purely conventional and motivated
 mainly by aesthetic reasons. 
That is, with the above choice for $\widetilde{m}_{5_M}^2$, the field
 $D^c$ has the same type of soft mass before and after the breaking of
 {\it SU}(5).

The second example we give here is that of an {\it SU}(2) singlet of the 
 MSSM, $U^c$, with an anomalous dimension:
\begin{equation}
 \gamma_{U^c}           =  
 -2\left(\frac{4}{3}g^2_3+\frac{4}{15}g^2_1 \right)\!{\mathop{\bf 1}}
 +2 Y_U^\ast Y_U^T  . 
\label{eq:adUexample}
\end{equation} 
The corresponding RGE for $\widetilde{m}_{U^c}^2$, defined in 
 Eq.~(\ref{eq:SOFTmassSINGL}), is given by 
\begin{equation}
 \dot{\widetilde{m}}^2_{U^c}     =    M_{G,U^c}^2
+2 {\cal F}(Y_U,
            \widetilde{m}_{U^c}^2,\widetilde{m}_{Q}^2,
            \widetilde{m}_{H_u}^2,A_U),
\label{eq:tildemUexample}
\end{equation}
 with an obvious definition of $M_{G,U^c}^2$.

The algorithm presented here is sufficient to deduce the RGEs for the
 soft masses of all scalar fields from the expression of their anomalous
 dimensions. 
As in the case of the bilinear and trilinear soft parameters, these
 RGEs could also be omitted. 
Nevertheless, since this algorithm is a little involved, we list them as
 in Eq.~(\ref{eq:tildemQexample}) or Eq.~(\ref{eq:tildemUexample}), with
 a further abbreviation of our notation, which consists in writing  
${\cal F}(Y_h^\dagger,
         \widetilde{m}^2_j,\widetilde{m}^2_k,\widetilde{m}^2_l,
         A_h^\dagger)$
 as
${\cal F}_{(Y_h^\dagger,
         \widetilde{j},\widetilde{k},\widetilde{l},
         A_h^\dagger)}$.
\end{list}

\subsection{\boldmath{$Q< M_{\rm ssw}$}}
\label{sec:RGEbelowSSWSCALE}  
%
When $Q< M_{\rm ssw}$, the typical scale of the heavy fields
 realizing the seesaw mechanism, the superpotential is that of the 
 MSSM plus an additional dimension-five operator obtained after
 integrating out the heavy fields needed to implement the seesaw of
 types~I,~II, or~III:  
\begin{equation}
 W^{{\rm MSSM},\nu} \ = \ W^{\rm MSSM} +  W_{\nu}, 
\label{eq:MSSM-SUPERPtoISC}
\end{equation}
 with 
\begin{equation}
 W^{\rm MSSM} \ = \
 U^c Y_U Q H_u -D^c Y_D Q H_d -E^c Y_E L H_d +\mu H_u H_d, 
\label{eq:MSSM}
\end{equation}
 and $W_{\nu}$ given in Eq.~(\ref{eq:nu}).
The RGEs for the superpotential parameters are known, but we report them
 for completeness.
\begin{list}{$\bullet$}{}
\item Beta function coefficients:
   $\displaystyle{\quad  b_i = \left\{\frac{33}{5},1,-3\right\}}.$
\item Yukawa couplings:
\begin{eqnarray}
&\dot{Y}_U    &=
\gamma^T_{U^c} Y_U + Y_U \gamma_Q + \gamma_{H_u} Y_U ,
\nonumber\\
&\dot{Y}_D    &=
\gamma^T_{D^c} Y_D + Y_D \gamma_Q + \gamma_{H_d} Y_D ,
\nonumber\\
&\dot{Y}_E    &=
\gamma^T_{E^c} Y_E + Y_E \gamma_L + \gamma_{H_d} Y_E ,
\label{eq:MSSM-YUKtoISC}
\end{eqnarray}
 where the anomalous dimension matrices $\gamma_Q$, $\gamma_{U^c}$,
 etc. are
\begin{eqnarray}
&\gamma_Q                          & =  
 -2\left(\displaystyle{\frac{4}{3}}g^2_3 
                      +\frac{3}{4}g^2_2 +\frac{1}{60}g^2_1 
   \right) \!{\mathop{\bf 1}}
 +Y_U^\dagger Y_U +Y_D^\dagger Y_D , 
\nonumber\\
&\gamma_{U^c}                      & =  
 -2\left(\displaystyle{\frac{4}{3}}g^2_3
                      +\frac{4}{15}g^2_1 \right) \!{\mathop{\bf 1}}
 + 2 Y_U^\ast Y_U^T ,
\nonumber\\
&\gamma_{D^c}                      & =  
 -2\left(\displaystyle{\frac{4}{3}}g^2_3
                      +\frac{1}{15}g^2_1 \right) \!{\mathop{\bf 1}}
 + 2 Y_D^\ast Y_D^T ,
\nonumber\\
&\gamma_L                          & =  
 -2\left(\displaystyle{\frac{3}{4}}g^2_2
                      +\frac{3}{20}g^2_1 \right) \!{\mathop{\bf 1}}  
 +Y_E^\dagger Y_E ,
\nonumber\\
&\gamma_{E^c}                      & =  
 -2\left(\displaystyle{\frac{3}{5}}g^2_1 \right) \!{\mathop{\bf 1}}  
 + 2 Y_E^\ast Y_E^T , 
\nonumber\\[1.001ex]
&\gamma_{H_u}                      & =  
 -2\left(\displaystyle{\frac{3}{4}}g^2_2 +\frac{3}{20}g^2_1 \right)
 + {\rm Tr} \left(3 Y_U^\dagger Y_U\right),
\nonumber\\
&\gamma_{H_d}                      & =  
 -2\left(\displaystyle{\frac{3}{4}}g^2_2 +\frac{3}{20}g^2_1 \right)
 + {\rm Tr} \left(3 Y_D^\dagger Y_D + Y_E^\dagger Y_E\right).
\label{eq:MSSM-anomDIMtoISC}
\end{eqnarray}
\item Superpotential dimensionful parameters:
\begin{eqnarray}
&\dot{\mu}    &=  \left( \gamma_{H_u} + \gamma_{H_d} \right)\mu ,
\nonumber\\
&\dot{\kappa} &= 
 \gamma^T_L \kappa + \kappa \gamma_L + 2 \gamma_{H_u} \kappa .
\label{eq:MSSM-muTERMandKAPPAtoISC}
\end{eqnarray}
\end{list}
The SUSY-breaking terms completing the description of this model
 are collected in
\begin{equation}
 \widetilde{V}^{{\rm MSSM},\nu} = \widetilde{V}^{\rm MSSM} +
                                  \widetilde{V}_{\nu},
\label{eq:MSSM-SOFTPOTtoISC}
\end{equation}
 where $\widetilde{V}^{\rm MSSM}$ is the usual
\begin{eqnarray}
  \widetilde{V}^{\rm MSSM}           & = & 
\left\{\,
 \widetilde{U}^c A_U \,\widetilde{Q} H_u 
-\widetilde{D}^c A_D \,\widetilde{Q} H_d
-\widetilde{E}^c A_E \,\widetilde{L} H_d 
+ B \, H_u H_d        
\ \ + \ {\rm H.c.} \right\}
\nonumber\\                        &   & 
+\widetilde{Q}^\ast \, \widetilde{m}^2_Q     \, \widetilde{Q} 
+\widetilde{U}^c    \, \widetilde{m}^2_{U^c} \, \widetilde{U}^{c\,\ast}
+\widetilde{D}^c    \, \widetilde{m}^2_{D^c} \, \widetilde{D}^{c\,\ast}
\nonumber\\                        &   & 
+\widetilde{L}^\ast \, \widetilde{m}^2_L \,     \widetilde{L} 
+\widetilde{E}^c    \, \widetilde{m}^2_{E^c} \, \widetilde{E}^{c\,\ast} 
\phantom{\frac{1}{2}}
\nonumber\\                        &   & 
+\widetilde{m}^2_{H_u} \, \widetilde{H}_u^\ast  \widetilde{H}_u 
+\widetilde{m}^2_{H_d} \, \widetilde{H}_d^\ast  \widetilde{H}_d 
\nonumber\\                        &   & 
  +  \frac{1}{2} \left(
    M_3\, \widetilde{g}\widetilde{g} + M_2\, \widetilde{W}\widetilde{W} 
  + M_1\, \widetilde{B}\widetilde{B} \right) ,
\label{eq:MSSMsoft}
\end{eqnarray}
 and $\widetilde{V}_{\nu}$ is the nonrenormalizable 
 lepton-number-{{{}}}violating operator ($\Delta L=2$)
\begin{equation}
 \widetilde{V}_{\nu} = -\frac{1}{2} \widetilde{L} H_u \widetilde{\kappa}
                        \widetilde{L} H_u.
\label{eq:nusoft}
\end{equation}
The symmetric matrix $\widetilde{\kappa}$ has elements of 
 ${\cal O}(\widetilde{m}/M_{\rm ssw})$, where $\widetilde{m}$ is a
 typical soft SUSY-breaking mass. 
We list in the following the RGEs for the parameters in this potential, 
 except for those for the bi- and trilinear scalar terms, which are 
 easily obtained using the algorithm mentioned above. 
We omit that for the dimensionless parameter $\widetilde{\kappa}$, which
 gives rise to a very suppressed bilinear term for the neutral
 component of $\widetilde{L}$. 
\begin{list}{$\bullet$}{}
\item Soft sfermion masses:
\begin{eqnarray}
&\dot{\widetilde{m}}^2_Q  \,      
&=
   M_{G,Q}^2
+{\cal F}_{(Y_U^\dagger,\widetilde{Q},\widetilde{U}^c,H_u,A_U^\dagger)}
+{\cal F}_{(Y_D^\dagger,\widetilde{Q},\widetilde{D}^c,H_d,A_D^\dagger)},
\nonumber\\
&\dot{\widetilde{m}}_{U^c}^2      
&=
   M_{G,U^c}^2
+2{\cal F}_{(Y_U,\widetilde{U}^c,\widetilde{Q},H_u,A_U)} ,
\nonumber\\
&\dot{\widetilde{m}}_{D^c}^2     
&=
   M_{G,D^c}^2
+2{\cal F}_{(Y_D,\widetilde{D}^c,\widetilde{Q},H_d,A_D)} ,
\nonumber\\
&\dot{\widetilde{m}}_L^2         
&=
   M_{G,L}^2
+{\cal F}_{(Y_E^\dagger,\widetilde{L},\widetilde{E}^c,H_d,A_E^\dagger)},
\nonumber\\
&\dot{\widetilde{m}}_{E^c}^2      
&=
   M_{G,E^c}^2
+2{\cal F}_{(Y_E,\widetilde{E}^c,\widetilde{L},H_d,A_E)} .
\label{eq:MSSM-softMASSEStoISC}
\end{eqnarray}
\item Soft Higgs masses:
\begin{eqnarray}
&\dot{\widetilde{m}}_{H_d}^2   
& =
  M_{G,{H_d}}^2
+3{\rm Tr}{\cal F}_{(Y_D^\dagger,H_d,\widetilde{D}^c,\widetilde{Q},
                     A_D^\dagger)} 
+\ {\rm Tr}{\cal F}_{(Y_E^\dagger,H_d,\widetilde{E}^c,\widetilde{L},
                      A_E^\dagger)},
\nonumber\\[1.1ex]
&\dot{\widetilde{m}}_{H_u}^2   
& =
   M_{G,{H_u}}^2
+3{\rm Tr}{\cal F}_{(Y_U^\dagger,H_u,\widetilde{U}^c,\widetilde{Q},
                     A_U^\dagger)}.
\label{eq:MSSM-softHIGGStoISC}
\end{eqnarray}
\end{list}

\subsection{\boldmath{$M_{\rm ssw}< Q < M_{\rm GUT}$}}
\label{sec:RGESSWtoGUT}
%
We distinguish the three cases in which the dimension-five operator in 
 Eq.~(\ref{eq:MSSM-SUPERPtoISC}) is induced by heavy right-handed
 neutrino singlets, $N^c$, or by Higgs fields in the 15 and $\bar{15}$
 representations of {\it SU}(5), $15_H$ and $\bar{15}_H$, or by three 
 matter fields in the adjoint representation of {\it SU}(5), $24_M$. 
Those presented here are not the most minimal implementations of the
 three types of seesaw mechanism, which, strictly speaking, require only
 two {\it SU}(2) triplet Higgs fields $T$ and $\bar{T}$ for the seesaw
 of type~II and three {\it SU}(2) fermion triplets $W_M$ for the 
 type~III, if no embedding in {\it SU}(5) is needed. 
(See \S2, where the relevant seesaw
 superpotentials are denoted by $W_{\rm sssw\, i}^\prime$, with
 $i=$I, II, III to differentiate them from those presented here.)
Here, we introduce all fields in the $15_H$, $\bar{15}_H$ and in the
 $24_M$ representation of {\it SU}(5) to which $T$, $\bar{T}$ and $W_M$
 belong, respectively, and we denote the corresponding seesaw
 superpotentials by $W_{\rm sssw\, i}$ ($i=$I, II, III).
Thus, the complete superpotentials in the three cases are 
\begin{equation}
 W^{{\rm MSSM},i} = W^{\rm MSSM} + W_{{\rm ssw}\,i} ,
\quad\quad (i={\rm I,\,II,\,III})  
\label{eq:i-SUPERPfromISCtoGUT}
\end{equation}
 with $W^{\rm MSSM}$ given in Eq.~(\ref{eq:MSSM}) and
 $W_{\rm ssw\,i}$ $(i= {\rm I,II,III})$ to be specified in the
 following. 
Similarly, the soft SUSY-breaking part of the scalar potential is, in 
 the three cases, 
\begin{equation}
 \widetilde{V}^{{\rm MSSM},i} = 
 \widetilde{V}^{\rm MSSM} + \widetilde{V}_{{\rm ssw}\,i},
 \quad\quad              (i={\rm I,\,II,\,III})  
\label{eq:i-SOFTPOTfromISCtoGUT}
\end{equation}
 with $\widetilde{V}^{\rm MSSM}$ given in Eq.~(\ref{eq:MSSMsoft}), and
 $\widetilde{V}_{{\rm ssw}\,i}$ also to be specified.

\subsubsection{\underline{MSSM, I}}
\label{sec:I-ISCALEtoGUT}  
%
The RGEs for this model have been widely studied in this context after
 Ref.~\citen{NuY->QFV} appeared (see for example Ref.~\citen{BGOO}).
Nevertheless, we list them here for completeness. 
The model is described using{{{}}} the superpotential $W^{\rm MSSM,I}$ with
\begin{equation}
  W_{\rm ssw\,I} = N^c Y_N^{\rm I} L H_u +\frac{1}{2} N^c M_N N^c .
\label{eq:I-SUPERPsswfromISCtoGUT}
\end{equation}
The RGEs for the superpotential parameters are as follows.
\begin{list}{$\bullet$}{}
\item Beta function coefficients:
 $\displaystyle{\quad  b_i^{\rm I} = \left\{\frac{33}{5},1,-3\right\}}.$
\item Yukawa couplings: 
To those in Eq.~(\ref{eq:MSSM-YUKtoISC}), the following RGE
\begin{equation}
\dot{Y}_{N}^{\rm I} = \gamma^T_{N^c} Y_N^{\rm I}
                    +Y_N^{\rm I}\gamma_L+\gamma_{H_u} Y_N^{\rm I},
\label{eq:I-YUKfromISCtoGUT}
\end{equation}
 must be added. Note, however, that $\gamma_{H_u}$ and $\gamma_L$ are
 now modified by the presence of the operator $N^c Y_N^{\rm I} L H_u$. 
Therefore, the following anomalous dimensions 
\begin{eqnarray}
&  \gamma_L                        & =  
  -2\left(\displaystyle{\frac{3}{4}}g^2_2
                       +\frac{3}{20}g^2_1 \right)\!{\mathop{\bf 1}}  
  +Y_E^\dagger Y_E + Y_N^{{\rm I}\,\dagger} Y_N^{\rm I},
\nonumber\\
&  \gamma_{H_u}                    & = 
  -2\left(\displaystyle{\frac{3}{4}}g^2_2 +\frac{3}{20}g^2_1 \right)
  + {\rm Tr}\left(
     3 Y_U^\dagger Y_U +Y_N^{{\rm I}\,\dagger} Y_N^{\rm I}\right),
\nonumber\\
&  \gamma_{N^c}                    & =  
 \ \  2 Y_N^{{\rm I}\,\ast} Y_N^{{\rm I}\,T} ,
\phantom{\displaystyle{\frac{1}{2}}}
\label{eq:I-anomDIMfromISCtoGUT}
\end{eqnarray}
 must be specified, whereas $\gamma_Q$, $\gamma_{U^c}$, $\gamma_{D^c}$,
 $\gamma_{E^c}$, $\gamma_{H_d}$ remain as in
 Eq.~(\ref{eq:MSSM-anomDIMtoISC}).
\item Superpotential dimensionful parameters:
\begin{equation}
\dot{M}_N  = \gamma_{N^c}^T M_N + M_N \gamma_{N^c}.
\label{eq:I-MNfromISCtoGUT}
\end{equation}
The RGE for $\mu$ is as in Eq.~(\ref{eq:MSSM-muTERMandKAPPAtoISC}), 
 with $\gamma_{H_u}$ given in Eq.~(\ref{eq:I-anomDIMfromISCtoGUT}). 
\end{list}
The soft SUSY-breaking part of the scalar potential,
 $\widetilde{V}_{\rm ssw\,I}$, is given by
\begin{equation}
  \widetilde{V}_{\rm ssw\,I}                    \ = \ 
\left\{
  \widetilde{N}^c A_N^{\rm I} \,\widetilde{L} H_u 
+ \frac{1}{2} \widetilde{N}^c B_N \widetilde{N}^c
+ \ {\rm H.c.} \right\}
+ \widetilde{N}^c \, \widetilde{m}^2_{N}\, \widetilde{N}^{c\,\ast},
\label{eq:I-SOFTPOTsswfromISCtoGUT}
\end{equation}
 and the parameters in $\widetilde{V}^{\rm MSSM,I}$ obey the following RGEs.
\begin{list}{$\bullet$}{}
\item Soft sfermion masses:
Of the RGEs in Eq.~(\ref{eq:MSSM-softMASSEStoISC}), only that for 
 $\widetilde{m}_L^2$ gets modified by the presence of the operator
 $N^c Y_N^{\rm I} L H_u$, whereas the RGE for the soft contribution to 
 the mass of $\widetilde{N}^c$ needs to be added:
\begin{eqnarray}
&\dot{\widetilde{m}}_L^2       
&=
   M_{G,L}^2
+ {\cal F}_{(Y_E^\dagger,\widetilde{L},\widetilde{E}^c,H_d,
              A_E^\dagger)}
+ {\cal F}_{(Y_N^{{\rm I}\,\dagger},\widetilde{L},\widetilde{N}^c,H_u,
              A_N^{{\rm I}\,\dagger})},
\nonumber\\
&\dot{\widetilde{m}}_{N^c}^2   
&= 
 2{\cal F}_{(Y_N^{\rm I},\widetilde{N}^c,\widetilde{L},H_u,
             A_N^{\rm I})} .
\label{eq:I-softMASSESfromISCtoGUT}
\end{eqnarray}
\item Soft Higgs masses: 
\begin{equation}
 \dot{\widetilde{m}}_{H_u}^2  \ =\  M_{G,{H_u}}^2
+3{\rm Tr}
  {\cal F}_{(Y_U^\dagger,H_u,\widetilde{U}^c,\widetilde{Q},A_U^\dagger)}
+\ {\rm Tr}
  {\cal F}_{(Y_N^{{\rm I}\,\dagger},H_u,\widetilde{N}^c,\widetilde{L},
             A_N^{{\rm I}\,\dagger})}.
\label{eq:I-softHIGGSfromISCtoGUT}
\end{equation}
The RGE for $\widetilde{m}_{H_d}^2$ is as in 
 Eq.~(\ref{eq:MSSM-softHIGGStoISC}).
\end{list}

\subsubsection{\underline{MSSM, II}}
\label{sec:II-ISCALEtoGUT}  
%
The RGEs for this model can also be found in Ref.~\citen{AROSSIorig}. 
Some of our equations, however, differ from those reported there.
We give a list of these equations at the end of this section.
The part of the superpotential needed to specify this model at these
 energies is 
\begin{eqnarray}
 W_{\rm ssw\,II}                    & = & 
 \ \frac{1}{\sqrt{2}} LY_N^{\rm II} T L 
 - D^c Y_{Q_{15}} L Q_{15}  
 +\frac{1}{\sqrt{2}} D^c Y_S S D^c 
\nonumber\\
&& 
 +\frac{1}{\sqrt{2}} \lambda_{\bar{T}} H_u \bar{T} H_u 
 +\frac{1}{\sqrt{2}} \lambda_T H_d  T  H_d 
\nonumber\\[1.1ex]
&&
 +M_T T\bar{T} +\!M_{Q_{15}}Q_{15}\bar{Q}_{15}
 +M_S S\bar{S},
\label{eq:II-SUPERPsswfromISCtoGUT}
\end{eqnarray}
 where $Y_N^{\rm II}$ and $Y_S$ are symmetric matrices,
 $\lambda_{\bar{T}}$ and $\lambda_T$, and
 $M_{T}$, $M_{Q}$, and $M_{S}$ are complex numbers.  
The RGEs for the superpotential parameters are as follows.
\begin{list}{$\bullet$}{}
\item Beta function coefficients:
 $\displaystyle{\quad  b^{\rm II}_i = \left\{\frac{68}{5},8,4\right\}}.$
\item Yukawa couplings: 
The RGEs for $Y_U$, $Y_D$, and $Y_E$ are as in
 Eq.~(\ref{eq:MSSM-YUKtoISC}). 
Those for $Y_N^{\rm II}$, $Y_S$, and $Y_{Q_{15}}$, and the
 flavour-independent couplings $\lambda_{\bar{T}}$ and $\lambda_T$ are
\begin{eqnarray}
& \dot{Y}_N^{\rm II}               & =  
 \gamma^T_LY_N^{\rm II} +Y_N^{\rm II}\gamma_L +\gamma_TY_N^{\rm II},
\nonumber\\
& \dot{Y}_{Q_{15}}                 & =  
  \gamma^T_{D^c}Y_{Q_{15}}+Y_{Q_{15}}\gamma_L+\gamma_{Q_{15}}Y_{Q_{15}},
\nonumber\\
& \dot{Y}_S                        & =  
  \gamma^T_{D^c} Y_S + Y_S \gamma_{D^c} + \gamma_S Y_S ,
\nonumber\\
& \dot{\lambda}_{\bar{T}}          & =  
 \left( 2 \gamma_{H_u} +\gamma_{\bar{T}} \right) \lambda_{\bar{T}}, 
\nonumber\\
& \dot{\lambda}_T                  & =  
 \left( 2 \gamma_{H_d} + \gamma_T \right) \lambda_T. 
\label{eq:II-YUKfromISCtoGUT}
\end{eqnarray}
{}For the anomalous dimensions $\gamma_Q$, $\gamma_{U^c}$, $\gamma_{E^c}$,  
 see Eq.~(\ref{eq:MSSM-anomDIMtoISC}); the others are 
\begin{eqnarray}
& \gamma_{D^c}                      & =  
  -2\left(\displaystyle{\frac{4}{3}}g^2_3
                       +\frac{1}{15}g^2_1 \right)\!{\mathop{\bf 1}}
  + 2 Y_D^\ast Y_D^T 
  + 2 Y_{Q_{15}}^\ast Y_{Q_{15}}^T 
  + 4 Y_S^\ast Y_S^T,
\nonumber\\
& \gamma_L                          & =  
  -2\left(\displaystyle{\frac{3}{4}}g^2_2
                       +\frac{3}{20}g^2_1 \right)\!{\mathop{\bf 1}}  
  +Y_E^\dagger Y_E
  +3 Y_{Q_{15}}^\dagger Y_{Q_{15}}
  + 3 Y_N^{{\rm II}\,\dagger} Y_N^{\rm II} ,
\nonumber\\[1.001ex]
& \gamma_{H_u}                      & =  
  -2\left(\displaystyle{\frac{3}{4}}g^2_2 +\frac{3}{20}g^2_1 \right)
  + {\rm Tr}\left(3 Y_U^\dagger Y_U\right)
           + 3 \vert \lambda_{\bar{T}} \vert^2  ,
\nonumber\\
& \gamma_{H_d}                      & =  
  -2\left(\displaystyle{\frac{3}{4}}g^2_2 +\frac{3}{20}g^2_1 \right)
 + {\rm Tr} \left(3 Y_D^\dagger Y_D + Y_E^\dagger Y_E\right)
          + 3 \vert \lambda_T \vert^2  ,
\nonumber\\
& \gamma_{Q_{15}}                   & =  
  -2\left(\displaystyle{\frac{4}{3}}g^2_3
                       +\frac{3}{4}g^2_2 +\frac{1}{60}g^2_1 
    \right)
  +{\rm Tr} \left(Y_{Q_{15}}^\dagger Y_{Q_{15}}\right) , 
\nonumber\\
& \gamma_{\bar{Q}_{15}}             & =  
  -2\left(\displaystyle{\frac{4}{3}}g^2_3
                       +\frac{3}{4}g^2_2 +\frac{1}{60}g^2_1 
    \right) ,
\nonumber\\
& \gamma_T                          & =  
 -2\left(2 g^2_2 +\displaystyle{\frac{3}{5}}g^2_1 \right)
 + {\rm Tr} \left(Y_N^{{\rm II}\,\dagger} Y_N^{\rm II}\right)
 + \vert \lambda_T \vert^2 ,
\nonumber\\
& \gamma_{\bar{T}}                  & =  
 -2\left(2 g^2_2 +\displaystyle{\frac{3}{5}}g^2_1 \right)
 + \vert \lambda_{\bar{T}} \vert^2 ,
\nonumber\\
& \gamma_S                          & =  
 -2\left(\displaystyle{\frac{10}{3}}g^2_3 +\frac{4}{15}g^2_1 \right)
 + {\rm Tr}\left(Y_S^\dagger Y_S \right) ,
\nonumber\\
& \gamma_{\bar{S}}                  & =  
 -2\left(\displaystyle{\frac{10}{3}}g^2_3 +\frac{4}{15}g^2_1 \right) .
\label{eq:II-anomDIMfromISCtoGUT}
\end{eqnarray}
\item Superpotential dimensionful parameters:
\begin{eqnarray}
& \dot{M}_T                        & =  
 \left(\gamma_T + \gamma_{\bar{T}} \right) M_T ,
\nonumber\\
& \dot{M}_{Q_{15}}                 & =  
 \left(\gamma_{Q_{15}} + \gamma_{\bar{Q}_{15}} \right) M_{Q_{15}} , 
\nonumber\\
& \dot{M}_S                        & =  
 \left(\gamma_S + \gamma_{\bar{S}} \right) M_S .
\label{eq:II-hardMASSESfromISCtoGUT}
\end{eqnarray}
The RGE for $\mu$ is as in Eq.~(\ref{eq:MSSM-muTERMandKAPPAtoISC}).
\end{list}
The part of the scalar potential specific to this model,
 $\widetilde{V}_{\rm ssw\,II}$, to be specified in addition to
 $\widetilde{V}^{\rm MSSM}$ is
\begin{eqnarray}
  \widetilde{V}_{\rm ssw\,II}       &\!=\! & 
\Bigl\{
  \frac{1}{\sqrt{2}}\widetilde{L} A_N^{\rm II} \, T \widetilde{L} 
- \widetilde{D}^c A_{Q_{15}} \,\widetilde{L} Q_{15}
+ \frac{1}{\sqrt{2}}\widetilde{D}^c A_S \, S \widetilde{D}^c 
\nonumber\\                        &   & 
+\frac{1}{\sqrt{2}} A_{\lambda_{\bar{T}}} H_{u} \bar{T} H_u
+\frac{1}{\sqrt{2}} A_{\lambda_T}         H_{d}      T  H_d
\nonumber\\                        &   & 
+ B_T \, T \bar{T}
+ B_{Q_{15}} \, Q_{15} \bar{Q}_{15}
+ B_S \, S \bar{S}
+ {\rm H.c.} 
\Bigr\}
\nonumber\\                        &   & 
+\widetilde{m}^2_{Q_{15}}       \,      Q_{15}^\ast       Q_{15}
+\widetilde{m}^2_{\bar{Q}_{15}} \, \bar{Q}_{15}^\ast \bar{Q}_{15} 
+\widetilde{m}^2_S \,              S^\ast       S
+\widetilde{m}^2_{\bar{S}} \, \bar{S}^\ast \bar{S}  
\nonumber\\[1.001ex]               &   & 
+\widetilde{m}^2_T \,              T^\ast       T
+\widetilde{m}^2_{\bar{T}} \, \bar{T}^\ast \bar{T}.  
\label{eq:II-SOFTPOTsswfromISCtoGUT}
\end{eqnarray}
Here, $A_N^{\rm II}$ and $A_S$ are symmetric matrices,
 $A_{\lambda_{\bar{T}}}$, $A_{\lambda_T}$, and 
 $B_T$, $B_{Q_{15}}$, $B_S$ are complex numbers.
The parameters in $\widetilde{V}^{\rm MSSM,II}$ satisfy the following
 RGEs. 
\begin{list}{$\bullet$}{}
\item Soft sfermion masses:
\begin{eqnarray}
&\dot{\widetilde{m}}_{D^c}^2     
& = 
  M_{G,D^c}^2
+2{\cal F}_{(Y_D,\widetilde{D}^c,\widetilde{Q},H_d,
               A_D)}    
+2{\cal F}_{(Y_{Q_{15}},\widetilde{D}^c,\widetilde{L},Q_{15},
               A_{Q_{15}})}
\nonumber\\
&& 
\hspace*{5.1truecm}
+4{\cal F}_{(Y_S,\widetilde{D}^c,\widetilde{D}^{c\,\ast},S,
               A_S)} ,
\nonumber\\
&\dot{\widetilde{m}}_L^2         
& =
  M_{G,L}^2
+ {\cal F}_{(Y_E^\dagger,\widetilde{L},\widetilde{E}^c,H_d,
                A_E^\dagger)}
+3{\cal F}_{(Y_{Q_{15}}^\dagger,\widetilde{L},\widetilde{D}^c,Q_{15},
               A_{Q_{15}}^\dagger)} 
\nonumber\\
&& 
\hspace*{4.8truecm}
+3{\cal F}_{(Y_{N}^{{\rm II}\,\dagger},\widetilde{L},
                                       \widetilde{L}^{\ast},T,
               A_{N}^{{\rm II}\,\dagger})} .
\label{eq:II-softMASSESfromISCtoGUT}
\end{eqnarray}
The RGEs for $\widetilde{m}^2_Q$, $\widetilde{m}^2_{U^c}$,
 $\widetilde{m}^2_{E^c}$ are as
 in Eq.~(\ref{eq:MSSM-softMASSEStoISC}).  
\item Soft Higgs masses:
\begin{eqnarray}
&\dot{\widetilde{m}}^2_{H_d}          
&=
  M_{G,{H_d}}^2
+3{\rm Tr}
 {\cal F}_{(Y_D^\dagger,H_d,\widetilde{D}^c,\widetilde{Q},A_D^\dagger)} 
+ {\rm Tr}
 {\cal F}_{(Y_E^\dagger,H_d,\widetilde{E}^c,\widetilde{L},A_E^\dagger)}
\nonumber\\
&&
\hspace*{5.5truecm}
+\,3{\cal F}_{(\lambda_T,H_d,H_d,T,A_{\lambda_T})}, 
\nonumber\\
&\dot{\widetilde{m}}_{H_u}^2          
&=
  M_{G,{H_u}}^2
+3{\rm Tr}
 {\cal F}_{(Y_U^\dagger,H_u,\widetilde{U}^c,\widetilde{Q},A_U^\dagger)}
+3{\cal F}_{(\lambda_{\bar{T}},H_u,H_u,\bar{T},A_{\lambda_{\bar{T}}})},
\nonumber\\
&\dot{\widetilde{m}}^2_{Q_{15}}       
&=
  M_{G,{Q_{15}}}^2
+ {\rm Tr}
 {\cal F}_{(Y_{Q_{15}}^\dagger,Q_{15},\widetilde{D}^c,\widetilde{L},
            A_{Q_{15}}^\dagger)} ,
\nonumber\\
&\dot{\widetilde{m}}^2_{\bar{Q}_{15}} 
&=
  M_{G,{\bar{Q}_{15}}}^2,
\nonumber\\
&\dot{\widetilde{m}}^2_T              
&=
  M_{G,T}^2
+ {\rm Tr}{\cal F}_{(Y_{N}^{{\rm II}\,\dagger},
                      T,\widetilde{L}^{\ast},\widetilde{L},
                        A_{N}^{{\rm II}\,\dagger})} 
+ {\cal F}_{(\lambda_T,T,H_d,H_d,A_{\lambda_T})},
\nonumber\\
&\dot{\widetilde{m}}^2_{\bar{T}}      
&=
  M_{G,{\bar{T}}}^2
+ {\cal F}_{(\lambda_{\bar{T}},\bar{T},H_u,H_u,A_{\lambda_{\bar{T}}})},
\nonumber\\
&\dot{\widetilde{m}}^2_S              
&=
  M_{G,S}^2
+ {\rm Tr}{\cal F}_{(Y_S,S,\widetilde{D}^{c\,\ast},\widetilde{D}^c,
                      A_S)} ,
\nonumber\\
&\dot{\widetilde{m}}^2_{\bar{S}}      
&=
  M_{G,{\bar{S}}}^2.
\label{eq:II-softHIGGSfromISCtoGUT}
\end{eqnarray}
Note that in the contributions originated by flavour-independent 
 operators,{{{}}} the order in which the three masses squared appear in the 
 function ${\cal F}$ is irrelevant. 
{}Both terms
 ${\cal F}_{(\lambda_T, H_d,H_d,T,A_{\lambda_T})}$ and 
 ${\cal F}_{(\lambda_T, T,H_d,H_d,A_{\lambda_T})}$, indeed, reduce to 
 $2 \vert \lambda_T\vert^2  
  \left( 2 \widetilde{m}_{H_d}^2 + \widetilde{m}_T^2  \right) 
  + 2 \vert A_{\lambda_T}\vert^2 $.
\end{list}
Our RGEs for this model differ from those reported in 
 Ref.~\citen{AROSSIorig} for the following parameters:
  $ Y_{N}^{\rm II}$, $ M_T$, $\widetilde{m}^2_T$, $\widetilde{m}^2_S$,
  $\widetilde{m}^2_{Q_{15}}$, $A_N^{\rm II}$, and $A_{Q_{15}}$.

\subsubsection{\underline{MSSM, III}}
\label{sec:III-ISCALEtoGUT}  
%
The superpotential is still that of 
 Eq.~(\ref{eq:i-SUPERPfromISCtoGUT}) with $i={\rm III}$ and 
 $W_{\rm ssw\,III}$ is
\begin{eqnarray}
 W_{\rm ssw\,III}           & = & 
  H_u  W_M Y_N^{\rm III} L   
-\sqrt{\frac{6}{5}}\frac{1}{2} H_u B_M Y_B^{\rm III} L   
+ H_u \bar{X}_M Y_{\bar{X}_M} D^c 
\nonumber\\                 &   & 
+ \frac{1}{2} B_M M_{B_M} B_M +\frac{1}{2} G_M M_{G_M} G_M
+ \frac{1}{2} W_M M_{W_M} W_M 
\nonumber\\[1.001ex]        &   & 
+ X_M M_{X_M} \bar{X}_M ,
\label{eq:III-SUPERPsswfromISCtoGUT}
\end{eqnarray}
 where $M_{B_M}$, $M_{G_M}$ and $ M_{W_M}$ are symmetric matrices.
The RGEs for the superpotential parameters are as follows.
\begin{list}{$\bullet$}{}
\item Beta function coefficients:
 $\displaystyle{\quad b^{\rm III}_i =
       \left\{\frac{108}{5},16,12\right\}}.$
\item Yukawa couplings: 
 $Y_U$, $Y_D$, and $Y_E$ have RGEs listed in 
 Eq.~(\ref{eq:MSSM-YUKtoISC}). 
Those for $Y_N^{\rm III}$, $Y_B^{\rm III}$, and $Y_{\bar{X}_M}$ are 
\begin{eqnarray}
& \dot{Y}_N^{\rm III}              & =  
   \gamma^T_{W_M} Y_N^{\rm III}+ Y_N^{\rm III}\gamma_L
 + \gamma_{H_u}   Y_N^{\rm III} ,
\nonumber\\
& \dot{Y}_B^{\rm III}              & =  
   \gamma^T_{B_M} Y_B^{\rm III}+ Y_B^{\rm III}\gamma_L
 + \gamma_{H_u}   Y_B^{\rm III} ,
\nonumber\\
& \dot{Y}_{\bar{X}_M}\!\!          & =  
   \gamma^T_{\bar{X}_M} Y_{\bar{X}_M}+ Y_{\bar{X}_M} \gamma_{D^c} 
 + \gamma_{H_u} Y_{\bar{X}_M} .
\label{eq:III-YUKfromISCtoGUT}
\end{eqnarray}
{}For the anomalous dimensions $\gamma_Q$, $\gamma_{U^c}$,
 $\gamma_{E^c}$, $\gamma_{H_d}$, see Eq.~(\ref{eq:MSSM-anomDIMtoISC}).
The remaining ones are 
\begin{eqnarray}
&\gamma_{D^c}                      & =  
-2\left(\displaystyle{\frac{4}{3}}g^2_3
                     +\frac{1}{15}g^2_1 \right) \!{\mathop{\bf 1}}
  + 2 Y_D^\ast Y_D^T 
  + 2 Y_{\bar{X}_M}^\dagger Y_{\bar{X}_M} ,
\nonumber\\
&\gamma_L                          & =  
-2\left(\displaystyle{\frac{3}{4}}g^2_2
                     +\frac{3}{20}g^2_1 \right) \!{\mathop{\bf 1}}  
  + Y_E^\dagger Y_E
  + \frac{3}{2} Y_N^{{\rm III}\,\dagger} Y_N^{\rm III}
  + \frac{3}{10} Y_B^{{\rm III}\,\dagger} Y_B^{\rm III} ,
\nonumber\\
&\gamma_{W_M}                      & =  
-2\left(2 g^2_2 \right) \!{\mathop{\bf 1}}
  + Y_N^{{\rm III}\,\ast} Y_N^{{\rm III}\,T} ,
\phantom{\displaystyle{\frac{3}{2}} Y_N^{{\rm III}\,\dagger}
                                    Y_N^{\rm III}}
\nonumber\\
&\gamma_{B_M}                      & =  
  \ \displaystyle{\frac{3}{5}} Y_B^{{\rm III}\,\ast} 
                               Y_B^{{\rm III}\,T} ,
\nonumber\\
&\gamma_{\bar{X}_M}                & =  
-2\left(\displaystyle{\frac{4}{3}}g^2_3 +\frac{3}{4}g^2_2 
         +\frac{5}{12}g^2_1 \right) \!{\mathop{\bf 1}}
  +  Y_{\bar{X}_M}^\ast Y_{\bar{X}_M}^T ,
\nonumber\\
&\gamma_{X_M}                      & =  
-2\left(\displaystyle{\frac{4}{3}}g^2_3 +\frac{3}{4}g^2_2 
         +\frac{5}{12}g^2_1 \right) \!{\mathop{\bf 1}} ,
\nonumber\\
&\gamma_{G_M}                      & =  
  -2\left(3 g^2_3 \right) \!{\mathop{\bf 1}} ,
\phantom{\displaystyle{\frac{3}{2}} Y_N^{{\rm III}\,\dagger}
                                    Y_N^{\rm III}}
\nonumber\\[1.001ex]
&\gamma_{H_u\ }                    & =  
  -2\left(\displaystyle{\frac{3}{4}}g^2_2 +\frac{3}{20}g^2_1 \right)
  +{\rm Tr} \left(
  3Y_U^\dagger Y_U +\frac{3}{2} Y_N^{{\rm III}\,\dagger}Y_N^{\rm III}
\right.
\nonumber\\
&&
\left. \hspace*{5truecm}
                +\frac{3}{10}Y_B^{{\rm III}\,\dagger}Y_B^{\rm III}  
 +3Y_{\bar{X}_M}^\dagger Y_{\bar{X}_M} 
           \right) .
\label{eq:III-anomDIMfromISCtoGUT}
\end{eqnarray}
\item Superpotential dimensionful parameters: 
\begin{eqnarray}
\dot{M}_{B_M}           & = & 
 \gamma_{B_M}^T M_{B_M} + M_{B_M} \gamma_{B_M} ,
\nonumber\\
\dot{M}_{G_M}           & = & 
 \gamma_{G_M}^T M_{G_M} + M_{G_M} \gamma_{G_M} ,
\nonumber\\
\dot{M}_{W_M}           & = & 
 \gamma_{W_M}^T M_{W_M} + M_{W_M} \gamma_{W_M} ,
\nonumber\\
\dot{M}_{X_M}           & = & 
 \gamma_{X_M}^T M_{X_M} + M_{X_M} \gamma_{\bar{X}_M} .
\label{eq:III-hardMASSESfromISCtoGUT}
\end{eqnarray}
\end{list}
The term $\widetilde{V}_{\rm ssw\,III}$ is 
\begin{eqnarray}
  \widetilde{V}_{\rm ssw\,III}\!  & = &  \!
\Bigl\{
  H_u  \widetilde{W}_M  A_N^{\rm III}  \widetilde{L}    
-\sqrt{\frac{6}{5}} \frac{1}{2}
  H_u \widetilde{B}_M A_B^{\rm III}\widetilde{L}
+ H_u \widetilde{\bar{X}}_M A_{\bar{X}_M} \widetilde{D}^c 
\nonumber\\                       &   &
+ \frac{1}{2} \widetilde{B}_M B_{B_M} \widetilde{B}_M 
+ \frac{1}{2} \widetilde{G}_M B_{G_M} \widetilde{G}_M
+ \frac{1}{2} \widetilde{W}_M B_{W_M} \widetilde{W}_M 
\nonumber\\                       &   &
+\widetilde{X}_M B_{X_M} \widetilde{\bar{X}}_M 
+ {\rm H.c.}
\Bigr\}
\nonumber\\                       &   & 
+\widetilde{W}_M\, \widetilde{m}^2_{W_M}\, \widetilde{W}_M^{\ast}
+\widetilde{B}_M\, \widetilde{m}^2_{B_M}\, \widetilde{B}_M^{\ast}
+\widetilde{G}_M\, \widetilde{m}^2_{G_M}\, \widetilde{G}_M^{\ast}
\nonumber\\                       &   & 
+\widetilde{X}_M\, \widetilde{m}^2_{X_M}\, \widetilde{X}_M^{\ast}
+\widetilde{\bar{X}}_M \, \widetilde{m}^2_{\bar{X}_M} \,
                        \widetilde{\bar{X}}_M^{\ast}.
\phantom{\frac{1}{2}}          
\label{eq:III-SOFTPOTsswfromISCtoGUT}
\end{eqnarray}
 The RGEs for the parameters in the complete soft scalar
  potential $\widetilde{V}^{\rm MSSM,III}$ are as follows.
\begin{list}{$\bullet$}{}
\item Soft sfermion masses:
\begin{eqnarray}
&\dot{\widetilde{m}}_{D^c}^2          
&=
  M_{G,D^c}^2
+2{\cal F}_{(Y_D,\widetilde{D}^c,\widetilde{Q},H_d, A_D)}    
+2{\cal F}_{(Y_{\bar{X}_M}^T,
              \widetilde{D}^c,\widetilde{\bar{X}}_M^{\ast},H_u,
              A_{\bar{X}_M}^T)},
\nonumber\\
&\dot{\widetilde{m}}_L^2              
&=
  M_{G,L}^2
+ {\cal F}_{(Y_E^\dagger,\widetilde{L},\widetilde{E}^c,H_d,A_E^\dagger)}
+\displaystyle{\frac{3}{2}}
           {\cal F}_{(Y_N^{{\rm III}\,\dagger},
                     \widetilde{L},\widetilde{W}_M,H_u,
                      A_N^{{\rm III}\,\dagger})} 
\nonumber\\ 
&&
\hspace*{5truecm}
+\displaystyle{\frac{3}{10}}
           {\cal F}_{(Y_B^{{\rm III}\,\dagger},
                     \widetilde{L},\widetilde{B}_M,H_u,
                      A_B^{{\rm III}\,\dagger})} , 
\qquad
\nonumber\\
&\dot{\widetilde{m}}_{W_M}^2          
&=
  M_{G,{W_M}}^2
+ {\cal F}_{(Y_N^{\rm III},\widetilde{W}_M,\widetilde{L},H_u,
              A_N^{\rm III})} ,
\nonumber\\
&\dot{\widetilde{m}}_{B_M}^2          
&=
\ \displaystyle{\frac{3}{5}}
            {\cal F}_{(Y_B^{\rm III},
                      \widetilde{B}_M,\widetilde{L},{H_u},
                       A_B^{\rm III})} ,
\nonumber\\
&\dot{\widetilde{m}}_{\bar{X}_M}^2    
&=
  M_{G,{\bar{X}_M}}^2
 + {\cal F}_{(Y_{\bar{X}_M},
              \widetilde{\bar{X}}_M,\widetilde{D}^{c\,\ast}, H_u,
              A_{\bar{X}_M})} ,
\nonumber\\
&\dot{\widetilde{m}}_{X_M}^2          
&=
   M_{G,{X_M}}^2,
\nonumber\\
&\dot{\widetilde{m}}_{G_M}^2          
&=
   M_{G,{G_M}}^2.
\label{eq:III-softMASSESfromISCtoGUT}
\end{eqnarray}
{}For $\dot{\widetilde{m}}^2_Q$, $\dot{\widetilde{m}}^2_{U^c}$, 
 $\dot{\widetilde{m}}^2_{E^c}$, see Eq.~(\ref{eq:MSSM-softMASSEStoISC}).
\item Soft Higgs masses: 
\begin{eqnarray}
 \dot{\widetilde{m}}_{H_u}^2         & = & M_{G,{H_u}}^2
 +3{\rm Tr}
 {\cal F}_{(Y_U^\dagger,H_u,\widetilde{U}^c,\widetilde{Q},
            A_U^\dagger)}
 +\frac{3}{2}{\rm Tr}
 {\cal F}_{(Y_N^{{\rm III}\,\dagger},
            H_u,\widetilde{W}_M,\widetilde{L},
            A_N^{{\rm III}\,\dagger})} 
\nonumber\\                          & &  
\hspace*{0.5truecm}
 +\frac{3}{10}{\rm Tr}
 {\cal F}_{(Y_B^{{\rm III}\,\dagger},
            H_u,\widetilde{B}_M,\widetilde{L},
            A_B^{{\rm III}\,\dagger})} 
 + 3{\rm Tr}
 {\cal F}_{(Y_{\bar{X}_M}^\dagger,
            H_u,\widetilde{\bar{X}}_M,\widetilde{D}^{c\,\ast},
            A_{\bar{X}_M}^\dagger)} . \qquad
\label{eq:III-softHIGGSfromISCtoGUT}
\end{eqnarray}
The RGE for $\widetilde{m}_{H_d}^2$ is as in 
 Eq.~(\ref{eq:MSSM-softHIGGStoISC}).
\end{list}

\subsection{\boldmath{$M_{\rm GUT}< Q$}}
\label{sec:RGEGUTtoPL} 
%
We give here the RGEs for the minimal SUSY {\it SU}(5) model without 
 NROs, with the three possible implementations of the seesaw mechanism.
The mechanism's mediators are schematically denoted as RHNs,
 ${\rm 15_H}$, and ${\rm 24_M}$ in three cases. 
The superpotentials and the soft SUSY-breaking scalar potentials
 for the three resulting models can be decomposed as 
\begin{eqnarray}
  W^{{\rm MSSU(5)},i}               &=& W^{\rm MSSU(5)} + W_{\rm i},
\nonumber\\
  \widetilde{V}^{{\rm MSSU(5)},i} \ &=&
  \widetilde{V}^{\rm MSSU(5)} \, + \,\widetilde{V}_{\rm i},
\quad\quad ({\rm i=RHN,\,15H,\,24M})
\label{eq:i-POTENTIALSfromGUT}
\end{eqnarray}
 where $W^{\rm MSSU(5)}$ and $\widetilde{V}^{\rm MSSU(5)}$ are given in
 \S\ref{sec:minSU5} (see
 Eqs.~(\ref{eq:WminSU5}),~(\ref{eq:WminSU5m}),~(\ref{eq:WminSU5H}), and
 Eqs.~(\ref{eq:VminSU5}),~(\ref{eq:VminSU5terms}), respectively), and 
 $W_{\rm i}$ and $\widetilde{V}_{\rm i}$ in \S\ref{sec:seesaw}
 (see Eqs.~(\ref{eq:SU5sswSUPERP}) and~(\ref{eq:SU5sswSOFTPOT})).

\subsubsection{\underline{MSSU(5), RHN}}
\label{sec:I-GUTtoPL}  
%
The RGEs for this model can be found also in Ref.~\citen{BGOO}.
The differences between our equations and those given in that paper are
 listed below.
We refer the reader to \S\ref{sec:minSU5} and~\S\ref{sec:seesaw}
 for the superpotential and soft SUSY-breaking potential of this model. 
\begin{list}{$\bullet$}{}
\item Beta function coefficient:
  $\displaystyle{ \quad b_5^{\rm I} = -3}.$
\item Yukawa couplings:
\begin{eqnarray}
& \dot{Y}^{10}       & = 
 \gamma_{10_M}^T Y^{10} +Y^{10}\gamma_{10_M} +\gamma_{5_H} Y^{10},  
\nonumber\\
& \dot{Y}^{5}        & = 
 \gamma_{\bar{5}_M}^T Y^5 +Y^5\gamma_{10_M}   
                         +\gamma_{\bar{5}_H} Y^5  ,
\nonumber\\
& \dot{Y}_N^{\rm I}  & =  
 \gamma_{N^c}^T Y_N^{\rm I} +Y_N^{\rm I}\gamma_{\bar{5}_M}   
                            +\gamma_{5_H} Y_N^{\rm I} ,
\nonumber\\
& \dot{\lambda}_5    & = 
 (\gamma_{5_H} + \gamma_{24_H} +\gamma_{\bar{5}_H})\lambda_5 ,
\nonumber\\
& \dot{\lambda}_{24} & =  3\gamma_{24_H} \lambda_{24} ,
\label{eq:I-YUKfromGUT}
\end{eqnarray}
 with anomalous dimensions:
\begin{eqnarray}
& \gamma_{10_M}                    & =  
  -2 \left(\displaystyle{\frac{18}{5}}g_5^2 \right) \!{\mathop{\bf 1}}
 + 3 Y^{10\,\dagger} Y^{10} + 2 Y^{5\,\dagger} Y^{5} ,
\phantom{\frac{12}{5}} 
\nonumber\\
& \gamma_{\bar{5}_M}               & =  
  -2 \left(\displaystyle{\frac{12}{5}}g_5^2 \right) \!{\mathop{\bf 1}}
 + 4 Y^{5\,\ast} Y^{5\,T} + Y_N^{{\rm I}\,\dagger} Y_N^{{\rm I}} ,
\nonumber\\
& \gamma_{N^c}                     & =  
 \ \  5 Y_N^{{\rm I}\,\ast} Y_N^{{\rm I}\,T} ,
 \phantom{\displaystyle{\frac12 }}
\nonumber\\
& \gamma_{5_H}                     & =  
  -2 \left(\displaystyle{\frac{12}{5}} g_5^2 \right)  
 + {\rm Tr}\left(3 Y^{10\,\dagger} Y^{10} + 
              Y_N^{\rm I\,\dagger} Y_N^{\rm I} \right) 
 + \displaystyle{\frac{24}{5}}\vert \lambda_5 \vert^2  ,
\nonumber\\
& \gamma_{\bar{5}_H}               & =  
  -2 \left(\displaystyle{\frac{12}{5}}g_5^2 \right)
 + {\rm Tr} \left(4 Y^{5\,\dagger} Y^{5}\right)
  + \displaystyle{\frac{24}{5}}\vert \lambda_5 \vert^2  ,
\nonumber\\
& \gamma_{24_H}                    & =  
  -2 \left(5  g_5^2 \right)
 + \vert \lambda_5 \vert^2 
 +\displaystyle{\frac{21}{20}} \vert \lambda_{24} \vert^2  .
\label{eq:I-anomDIMfromGUT}
\end{eqnarray}
The coefficient $21/20$ in $\gamma_{24_H}$ can be calculated by using
 the properties of the third-order Casimir for adjoint representations
 reported, for example in Ref.~\citen{CUI}. 
\item Superpotential dimensionful parameters:
\begin{eqnarray}
& \dot{M}_N                        & = 
     \gamma_{N^c}^T M_N + M_N \gamma_{N^c}, 
\nonumber\\
& \dot{M}_5                        & = 
    (\gamma_{5_H} + \gamma_{\bar{5}_H}) M_5,
\nonumber\\
& \dot{M}_{24}                     & = 
   2 \gamma_{24_H} M_{24}.
\label{eq:I-hardMASSESfromGUT}
\end{eqnarray}
\end{list}
\begin{list}{$\bullet$}{}
\item Soft sfermion masses:
\begin{eqnarray}
& \dot{\widetilde{m}}_{10_M}^2     
& =  
 M_{G,{10_M}}^2
 +3{\cal F}_{(Y^{10\,\dagger},
              \widetilde{10}_M,
              \widetilde{10}_M^{\ast},
              {5}_H,
              A^{10\,\dagger})}   
 +2{\cal F}_{(Y^{5\,\dagger},
              \widetilde{10}_M,
              \widetilde{\bar{5}}_M,
              {\bar{5}}_H,
              A^{5\,\dagger})} ,  
\nonumber\\
&\dot{\widetilde{m}}_{\bar{5}_M}^2 
& = 
  M_{G,{\bar{5}_M}}^2
 +4{\cal F}_{(Y^{5},
              \widetilde{\bar{5}}_M,
              \widetilde{10}_M,
              \bar{5}_H,
              A^{5})}   
 + {\cal F}_{(Y_N^{{\rm I}\,T},
              \widetilde{\bar{5}}_M,
              \widetilde{N}^{c\,\ast},
              5_H,
              A_N^{{\rm I}\,T})} ,  
\nonumber\\
&\dot{\widetilde{m}}_{N^c}^2       
& =  
\ 5{\cal F}_{(Y_N^{\rm I},
              \widetilde{N}^c,
              \widetilde{\bar{5}}_M^{\ast},
              5_H,
              A_N^{\rm I})}.
\label{eq:I-softMASSESfromGUT}
\end{eqnarray}
\item Soft Higgs masses:
\begin{eqnarray}
&\dot{\widetilde{m}}_{5_H}^2       
& =  
 M_{G,{5_H}}^2
+ 3{\rm Tr}
          {\cal F}_{(Y^{10\,\dagger},
                     5_H,\widetilde{10}_M^{\ast},
                     \widetilde{10}_M,
                     A^{10\,\dagger})}   
\nonumber\\
&   & 
\phantom{= M_{G,{5_H}}^2}
+ {\rm Tr} {\cal F}_{(Y_N^{{\rm I}\,\dagger},
                     5_H,\widetilde{N}^c,
                     \widetilde{\bar{5}}_M^{\ast},
                     A_N^{{\rm I}\,\dagger})}   
+ \displaystyle{\frac{24}{5}}
          {\cal F}_{(\lambda_5,
                     5_H,
                     24_H,
                     \bar{5}_H,
                     A_{\lambda_5})},  
\nonumber\\
&\dot{\widetilde{m}}_{\bar{5}_H}^2 
& = 
 M_{G,{\bar{5}_H}}^2
+ 4{\rm Tr}{\cal F}_{(Y^{5\,\dagger},
                     \bar{5}_H,
                     \widetilde{\bar{5}}_M,
                     \widetilde{10}_M,
                     A^{5\,\dagger})}
+ \displaystyle{\frac{24}{5}}
          {\cal F}_{(\lambda_5,
                     \bar{5}_H,
                     5_H,
                     24_H,
                     A_{\lambda_5})},
\nonumber\\
&\dot{\widetilde{m}}_{24_H}^2      
& = 
 M_{G,{24_H}}^2
+          {\cal F}_{(\lambda_5,
                     24_H,
                     \bar{5}_H,
                     5_H,
                     A_{\lambda_5})} 
+\displaystyle{\frac{21}{20}}
          {\cal F}_{(\lambda_{24},
                     24_H,
                     24_H,
                     24_H,
                     A_{\lambda_{24}})}.
\label{eq:I-softHIGGSfromGUT}
\end{eqnarray}
\end{list}
Our RGEs for this model differ from those in
 Ref.~\citen{BGOO} for $\gamma_{5_H}$ and $\tilde{\gamma}_{5_H}$. 
Moreover, the parameters $\lambda_5$ and $\lambda_{24}$, as well as
 $A_{\lambda_5}$ and $A_{\lambda_{24}}$, are systematically taken as
 vanishing in the RGEs listed in that paper.

\subsubsection{\underline{MSSU(5), 15H}}
\label{sec:II-GUTtoPL}  
%
Also in this case, the expressions for the superpotential and the soft
 SUSY-breaking potential can be found in \S\S\ref{sec:minSU5}
 and \ref{sec:seesaw}.
\begin{list}{$\bullet$}{}
\item Beta function coefficient:
  $\displaystyle{ \quad b_5^{\rm II} = 4}.$
\item Yukawa couplings: 
The RGEs for $Y^{10}$, $Y^5$, $\lambda_5$, and $\lambda_{24}$ can be 
 read from Eq.~(\ref{eq:I-YUKfromGUT}), those for $Y_{N}^{\rm II}$, 
 $\lambda_{15}$, $\lambda_U$, $\lambda_D$ are
\begin{eqnarray}
& \dot{Y}_{N}^{\rm II} & =  
 \gamma_{\bar{5}_M}^T Y_{N}^{\rm II} 
                           +Y_{N}^{\rm II} \gamma_{\bar{5}_M}
                           +\gamma_{15_H} Y_{N}^{\rm II},
\nonumber\\
& \dot{\lambda}_{15}   & = 
 \left(\gamma_{15_H} +\gamma_{24_H}
                           +\gamma_{\bar{15}_H}\right)\lambda_{15},
\nonumber\\
& \dot{\lambda}_U      & = 
 \left(2 \gamma_{5_H} +\gamma_{\bar{15}_H}\right) \lambda_U,
\nonumber\\
& \dot{\lambda}_D      & = 
 \left(2 \gamma_{\bar{5}_H} +\gamma_{15_H}\right) \lambda_D   .
\label{eq:II-YUKfromGUT}
\end{eqnarray}
Among the anomalous dimensions, only $\gamma_{10_M}$ is as in
 Eq.~(\ref{eq:I-anomDIMfromGUT}). The others are
\begin{eqnarray}
& \gamma_{\bar{5}_M}\       & =  
  -2 \left(\displaystyle{\frac{12}{5}}g_5^2 \right) \!{\mathop{\bf 1}} 
 + 4 Y^{5\,\ast} Y^{5\,T} + 6 Y_N^{{\rm II}\,\ast} Y_N^{{\rm II}\,T} ,
\nonumber\\
& \gamma_{5_H}\             & =  
  -2 \left(\displaystyle{\frac{12}{5}} g_5^2 \right)
 + {\rm Tr} \left(3 Y^{10\,\dagger} Y^{10}\right)
 + \frac{24}{5}\vert \lambda_5 \vert^2  
 + 6 \vert \lambda_U \vert^2     ,
\nonumber\\
& \gamma_{\bar{5}_H}\       & =  
  -2 \left(\displaystyle{\frac{12}{5}} g_5^2 \right)
 + {\rm Tr} \left(4 Y^{5\,\dagger} Y^{5} \right)
 + \frac{24}{5}\vert \lambda_5 \vert^2
 + 6 \vert \lambda_D \vert^2    ,
\nonumber\\
& \gamma_{24_H}             & =  
  -2 \left(5  g_5^2 \right)
 +  \vert \lambda_5 \vert^2
 + \displaystyle{\frac{21}{20}}  \vert \lambda_{24} \vert^2
 +  7 \vert \lambda_{15} \vert^2   ,
\nonumber\\
& \gamma_{15_H}             & =  
  -2 \left(\displaystyle{\frac{28}{5}} g_5^2 \right)
 + {\rm Tr} \left(Y_N^{{\rm II}\,\dagger} Y_N^{{\rm II}} \right)
 + \vert \lambda_D \vert^2 
 + \frac{56}{5} \vert \lambda_{15} \vert^2 , 
\nonumber\\
& \gamma_{\bar{15}_H}       & = 
  -2 \left(\displaystyle{\frac{28}{5}} g_5^2 \right) 
 +  \vert \lambda_U \vert^2 
 + \frac{56}{5} \vert \lambda_{15} \vert^2  .
\label{eq:II-anomDIMfromGUT}
\end{eqnarray}
\item Superpotential dimensionful parameters: 
\begin{eqnarray}
 \dot{M}_{15}\!    & = &  (\gamma_{15_H} + \gamma_{\bar{15}_H}) M_{15} .
\label{eq:II-hardMASSESfromGUT}
\end{eqnarray}
The RGEs for ${M}_5$ and ${M}_{24}$ are as in
 Eq.~(\ref{eq:I-hardMASSESfromGUT}).
\end{list}
\begin{list}{$\bullet$}{}
\item Soft sfermion masses:
The RGE for $\widetilde{m}_{10_M}^2$ is still as in 
 Eq.~(\ref{eq:I-softMASSESfromGUT}), that for
 $\widetilde{m}_{\bar{5}_M}^2$ is
\begin{eqnarray}
\dot{\widetilde{m}}_{\bar{5}_M}^2      &=&  M_{G,{\bar{5}_M}}^2
 +4{\cal F}_{(Y^5,
              \widetilde{\bar{5}}_M,
              \widetilde{10}_M,
              {\bar{5}}_H,
              A^5)} 
 +6{\cal F}_{(Y_N^{\rm II},
              \widetilde{\bar{5}}_M,
              \widetilde{\bar{5}}_M^{\ast},
              15_H,
              A_N^{\rm II})} .
\label{eq:II-softMASSESfromGUT}
\end{eqnarray}
\item Soft Higgs masses:
\begin{eqnarray}
&\dot{\widetilde{m}}_{5_H}^2        
&=
  M_{G,{5_H}}^2
 +3{\rm Tr}{\cal F}_{(Y^{10\,\dagger},
                     5_H,\widetilde{10}_M^{\ast},
                     \widetilde{10}_M,
                     A^{10\,\dagger})}
\nonumber\\
& &
\phantom{=  M_{G,{5_H}}^2}
+\displaystyle{\frac{24}{5}}
          {\cal F}_{(\lambda_5,
                     5_H,24_H,
                     \bar{5}_H,
                     A_{\lambda_5})}   
 +6       {\cal F}_{(\lambda_U,
                     5_H,
                     \bar{15}_H,
                     5_H,
                     A_{\lambda_U})} ,
\nonumber\\
&\dot{\widetilde{m}}_{\bar{5}_H}^2  
&=
  M_{G,{\bar{5}_H}}^2
 +4{\rm Tr}{\cal F}_{(Y^{5\,\dagger},
                      \bar5_H,
                      \widetilde{\bar{5}}_M,
                      \widetilde{10}_M,
                      A^{5\,\dagger})}
\nonumber\\
& &
\phantom{=  M_{G,{5_H}}^2}
+\displaystyle{\frac{24}{5}}
          {\cal F}_{(\lambda_5,
                     \bar{5}_H,
                     5_H,
                     24_H,
                     A_{\lambda_5})} 
 +6       {\cal F}_{(\lambda_D,
                     \bar{5}_H,
                     15_H,
                     \bar{5}_H,
                     A_{\lambda_D})} ,
\nonumber\\
&\dot{\widetilde{m}}_{24_H}^2       
&=  
 M_{G,{24_H}}^2
  +        {\cal F}_{(\lambda_5,
                     24_H,
                     \bar{5}_H,
                     5_H,
                     A_{\lambda_5})} 
\nonumber\\
& &
\phantom{=  M_{G,{5_H}}^2}
+\displaystyle{\frac{21}{20}}
          {\cal F}_{(\lambda_{24},
                     24_H,
                     24_H,
                     24_H,
                     A_{\lambda_{24}})} 
 +7       {\cal F}_{(\lambda_{15},
                     24_H,
                     \bar{15}_H,
                     15_H,
                     A_{\lambda_{15}})} ,
\nonumber\\
&\dot{\widetilde{m}}_{15_H}^2       
&= 
 M_{G,{15_H}}^2
 + {\rm Tr}
          {\cal F}_{(Y_N^{{\rm II}\,\dagger},
                     15_H,
                     \widetilde{\bar{5}}_M,
                     \widetilde{\bar{5}}_M^{\ast},
                     A_N^{{\rm II}\,\dagger})}
\nonumber\\
& &
\phantom{=  M_{G,{5_H}}^2}
 +         {\cal F}_{(\lambda_D,
                     15_H,
                     \bar{5}_H,
                     \bar{5}_H,
                     A_{\lambda_D})}
 +\displaystyle{\frac{56}{5}}
          {\cal F}_{(\lambda_{15},
                     15_H,
                     24_H,
                     \bar{15}_H,
                     A_{\lambda_{15}})} ,
\nonumber\\
&\dot{\widetilde{m}}_{\bar{15}_H}^2 
&=
 M_{G,{\bar{15}_H}}^2
 +         {\cal F}_{(\lambda_U,
                     \bar{15}_H,
                     5_H,
                     5_H,
                     A_{\lambda_U})}
 +\displaystyle{\frac{56}{5}}
          {\cal F}_{(\lambda_{15},
                     \bar{15}_H,
                     15_H,
                     24_H,
                     A_{\lambda_{15}})} .
\label{eq:II-softHIGGSfromGUT}
\end{eqnarray}
\end{list}

\subsubsection{\underline{MSSU(5), 24M}}
\label{sec:III-GUTtoPL}  
%
The definitions of the superpotential and soft SUSY-breaking scalar
 potential can be found in \S\S\ref{sec:minSU5} and \ref{sec:seesaw}.
\begin{list}{$\bullet$}{}
\item Beta function coefficient:
  $\displaystyle{ \quad b_5^{\rm III} = 12}.$
\item Yukawa couplings: 
\begin{eqnarray}
& \dot{Y}_N^{\rm III}              & =  
 \gamma_{24_M}^T Y_N^{\rm III} +Y_N^{\rm III}\gamma_{\bar{5}_M}   
+\gamma_{5_H} Y_N^{\rm III} ,
\nonumber\\
& \dot{Y}^x_{24_M}                 & =  
 \gamma_{24_M}^T Y^x_{24_M}  +Y^x_{24_M} \gamma_{24_M}
+ \gamma_{24_H}  Y^x_{24_M} . 
\label{eq:III-YUKfromGUT}
\end{eqnarray}
The RGEs for $Y_5$, $Y_{10}$, $\lambda_5$ and $\lambda_{24}$ are
 formally as in Eq.~(\ref{eq:I-YUKfromGUT}). 
The anomalous dimensions are now
\begin{eqnarray}
& \gamma_{\bar{5}_M}              & =  
  -2 \left(\displaystyle{\frac{12}{5}} g_5^2 \right) 
   \!{\mathop{\bf 1}}
 + 4 Y^{5\,\ast} Y^{5\,T} 
 + \frac{24}{5} Y_N^{{\rm III}\,\dagger} Y_N^{{\rm III}} ,
 \phantom{
 +\frac{5}{2}  \left( Y_{24_M}^{A\,\dagger} Y_{24_M}^A\right) ,
         }
\nonumber\\
& \gamma_{24_M}                   & =  
  -2 \left(5 g_5^2 \right) \!{\mathop{\bf 1}}
   +Y_N^{{\rm III}\,\ast} Y_N^{{\rm III}\,T}  
   +\displaystyle{\frac{21}{10}}
                 \left( Y_{24_M}^{S\,\dagger} Y_{24_M}^S\right)
   +\frac{5}{2}  \left( Y_{24_M}^{A\,\dagger} Y_{24_M}^A\right) ,
\nonumber\\
& \gamma_{5_H}                    & =  
  -2 \left(\displaystyle{\frac{12}{5}} g_5^2 \right)  
   +{\rm Tr}\left(3 Y^{10\,\dagger} Y^{10} 
   +\frac{24}{5} Y_N^{\rm III\,\dagger} Y_N^{\rm III} \right) 
   +\frac{24}{5}\vert \lambda_5 \vert^2  ,
\nonumber\\
& \gamma_{24_H}                   & = 
  -2 \left(5  g_5^2 \right)
   +\vert \lambda_5 \vert^2 
   +\displaystyle{\frac{21}{20}}  \vert \lambda_{24} \vert^2 
   +{\rm Tr}\!\left(\! \frac{21}{20} Y_{24_M}^{S\,\dagger} Y_{24_M}^S
   +\!\frac{5}{4}   Y_{24_M}^{A\,\dagger} Y_{24_M}^A \!\right) ,
\nonumber\\
& & 
\label{eq:III-anomDIMfromGUT}
\end{eqnarray}
 and $\gamma_{10_M}$ and $\gamma_{\bar{5}_H}$ are listed in 
 Eq.~(\ref{eq:I-anomDIMfromGUT}).
\item Superpotential dimensionful parameters: 
\begin{eqnarray}
 \dot{M}_{24_M} & = & \gamma_{24_M}^T M_{24_M} +M_{24_M} \gamma_{24_M}. 
\label{eq:III-hardMASSESfromGUT}
\end{eqnarray}
The RGEs for $M_5$ and $M_{24}$ can be found in 
 Eq.~(\ref{eq:I-hardMASSESfromGUT}). 
\item Soft sfermion masses:
\begin{eqnarray}
&\dot{\widetilde{m}}_{\bar{5}_M}^2      
&=  
 M_{G,{\bar{5}_M}}^2
 +4  {\cal F}_{(Y^{5},
               \widetilde{\bar{5}}_M,
               \widetilde{10}_M,
               \bar{5}_H, 
               A^{5})}   
+\displaystyle{\frac{24}{5}}
    {\cal F}_{(Y_N^{{\rm III}\,T},
               \widetilde{\bar{5}}_M,
               \widetilde{24}_M^{\ast},
               5_H,
               A_N^{{\rm III}\,T})} ,  
\nonumber\\
&\dot{\widetilde{m}}_{24_M}^2           
&=
 M_{G,{24_M}}^2
 +\  {\cal F}_{(Y_N^{\rm III},
               \widetilde{24}_M,
               \widetilde{\bar{5}}_M^{\ast},
               5_H,
               A_N^{\rm III})} 
+\displaystyle{\frac{21}{10}}
    {\cal F}_{(Y_{24_M}^S,
               \widetilde{24}_M,
               \widetilde{24}_M^{\ast},
               24_H,
               A_{24_M}^S)}   
\nonumber\\
& &
\phantom{=  M_{G,{24_M}}^2} 
+\displaystyle{\frac{5}{2}}
    {\cal F}_{(Y_{24_M}^A,
               \widetilde{24}_M,
               \widetilde{24}_M^{\ast},
               24_H,
               A_{24_M}^A)} .
\label{eq:III-softMASSESfromGUT}
\end{eqnarray}
The RGE for $\widetilde{m}_{10_M}^2$ is as in 
  Eq.~(\ref{eq:I-softMASSESfromGUT}).
\item Soft Higgs masses: 
\begin{eqnarray}
&  \dot{\widetilde{m}}_{5_H}^2        
&=
  M_{G,{5_H}}^2
 +3{\rm Tr}
    {\cal F}_{(Y^{10\,\dagger},
               5_H,
               \widetilde{10}_M^{\ast},
               \widetilde{10}_M,
               A^{10\,\dagger})}   
 +\displaystyle{\frac{24}{5}}{\rm Tr}
    {\cal F}_{(Y_N^{{\rm III}\,\dagger}\!,
               5_H,
               \widetilde{24}_M,
               \widetilde{\bar{5}}_M^{\ast},
               A_N^{{\rm III}\,\dagger})}   
\nonumber\\
& & 
 \phantom{=   M_{G,{5_H}}^2}
 +\displaystyle{\frac{24}{5}}
    {\cal F}_{(\lambda_5,
               5_H,
               24_H,
               \bar{5}_H,
               A_{\lambda_5})} ,  
\nonumber\\
&\dot{\widetilde{m}}_{24_H}^2       
&= 
 M_{G,{24_H}}^2                    
 +\  {\cal F}_{(\lambda_5, 24_H,\bar{5}_H,5_H, A_{\lambda_5})} 
 +\displaystyle{\frac{21}{20}}{\rm Tr}
    {\cal F}_{(Y_{24_M}^{S\,\dagger},
               24_H,
               \widetilde{24}_M,
               \widetilde{24}_M^{\ast},
               A_{24_M}^{S\,\dagger})}
\nonumber\\
& &                      
 \phantom{=   M_{G,{5_H}}^2}
 +\displaystyle{\frac{5}{4}{\rm Tr}}
    {\cal F}_{(Y_{24_M}^{A\,\dagger},
               24_H,
               \widetilde{24}_M,
               \widetilde{24}_M^{\ast},
               A_{24_M}^{A\,\dagger})}
\nonumber \\
& &                      
 \phantom{=   M_{G,{5_H}}^2}
 +\displaystyle{\frac{21}{20}}
    {\cal F}_{(\lambda_{24},
               24_H,
               24_H,
               24_H, 
               A_{\lambda_{24}})} .\qquad
\label{eq:III-softHIGGSfromGUT}
\end{eqnarray}
The RGE for $\widetilde{m}_{\bar{5}_H}^2$ is as in 
 Eq.~(\ref{eq:I-softHIGGSfromGUT}).
\end{list}

\subsection{\boldmath{$M_{\rm GUT}< Q$} - nonvanishing NROs}
\label{sec:RGEGUTtoPL-NRO} 
%
As explained in the text, we express the superpotential and scalar
 potential in terms of effective operators, with effective parameters
 indicated in boldface. 
A decomposition similar to that of Eq.~(\ref{eq:i-POTENTIALSfromGUT})
 still holds:
\begin{eqnarray}
  \mbox{\boldmath{$W$}}^{{\rm MSSU(5)},i}        &=&
  \mbox{\boldmath{$W$}}^{\rm MSSU(5)} +
  \mbox{\boldmath{$W$}}_{\rm i},
\nonumber\\
  \widetilde{\mbox{\boldmath{$V$}}}^{{\rm MSSU(5)},i} \ &=&
  \widetilde{\mbox{\boldmath{$V$}}}^{\rm MSSU(5)}  \, + \,
  \widetilde{\mbox{\boldmath{$V$}}}_{\rm i}.
\quad\quad ({\rm i=RHN,\,15H,\,24M})
\label{eq:i-NROPOTSfromGUT}
\end{eqnarray}
The various seesaw potentials 
 $\mbox{\boldmath{$W$}}_{\rm i}$ and
 $\widetilde{\mbox{\boldmath{$V$}}}_{\rm i}$,
 $({\rm i = RHN,\,15H,\,24M})$ will
 be given explicitly in the following subsections.
The two potentials
 $\mbox{\boldmath{$W$}}^{\rm MSSU(5)}$ and 
 $\widetilde{\mbox{\boldmath{$V$}}}^{\rm MSSU(5)}$, now decomposed 
 as
\begin{eqnarray}
  \mbox{\boldmath{$W$}}^{\rm MSSU(5)} &=&
  \mbox{\boldmath{$W$}}^{\rm MSSU(5)}_{\rm M} +
   W^{\rm MSSU(5)}_{\rm H},
\nonumber \\
 \widetilde{\mbox{\boldmath{$V$}}}^{\rm MSSU(5)} &=& 
 \widetilde{\mbox{\boldmath{$V$}}}^{\rm MSSU(5)}_{\rm M} +
 \widetilde{V}^{\rm MSSU(5)}_{\rm H} +
 \widetilde{V}^{\rm MSSU(5)}_{\rm gaug},
\label{eq:WandVdecomp}
\end{eqnarray}
 include NROs only in the matter parts, which are now
\begin{eqnarray}
  \mbox{\boldmath{$W$}}^{\rm MSSU(5)}_{M}  \!    & = & \!
-  D^c \mbox{\boldmath{$Y$}}^5_{D\phantom u} Q H_d     
-\!E^c(\mbox{\boldmath{$Y$}}^5_{E\phantom q})^T L H_d 
-\!D^c \mbox{\boldmath{$Y$}}^5_{DU} U^c H_D^C        
-\!L\, \mbox{\boldmath{$Y$}}^5_{LQ} Q H_D^C
\nonumber\\                                      &   & 
+  U^c \mbox{\boldmath{$Y$}}^{10}_U    Q H_u     
+\!U^c \mbox{\boldmath{$Y$}}^{10}_{UE} E^c H_U^C 
+\!\frac{1}{2}Q \,\mbox{\boldmath{$Y$}}^{10}_{QQ} Q H_U^C,
\label{eq:WSU5Mdecomp}
\end{eqnarray}
 and
\begin{eqnarray}
  \widetilde{\mbox{\boldmath{$V$}}}^{\rm MSSU(5)}_{M} 
                                 \!& = & \!
\Bigl\{
-\widetilde{D}^c \mbox{\boldmath{$A$}}^5_{D\phantom u}
 \widetilde{Q} H_d     
-\widetilde{E}^c(\mbox{\boldmath{$A$}}^5_{E\phantom q})^T 
 \widetilde{L} H_d 
-\widetilde{D}^c \mbox{\boldmath{$A$}}^5_{DU} 
 \widetilde{U}^c H_D^C        
-\widetilde{L}\, \mbox{\boldmath{$A$}}^5_{LQ} 
 \widetilde{Q} H_D^C
\nonumber\\                        &   &
+\widetilde{U}^c \mbox{\boldmath{$A$}}^{10}_U    
 \widetilde{Q} H_u     
+\widetilde{U}^c \mbox{\boldmath{$A$}}^{10}_{UE} 
 \widetilde{E}^c H_U^C 
+\!\frac{1}{2}\widetilde{Q} \,\mbox{\boldmath{$A$}}^{10}_{QQ} 
              \widetilde{Q} H_U^C  + {\rm H.c.}
\Bigr\}
\nonumber\\[1.001ex]               &   & 
+\widetilde{Q}^\ast \, \widetilde{\mbox{\boldmath{$m$}}}^2_Q \, 
 \widetilde{Q} 
+\widetilde{U}^c    \, \widetilde{\mbox{\boldmath{$m$}}}^2_{U^c}\, 
 \widetilde{U}^{c\,\ast}\!
+\widetilde{D}^c    \, \widetilde{\mbox{\boldmath{$m$}}}^2_{D^c}\, 
 \widetilde{D}^{c\,\ast}\!
\nonumber\\[1.001ex]                &   & 
+\widetilde{L}^\ast \, \widetilde{\mbox{\boldmath{$m$}}}^2_L \,     
 \widetilde{L} 
+\widetilde{E}^c    \, \widetilde{\mbox{\boldmath{$m$}}}^2_{E^c}\, 
 \widetilde{E}^{c\,\ast} .
\quad 
\label{eq:VSU5Mdecomp}
\end{eqnarray}
No boldface type is used for $\widetilde{V}^{\rm MSSU(5)}_{\rm gaug}$,
 which is still that of Eq.~(\ref{eq:VminSU5terms}). 
Normal character types are also used 
 for the Higgs potentials, $W^{\rm MSSU(5)}_{H}$ and
 $\widetilde{V}^{\rm MSSU(5)}_{H}$, because we neglect NROs in the
 Higgs sector. 
These potentials are then those in Eqs.~(\ref{eq:WminSU5H})
 and~(\ref{eq:VminSU5terms}).
Similarly, the purely Higgs parts of $\mbox{\boldmath{$W$}}_{15H}$
 and $\widetilde{\mbox{\boldmath{$V$}}}_{15H}$ also remain unchanged.
The corresponding RGEs are also those reported in the case of 
 vanishing NROs, with undecomposed anomalous dimensions, soft mass 
 squared, trilinear and bilinear couplings. 
Nevertheless, we need, at least formally, to decompose these last
 quantities when they enter in RGEs of effective flavoured fields. 
In this case, we use distinct symbols, for example,
 $\mbox{\boldmath{$\gamma$}}_{H_u}$ and
 $\mbox{\boldmath{$\gamma$}}_{H_U^C}$ for the same $\gamma_{5_H}$,
 as well as 
 $\mbox{\boldmath{$\widetilde{m}$}}_{H_u}^2$ and 
 $\mbox{\boldmath{$\widetilde{m}$}}_{H_U^C}^2$ for 
 $\widetilde{m}_{5_H}^2$, but we make the identifications
 $\mbox{\boldmath{$\gamma$}}_{H_u}   \!=\! 
  \mbox{\boldmath{$\gamma$}}_{H_U^C} \!=\! \gamma_{5_H}$ and
 $\mbox{\boldmath{$\widetilde{m}$}}_{H_u}^2   \!=\! 
  \mbox{\boldmath{$\widetilde{m}$}}_{H_U^C}^2 \!=\!
                   \widetilde{m}_{5_H}^2$.
As said, these are only formal decompositions, and we approximate the
 Yukawa couplings of the renormalizable operators,
 $Y^{10}$, $Y^{5}$, $Y_N^{\rm I}$, which enter in their definitions or
 in the expressions of their RGEs, with the effective couplings
 $\mbox{\boldmath{$Y$}}^{10}_U$,
 $\mbox{\boldmath{$Y$}}^{5}_D$, and 
 $\mbox{\boldmath{$Y$}}^{\rm i}_N$ (i = I, II, III), respectively.

We also neglect NROs for terms that involve only heavy fields, even when
 these have flavour, such as the fields $24_M$.
In the expressions for the anomalous dimensions of these fields, we use 
 a ``hybrid'' form, with contributions from the undecomposed couplings
 of interactions for which we have omitted NROs (Higgs couplings), and
 from the decomposed effective couplings of interactions for which NROs
 are nonvanishing 
(see for example Eq.~(\ref{eq:III-HYBRIDanomDIMfromGUT-NRO})).
A similar treatment in this section is reserved to the RGEs for the soft
 masses of such superheavy flavour fields.  
We keep undecomposed the contributions from the interactions that do not
 involve MSSM fields, but we do decompose the contributions from 
 the flavour interactions to which the MSSM fields take part
(see for example Eq.~(\ref{eq:III-softM24fromGUT-NRO})).
This is to allow {\it SU}(5)-violating field rotations needed to embed the 
 light fields in the {\it SU}(5) multiplets.

\subsubsection{\underline{nrMSSU(5), RHN}}
\label{sec:I-GUTtoPL-NRO}  
%
The form of the superpotential \mbox{\boldmath{$W$}}$_{\rm RHN}$ is
 now
\begin{equation}
  \mbox{\boldmath{$W$}}_{\rm RHN} =
   N^c \mbox{\boldmath{$Y$}}_{N}^{\rm I} L H_u
  -N^c \mbox{\boldmath{$Y$}}_{ND}^{\rm I}  D^c H_U^C 
  + \frac{1}{2} N^c M_N N^c.    
\label{eq:I-SU5RHNdec}
\end{equation}   
The RGEs for the superpotential parameters in this class of models
 are as follows.
\begin{list}{$\bullet$}{}
\item Effective Yukawa couplings: 
\begin{eqnarray}
& \dot{\mbox{\boldmath{$Y$}}}^{10}_U       &=
 \mbox{\boldmath{$\gamma$}}_{U^c}^T \,
      \mbox{\boldmath{$Y$}}^{10}_U         +
      \mbox{\boldmath{$Y$}}^{10}_U         
  \mbox{\boldmath{$\gamma_Q$}}             + 
  \mbox{\boldmath{$\gamma$}}_{H_u} 
      \mbox{\boldmath{$Y$}}^{10}_U,
\nonumber\\
& \dot{\mbox{\boldmath{$Y$}}}^{10}_{\!QQ}  &=
 \mbox{\boldmath{$\gamma$}}_Q^T \,
      \mbox{\boldmath{$Y$}}^{10}_{\!QQ}    +
      \mbox{\boldmath{$Y$}}^{10}_{\!QQ} 
 \mbox{\boldmath{$\gamma$}}_Q              +
 \mbox{\boldmath{$\gamma$}}_{H_U^C} 
      \mbox{\boldmath{$Y$}}^{10}_{\!QQ}, 
\nonumber\\
& \dot{\mbox{\boldmath{$Y$}}}^{10}_{UE}    &=
 \mbox{\boldmath{$\gamma$}}_{U^c}^T \,
      \mbox{\boldmath{$Y$}}^{10}_{\!UE}    +
      \mbox{\boldmath{$Y$}}^{10}_{\!UE}
 \mbox{\boldmath{$\gamma$}}_{E^c}          +
 \mbox{\boldmath{$\gamma$}}_{H_U^C} 
      \mbox{\boldmath{$Y$}}^{10}_{\!UE},
\nonumber\\
& \dot{\mbox{\boldmath{$Y$}}}^{5}_D        &=  
 \mbox{\boldmath{$\gamma$}}_{D^c}^T \,
      \mbox{\boldmath{$Y$}}^{5}_D          + 
      \mbox{\boldmath{$Y$}}^{5}_D
 \mbox{\boldmath{$\gamma$}}_Q              +
 \mbox{\boldmath{$\gamma$}}_{H_d} 
      \mbox{\boldmath{$Y$}}^{5}_D,
\nonumber\\
& \dot{\mbox{\boldmath{$Y$}}}^{5}_E        &=  
 \mbox{\boldmath{$\gamma$}}_L^T \,
      \mbox{\boldmath{$Y$}}^{5}_E          +   
      \mbox{\boldmath{$Y$}}^{5}_E
 \mbox{\boldmath{$\gamma$}}_{E^c}          + 
 \mbox{\boldmath{$\gamma$}}_{H_d} 
      \mbox{\boldmath{$Y$}}^{5}_E,
\nonumber\\
& \dot{\mbox{\boldmath{$Y$}}}^{5}_{DU}     &= 
 \mbox{\boldmath{$\gamma$}}_{D^c}^T \,
      \mbox{\boldmath{$Y$}}^{5}_{\!DU}     +
      \mbox{\boldmath{$Y$}}^{5}_{\!DU}     
 \mbox{\boldmath{$\gamma$}}_{U^c}          + 
 \mbox{\boldmath{$\gamma$}}_{H_D^C} 
      \mbox{\boldmath{$Y$}}^{5}_{\!DU},
\nonumber\\
& \dot{\mbox{\boldmath{$Y$}}}^{5}_{LQ}     &= 
 \mbox{\boldmath{$\gamma$}}_L^T \,
      \mbox{\boldmath{$Y$}}^{5}_{\!LQ}     +
      \mbox{\boldmath{$Y$}}^{5}_{\!LQ}
 \mbox{\boldmath{$\gamma$}}_Q              +
 \mbox{\boldmath{$\gamma$}}_{H_D^C} 
      \mbox{\boldmath{$Y$}}^{5}_{\!LQ},
\label{eq:I-YUK5and10fromGUT-NRO}
\end{eqnarray}
 and
\begin{eqnarray}
& \dot{\mbox{\boldmath{$Y$}}}^{\rm I}_{N{\phantom{L}}} &=
 \mbox{\boldmath{$\gamma$}}_{N^c}^T\,
      \mbox{\boldmath{$Y$}}^{\rm I}_{\!N}              +
      \mbox{\boldmath{$Y$}}^{\rm I}_{\!N}
 \mbox{\boldmath{$\gamma$}}_L                          +
 \mbox{\boldmath{$\gamma$}}_{H_u}
      \mbox{\boldmath{$Y$}}^{\rm I}_{\!N},
\nonumber\\
& \dot{\mbox{\boldmath{$Y$}}}^{\rm I}_{ND}             &= 
 \mbox{\boldmath{$\gamma$}}_{N^c}^T\,
      \mbox{\boldmath{$Y$}}^{\rm I}_{\!ND}             +
      \mbox{\boldmath{$Y$}}^{\rm I}_{\!ND}
 \mbox{\boldmath{$\gamma$}}_{D^c}                      +
 \mbox{\boldmath{$\gamma$}}_{H_U^C}
      \mbox{\boldmath{$Y$}}^{\rm I}_{\!ND}.
\label{eq:I-YUKNfromGUT-NRO}
\end{eqnarray}
Given the approximation made here, the RGEs for
 $\lambda_5$ and $\lambda_{24}$, not decomposed, are as in
 Eq.~(\ref{eq:I-YUKfromGUT}).
{}For the same reason, $\mbox{\boldmath{$\gamma $}}_{24_H}$,
 $\mbox{\boldmath{$\gamma $}}_{5_H}$ and
 $\mbox{\boldmath{$\gamma $}}_{\bar{5}_H}$ are those of
 Eq.~(\ref{eq:I-anomDIMfromGUT}),
 with $Y^5$, $Y^{10}$, and $Y^{\rm I}_N$ approximated by
 $\mbox{\boldmath{$Y$}}^{5}_D$,
 $\mbox{\boldmath{$Y$}}^{10}_U$, and
 $\mbox{\boldmath{$Y$}}^{\rm I}_N$,
 respectively.
The anomalous dimensions $\mbox{\boldmath{$\gamma $}}_{H_u}$ and
 $\mbox{\boldmath{$\gamma $}}_{H_U^C}$ are taken to be equal to
 $\mbox{\boldmath{$\gamma $}}_{5_H}$;
 $\mbox{\boldmath{$\gamma $}}_{H_d}$ and
 $\mbox{\boldmath{$\gamma $}}_{H_D^C}$ to
 $\mbox{\boldmath{$\gamma $}}_{\bar{5}_H}$.  
The remaining anomalous dimensions needed for the evaluation 
 of the above RGEs are
\begin{eqnarray}
& \gamma_Q                         & = 
-2 \left(\displaystyle{\frac{18}{5}} g_5^2 \right) \!{\mathop{\bf 1}}
+ \mbox{\boldmath{$Y$}}^{10\,\dagger}_U 
  \mbox{\boldmath{$Y$}}^{10}_U 
+2\mbox{\boldmath{$Y$}}^{10\,\dagger}_{\!QQ}
  \mbox{\boldmath{$Y$}}^{10}_{\!QQ} 
+ \mbox{\boldmath{$Y$}}^{5\,\dagger}_D
  \mbox{\boldmath{$Y$}}^{5}_D 
+ \mbox{\boldmath{$Y$}}^{5\,\dagger}_{\!LQ}
  \mbox{\boldmath{$Y$}}^{5}_{\!LQ}, 
\nonumber\\
& \gamma_{U^c}                     & = 
-2 \left(\displaystyle{\frac{18}{5}} g_5^2 \right) \!{\mathop{\bf 1}}
+2\mbox{\boldmath{$Y$}}^{10\,\ast}_U 
  \mbox{\boldmath{$Y$}}^{10\,T}_U 
+ \mbox{\boldmath{$Y$}}^{10\,\ast}_{\!UE} 
  \mbox{\boldmath{$Y$}}^{10\,T}_{\!UE} 
+2\mbox{\boldmath{$Y$}}^{5\,\dagger}_{\!DU} 
  \mbox{\boldmath{$Y$}}^{5}_{\!DU}, 
\nonumber\\
& \gamma_{E^c}                     & = 
-2 \left(\displaystyle{\frac{18}{5}} g_5^2 \right) \!{\mathop{\bf 1}}
+3\mbox{\boldmath{$Y$}}^{10\,\dagger}_{\!UE}
  \mbox{\boldmath{$Y$}}^{10}_{\!UE} 
+2\mbox{\boldmath{$Y$}}^{5\,\dagger}_E
  \mbox{\boldmath{$Y$}}^{5}_E, 
\nonumber\\
& \gamma_L                         & =   
-2 \left(\displaystyle{\frac{12}{5}} g_5^2 \right) \!{\mathop{\bf 1}}
+ \mbox{\boldmath{$Y$}}^{5\,\ast}_E
  \mbox{\boldmath{$Y$}}^{5\,T}_E 
+3\mbox{\boldmath{$Y$}}^{5\,\ast}_{\!LQ} 
  \mbox{\boldmath{$Y$}}^{5\,T}_{\!LQ} 
+ \mbox{\boldmath{$Y$}}^{{\rm I}\,\dagger}_{\!N} 
  \mbox{\boldmath{$Y$}}^{\rm I}_{\!N}, 
\nonumber\\
& \gamma_{D^c}                     & =  
-2 \left(\displaystyle{\frac{12}{5}} g_5^2 \right) \!{\mathop{\bf 1}}
+2\mbox{\boldmath{$Y$}}^{5\rm \,\ast}_D
  \mbox{\boldmath{$Y$}}^{5\,T}_D 
+2\mbox{\boldmath{$Y$}}^{5\rm \,\ast}_{\!DU}
  \mbox{\boldmath{$Y$}}^{5\,T}_{\!DU} 
+ \mbox{\boldmath{$Y$}}^{{\rm I}\,\dagger}_{\!ND}
  \mbox{\boldmath{$Y$}}^{\rm I}_{\!ND}, 
\nonumber\\[1.001ex]
& \gamma_{N^c}                     & = \ 
 2\mbox{\boldmath{$Y$}}^{{\rm I}\,\ast}_{\!N}
  \mbox{\boldmath{$Y$}}^{{\rm I}\,T}_{\!N} 
+3\mbox{\boldmath{$Y$}}^{{\rm I}\,\ast}_{\!ND}
  \mbox{\boldmath{$Y$}}^{{\rm I}\,T}_{\!ND}. 
 \phantom{
-2 \left(\displaystyle{\frac{12}{5}} g_5^2 \right) \!{\mathop{\bf 1}}
+    }
\label{eq:I-anomDIMsfromGUT-NRO}
\end{eqnarray}
\item Superpotential dimensionful parameters: 
See Eq.~(\ref{eq:I-hardMASSESfromGUT}).
We are implicitly assuming that NROs involving only $24_H$ fields
 have small coefficients and/or that $\lambda_{24}$ is not too 
 small.
\end{list}
The term $\widetilde{\mbox{\boldmath{$V$}}}_{\rm RHN}$ specific of this
 model is 
\begin{equation}
  \widetilde{\mbox{\boldmath{$V$}}}_{\rm RHN} =
\left\{
   N^c \mbox{\boldmath{$A$}}_{N}^{\rm I} L H_u
  -N^c \mbox{\boldmath{$A$}}_{ND}^{\rm I}  D^c H_U^C 
  + \frac{1}{2} N^c B_N N^c
 +{\rm H.c.}
\right\}
 +\widetilde{N}^c\, \widetilde{\mbox{\boldmath{$m$}}}^2_{N^c} \, 
  \widetilde{N}^{c\,\ast} 
.    
\label{eq:I-SOFTPOTSU5RHNdec}
\end{equation}   
 The RGEs for the parameters appearing in 
 $\widetilde{\mbox{\boldmath{$V$}}}^{{\rm MSSU(5)},{\rm RHN}}$ are as
 follows.
\begin{list}{$\bullet$}{}
\item Soft sfermion masses:
\begin{eqnarray}
&\dot{\widetilde{\mbox{\boldmath{$m$}}}}_Q^{2}              
& =  
 M_{G,{10_M}}^2
  + {\cal F}_{(\mbox{\boldmath{$_Y$}}^{10\,\dagger}_U \!,
               \widetilde{\mbox{\boldmath{$_Q$}}},
               \widetilde{\mbox{\boldmath{$_{U^c}$}}},
               \mbox{\boldmath{$_{H_u}$}},
               \mbox{\boldmath{$_A$}}^{10\,\dagger}_U)}  
  +2{\cal F}_{(\mbox{\boldmath{$_Y$}}^{10\,\dagger}_{\!QQ}\!,
               \widetilde{\mbox{\boldmath{$_Q$}}},
               \widetilde{\mbox{\boldmath{$_Q$}}}^{\ast} \!,
               \mbox{\boldmath{$_{H_U^C}$}},
               \mbox{\boldmath{$_A$}}^{10\,\dagger}_{\!QQ})}  
\nonumber\\                                                          
&   & 
\phantom{= M_{G,{10_M}}^2}
  + {\cal F}_{(\mbox{\boldmath{$_Y$}}^{5\,\dagger}_D \!,
               \widetilde{\mbox{\boldmath{$_Q$}}},
               \widetilde{\mbox{\boldmath{$_{D^c}$}}},
               \mbox{\boldmath{$_{H_d}$}},
               \mbox{\boldmath{$_A$}}^{5\rm\,\dagger}_D)}  
 +\ {\cal F}_{(\mbox{\boldmath{$_Y$}}^{5\rm\,\dagger}_{\!LQ},
               \widetilde{\mbox{\boldmath{$_Q$}}},
               \widetilde{\mbox{\boldmath{$_L$}}}^{\ast} \!,
               \mbox{\boldmath{$_{H_D^C}$}},
               \mbox{\boldmath{$_A$}}^{5\,\dagger}_{\!LQ})},  
\nonumber\\
&\dot{\widetilde{\mbox{\boldmath{$m$}}}}_{U^c}^{2}          
& =  
 M_{G,{10_M}}^2
 +2{\cal F}_{(\mbox{\boldmath{$_Y$}}^{10}_U \!,
              \widetilde{\mbox{\boldmath{$_{U^c}$}}},
              \widetilde{\mbox{\boldmath{$_Q$}}},
              \mbox{\boldmath{$_{H_u}$}},
              \mbox{\boldmath{$_A$}}^{10}_U)}  
 + {\cal F}_{(\mbox{\boldmath{$_Y$}}^{10}_{\!UE},
              \widetilde{\mbox{\boldmath{$_{U^c}$}}},
              \widetilde{\mbox{\boldmath{$_{E^c}$}}}^{\ast}\!,
              \mbox{\boldmath{$_{H_U^C}$}},
              \mbox{\boldmath{$_A$}}^{10}_{\!UE})}  
\nonumber\\                                                          
&   & 
\phantom{= M_{G,{10_M}}^2}
 +2{\cal F}_{(\mbox{\boldmath{$_Y$}}^{5\,T}_{DU},
              \widetilde{\mbox{\boldmath{$_{U^c}$}}},
              \widetilde{\mbox{\boldmath{$_{D^c}$}}}^{\ast} \!,
              \mbox{\boldmath{$_{H_D^C}$}},
              \mbox{\boldmath{$_A$}}^{5\,T}_{DU})},  
\nonumber\\
&\dot{\widetilde{\mbox{\boldmath{$m$}}}}_{E^c}^{2}          
& =  
 M_{G,{10_M}}^2
 +3{\cal F}_{(\mbox{\boldmath{$_Y$}}^{10\,T}_{\!UE},
              \widetilde{\mbox{\boldmath{$_{E^c}$}}},
              \widetilde{\mbox{\boldmath{$_{U^c}$}}}^{\,\ast} \!,
              \mbox{\boldmath{$_{H_U^C}$}},
              \mbox{\boldmath{$_A$}}^{10\,T}_{\!UE})}  
 +2{\cal F}_{(\mbox{\boldmath{$_Y$}}^{5\,T}_E \!,
              \widetilde{\mbox{\boldmath{$_{E^c}$}}},
              \widetilde{\mbox{\boldmath{$_L$}}},
              \mbox{\boldmath{$_{H_d}$}},
              \mbox{\boldmath{$_A$}}^{5\,T}_E)},  
\nonumber\\
&\dot{\widetilde{\mbox{\boldmath{$m$}}}}_{L}^{2}             
& =    
 M_{G,{\bar{5}_M}}^2
 + {\cal F}_{(\mbox{\boldmath{$_Y$}}^{5\,\ast}_E \!,
              \widetilde{\mbox{\boldmath{$_L$}}},
              \widetilde{\mbox{\boldmath{$_{E^c}$}}},
              \mbox{\boldmath{$_{H_d}$}},
              \mbox{\boldmath{$_A$}}^{5\,\ast}_E)} 
 +3{\cal F}_{(\mbox{\boldmath{$_Y$}}^{5\,\ast}_{\!LQ},
              \widetilde{\mbox{\boldmath{$_L$}}},
              \widetilde{\mbox{\boldmath{$_Q$}}}^{\ast}\!,
              \mbox{\boldmath{$_{H_D^C}$}},
              \mbox{\boldmath{$_A$}}^{5\,\ast}_{\!LQ})}  
\nonumber\\                                                          
&   & 
\phantom{= M_{G,{\bar{5}_M}}^2}
  + {\cal F}_{(\mbox{\boldmath{$_Y$}}^{{\rm I}\,\dagger}_{N},
               \widetilde{\mbox{\boldmath{$_L$}}},
               \widetilde{\mbox{\boldmath{$_{N^c}$}}},
               \mbox{\boldmath{$_{H_u}$}},
               \mbox{\boldmath{$_A$}}^{{\rm I}\,\dagger}_{N})},  
\nonumber\\
&\dot{\widetilde{\mbox{\boldmath{$m$}}}}_{D^c}^{2}           
& =  
 M_{G,{\bar{5}_M}}^2
 +2{\cal F}_{(\mbox{\boldmath{$_Y$}}^{5}_D,
              \widetilde{\mbox{\boldmath{$_{D^c}$}}},
              \widetilde{\mbox{\boldmath{$_Q$}}},
              \mbox{\boldmath{$_{H_d}$}},
              \mbox{\boldmath{$_A$}}^{5}_D)}  
 +2{\cal F}_{(\mbox{\boldmath{$_Y$}}^{5}_{\!DU},
              \widetilde{\mbox{\boldmath{$_{D^c}$}}},
              \widetilde{\mbox{\boldmath{$_{U^c}$}}}^{\ast} \!,
              \mbox{\boldmath{$_{H_D^C}$}},
              \mbox{\boldmath{$_A$}}^{5}_{\!DU})}  
\nonumber\\                                                          
&   & 
\phantom{= M_{G,{\bar{5}_M}}^2}
+\ {\cal F}_{(\mbox{\boldmath{$_Y$}}^{{\rm I}\,T}_{ND},
              \widetilde{\mbox{\boldmath{$_{D^c}$}}},
              \widetilde{\mbox{\boldmath{$_{N^c}$}}}^{\ast}\!,
              \mbox{\boldmath{$_{H_U^C}$}},
              \mbox{\boldmath{$_A$}}^{{\rm I} \,T}_{ND})}, 
\nonumber\\
&\dot{\widetilde{\mbox{\boldmath{$m$}}}}_{N^c}^2             
& =  
 \ 2{\cal F}_{(\mbox{\boldmath{$_Y$}}^{\rm I}_{N},
               \widetilde{\mbox{\boldmath{$_{N^c}$}}},
               \widetilde{\mbox{\boldmath{$_L$}}},
              {\mbox{\boldmath{$_{H_u}$}}},
               \mbox{\boldmath{$_A$}}^{\rm I}_{N})} 
 + 3{\cal F}_{(\mbox{\boldmath{$_Y$}}^{\rm I}_{ND},
               \widetilde{\mbox{\boldmath{$_{N^c}$}}},
               \widetilde{\mbox{\boldmath{$_{D^c}$}}}^{\ast} \!,
              {\mbox{\boldmath{$_{H_U^C}$}}},
               \mbox{\boldmath{$_A$}}^{\rm I}_{ND})} ,   
\label{eq:I-softTildemfromGUT-NRO}
\end{eqnarray}
 with $M_{G,{10_M}}^2$ and $ M_{G,{\bar{5}_M}}^2$ as in the case of
 vanishing NROs.
\item Soft Higgs masses: 
 Since no decomposition is needed for $5_H$, $\bar{5}_H$, and $24_H$,
 the RGEs for
 $\widetilde{\mbox{\boldmath{$m$}}}_{5_H}^2$,
 $\widetilde{\mbox{\boldmath{$m$}}}_{\bar{5}_H}^2$, and
 $\widetilde{\mbox{\boldmath{$m$}}}_{24_H}^2$ are as in
 Eq.~(\ref{eq:I-softHIGGSfromGUT}), with the already mentioned
 approximation taken for $Y^5$, $Y^{10}$, and $Y^{\rm I}_N$.  
\end{list}

\subsubsection{\underline{nrMSSU(5), 15H}}
\label{sec:II-GUTtoPL-NRO}  
%
The superpotential term $W_{\rm 15H}$ now has the form 
\begin{eqnarray}
 \mbox{\boldmath{$W$}}_{\rm 15H}    & = & 
  \ \frac{1}{\sqrt{2}} L \mbox{\boldmath{$Y$}}_N^{\rm II} T L 
  -\! D^c \mbox{\boldmath{$Y$}}_{DL}^{\rm II} L Q_{15}  
  +\!\frac{1}{\sqrt{2}}
    D^c \mbox{\boldmath{$Y$}}_{DD}^{\rm II} S D^c 
\nonumber\\                      &   & 
  +\frac{1}{\sqrt{2}} \lambda_{D\,} \bar{5}_H 15_H \bar{5}_H
  +\!\frac{1}{\sqrt{2}} \lambda_{U\,}  5_H \bar{15}_H 5_H 
  +\!\lambda_{15} 15_H 24_H \bar{15}_H 
\nonumber\\                      &   & 
\hspace*{6.6truecm}
  + M_{15} 15_H \bar{15}_H, 
\label{eq:II-SU515Hdec}
\end{eqnarray}
 where only the first term in $W_{\rm 15H}$ of
 Eq.~(\ref{eq:SU5sswSUPERP}) was decomposed, with $Y_N^{\rm II}$ split
 in the three couplings $\mbox{\boldmath{$Y$}}_N^{\rm II}$,
 $\mbox{\boldmath{$Y$}}_{DL}^{\rm II}$, and
 $\mbox{\boldmath{$Y$}}_{DD}^{\rm II}$.    
The RGEs for the parameters in the superpotential are as follows.
\begin{list}{$\bullet$}{}
\item Effective Yukawa couplings: 
\begin{eqnarray}
& \dot{\mbox{\boldmath{$Y$}}}^{\rm II}_N \       
&= 
   \mbox{\boldmath{$\gamma$}}_L^T   \,
      \mbox{\boldmath{$Y$}}^{\rm II}_N      + 
      \mbox{\boldmath{$Y$}}^{\rm II}_N
   \mbox{\boldmath{$\gamma$}}_L           +
   \mbox{\boldmath{$\gamma$}}_T
      \mbox{\boldmath{$Y$}}^{\rm II}_N, 
\nonumber\\
& \dot{\mbox{\boldmath{$Y$}}}^{\rm II}_{DL}       
&=
   \mbox{\boldmath{$\gamma$}}_{D^c}^T \,
      \mbox{\boldmath{$Y$}}^{\rm II}_{DL}    +
      \mbox{\boldmath{$Y$}}^{\rm II}_{DL}
   \mbox{\boldmath{$\gamma$}}_L           +    
   \mbox{\boldmath{$\gamma$}}_{Q_{15}}
      \mbox{\boldmath{$Y$}}^{\rm II}_{DL}, 
\nonumber\\
& \dot{\mbox{\boldmath{$Y$}}}^{\rm II}_{DD}       
&=
   \mbox{\boldmath{$\gamma$}}_{D^c}^T \,
      \mbox{\boldmath{$Y$}}^{\rm II}_{DD}    + 
      \mbox{\boldmath{$Y$}}^{\rm II}_{DD}
   \mbox{\boldmath{$\gamma$}}_{D^c}        + 
   \mbox{\boldmath{$\gamma$}}_S \,
      \mbox{\boldmath{$Y$}}^{\rm II}_{DD}.
\label{eq:II-YUKfromGUT-NRO}
\end{eqnarray}
The RGEs for  $\lambda_{15}$, $\lambda_D$ and $\lambda_U$ can be found
 in Eq.~(\ref{eq:II-YUKfromGUT}), the remaining ones in 
 Eqs.~(\ref{eq:I-YUK5and10fromGUT-NRO}). 
The anomalous dimensions
 $\mbox{\boldmath{$\gamma $}}_Q$,
 $\mbox{\boldmath{$\gamma $}}_{U^c}$, and
 $\mbox{\boldmath{$\gamma $}}_{E^c}$
 are as in Eq.~(\ref{eq:I-anomDIMsfromGUT-NRO}),
 $\mbox{\boldmath{$\gamma $}}_L$ and 
 $\mbox{\boldmath{$\gamma $}}_{D^c}$ are
\begin{eqnarray}
&\gamma_L  \   & =\! 
 -\!2 \left(\displaystyle{\frac{12}{5}} g_5^2 \right)
  \!{\mathop{\bf 1}}
+ \mbox{\boldmath{$Y$}}^{5\rm\,*}_E \mbox{\boldmath{$Y$}}^{5\,T}_E 
+\!3 \mbox{\boldmath{$Y$}}^{5\,*}_{\!LQ}
                                \mbox{\boldmath{$Y$}}^{5\,T}_{\!LQ} 
+\!3 \mbox{\boldmath{$Y$}}^{{\rm II}\,\dagger}_N
                                \mbox{\boldmath{$Y$}}_N^{\rm II} 
+\!3 \mbox{\boldmath{$Y$}}^{{\rm II}\,\dagger}_{DL}
                                \mbox{\boldmath{$Y$}}^{\rm II}_{DL}, 
\nonumber\\
&\gamma_{D^c}  & =\! 
-\!2 \left(\displaystyle{\frac{12}{5}} g_5^2 \right)
 \!{\mathop{\bf 1}}
+\!2 \mbox{\boldmath{$Y$}}^{5\rm\,*}_D
                           \mbox{\boldmath{$Y$}}^{5\,T}_D 
+\!2 \mbox{\boldmath{$Y$}}^{5\,*}_{\!DU}
                           \mbox{\boldmath{$Y$}}^{5\,T}_{\!DU} 
+\!4 \mbox{\boldmath{$Y$}}^{{\rm II}\,*}_{DD}
                           \mbox{\boldmath{$Y$}}^{{\rm II}\,T}_{DD} 
\nonumber\\         & & 
\hspace*{7.5truecm}
+\!2 \mbox{\boldmath{$Y$}}^{{\rm II}\,*}_{DL}
                          \mbox{\boldmath{$Y$}}^{{\rm II} \,T}_{DL} .
\qquad 
\label{eq:II-anomDIMfromGUT-NRO}
\end{eqnarray}
As mentioned above,
 $\mbox{\boldmath{$\gamma $}}_T$,
 $\mbox{\boldmath{$\gamma $}}_{Q_{15}}$, and
 $\mbox{\boldmath{$\gamma $}}_S$ are identified,
 $\mbox{\boldmath{$\gamma $}}_T =\mbox{\boldmath{$\gamma $}}_{Q_15}=
  \mbox{\boldmath{$\gamma $}}_S=
  \mbox{\boldmath{$\gamma $}}_{15_H}$, and the expression for
 $\mbox{\boldmath{$\gamma $}}_{15_H}$,
 $\mbox{\boldmath{$\gamma $}}_{\bar{15}_H}$,
 $\mbox{\boldmath{$\gamma $}}_{5_H}$,
 $\mbox{\boldmath{$\gamma $}}_{\bar{5}_H}$, and
 $\mbox{\boldmath{$\gamma $}}_{24_H}$ are as in 
 Eq.~(\ref{eq:II-anomDIMfromGUT}). 
\item Superpotential dimensionful parameters: 
{}For $\dot{M}_{15}$, see Eq.~(\ref{eq:II-hardMASSESfromGUT}); 
 for  $\dot{M}_5$ and $\dot{M}_{24}$,
 Eq.~(\ref{eq:I-hardMASSESfromGUT}).
\end{list}
The term $\widetilde{\mbox{\boldmath{$V$}}}_{\rm 15H}$ is
\begin{eqnarray}
 \widetilde{\mbox{\boldmath{$V$}}}_{\rm 15H} 
                                  & = & 
 \Bigl\{
  \ \frac{1}{\sqrt{2}} L \mbox{\boldmath{$A$}}_N^{\rm II} T L 
  - D^c \mbox{\boldmath{$A$}}_{DL}^{\rm II} L Q_{15}  
  +\frac{1}{\sqrt{2}}
    D^c \mbox{\boldmath{$A$}}_{DD}^{\rm II} S D^c 
\nonumber\\                       &   & 
  +\frac{1}{\sqrt{2}} A_{\lambda_{D\,}} \bar{5}_H 15_H \bar{5}_H
  +\frac{1}{\sqrt{2}} A_{\lambda_{U\,}}  5_H \bar{15}_H 5_H 
  + A_{\lambda_{15}} 15_H 24_H \bar{15}_H 
\nonumber\\                       &   & 
  + B_{15} 15_H \bar{15}_H
  +{\rm H.c.}
 \Bigr\}
\nonumber\\[1.001ex]              &   & 
+\widetilde{m}^2_{15H}        \,      15_H^\ast       15_H
+\widetilde{m}^2_{\bar{15}_H} \, \bar{15}_H^\ast \bar{15}_H. 
\quad \quad 
\label{eq:II-SOFTPOTSU515Hdec}
\end{eqnarray}   
 The RGEs for the parameters appearing in 
 $\widetilde{\mbox{\boldmath{$V$}}}^{{\rm MSSU(5)},{\rm 15H}}$ are as
 follows.
\begin{list}{$\bullet$}{}
\item Soft sfermion masses:
\begin{eqnarray}
&\dot{\widetilde{\mbox{\boldmath{$m$}}}}_{L}^{2}  
& = 
 M_{G,{\bar{5}_M}}^2
  + {\cal F}_{(\mbox{\boldmath{$_Y$}}^{5\,\ast}_E\!,
             \widetilde{\mbox{\boldmath{$_L$}}},
             \widetilde{\mbox{\boldmath{$_{E^c}$}}},
             \mbox{\boldmath{$_{H_d}$}},
             \mbox{\boldmath{$_A$}}^{5\rm,\ast}_E)}  
+3{\cal F}_{(\mbox{\boldmath{$_Y$}}^{5\,\ast}_{\!LQ},
             \widetilde{\mbox{\boldmath{$_L$}}},
             \widetilde{\mbox{\boldmath{$_Q$}}}^{\ast}\!,
             \mbox{\boldmath{$_{H_D^C}$}},
             \mbox{\boldmath{$_A$}}^{5\,\ast}_{\!LQ})}  
\nonumber\\                                               
&   &
\phantom{= M_{G,{\bar{5}_M}}^2}
+3{\cal F}_{(\mbox{\boldmath{$_Y$}}^{{\rm II}\,\dagger}_N\!,
             \widetilde{\mbox{\boldmath{$_L$}}},
             \widetilde{\mbox{\boldmath{$_L$}}}^{\ast}\!,
             \mbox{\boldmath{$_T$}},
             \mbox{\boldmath{$_A$}}^{{\rm II}\,\dagger}_N)}  
+3{\cal F}_{(\mbox{\boldmath{$_Y$}}^{{\rm II}\,\dagger}_{DL},
             \widetilde{\mbox{\boldmath{$_L$}}},
             \widetilde{\mbox{\boldmath{$_{D^c}$}}},
             \mbox{\boldmath{$_{Q_{15}}$}},
             \mbox{\boldmath{$_A$}}^{{\rm II}\,\dagger}_{DL})} , 
\nonumber\\
&\dot{\widetilde{\mbox{\boldmath{$m$}}}}_{D^c}^{2}  
& = 
 M_{G,{\bar{5}_M}}^2
+2{\cal F}_{(\mbox{\boldmath{$_Y$}}^{5}_D,
             \widetilde{\mbox{\boldmath{$_{D^c}$}}},
             \widetilde{\mbox{\boldmath{$_Q$}}},
             \mbox{\boldmath{$_{H_d}$}},
             \mbox{\boldmath{$_A$}}^{5}_D)}  
+2{\cal F}_{(\mbox{\boldmath{$_Y$}}^{5}_{\!DU},
             \widetilde{\mbox{\boldmath{$_{D^c}$}}},
             \widetilde{\mbox{\boldmath{$_{U^c}$}}}^{\ast}\!,
             \mbox{\boldmath{$_{H_D^C}$}},
             \mbox{\boldmath{$_A$}}^{5}_{\!DU})}  
\nonumber\\                                               
&   &
\phantom{= M_{G,{\bar{5}_M}}^2}
+4{\cal F}_{(\mbox{\boldmath{$_Y$}}^{\rm II}_{DD},
             \widetilde{\mbox{\boldmath{$_{D^c}$}}},
             \widetilde{\mbox{\boldmath{$_{D^c}$}}}^{\ast}\!,
             \mbox{\boldmath{$_S$}},
             \mbox{\boldmath{$_A$}}^{\rm II}_{DD})}  
+2{\cal F}_{(\mbox{\boldmath{$_Y$}}^{\rm II}_{DL},
             \widetilde{\mbox{\boldmath{$_{D^c}$}}},
             \widetilde{\mbox{\boldmath{$_L$}}},
             \mbox{\boldmath{$_{Q_{15}}$}},
             \mbox{\boldmath{$_A$}}^{\rm II}_{DL})} ,
\nonumber \\
&& 
\label{eq:II-softM5fromGUT-NRO}
\end{eqnarray}
 with $ M_{G,{\bar{5}_M}}^2$ as in the case of vanishing NROs.
The RGEs for
 $\widetilde{\mbox{\boldmath{$m$}}}_{Q}^2$,
 $\widetilde{\mbox{\boldmath{$m$}}}_{U^c}^2$, and
 $\widetilde{\mbox{\boldmath{$m$}}}_{E^c}^2$ are as in
 Eq.~(\ref{eq:I-softTildemfromGUT-NRO}).
\item Soft Higgs masses: 
See RGEs in \S\ref{sec:II-GUTtoPL}, with 
 $\mbox{\boldmath{$Y$}}^{5}_D$,
 $\mbox{\boldmath{$Y$}}^{10}_U$ and 
 $\mbox{\boldmath{$Y$}}_{N}^{\rm II}$ used for
 $Y^5$, $Y^{10}$ and $Y_{N}^{\rm II}$, respectively.   
\end{list}

\subsubsection{\underline{nrMSSU(5), 24M}}
\label{sec:III-GUTtoPL-NRO}  
%
The missing ingredient needed to specify the superpotential of
 this class of models is
\begin{eqnarray}
 \mbox{\boldmath{$W$}}_{\rm 24M}              & = & 
  H_U^C \Bigl(
   \sqrt{2} G_M  \mbox{\boldmath{$Y$}}_{GD}^{\rm III}
  +\sqrt{\frac{6}{5}}\frac{1}{3}B_M \mbox{\boldmath{$Y$}}_{BD}^{\rm III}
        \!\Bigr) D^c 
 + H_U^C     X_M   \mbox{\boldmath{$Y$}}_{XL}^{\rm III} L
\nonumber\\                  &   & 
 + H_u  \Bigl(
  \sqrt{2} W_M \mbox{\boldmath{$Y$}}_N^{\rm III} 
 -\sqrt{\frac{6}{5}}\frac{1}{2}B_M \mbox{\boldmath{$Y$}}_{BL}^{\rm III} 
        \!\Bigr) L   
 + H_u   \bar{X}_M \mbox{\boldmath{$Y$}}_{\bar{X}D}^{\rm III} D^c 
\nonumber\\                  &   & 
    + \frac{1}{2} 24_M M_{24_M} 24_M
    + \frac{1}{2} \sum_{x=S,A} 
       \left(24_M Y_{24_M}^x 24_M \right)_{x} 24_H,  
\label{eq:III-SU524Mdec}
\end{eqnarray}
 where only the first term in $W_{\rm 24M}$ of
 Eq.~(\ref{eq:SU5sswSUPERP}) was decomposed, with $Y^{\rm III}_N$ split
 in the six couplings
 $\mbox{\boldmath{$Y$}}^{\rm III}_N$,
 $\mbox{\boldmath{$Y$}}^{\rm III}_{BL}$,
 $\mbox{\boldmath{$Y$}}^{\rm III}_{\bar{X}D}$, and 
 $\mbox{\boldmath{$Y$}}^{\rm III}_{GD}$,
 $\mbox{\boldmath{$Y$}}^{\rm III}_{BD}$, 
 $\mbox{\boldmath{$Y$}}^{\rm III}_{XL}$.     
The index $x$, as usual, labels the symmetric and antisymmetric
 products of the two $24_M$. 
\begin{list}{$\bullet$}{}
\item Effective Yukawa couplings: 
Also in this case, the RGEs in Eq.~(\ref{eq:I-YUK5and10fromGUT-NRO}) 
 hold. 
The RGEs for $\lambda_5$, $\lambda_{24}$ are in
 Eq.~(\ref{eq:I-YUKfromGUT}); those for the couplings $Y^x_{24_M}$ in 
 Eq.~(\ref{eq:III-YUKfromGUT}).
We have, in addition,
\begin{eqnarray}
&\dot{\mbox{\boldmath{$Y$}}}^{\rm III}_N          
&=
  \mbox{\boldmath{$\gamma$}}_{W_M}^T \,
     \mbox{\boldmath{$Y$}}^{\rm III}_N  
 +   \mbox{\boldmath{$Y$}}^{\rm III}_N
  \mbox{\boldmath{$\gamma$}}_L 
 +\mbox{\boldmath{$\gamma$}}_{H_u} \,
     \mbox{\boldmath{$Y$}}^{\rm III}_N ,
\nonumber\\
&\dot{\mbox{\boldmath{$Y$}}}^{\rm III}_{BL}          
&=
  \mbox{\boldmath{$\gamma$}}_{B_M}^T \,
     \mbox{\boldmath{$Y$}}^{\rm III}_{BL}  
 +   \mbox{\boldmath{$Y$}}^{\rm III}_{BL}
  \mbox{\boldmath{$\gamma$}}_L 
 +\mbox{\boldmath{$\gamma$}}_{H_u} \,
     \mbox{\boldmath{$Y$}}^{\rm III}_{BL}, 
\nonumber\\
&\dot{\mbox{\boldmath{$Y$}}}^{\rm III}_{BD}      
&=
  \mbox{\boldmath{$\gamma$}}_{B_M}^T \,
     \mbox{\boldmath{$Y$}}^{\rm III}_{BD}  
 +   \mbox{\boldmath{$Y$}}^{\rm III}_{BD}
  \mbox{\boldmath{$\gamma$}}_{D^c} 
 +\mbox{\boldmath{$\gamma$}}_{H_U^C} \,
     \mbox{\boldmath{$Y$}}^{\rm III}_{BD}, 
\nonumber\\
&\dot{\mbox{\boldmath{$Y$}}}^{\rm III}_{GD}      
&=
  \mbox{\boldmath{$\gamma$}}_{G_M}^T \,
     \mbox{\boldmath{$Y$}}^{\rm III}_{GD}  
 +   \mbox{\boldmath{$Y$}}^{\rm III}_{GD}
  \mbox{\boldmath{$\gamma$}}_{D^c} 
 +\mbox{\boldmath{$\gamma$}}_{H_U^C} \,
     \mbox{\boldmath{$Y$}}^{\rm III}_{GD} ,
\nonumber\\
&\dot{\mbox{\boldmath{$Y$}}}^{\rm III}_{\bar{X}D}    
&=
  \mbox{\boldmath{$\gamma$}}_{\bar X_M}^T\,
     \mbox{\boldmath{$Y$}}^{\rm III}_{\bar{X}D}
 +   \mbox{\boldmath{$Y$}}^{\rm III}_{\bar{X}D}
  \mbox{\boldmath{$\gamma$}}_{D^c} 
 +\mbox{\boldmath{$\gamma$}}_{H_u} \,
     \mbox{\boldmath{$Y$}}^{\rm III}_{\bar{X}D} ,
\nonumber\\
&\dot{\mbox{\boldmath{$Y$}}}^{\rm III}_{XL}      
&=
  \mbox{\boldmath{$\gamma$}}_{X_M}^T \,
     \mbox{\boldmath{$Y$}}^{\rm III}_{XL}  
 +   \mbox{\boldmath{$Y$}}^{\rm III}_{XL}
  \mbox{\boldmath{$\gamma$}}_L 
 +\mbox{\boldmath{$\gamma$}}_{H_U^C} \,
     \mbox{\boldmath{$Y$}}^{\rm III}_{XL}. 
\label{eq:III-YUKfromGUT-NRO}
\end{eqnarray}
The anomalous dimensions
 $\mbox{\boldmath{$\gamma $}}_Q$,
 $\mbox{\boldmath{$\gamma $}}_{U^c}$ and
 $\mbox{\boldmath{$\gamma $}}_{E^c}$ are as in
 Eq.~(\ref{eq:I-anomDIMsfromGUT-NRO}); 
 $\mbox{\boldmath{$\gamma $}}_L$ and
 $\mbox{\boldmath{$\gamma $}}_{D^c}$ are
\begin{eqnarray}
 \mbox{\boldmath{$\gamma$}}_L               \       & = & 
 -2\!\left(\!\displaystyle{\frac{12}{5}} g_5^2 \!\right)
   \!{\mathop{\bf 1}} 
 +   \mbox{\boldmath{$Y$}}^{5\rm\,\ast}_E
     \mbox{\boldmath{$Y$}}^{5\,T}_E 
 +\!3\mbox{\boldmath{$Y$}}^{5\,\ast}_{\!LQ}
     \mbox{\boldmath{$Y$}}^{5\,T}_{\!LQ} 
 +\!\displaystyle{\frac32}
     \mbox{\boldmath{$Y$}}^{{\rm III}\,\dagger}_N
     \mbox{\boldmath{$Y$}}^{\rm III}_N 
\nonumber\\                                        &   & 
 \hspace*{4truecm}
 +\!\displaystyle{\frac3{10}}
     \mbox{\boldmath{$Y$}}^{{\rm III}\,\dagger}_{BL} 
     \mbox{\boldmath{$Y$}}^{\rm III}_{BL} 
+ 3 \mbox{\boldmath{$Y$}}^{{\rm III}\,\dagger}_{XL}
    \mbox{\boldmath{$Y$}}^{\rm III}_{XL}, 
\nonumber\\
 \mbox{\boldmath{$\gamma$}}_{D^c}                     & = &  
 -2\!\left(\!\displaystyle{\frac{12}{5}} g_5^2 \!\right)
   \!{\mathop{\bf 1}} 
 +\!2\mbox{\boldmath{$Y$}}^{5\,\ast}_D
     \mbox{\boldmath{$Y$}}^{5\,T}_D 
 +\!2\mbox{\boldmath{$Y$}}^{5\,\ast}_{\!DU}
     \mbox{\boldmath{$Y$}}^{5\,T}_{\!DU} 
 +\!\displaystyle{\frac83}
     \mbox{\boldmath{$Y$}}^{{\rm III}\,\dagger}_{GD}
     \mbox{\boldmath{$Y$}}^{\rm III}_{GD} 
\nonumber\\                                &   &
 \hspace*{4truecm}
 +\!\displaystyle{\frac2{15}}
     \mbox{\boldmath{$Y$}}^{{\rm III}\,\dagger}_{BD}
     \mbox{\boldmath{$Y$}}^{\rm III}_{BD} 
+ 2 \mbox{\boldmath{$Y$}}^{{\rm III}\,\dagger}_{\bar{X}D}
    \mbox{\boldmath{$Y$}}^{\rm III}_{\bar{X}D}     ;
\label{eq:III-anomDIMfromGUT-NRO}
\end{eqnarray}
 $\mbox{\boldmath{$\gamma $}}_{H_u}$ ,
 $\mbox{\boldmath{$\gamma $}}_{H_U^C}$ ,
 $\mbox{\boldmath{$\gamma $}}_{H_d}$ , and 
 $\mbox{\boldmath{$\gamma $}}_{H_D^C}$ are approximated as mentioned 
 earlier.
The anomalous dimensions for the fields that are components of $24_M$
 can be written in the ``hybrid'' form: 
\begin{eqnarray}
 \mbox{\boldmath{$\gamma$}}_{G_M}                    & = & 
 -2 \left(5 g_5^2 \right) \!{\mathop{\bf 1}}
    +\frac{21}{10} Y_{24_M}^{S\,\dagger} Y_{24_M}^S
    +\frac{5}{2}  Y_{24_M}^{A\,\dagger} Y_{24_M}^A 
   + \mbox{\boldmath{$Y$}}^{{\rm III}\,\ast}_{GD}
     \mbox{\boldmath{$Y$}}^{{\rm III}\,T}_{GD} ,
\nonumber\\ 
 \mbox{\boldmath{$\gamma$}}_{W_M}                    & = & 
 -2 \left(5 g_5^2 \right) \!{\mathop{\bf 1}}
    +\frac{21}{10} Y_{24_M}^{S\,\dagger} Y_{24_M}^S
    +\frac{5}{2}   Y_{24_M}^{A\,\dagger} Y_{24_M}^A 
   + \mbox{\boldmath{$Y$}}^{{\rm III}\,\ast}_N
     \mbox{\boldmath{$Y$}}^{{\rm III}\,T}_N ,
\nonumber\\ 
 \mbox{\boldmath{$\gamma$}}_{B_M}                    & = & 
 -2 \left(5 g_5^2 \right) \!{\mathop{\bf 1}}
    +\frac{21}{10} Y_{24_M}^{S\,\dagger} Y_{24_M}^S
    +\frac{5}{2}  Y_{24_M}^{A\,\dagger} Y_{24_M}^A 
    +\frac35  \mbox{\boldmath{$Y$}}^{{\rm III}\,\ast}_{BL}
              \mbox{\boldmath{$Y$}}^{{\rm III}\,T}_{BL}
\nonumber\\                                        & & 
\hspace*{7.2truecm}
   + \frac25  \mbox{\boldmath{$Y$}}^{{\rm III}\,\ast}_{BD}
              \mbox{\boldmath{$Y$}}^{{\rm III}\,T}_{BD} ,
\nonumber\\ 
 \mbox{\boldmath{$\gamma$}}_{X_M}                    & = & 
 -2 \left(5 g_5^2 \right) \!{\mathop{\bf 1}}
    +\frac{21}{10} Y_{24_M}^{S\,\dagger} Y_{24_M}^S
    +\frac{5}{2}  Y_{24_M}^{A\,\dagger} Y_{24_M}^A 
   + \mbox{\boldmath{$Y$}}^{{\rm III}\,\ast}_{XL}
     \mbox{\boldmath{$Y$}}^{{\rm III}\,T}_{XL} ,
\nonumber\\ 
 \mbox{\boldmath{$\gamma$}}_{{\bar X}_M}             & = &
 -2 \left(5 g_5^2 \right) \!{\mathop{\bf 1}}
    +\frac{21}{10} Y_{24_M}^{S\,\dagger} Y_{24_M}^S
    +\frac{5}{2}   Y_{24_M}^{A\,\dagger} Y_{24_M}^A 
   + \mbox{\boldmath{$Y$}}^{{\rm III}\,\ast}_{\bar{X}D}
     \mbox{\boldmath{$Y$}}^{{\rm III}\,T}_{\bar{X}D} .
\label{eq:III-HYBRIDanomDIMfromGUT-NRO}
\end{eqnarray}
\item Superpotential dimensionful parameters: 
See Eq.~(\ref{eq:III-hardMASSESfromGUT}) for 
 $\dot{M}_{24_M}$,  Eq.~(\ref{eq:I-hardMASSESfromGUT}) for 
 $\dot{M}_5$ and $\dot{M}_{24}$.
\end{list}
The term $\widetilde{\mbox{\boldmath{$V$}}}_{\rm 24M}$ is
\begin{eqnarray}
 \widetilde{\mbox{\boldmath{$V$}}}_{\rm 24_M}  &\! = \!& 
 \left\{
  H_U^C \Bigl(
   \sqrt{2} G_M  \mbox{\boldmath{$A$}}_{GD}^{\rm III}
  +\sqrt{\frac{6}{5}}\frac{1}{3}B_M \mbox{\boldmath{$A$}}_{BD}^{\rm III}
        \!\Bigr) D^c 
 + H_U^C     X_M   \mbox{\boldmath{$A$}}_{XL}^{\rm III} L
 \right.
\nonumber\\                      &   & 
 \left.
 + H_u  \Bigl(
  \sqrt{2} W_M \mbox{\boldmath{$A$}}_N^{\rm III} 
 -\sqrt{\frac{6}{5}}\frac{1}{2}B_M \mbox{\boldmath{$A$}}_{BL}^{\rm III} 
        \!\Bigr) L   
 + H_u   \bar{X}_M \mbox{\boldmath{$A$}}_{\bar{X}D}^{\rm III} D^c 
 \right.
\nonumber\\                      &   & 
 \left.
    + \frac{1}{2} 24_M B_{24_M} 24_M
     + \frac{1}{2} \sum_{x=S,A} 
       \left(24_M A_{24_M}^x 24_M \right)_{x} 24_H
 +{\rm H.c.}
\right\}
\nonumber\\                  &   & 
+\widetilde{W}_M\, \widetilde{\mbox{\boldmath{$m$}}}^2_{W_M}\, 
 \widetilde{W}_M^{\ast}
+\widetilde{B}_M\, \widetilde{\mbox{\boldmath{$m$}}}^2_{B_M}\, 
 \widetilde{B}_M^{\ast}
\nonumber\\                  &   & 
+\widetilde{G}_M\, \widetilde{\mbox{\boldmath{$m$}}}^2_{G_M}\, 
 \widetilde{G}_M^{\ast}
+\widetilde{X}_M\, \widetilde{\mbox{\boldmath{$m$}}}^2_{X_M}\, 
 \widetilde{X}_M^{\ast}
+\widetilde{\bar{X}}_M \, 
 \widetilde{\mbox{\boldmath{$m$}}}^2_{\bar{X}_M} \,
 \widetilde{\bar{X}}_M^{\ast}.
\label{eq:II-SOFTPOTSU524Mdec}
\end{eqnarray}   
 The RGEs for the parameters appearing in 
 $\widetilde{\mbox{\boldmath{$V$}}}^{{\rm MSSU(5)},{\rm 24M}}$ are as{{}}
 follows.
\begin{list}{$\bullet$}{}
\item Soft sfermion masses:
\begin{eqnarray}
& \dot{\widetilde{\mbox{\boldmath{$m$}}}}_{L}^{2}      
& = 
 M_{G,{\bar{5}_M}}^2                                  
+\  {\cal F}_{(\mbox{\boldmath{$_Y$}}^{5\,\ast}_E\!,
               \widetilde{\mbox{\boldmath{$_L$}}},
               \widetilde{\mbox{\boldmath{$_{E^c}$}}},
               \mbox{\boldmath{$_{H_d}$}},
               \mbox{\boldmath{$_A$}}^{5\,\ast}_E)}  
+3\,{\cal F}_{(\mbox{\boldmath{$_Y$}}^{5\rm\,\ast}_{\!LQ},
               \widetilde{\mbox{\boldmath{$_L$}}},
               \widetilde{\mbox{\boldmath{$_Q$}}}^{\ast}\!,
               \mbox{\boldmath{$_{H_D^C}$}},
               \mbox{\boldmath{$_A$}}^{5\,\ast}_{\!LQ})}  
\phantom{\displaystyle{\frac83}} 
\nonumber\\                                                    
&  & \
  +\displaystyle{\frac32}
    {\cal F}_{(\mbox{\boldmath{$_Y$}}^{{\rm III}\,\dagger}_N\!\!,
               \widetilde{\mbox{\boldmath{$_L$}}},
               \widetilde{\mbox{\boldmath{$_{W_M}$}}},
               \mbox{\boldmath{$_{H_u}$}},
               \mbox{\boldmath{$_A$}}^{{\rm III}\,\dagger}_N)}  
  +\displaystyle{\frac3{10}}
    {\cal F}_{(\mbox{\boldmath{$_Y$}}^{{\rm III}\,\dagger}_{BL}\!,
               \widetilde{\mbox{\boldmath{$_L$}}},
               \widetilde{\mbox{\boldmath{$_{B_M}$}}},
               \mbox{\boldmath{$_{H_u}$}},
               \mbox{\boldmath{$_A$}}^{{\rm III}\,\dagger}_{BL})}  
\nonumber\\                                                    
&  & \
+3\,{\cal F}_{(\mbox{\boldmath{$_Y$}}^{{\rm III}\,\dagger}_{XL}\!,
               \widetilde{\mbox{\boldmath{$_L$}}},
               \widetilde{\mbox{\boldmath{$_{X_M}$}}},
               \mbox{\boldmath{$_{H_U^C}$}},
               \mbox{\boldmath{$_A$}}^{{\rm III}\,\dagger}_{XL})} , 
\phantom{\displaystyle{\frac83}} 
\nonumber\\
& \dot{\widetilde{\mbox{\boldmath{$m$}}}}_{D}^{2}      
& = 
 M_{G,{\bar{5}_M}}^2                                  
+2\,{\cal F}_{(\mbox{\boldmath{$_Y$}}^{5}_D,
               \widetilde{\mbox{\boldmath{$_{D^c}$}}},
               \widetilde{\mbox{\boldmath{$_Q$}}},
               \mbox{\boldmath{$_{H_d}$}},
               \mbox{\boldmath{$_A$}}^{5}_D)}  
+2\,{\cal F}_{(\mbox{\boldmath{$_Y$}}^{5}_{\!DU},
               \widetilde{\mbox{\boldmath{$_{D^c}$}}},
               \widetilde{\mbox{\boldmath{$_{U^c}$}}}^{\ast}\!,
               \mbox{\boldmath{$_{H_D^C}$}},
               \mbox{\boldmath{$_A$}}^{5}_{\!DU})}  
\nonumber\\                                                    
&  & \
 +\displaystyle{\frac{8}{3}}
    {\cal F}_{(\mbox{\boldmath{$_Y$}}^{{\rm III}\,T}_{GD}\!,
               \widetilde{\mbox{\boldmath{$_{D^c}$}}},
               \widetilde{\mbox{\boldmath{$_{G_M}$}}}^{\ast}\!,
               \mbox{\boldmath{$_{H_U^C}$}},
               \mbox{\boldmath{$_A$}}^{{\rm III}\,T}_{GD})}  
 +\displaystyle{\frac{2}{15}}
    {\cal F}_{(\mbox{\boldmath{$_Y$}}^{{\rm III}\,T}_{BD}\!,
               \widetilde{\mbox{\boldmath{$_{D^c}$}}},
               \widetilde{\mbox{\boldmath{$_{B_M}$}}}^{\ast}\!,
               \mbox{\boldmath{$_{H_U^C}$}},
               \mbox{\boldmath{$_A$}}^{{\rm III}\,T}_{BD})}  
\nonumber\\                                                    
&  & \
+2\,{\cal F}_{(\mbox{\boldmath{$_Y$}}^{{\rm III}\,T}_{\bar{X}D}\!,
               \widetilde{\mbox{\boldmath{$_{D^c}$}}},
               \widetilde{\mbox{\boldmath{$_{\bar{X}_M}$}}}^{\ast}\!,
               \mbox{\boldmath{$_{H_u}$}},
               \mbox{\boldmath{$_A$}}^{{\rm III}\,T}_{\bar{X}D})} .
\phantom{\displaystyle{\frac83}} 
\label{eq:III-softM5DfromGUT-NRO}
\end{eqnarray}
The RGEs for
 $\widetilde{\mbox{\boldmath{$m$}}}_{Q}^2$,
 $\widetilde{\mbox{\boldmath{$m$}}}_{U^c}^2$, and
 $\widetilde{\mbox{\boldmath{$m$}}}_{E^c}^2$ are as in
 Eq.~(\ref{eq:I-softTildemfromGUT-NRO}).
As for the scalar components of the $24_M$, their soft-mass RGEs 
 can be given in the ``hybrid'' form:
\begin{eqnarray}
& \dot{\widetilde{\mbox{\boldmath{$m$}}}}_{G_M}^{2}      & = 
 ({\widetilde m}^{\rm com}_{24_M})^2  
  +\ {\cal F}_{(\mbox{\boldmath{$_Y$}}^{\rm III}_{GD},
                \widetilde{\mbox{\boldmath{$_{G_M}$}}},
                \widetilde{\mbox{\boldmath{$_{D^c}$}}}^{\ast}\!,
                \mbox{\boldmath{$_{H_U^C}$}},
                \mbox{\boldmath{$_A$}}^{\rm III}_{GD})} ,
\nonumber\\
& \dot{\widetilde{\mbox{\boldmath{$m$}}}}_{W_M}^{2}      & = 
 ({\widetilde m}^{\rm com}_{24_M})^2  
  +\ {\cal F}_{(\mbox{\boldmath{$_Y$}}^{\rm III}_N\!,
                \widetilde{\mbox{\boldmath{$_{W_M}$}}},
                \widetilde{\mbox{\boldmath{$_L$}}},
                \mbox{\boldmath{$_{H_u}$}},
                \mbox{\boldmath{$_A$}}^{\rm III}_N)} ,
\nonumber\\
& \dot{\widetilde{\mbox{\boldmath{$m$}}}}_{B_M}^{2}      & = 
 ({\widetilde m}^{\rm com}_{24_M})^2  
  +\displaystyle{\frac{3}{5}}
     {\cal F}_{(\mbox{\boldmath{$_Y$}}^{\rm III}_{BL}\!,
                \widetilde{\mbox{\boldmath{$_{B_M}$}}},
                \widetilde{\mbox{\boldmath{$_L$}}},
                \mbox{\boldmath{$_{H_u}$}},
                \mbox{\boldmath{$_A$}}^{\rm III}_{BL})} 
 +\displaystyle{\frac{2}{5}}
     {\cal F}_{(\mbox{\boldmath{$_Y$}}^{\rm III}_{BD},
                \widetilde{\mbox{\boldmath{$_{B_M}$}}},
                \widetilde{\mbox{\boldmath{$_{D^c}$}}}^{\ast}\!,
                \mbox{\boldmath{$_{H_U^C}$}},
                \mbox{\boldmath{$_A$}}^{\rm III}_{BD})} ,
\nonumber\\
& \dot{\widetilde{\mbox{\boldmath{$m$}}}}_{X_M}^{2}      & = 
 ({\widetilde m}^{\rm com}_{24_M})^2  
  +\ {\cal F}_{(\mbox{\boldmath{$_Y$}}^{\rm III}_{XL},
                \widetilde{\mbox{\boldmath{$_{X_M}$}}},
                \widetilde{\mbox{\boldmath{$_L$}}},
                \mbox{\boldmath{$_{H_U^C}$}},
                \mbox{\boldmath{$_A$}}^{\rm III}_{XL})} ,
\nonumber\\
& \dot{\widetilde{\mbox{\boldmath{$m$}}}}_{\bar X_M}^{2} & = 
 ({\widetilde m}^{\rm com}_{24_M})^2  
  +\ {\cal F}_{(\mbox{\boldmath{$_Y$}}^{\rm III}_{\bar{X}D},
                \widetilde{\mbox{\boldmath{$_{\bar{X}_M}$}}},
                \widetilde{\mbox{\boldmath{$_{D^c}$}}}^{\ast}\!,
                \mbox{\boldmath{$_{H_u}$}},
                \mbox{\boldmath{$_A$}}^{\rm III}_{\bar{X}D})} ,
\label{eq:III-softM24fromGUT-NRO}
\end{eqnarray}
 where the common term $({\widetilde m}^{\rm com}_{24_M})^2$  
 is given by 
\begin{eqnarray}
 ({\widetilde m}^{\rm com}_{24_M})^2   & = & 
 M_{G,{24_M}}^2                          +
 \displaystyle{\frac{21}{10}}
    {\cal F}_{(Y_{24_M}^S,
               \widetilde{24}_M^2,
               \widetilde{24}_M^{2\,\ast},
               \widetilde{24}_H^2,
               A_{24_M}^S)}            
\nonumber \\                      & & 
\hspace*{2.7truecm}
+ \displaystyle{\frac{5}{2}}\,
    {\cal F}_{(Y_{24_M}^A,
               \widetilde{24}_M^2,
               \widetilde{24}_M^{2\,\ast},
               \widetilde{24}_H^2,
               A_{24_M}^A)} .
\label{eq:shortM}
\end{eqnarray}
The quantities $M_{G,{\bar{5}_M}}^2$ and $M_{G,{24_M}}^2$ 
 are as in the case with vanishing NROs. 
\item Soft Higgs masses: 
The RGEs are as those given in Appendix~\ref{sec:III-GUTtoPL},
 with $Y^5$, $Y^{10}$ and $Y^{\rm III}_N$ replaced by 
 $\mbox{\boldmath{$Y$}}^{5}_D$,
 $\mbox{\boldmath{$Y$}}^{10}_U$ and
 $\mbox{\boldmath{$Y$}}^{\rm III}_N$, respectively.   
\end{list}


\begin{thebibliography}{99}
  
\bibitem{FLVreview}
M.~Artuso et al., Eur.\ Phys.\ J.\ C {\bf 57} (2008), 309.

\bibitem{sFVsmall}
J.~R.~Ellis and D.~V.~Nanopoulos, Phys.\ Lett.\ B {\bf 110} (1982), 44.\\
R.~Barbieri and R.~Gatto, Phys.\ Lett.\ B {\bf 110} (1982), 211.

\bibitem{DNW}
J.~F.~Donoghue, H.~P.~Nilles and D.~Wyler, Phys.\ Lett.\ B {\bf 128} (1983), 55.

\bibitem{BKS}
A.~Bouquet, J.~Kaplan and C.~A.~Savoy, Phys.\ Lett.\ B {\bf 148} (1984), 69.
 
\bibitem{DGH}
M.~Dugan, B.~Grinstein and L.~J.~Hall, Nucl.\ Phys.\ B {\bf 255} (1985), 413.

\bibitem{BBMR}
S.~Bertolini, F.~Borzumati, A.~Masiero and G.~Ridolfi,
Nucl.\ Phys.\ B {\bf 353} (1991), 591.
 
\bibitem{MFV}
L.~J.~Hall and L.~Randall, Phys.\ Rev.\ Lett.\ {\bf 65} (1990), 2939. \\
G.~D'Ambrosio, G.~F.~Giudice, G.~Isidori and A.~Strumia,
Nucl.\ Phys.\ B {\bf 645} (2002), 155.

\bibitem{FB}
N.~Oshimo, Nucl.\ Phys.\ B {\bf 404} (1993), 20. \\
F.~Borzumati, Z.\ Phys.\ C {\bf 63} (1994), 291.\\
F.~Borzumati, M.~Olechowski and S.~Pokorski, Phys.\ Lett.\ B {\bf 349} (1995), 311.\\
R.~Rattazzi and U.~Sarid, Nucl.\ Phys.\ B {\bf 501} (1997), 297.\\
F.~Borzumati, hep-ph/9702307.

\bibitem{SEESAW1}
P.~Minkowski, Phys.\ Lett.\ B {\bf 67} (1977), 421.\\
T.~Yanagida,  in {\it Proceedings of the Workshop on the Unified Theory and 
Baryon Number in the Universe}, KEK report 79-18, ed. O.~Sawada and A.~Sugamoto
(KEK, 1979), p.~95.\\ 
M.~Gell-Mann, P.~Ramond, R.~Slansky, 
in {\it Supergravity}, ed. P.~van Nieuwenhuizen and D.~Z.~Freedman 
(North Holland, Amsterdam, 1979), p.~315.

\bibitem{SEESAW2}
W.~Konetschny and W.~Kummer, Phys.\ Lett.\ B {\bf 70} (1977), 433.\\
R.~Barbieri, D.~V.~Nanopoulos, G.~Morchio and F.~Strocchi,
Phys.\ Lett.\ B {\bf 90} (1980), 91.\\
R.~N.~Mohapatra and G.~Senjanovic, Phys.\ Rev.\ Lett.\ {\bf 44} (1980), 912.\\
M.~Magg and C.~Wetterich, Phys.\ Lett.\ B {\bf 94} (1980), 61.\\
J.~Schechter and J.~W.~F.~Valle, Phys.\ Rev.\ D {\bf 22} (1980), 2227.\\
T.~P.~Cheng and L.~F.~Li, Phys.\ Rev.\ D {\bf 22} (1980), 2860.\\
G.~Lazarides, Q.~Shafi and C.~Wetterich, Nucl.\ Phys.\ B {\bf 181} (1981), 287.\\
R.~N.~Mohapatra and G.~Senjanovic, Phys.\ Rev.\ D {\bf 23} (1981), 165.

\bibitem{SEESAW3}
R.~Foot, H.~Lew, X.~G.~He and G.~C.~Joshi, Z.\ Phys.\ C {\bf 44} (1989), 441.\\
E.~Ma, Phys.\ Rev.\ Lett.\ {\bf 81} (1998), 1171.

\bibitem{STRUMIAVISSANI}
For a thorough review of the three types of seesaw mechanism, see \\
A.~Strumia and F.~Vissani, hep-ph/0606054.

\bibitem{NuY->LFV}
F.~Borzumati and A.~Masiero, Phys.\ Rev.\ Lett.\ {\bf 57} (1986), 961.

\bibitem{YuMEDIAT}
M.~Dine, Y.~Nir and Y.~Shirman, Phys.\ Rev.\ D {\bf 55} (1997), 1501.\\
G.~R.~Dvali and M.~A.~Shifman, Phys.\ Lett.\ B {\bf 399} (1997), 60.

\bibitem{BFS}
R.~Barbieri, S.~Ferrara and C.~A.~Savoy, Phys.\ Lett.\ B {\bf 119} (1982), 343.

\bibitem{HLW}
L.~J.~Hall, J.~D.~Lykken and S.~Weinberg, Phys.\ Rev.\ D {\bf 27} (1983), 2359.

\bibitem{BGOO1}
S.~Baek, T.~Goto, Y.~Okada and K.~I.~Okumura,
Phys.\ Rev.\ D {\bf 63} (2001), 051701.

\bibitem{NuY->QFV}
T.~Moroi, J.~High~Energy~Phys. {\bf 03} (2000), 019;
Phys.\ Lett.\ B {\bf 493} (2000), 366.

\bibitem{BGOO}
S.~Baek, T.~Goto, Y.~Okada and K.~I.~Okumura, Phys.\ Rev.\ D {\bf 64} (2001), 095001.

\bibitem{ETbefMEAS}
N.~Akama, Y.~Kiyo, S.~Komine and T.~Moroi, Phys.\ Rev.\ D {\bf 64} (2001), 095012.\\
R.~Harnik, D.~T.~Larson, H.~Murayama and A.~Pierce,
Phys.\ Rev.\ D {\bf 69} (2004), 094024.\\
J.~Hisano and Y.~Shimizu, Phys.\ Lett.\ B {\bf 565} (2003), 183;
Phys.\ Lett.\ B {\bf 581} (2004), 224.\\
M.~Ciuchini, A.~Masiero, L.~Silvestrini, S.~K.~Vempati and O.~Vives,
Phys.\ Rev.\ Lett.\ {\bf 92} (2004), 071801.\\
D.~T.~Larson, H.~Murayama and G.~Perez,
J.~High~Energy~Phys. {\bf 07} (2005), 057.

\bibitem{ETaftMEAS}
K.~Cheung, S.~K.~Kang, C.~S.~Kim and J.~Lee,
Phys.\ Lett.\ B {\bf 652} (2007), 319.\\
M.~Ciuchini, A.~Masiero, P.~Paradisi, L.~Silvestrini, S.~K.~Vempati
and O.~Vives, Nucl.\ Phys.\ B {\bf 783} (2007), 112.\\
L.~Calibbi, Y.~Mambrini and S.~K.~Vempati,
J.~High~Energy~Phys. {\bf 09} (2007), 081.\\
J.~K.~Parry and H.~H.~Zhang, Nucl.\ Phys.\ B {\bf 802} (2008), 63.\\
T.~Goto, Y.~Okada, T.~Shindou and M.~Tanaka,
Phys.\ Rev.\ D {\bf 77} (2008), 095010.\\
B.~Dutta and Y.~Mimura, Phys.\ Rev.\ D {\bf 78} (2008), 071702;
Phys.\ Lett.\ B {\bf 677} (2009), 164.

\bibitem{ParkEt}
J.~H.~Park, arXiv:0710.4529; arXiv:0809.3004\\
J.~H.~Park and M.~Yamaguchi, Phys.\ Lett.\ B {\bf 670} (2009), 356.\\
P.~Ko and J.~H.~Park, Phys.\ Rev.\ D {\bf 80} (2009), 035019.\\
P.~Ko, J.~H.~Park and M.~Yamaguchi, J.~High~Energy~Phys. {\bf 11} (2008), 051.

\bibitem{AROSSIorig}
A.~Rossi, Phys.\ Rev.\ D {\bf 66} (2002), 075003.

\bibitem{JoaROSSI}
F.~R.~Joaquim and A.~Rossi, Phys.\ Rev.\ Lett.\ {\bf 97} (2006), 181801;
Nucl.\ Phys.\ B {\bf 765} (2007), 71.

\bibitem{MOHOKA}
R.~N.~Mohapatra, N.~Okada and H.~B.~Yu, Phys.\ Rev.\ D {\bf 78} (2008), 075011.

\bibitem{PatEmiShab}
For the relevance of sQFVs in the right--right down-squark sector for 
$\Delta M_S$, see for example: \\
P.~Ball, S.~Khalil and E.~Kou, Phys.\ Rev.\ D {\bf 69} (2004), 115011.

\bibitem{TEVmix}
V.~M.~Abazov et al. (D0 Collaboration), Phys.\ Rev.\ Lett.\ {\bf 97} (2006), 021802.\\
A.~Abulencia et al. (CDF Collaboration), Phys.\ Rev.\ Lett.\ {\bf 97} (2006), 062003; Phys.\ Rev.\ Lett.\ {\bf 97} (2006), 242003.

\bibitem{AlexUlli}
A.~Lenz and U.~Nierste, J.~High~Energy~Phys. {\bf 06} (2007), 072.

\bibitem{RFleischer}
R.~Fleischer, arXiv:0802.2882.

\bibitem{TEVphase}
V.~M.~Abazov et al. (D0 Collaboration), Phys.\ Rev.\ Lett.\ {\bf 98} (2007), 121801; Phys.\ Rev.\ Lett.\ {\bf 101} (2008), 241801. \\
T.~Aaltonen et al. (CDF Collaboration), Phys.\ Rev.\ Lett.\ {\bf 100} (2008), 161802.

\bibitem{UTfitnew}
M.~Bona et al. (UTfit Collaboration), arXiv:0909.5065.

\bibitem{HFavG}
E.~Barberio et al. (Heavy Flavour Averaging Group), arXiv:0808.1297.

\bibitem{CDFDOcomb}
CDF/D0 $\Delta \Gamma_s$, $\beta_S$ Combination Working Group, 
CDF/PHYS/BOTTOM/CDFR/9787 (D0 Note 5928-CONF)

\bibitem{SakaiYanag}
N.~Sakai and T.~Yanagida, Nucl.\ Phys.\ B {\bf 197} (1982), 533.\\
S.~Weinberg, Phys.\ Rev.\ D {\bf 26} (1982), 287.\\
S.~Dimopoulos, S.~Raby and F.~Wilczek, Phys.\ Lett.\ B {\bf 112} (1982), 133.

\bibitem{FERMASSES}
P.~H.~Frampton, S.~Nandi and J.~J.~G.~Scanio, Phys.\ Lett.\ B {\bf 85} (1979), 225.\\
H.~Georgi and C.~Jarlskog, Phys.\ Lett.\ B {\bf 86} (1979), 297.

\bibitem{missingPARTN}
A.~Masiero, D.~V.~Nanopoulos, K.~Tamvakis and T.~Yanagida,
Phys.\ Lett.\ B {\bf 115} (1982), 380.\\
B.~Grinstein, Nucl.\ Phys.\ B {\bf 206} (1982), 387.

\bibitem{AFM}
G.~Altarelli, F.~Feruglio and I.~Masina, J.~High~Energy~Phys. {\bf 11} (2000), 040.\\
P.~Fileviez Perez, Phys.\ Rev.\ D {\bf 76} (2007), 071701.

\bibitem{NROsFLsAHCH}
N.~Arkani-Hamed, H.~C.~Cheng and L.~J.~Hall,
Phys.\ Rev.\ D {\bf 53} (1996), 413.

\bibitem{NROsFLsHNOST}
J.~Hisano, D.~Nomura, Y.~Okada, Y.~Shimizu and M.~Tanaka,
Phys.\ Rev.\ D {\bf 58} (1998), 116010.

\bibitem{TWW}
S.~Trine, S.~Westhoff and S.~Wiesenfeldt, J.~High~Energy~Phys. {\bf 08} (2009), 002.

\bibitem{ZURAB}
Z.~Berezhiani, Z.~Tavartkiladze and M.~Vysotsky, hep-ph/9809301.

\bibitem{GORAN}
B.~Bajc, P.~Fileviez Perez and G.~Senjanovic,
Phys.\ Rev.\ D {\bf 66} (2002), 075005; hep-ph/0210374.
 
\bibitem{DESYpeople}
D.~Emmanuel-Costa and S.~Wiesenfeldt, Nucl.\ Phys.\ B {\bf 661} (2003), 62.

\bibitem{ExperimentPD}
For an update of the experimental results on proton decay 
 rates, see for example \\
M.~Shiozawa, talk at SUSY08, Seoul, Jun. 16--21, 2008.

\bibitem{KMY}
Y.~Kawamura, H.~Murayama and M.~Yamaguchi, Phys.\ Rev.\ D {\bf 51} (1995), 1337.

\bibitem{BMYtalks} 
F.~Borzumati, S.~Mishima and T.~Yamashita, arXiv:0705.2664.

\bibitem{BY-SUSY09}
See also: F.~Borzumati and T.~Yamashita, arXiv:0910.0372.

\bibitem{SUSYSU5}
S.~Dimopoulos, S.~Raby and F.~Wilczek, Phys.\ Rev.\ D {\bf 24} (1981),
1681. \\
S.~Dimopoulos and H.~Georgi, Nucl.\ Phys.\ B {\bf 193} (1981), 150.\\
N.~Sakai, Z.\ Phys.\ C {\bf 11} (1981), 153.

\bibitem{SU5FIRST}
H.~Georgi and S.~L.~Glashow, Phys.\ Rev.\ Lett.\ {\bf 32} (1974), 438.

\bibitem{EllGaill}
J.~R.~Ellis and M.~K.~Gaillard, Phys.\ Lett.\ B {\bf 88} (1979), 315.

\bibitem{MPnonMSSU5}
Z.~Berezhiani and Z.~Tavartkiladze, Phys.\ Lett.\ B {\bf 396} (1997), 150.

\bibitem{SS}
E.~Witten, Phys.\ Lett.\ B {\bf 105} (1981), 267.\\
H.~Georgi, Phys.\ Lett.\ B {\bf 108} (1982), 283.\\
A.~Sen, Phys.\ Lett.\ B {\bf 148} (1984), 65.\\
N.~Maekawa and T.~Yamashita, Phys.\ Rev.\ D {\bf 68} (2003), 055001.

\bibitem{flipped}
S.~M.~Barr, Phys.\ Lett.\ B {\bf 112} (1982), 219.\\
N.~Maekawa and T.~Yamashita, Phys.\ Lett.\ B {\bf 567} (2003), 330.

\bibitem{DW}
S.~Dimopoulos and F.~Wilczek, Print-81-0600 (SANTA BARBARA), NSF-ITP-82-07.\\
K.~S.~Babu and S.~M.~Barr, Phys.\ Rev.\ D {\bf 48} (1993), 5354.\\
S.~M.~Barr and S.~Raby, Phys.\ Rev.\ Lett.\ {\bf 79} (1997), 4748.\\
N.~Maekawa, Prog.\ Theor.\ Phys.\ {\bf 106} (2001), 401.\\
N.~Maekawa and T.~Yamashita, Prog.\ Theor.\ Phys.\ {\bf 107} (2002), 1201; 
Prog.\ Theor.\ Phys.\ {\bf 110} (2003), 93.

\bibitem{pseudoNG}
K.~Inoue, A.~Kakuto and H.~Takano, Prog.\ Theor.\ Phys.\ {\bf 75} (1986), 664.

\bibitem{5dDTS}
Y.~Kawamura, Prog.\ Theor.\ Phys.\ {\bf 103} (2000), 613;
Prog.\ Theor.\ Phys.\ {\bf 105} (2001), 691;
Prog.\ Theor.\ Phys.\ {\bf 105} (2001), 999.

\bibitem{EMO}
J.~Ellis, A.~Mustafayev and K.~A.~Olive, arXiv:1004.5399.
 
\bibitem{CASAS}
J.~A.~Casas and A.~Ibarra, Nucl.\ Phys.\ B {\bf 618} (2001), 171.

\bibitem{ADDGHR}
A.~Arvanitaki, S.~Dimopoulos, S.~Dubovsky, P.~W.~Graham,
R.~Harnik and S.~Rajendran, Phys.\ Rev.\ D {\bf 79} (2009), 105022;
Phys.\ Rev.\ D {\bf 80} (2009), 055011.

\bibitem{TheoryPD}
T.~Goto and T.~Nihei, Phys.\ Rev.\ D {\bf 59} (1999), 115009.

\bibitem{MP}
H.~Murayama and A.~Pierce, Phys.\ Rev.\ D {\bf 65} (2002), 055009.

\bibitem{GUTeLFV}
L.~J.~Hall, V.~A.~Kostelecky and S.~Raby, Nucl.\ Phys.\ B {\bf 267} (1986), 415.\\
R.~Barbieri, L.~J.~Hall and A.~Strumia, Nucl.\ Phys.\ B {\bf 445} (1995), 219.

\bibitem{RGES}
S.~P.~Martin, hep-ph/9709356.

\bibitem{RGESfalck}
N.~K.~Falck, Z.\ Phys.\ C {\bf 30} (1986), 247.

\bibitem{RGEs2loopYY}
Y.~Yamada, Phys.\ Rev.\ Lett.\ {\bf 72} (1994), 25;
Phys.\ Rev.\ D {\bf 50} (1994), 3537.

\bibitem{RGEs2loop}
S.~P.~Martin and M.~T.~Vaughn, Phys.\ Rev.\ D {\bf 50} (1994), 2282
[Errata; {\bf 78} (2008), 039903].\\
I.~Jack, D.~R.~T.~Jones, S.~P.~Martin, M.~T.~Vaughn and Y.~Yamada,
Phys.\ Rev.\ D {\bf 50} (1994), 5481.

\bibitem{SLANSKY}
R.~Slansky, Phys.\ Rep.\ {\bf 79} (1981), 1.

\bibitem{CUI}
Y.~Cui, Phys.\ Rev.\ D {\bf 74} (2006), 075010.

\bibitem{RGEsfromWaveFncRen}
G.~F.~Giudice and R.~Rattazzi, Nucl.\ Phys.\ B {\bf 511} (1998), 25.\\
N.~Arkani-Hamed, G.~F.~Giudice, M.~A.~Luty and R.~Rattazzi,
Phys.\ Rev.\ D {\bf 58} (1998), 115005.

\end{thebibliography}
\end{document}